\documentclass[12pt]{article}    

\usepackage{graphicx}
\usepackage{amsfonts}
\usepackage{amsmath}
\usepackage{amssymb}
\usepackage{amsmath,amscd}
\usepackage{amsthm}
\usepackage[T1]{fontenc}
\usepackage[hidelinks]{hyperref}
\usepackage{lmodern}
\hypersetup{
     colorlinks   = true,
     citecolor    = gray
}

\usepackage{xypic}
\usepackage{mathrsfs}
\usepackage{upgreek}
\usepackage{esint}
\usepackage[usenames,dvipsnames,svgnames,table]{xcolor}
\usepackage[none]{hyphenat}
\usepackage{eucal} 
\usepackage{amsthm}
\usepackage{bbm}
\usepackage{textcomp}
\usepackage{wrapfig}

\newtheorem{theorem}{Theorem}[subsection]
\newtheorem{corollary}{Corollary}[subsection]
\newtheorem{lemma}{Lemma}[subsection]

\newtheorem{proposition}{Proposition}[subsection]

\newtheorem{definition}{Definition}[subsection]

\numberwithin{equation}{subsection}

\makeatletter
\renewcommand\tableofcontents{%
    \@starttoc{toc}%
}
\makeatother

\newcommand{\ebar}
{\mathchar'26\mkern-9mu \varepsilon}

\DeclareMathOperator*{\R}{\mathbb{R}}
\DeclareMathOperator*{\N}{\mathbb{N}}

\DeclareMathOperator*{\Z}{\mathbb{Z}}

\DeclareMathOperator*{\C}{\mathbb{C}}

\DeclareMathOperator*{\e}{\textcolor{purple}{\varepsilon_1}}
\DeclareMathOperator*{\ee}{\textcolor{purple}{\varepsilon_2}}
\DeclareMathOperator*{\ered}{\textcolor{red}{--\varepsilon_1 \varepsilon_2}}
\DeclareMathOperator*{\eblue}{\textcolor{blue}{\varepsilon_1 + \varepsilon_2}}
\DeclareMathOperator*{\eebrown}{\textcolor{brown}{-- \varepsilon}}
\DeclareMathOperator*{\ebrown}{\textcolor{brown}{\varepsilon}}

\DeclareMathOperator*{\tealN}{\textit{\textcolor{teal}{N}}}

\DeclareMathOperator*{\vcurrent}{\widehat{\textit{\textbf{v}}}}

\begin{document}

\pagebreak

\hypersetup{linkcolor=black} 

\begin{center}{\large Random Partitions and the Quantum Benjamin-Ono Hierarchy} \end{center}
\begin{center} Alexander Moll \end{center}

\begin{abstract} {\small \noindent \underline{ABSTRACT} We derive exact and asymptotic results for random partitions from general results in the semi-classical analysis of coherent states.  In geometric quantization of Hermitian affine spaces $(\mathscr{M}, J, \mathsf{g} , \omega)$, a coherent state $\Upsilon_v ( \cdot | \hbar)$ around a classical state $v \in \mathscr{M}$ is the reproducing kernel in the Fock space $L^2_{\textnormal{J-hol}} (\mathscr{M}, d \rho_{\hbar, \mathsf{g}})$ of $J$-holomorphic functions on $\mathscr{M}$ square-integrable against the Segal-Bargmann Gaussian weight $d \rho_{\hbar, \mathsf{g}}$.  Under certain regularity assumptions, in any canonical quantization defined by an ordering $\eta$, we prove that in the semi-classical limit $\hbar \rightarrow 0$ \begin{equation} \widehat{O}^{\eta}(\hbar) |_{\Upsilon_v(\cdot | \hbar)} \sim O|_v + \hbar^{1/2} \mathbb{G}(O)|_v \nonumber \end{equation}

\noindent the random value $\widehat{O}^{\eta}(\hbar) |_{\Upsilon_v(\cdot | \hbar)}$ of any quantized observable $\widehat{O}^{\eta}(\hbar)$ in a coherent state $\Upsilon_v( \cdot | \hbar)$ around a classical state $v$ concentrates to leading order on the non-random value $O|_v$ of the classical observable $O$ at the classical state $v$, independent of $\eta$.  Moreover, quantum corrections are Gaussian $\mathbb{G}(O)|_v$ at scale $\hbar^{1/2}$ with mean $0$ and variance $|| (\nabla O)|_v ||_{\mathsf{g}}^2$ independent of $\eta$.  These results do not assume integrability of the Hamiltonian flow generated by $O$ but follow directly from the fact that at fixed $\hbar >0$ the Segal-Bargmann weight on $\mathscr{M}$ is already Gaussian with covariance kernel $\hbar \mathsf{g}^{-1}$ given by the inverse metric.\\
\\
\noindent The classical periodic Benjamin-Ono equation for real $2\pi$-periodic $v$ of mean $a \in \R$ is Hamiltonian in the leaf $(\mathscr{M}(a), J, \mathsf{g}_{-1/2}, \omega_{- 1/2})$ of the real $L^2$-Sobolev space on the circle $\mathbb{T}$ at critical regularity $s =-1/2$ with $J$ the spatial periodic Hilbert transform.  We find a classical conserved density $dF_{\star |v} (c| \ebar)$ on $c \in \R$ for this system with dispersion coefficient $\ebar$, extending Nazarov-Sklyanin (2013).  The authors also give an ordering $\eta_{NS}$ for an integrable canonical quantization, which we use to construct a quantum conserved density $d\widehat{F}^{\eta_{NS}}( c | \hbar, \ebar)|_{\Psi}$.  For quantum stationary states, we identify this conserved density with $dF_{\lambda}(c | \ee, \e)$ the Rayleigh measure of the profile of a partition $\lambda$ of anisotropy $(\ee, \e) \in \C^2$ for $\hbar = \ered$, $\ebar = \eblue$ invariant under $\ee \longleftrightarrow \e$.  As Jack polynomials are the quantum stationary states and Stanley's Cauchy kernel (1989) is the reproducing kernel, the random values of the quantum periodic Benjamin-Ono hierarchy in a coherent state $\Upsilon_v ( \cdot | \hbar)$ are a ``Jack measure'' on partitions, a dispersive generalization of Okounkov's Schur measures (1999).  By the above, \begin{equation} d \widehat{F}_{\lambda}^{\eta_{NS}} ( c | \hbar, \ebar ) \big |_{\Upsilon_v(\cdot | \hbar)} \sim d F_{\star |v} ( c | \ebar)+ \hbar^{1/2} \mathbb{G}(c|\ebar)|_v \nonumber \end{equation} \noindent we have concentration on a limit shape as $\hbar \rightarrow 0$, the classical conserved density at $v$, and quantum fluctuations are an explicit Gaussian field.  Our results follow from an enumerative asymptotic expansion in $\hbar$ and $\ebar$ of joint cumulants over new combinatorial objects we call ``ribbon paths''.  As above, our results reflect the fact that at fixed $\hbar>0$ the weight defining Fock space is already a fractional Brownian motion of variance $\hbar$ and Hurst index $(-s) - \tfrac{1}{2} \dim \mathbb{T} = + \tfrac{1}{2} - \tfrac{1}{2} = 0.$
}
 \end{abstract}

\pagebreak

\tableofcontents

\section{\textcolor{black}{Introduction}} \label{secIntroduction}

\subsection{\textcolor{black}{Overview}} \label{subsecOverview}

\noindent This paper is devoted to the semi-classical $\hbar \rightarrow 0$ and dispersionless $\ebar \rightarrow 0$ limits of Nazarov-Sklyanin's integrable geometric quantization \cite{NaSk2} of the classical periodic Benjamin-Ono equation \cite{AblCla, Benj, Molinet, Ono, TaoBenjaminOno}.  In both classical and quantum cases, the periodic Benjamin-Ono wave equations arise as hydrodynamic limits $\tealN \rightarrow \infty$ of the Calogero-Sutherland $\tealN$-body problem on the circle \cite{Calog1, Suth1} taken in a \textit{chiral sector} so that the density field is \textit{approximately uniform} \cite{AbBeWi, Poly1995, StoneAnduagaXing, StoneGutman}.  In this regime, the $\tealN$ particle configuration is a vibrating periodic lattice, so both classical and quantum periodic Benjamin-Ono equations describe interacting dispersive \textit{phonons} \cite{LamPri2014, LamPri2015}.\\
\\
\noindent The quantum periodic Benjamin-Ono equation is of current interest in both pure and applied aspects of mathematics and physics, many of which we detail as the paper develops.  Most notably, in light of its hydrodynamic origin mentioned above, decades of research on abelian braid statistics of collective excitations of the quantum Calogero-Sutherland system \cite{AbWi1, CaLa1, Eti0, EsPaSaSe, Ha1995Fractional, OkounkovLogGas, Pasq1, Poly0, SerbanLesagePasquier, Suth0} culminate in Wiegmann's proposal \cite{Wieg1} that the quantum periodic Benjamin-Ono equation is an effective model of edge excitations in the abelian fractional quantum Hall effect.\\
\\
\noindent In this paper, we do not study the finite or long time evolution of either the classical nor of the quantum periodic Benjamin-Ono systems.  Instead, we lay a foundation for doing so by studying the infinitesimal time evolution of quantum \textit{coherent states} $\Upsilon_v (\cdot | \hbar)$ around a classical state $v$ \cite{GazeauCoherentBook, HallBook, Perelomov}.  Moreover, for quantum periodic Benjamin-Ono, knowledge of its quantum stationary states and spectrum reduces our infinitesimal study of coherent states to a model of random partitions we call \textit{Jack measures}, a dispersive generalization of Okounkov's Schur measures \cite{Ok1} at the crosssroads of modern probability \cite{BoGo0, BoPe0, Co0} and geometric representation theory \cite{KimuraPestun1, MaulOk, NekYI, NekOk, Ok2}.\\
\\
\noindent In section \textbf{[\ref{secPartitionsFromQuantumBenjaminOno}]}, we introduce classical and quantum periodic Benjamin-Ono systems and Jack measures and present our results: $\hbar$-expansions of joint cumulants of linear statistics, concentration of profiles around limit shapes, and Gaussian fluctuations at the global scale.  We discuss our results in relation to ongoing research and derive them as consequences of results we prove later in sections \textbf{[\ref{secColumn1}]}, \textbf{[\ref{secColumn2}]}, \textbf{[\ref{secColumn3}]}, and \textbf{[\ref{secConstructionsForPeriodicBenjaminOno}]}.\\
\\
\noindent To best frame our results, we review precise definitions of ``randomness'' in chaos theory \textbf{[\ref{subsecOrbits}]} and quantum theory \textbf{[\ref{subsecObservables}]}, as the qualitative behavior of the quantum periodic Benjamin-Ono equation depends on the fact that it is quantum but not chaotic.  To best understand our results, we present a complete account of a much simpler and fundamental example in \textbf{[\ref{subsecOscillations}]}, the semi-classical analysis $\hbar \rightarrow 0$ of coherent states of a single quantized harmonic oscillator and its reduction to the high intensity asymptotics of a single Poisson random variable, a random partition with one row.  We provide an outline of the paper in \textbf{[\ref{subsecOutline}]} and concluding remarks in \textbf{[\ref{subsecOutlook}]}.\\
\\
\noindent As Jack measures are of intrinsic interest in probability, asymptotic representation theory, enumerative combinatorics, and enumerative algebraic geometry, we emphasize that section \textbf{[\ref{subsecRandomPartitionsIntroduction}]} gives a definition of Jack measures and \textbf{[\ref{subsecAOEIntro}]}, \textbf{[\ref{subsecLLNIntro}]}, \textbf{[\ref{subsecCLTIntro}]} an account of our results that is self-contained and does not rely on the derivation of Jack measures via quantization.  That being said, our results are specializations of much more general semi-classical results we prove for coherent states in sections \textbf{[\ref{secColumn1}], [\ref{secColumn2}], [\ref{secColumn3}]} in the analytic setting of geometric quantizations of Hermitian affine spaces, a special class of infinite-dimensional homogeneous K\"{a}hler manifolds.  \underline{Although our results and} \underline{methods are both new in random partitions, it is our new approach to these random} \underline{partitions by the first principles of geometric quantization and semi-classical analysis}
\cite{Englis, FollandBook, GuilleminSternbergSemiClassicalAnalysis, HallBook, MartinezBook, Maslov0, Woodhouse, Zwor0} \underline{that we regard as the main contribution of this paper.}

\subsection{\textcolor{black}{Orbits: Integrability and Chaos}} \label{subsecOrbits}

\noindent In sections \textbf{[\ref{subsecCBOHIntroduction}]} and \textbf{[\ref{subsecCBOHConstruction}]}, we study the classical periodic Benjamin-Ono equation and the sense in which it is a \textit{classical integrable Hamiltonian system}.  It is hard to determine if a given equation is a classical integrable Hamiltonian system, let alone to exhibit its exact solution via model-dependent special functions.  However, once an equation is known to be a classical integrable Hamiltonian system, one automatically has qualitative knowledge of its flow.  In particular, {the flow is not \textit{chaotic}.}
  \begin{figure}[htb]
\centering
\includegraphics[width=1.0 \textwidth]{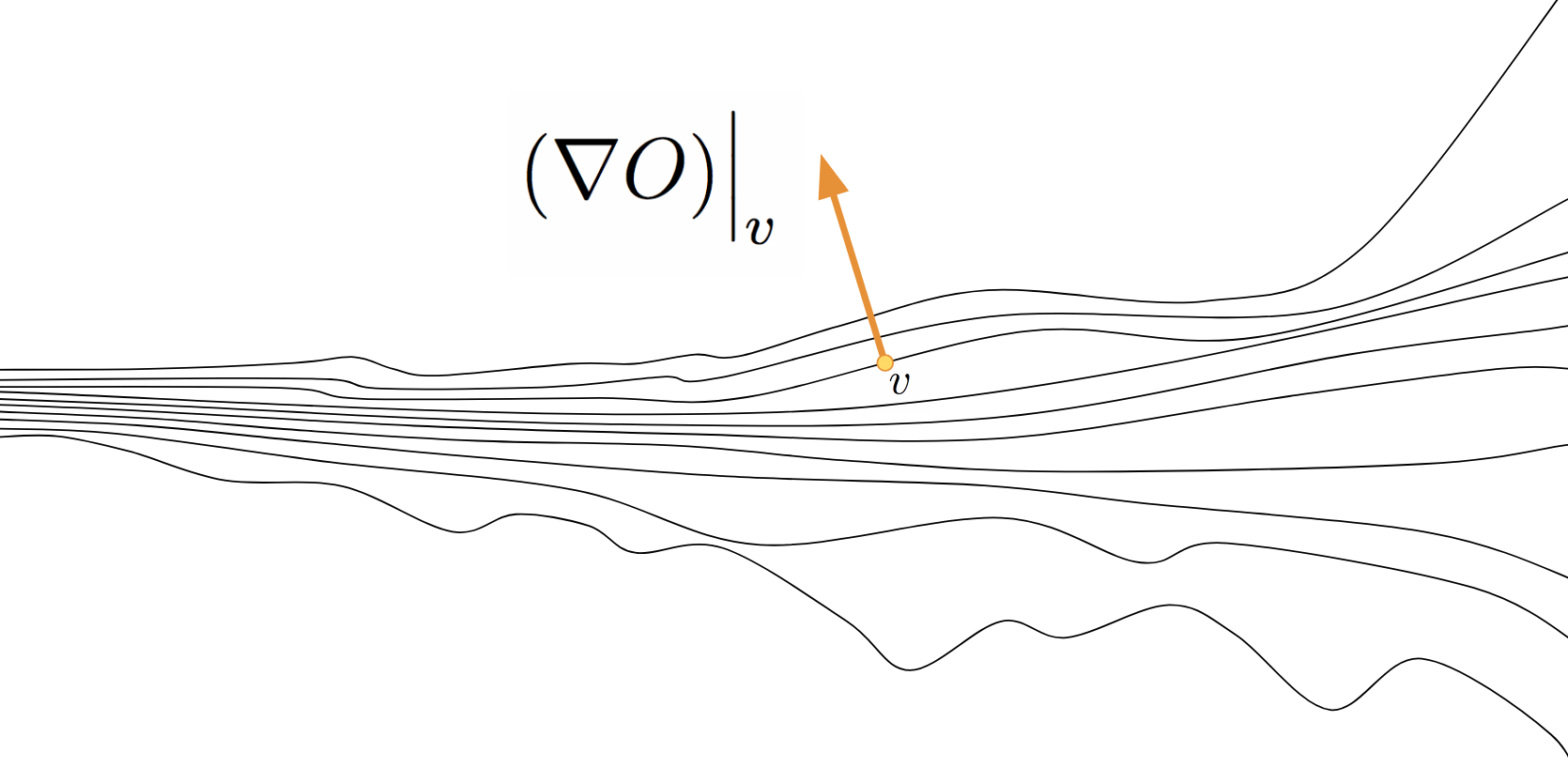}
\end{figure}

\noindent Let $(\mathscr{M}, \omega)$ be a symplectic manifold and $O: \mathscr{M} \rightarrow \R$ a classical observable.  The value $O|_v$ of $O$ at any $v \in \mathscr{M}$ is a conserved quantity for the $O$-Hamiltonian flow: $v$ must evolve within a level set of $O$.  In the figure above, we depict level sets of $O$ for different initial data on the left that are nearby when measured in a Riemannian metric $\mathsf{g}$ compatible with $\omega$.  That small changes in the initial condition on the left might lead to drastically different outcomes on the right is a defining property of {chaotic systems}, the positivity of the maximal Lyapunov exponent known as the ``{butterfly effect}''.  However, if one only knows the values of $O$ in an infinitesimal neighborhood around some $v \in \mathscr{M}$, one doesn't know the global geometry of level lines of $O$, hence cannot tell whether or not the system is chaotic.  Said again,

\begin{proposition} \label{RespectTheButterflyEffect} If a quantity $\mathbf{X}$ depends only on a classical state $v \in (\mathscr{M}, \omega)$, the value $O|_v$ of the classical Hamiltonian $O$ at $v$, and the gradient $(\nabla O)|_v$ of $O$ at $v$ defined through a compatible Riemannian metric $\mathsf{g}$, then $\mathbf{X}$ does not depend on whether or not the Hamiltonian flow generated by $O$ is integrable or chaotic.\end{proposition}

\noindent Only global knowledge of orbits - not the local knowledge of gradients - distinguishes between integrability and chaos.  The emergent objects in our Theorems [\ref{LLNColumn2}] and Theorems [\ref{CLTColumn2}], from which we derive our limit shape and Gaussan fluctuations for Jack measures, are quantities of the type in Proposition [\ref{RespectTheButterflyEffect}].  This fact will rear its head throughout, as it gives a precise sense in which the study of random partitions at the {global scale} does not fundamentally engage with the property of integrability.

\pagebreak

\subsection{\textcolor{black}{Observables: Randomness from Quantization}} \label{subsecObservables}

\noindent In the literature on classical chaos, the word ``random'' is often used colloquially to describe the ``seemingly unrelated'' outcomes of a chaotic system due to slight perturbations of the initial conditions.  For chaotic systems, observables of trajectories are \textit{unpredictable} but not actually random in the sense of probability theory.  By contrast, for quantum systems, observables of trajectories are intrinsically \textit{uncertain} and truly random.  To reduce our study of the infinitesimal time evolution of the quantum periodic Benjamin-Ono equation in a coherent state to the Jack measures of random partitions, we need to review the basic analysis necessary to extract random variables from any quantum system written in the operator formalism.\\
\\
\noindent Recall the spectral theorem for self-adjoint operators, as can be found e.g. in \cite{Lax0}.
\begin{theorem} \label{SpectralTheorem} \textnormal{[von Neumann 1932]} Let $\widehat{\mathcal{O}}$ be a possibly unbounded self-adjoint operator in a Hilbert space $(\mathscr{H}, \langle \cdot, \cdot \rangle)$.  For every $\Psi \in \mathscr{H}$ with $|| \Psi ||=1$, there exists a probability measure $\upmu_{\Psi, \Psi}( \cdot | \widehat{\mathcal{O}})$ on $\R$ so that for all bounded continuous $\phi : \R \rightarrow \C$ \begin{equation} \label{SpectralMeasureDefiningRelation} { \langle \Psi | \phi(\widehat{\mathcal{O}}) | \Psi \rangle } = \int_{- \infty}^{+\infty} \phi(E) d\upmu_{\Psi, \Psi}( E | \widehat{\mathcal{O}})
\end{equation}
\noindent  called the \underline{spectral measure {of} $\widehat{\mathcal{O}}$ {at} $\Psi$.}
\end{theorem}

\noindent $\Psi \in \mathscr{H}$ and self-adjoint $\widehat{\mathcal{O}}$ on $\mathscr{H}$ are called quantum states and quantum observables, respectively.  The next definition is the pivot upon which everything turns:

\begin{definition} \label{TheModelColumn1PreDefinition}The \underline{random value} $\widehat{\mathcal{O}}|_{\Psi}$ of a quantum observable $\widehat{\mathcal{O}}$ in a quantum state $\Psi \in \mathscr{H}$ is the random variable $\widehat{\mathcal{O}}|_{\Psi}$ whose law is $d \mu_{\Psi, \Psi}( \cdot | \widehat{\mathcal{O}})$ the spectral measure of $\widehat{\mathcal{O}}$ at $\Psi$.  For bounded continuous $\phi : \R \rightarrow \C$, $\mathbb{E} [ \phi( \widehat{O} |_{\Psi} )]$ is either side of \textnormal{(\ref{SpectralMeasureDefiningRelation})}.
\end{definition}
\noindent The random variables $\widehat{\mathcal{O}}(\hbar)|_{\Psi}$ of Definition [\ref{TheModelColumn1PreDefinition}] are physically significant:
\begin{proposition} \textnormal{[Born's Rule 1926]} The random variable $\widehat{\mathcal{O}} |_{\Psi}$ in \textnormal{Definition [\ref{TheModelColumn1PreDefinition}]} is the random outcome of the observation $\widehat{\mathcal{O}}$ of a quantum system in a state $\Psi$. \end{proposition}
\noindent Physically, Born's Rule explains how the uncertainty and randomness in observing the trajectories of a quantum system come from the deterministic Schr\"{o}dinger equation (\ref{SchrodingerEquation}).  The probabilistic behavior of quantizations of classical systems is qualitatively different from the result of adding a random forcing term to -- or taking random initial data in -- the equations of motion of a classical system.\\
\\
\noindent Mathematically, Born's Rule is a simple key that opens the door separating the pure probabilist from the vast and active research in geometry and representation theory devoted to quantum integrable systems in (1+1)-dimensions, such as quantum spin chains and conformal field theories.  In particular, Born's Rule is what will connect the two objects in the title of this paper.

\pagebreak

\subsection{\textcolor{black}{Oscillators: Poisson Measures at High Intensity}} \label{subsecOscillations}

\noindent In sections \textbf{[\ref{subsecRandomPartitionsIntroduction}]} and \textbf{[\ref{subsecColumn3Row4}]}, we use Born's Rule to reduce the quantum periodic Benjamin-Ono equation in a coherent state to the Jack measures of random partitions.  Doing so is an involved application of the simple, concise recipe of Born's Rule on the previous page, and so to best prepare for what is to come let us quantize the classical harmonic oscillator equation and use Born's Rule to arrive at Poisson measures, a model of random partitions with one row.  All core concepts in this paper are faithfully represented in miniature in this example, as the periodic Benjamin-Ono equation is a system of infinitely-many coupled harmonic oscillators when written in Fourier series.

\subsubsection{Classical Harmonic Oscillator Equation}

\noindent The \textit{classical harmonic oscillator equation} of mass $1$ and angular frequency $\sigma >0$ is
\begin{equation} \label{CHOE} q''(t) + \sigma^2 q(t) = 0 \end{equation} \noindent a second-order linear evolution equation.  In the language of symplectic geometry, \begin{wrapfigure}{r}{0.40 \textwidth} \begin{center}
    \includegraphics[width=0.40 \textwidth]{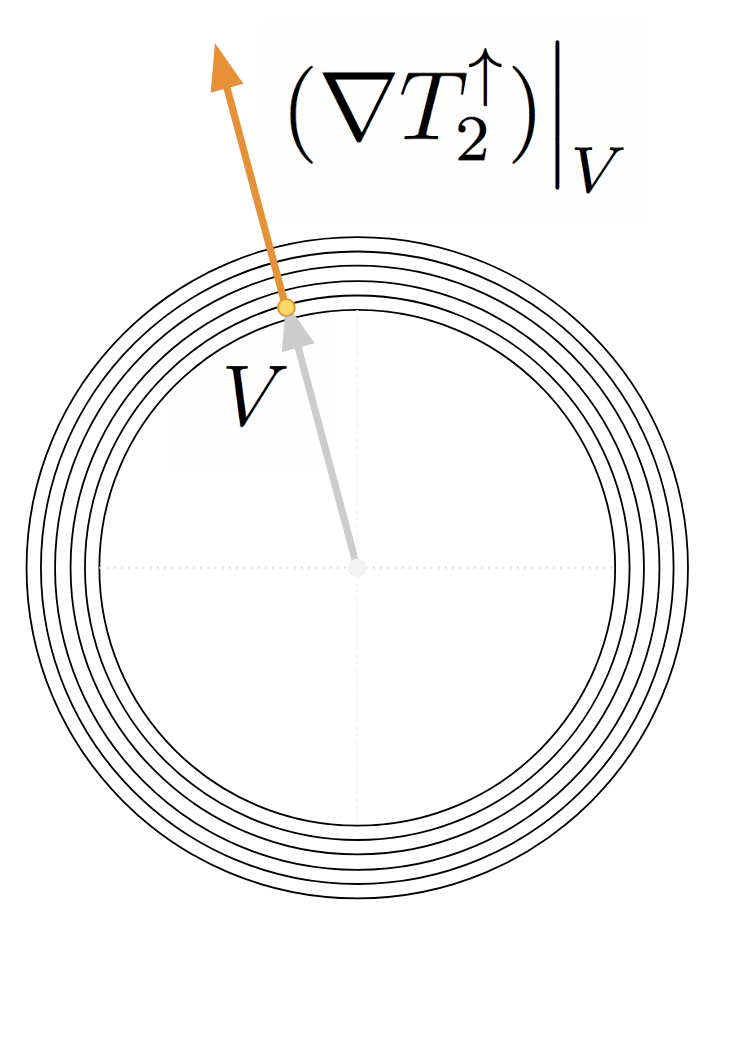}
  \end{center} 
\end{wrapfigure} \noindent \begin{theorem} \label{CHOHamiltonianSolution} In the complex variable \begin{equation} V = \frac{1}{ \sqrt{2}} \Big ( \sigma q + \textnormal{\textbf{i}} q'(t) \Big )  \end{equation} \noindent the equation \textnormal{(\ref{CHOE})} is a classical integrable Hamiltonian system in the sense of section \textnormal{\textbf{[\ref{subsecOrbits}]}} whose classical phase space is $\mathscr{M} \cong \C^1$ with the standard K\"{a}hler structure, whose classical Hamiltonian $T_2^{\uparrow} : \mathscr{M} \rightarrow \R$ is \begin{equation} T_2^{\uparrow} |_V  = |V|^2 \end{equation} \noindent for $\overline{V}, V$ coordinates in which the metric is the $\sigma>0$ scaling of the Euclidean metric, and whose exact solution at time $t$ with initial data $V(0)$ is
\begin{equation} V(t) = e^{\textnormal{\textbf{i}} \sigma t} V(0). \end{equation} \end{theorem}

\noindent In the special coordinates $\overline{V}, V$, the orbits of the classical harmonic oscillator flow are concentric circles around the origin, and the gradient $\nabla T_2^{\uparrow}|_V$ of the Hamiltonian $T_2^{\uparrow}$ is again $V$.  Unlike our depiction of the butterfly effect above, nearby initial data stay nearby for all time as they execute clockwise rotation by a phase at angular frequency $\sigma >0$.\\
\\
\noindent A classical mechanical particle in an arbitrary background potential $\Gamma(q)$ is always an integrable system due to the conservation of energy, but for the special background potential of the harmonic oscillator $\Gamma(q) = \frac{1}{2} \sigma^2 q^2$ the equation of motion is linear.  One regards a classical integrable Hamiltonian system with $K$ degrees of freedom to be ``solved'' if one can find \textit{action-angle variables} on its classical phase space $\mathscr{M}$ so that the original equations of motion are locally $K$ independent copies of (\ref{CHOE}) for variables $\overline{V_k}, V_k$ and angular frequencies $\sigma_k >0$ for $k=1,2,\ldots, K$ \cite{ArnoldMCM, DubrovinKricheverNovikov}.

\pagebreak

\subsubsection{Quantum Harmonic Oscillator Equation}

\noindent A priori unrelated to (\ref{CHOE}), the \textit{quantum harmonic oscillator equation} at quantum dimensionless scale $\hbar >0$ and angular frequency $\sigma$ is the deterministic, linear equation

\begin{equation} \label{QHOE} \begin{cases} \textnormal{\textbf{i}} \hbar \partial_t \Psi = \widehat{\mathcal{T}}_2^{\uparrow}(\hbar) \Psi  \\ \ \ \Psi \in (\mathscr{F}, \langle \cdot, \cdot \rangle_{\hbar, \sigma})\end{cases} \end{equation} 

\noindent posed for $\Psi$ in the Fock space with quantum Hamiltonian $\widehat{\mathcal{T}}_2^{\uparrow}(\hbar)$ defined as follows:

\begin{definition} The \underline{Fock space} of the classical harmonic oscillator phase space $\mathscr{M} \cong \C^1$ with its standard K\"{a}hler structure is the Hilbert space \begin{equation}( \mathscr{F}, \langle \cdot, \cdot \rangle_{\hbar, \sigma} )= L^2_{\textnormal{hol}} (\mathscr{M}, d \rho_{\hbar, \sigma}) \end{equation} \noindent of holomorphic functions on $\mathscr{M}$ square-integrable against the Segal-Bargmann weight \begin{equation} \label{PhaseSpaceFootprint} d \rho_{\hbar, \sigma} = e^{ - \frac{ |V|^2}{ \hbar \sigma^2} } dV d \overline{V}  \end{equation} \noindent a rotationally-invariant complex Gaussian of mean $0$ and complex variance $\hbar \sigma^2$. \end{definition}

\begin{definition} The \underline{Wick-quantized quantum harmonic oscillator Hamiltonian} is
\begin{equation} \label{WickT2} \widehat{\mathcal{T}}^{\uparrow}_2 ( \hbar) = \widehat{\mathcal{V}}_{+} \widehat{\mathcal{V}}_{-} \end{equation}

\noindent self-adjoint on Fock space for $\widehat{\mathcal{V}}_+$ the operator of multiplication by $V$ and 
\begin{equation} \widehat{\mathcal{V}}_- = \hbar \sigma^2 \frac{\partial}{\partial V}. \end{equation}

\end{definition}

\noindent $\widehat{\mathcal{V}}_{\pm}$, the creation and annihilation operators reappearing throughout this paper, yield an explicit description of eigenvalues and eigenvectors of the quantum Hamiltonian: \begin{theorem} \label{QHOSpectrumTheorem} The spectrum of the Wick-quantized quantum harmonic oscillator Hamiltonian $\widehat{\mathcal{T}}_2^{\uparrow}(\hbar)$ on Fock space is discrete whose solutions to the eigenvalue problem \begin{equation} \widehat{\mathcal{T}}_2^{\uparrow} (\hbar) \Psi_{d} ( \cdot | \hbar) = \widehat{T}_2^{\uparrow}(\hbar)|_{d} \Psi_d( \cdot  | \hbar) \end{equation} \noindent are indexed by $d=0,1,2,3,\ldots$ and given by
\begin{eqnarray} \Psi_d ( \cdot | \hbar ) &=& V^d \\  \widehat{T}_2^{\uparrow}(\hbar)|_{d}  &=& \hbar \sigma^2 d \end{eqnarray}

\end{theorem}

\noindent Quantum eigenstates $\Psi_d( \cdot | \hbar)$ do not depend on $\hbar, \sigma>0$ but their norms do.  

\subsubsection{Quantizations and Schr\"{o}dinger Operators}

\noindent We have not specified a \textit{quantization} associating to any $O: \mathscr{M} \rightarrow \R$ a self-adjoint operator $\widehat{\mathcal{O}}(\hbar)$ on Fock space.  We only declared $\widehat{\mathcal{T}}_2^{\uparrow}(\hbar)$ in (\ref{WickT2}) to be the quantum analog of the classical Hamiltonian $T_2^{\uparrow}|_V = |V|^2 = \overline{V} V$.  We could have made a different choice: to the numerically equivalent expression for the classical Hamiltonian
\begin{equation} T_2^{\uparrow}|_{V} = \frac{1}{2} \Big ( (q')^2 + (\sigma q)^2 \Big ) \end{equation} \noindent one associates a Schr\"{o}dinger operator $\frac{1}{2} ( - \hbar^2 \tfrac{\partial}{\partial q^2} + \sigma^2 q^2 )$ differing from $\widehat{\mathcal{T}}_2^{\uparrow}(\hbar)$ by an additive constant $\tfrac{1}{2} \hbar \sigma^2$.  To distinguish them, (\ref{WickT2}) is also called the \textit{number operator}.  

\pagebreak

\subsubsection{Random Partitions from Coherent States}

\noindent We now define a model of random partitions with one row.

\begin{definition} \label{PoissonMeasureDefinition} For parameters $V \in \C$, $\hbar>0$, and $\sigma >0$, define a \underline{Poisson measure} 

\begin{equation} \textnormal{Prob} (d) = \frac{1}{\textnormal{exp} \big ( \tfrac{|V|^2}{ \hbar \sigma^2} \big )} \cdot \frac{1}{d!} \Bigg ( \frac{|V|^2}{ \hbar \sigma^2} \Bigg )^d \end{equation}

\noindent a random $d=0,1,2,3,\ldots$ of \underline{intensity} $\tfrac{|V|^2}{\hbar \sigma^2}$. 

\end{definition}

\noindent It is well-known that the number distribution of \textit{coherent states} is Poisson:

\begin{definition} \label{BabyCoherentStatesDefinition} A \underline{coherent state $\Upsilon_{V^{\textnormal{out}}} ( \cdot | \hbar)$ around a classical state $V^{\textnormal{out}} \in \mathscr{M} \cong \C^1$} is the reproducing kernel in the Fock space $\mathscr{F}$ of $\mathscr{M}$, namely the holomorphic functions
\begin{equation} \Upsilon_{V^{\textnormal{out}}} ( V^{\textnormal{in}} | \hbar) = \textnormal{exp} \Bigg ( \frac{\overline{V^{\textnormal{out}}} \cdot V^{\textnormal{in}} }{\hbar \sigma^2} \Bigg ) \end{equation} \noindent of the complex coordinate $V^{\textnormal{in}} \in \mathscr{M}$.
\end{definition}

\noindent By Theorem [\ref{QHOSpectrumTheorem}] and the discussion of Born's Rule in section \textbf{[\ref{subsecObservables}]},
\begin{proposition} \label{PoissonDynamical} The random values of the Wick-quantized quantum harmonic oscillator Hamiltonian $\widehat{\mathcal{T}}^{\uparrow}_2(\hbar)$ in a coherent state $\Upsilon_V ( \cdot | \hbar)$ is the random variable \begin{equation} \widehat{T}^{\uparrow}_2(\hbar) |_{\Upsilon_V ( \cdot | \hbar)} = \hbar \sigma^2 d \end{equation} \noindent for $d$ sampled randomly from the Poisson measure of \textnormal{Definition [\ref{PoissonMeasureDefinition}]}.
\end{proposition}

\subsubsection{Asymptotic Expansions of Cumulants}

\noindent The behavior of a Poisson measure can be determined exactly:

\begin{theorem} \label{PoissonAOE} For random $d=0,1,2,\ldots$ from the Poisson measure of \textnormal{Definition [\ref{PoissonMeasureDefinition}]}, the random variable $T_2^{\uparrow}(\hbar) \big |_{\Upsilon_V(\cdot | \hbar)} = \hbar \sigma^2 d$ has characteristic function \begin{equation} \mathbb{E} \Bigg [ e^{ \textnormal{\textbf{i}} t T_2^{\uparrow}(\hbar) \big |_{\Upsilon_V(\cdot | \hbar)} } \Bigg ] = \textnormal{exp} \Bigg ( \frac{ |V|^2}{\hbar \sigma^2} \cdot \Big ( e^{ \textnormal{\textbf{i}} \hbar \sigma^2 t } -1 \Big ) \Bigg ) \end{equation} \noindent hence its $n$th cumulant is \begin{equation} \label{ExactYaYa} W_n \Big ( T_2^{\uparrow}(\hbar) \big |_{\Upsilon_V(\cdot | \hbar)} \Big ) = |V|^2 \sigma^{2(n-1)} \hbar^{n-1} \end{equation} \noindent a polynomial in $\hbar$, $\overline{V}$, and $V$ with order of vanishing $n-1$ in $\hbar$.
\end{theorem}
\noindent We give two proofs of Theorem [\ref{PoissonAOE}]:
\begin{itemize}
\item \textit{Computational Proof:} follows in one line using the power series for $e^x$. $\square$
\item \textit{Dynamical Proof:} by Proposition [\ref{PoissonDynamical}], use our Theorem [\ref{AOEColumn2}] for $\eta$-quantized observables $\widehat{O}^{\eta}$ for coherent states in Hermitian affine spaces, the specific form of the Wick-quantized classical harmonic oscillator Hamiltonian $\widehat{\mathcal{T}}_2^{\uparrow}(\hbar)$, and the special case of the Baker-Campbell-Hausdorff formula in which $[A,B] = sB$. $\square$
\end{itemize}

\pagebreak

\subsubsection{Concentration of Measure in Static Semi-Classical Limits}
\noindent The semi-classical limit $\hbar \rightarrow 0$ probes high intensity asymptotics of a Poisson measure.

\begin{theorem} \label{PoissonLLN} For random $d=0,1,2,\ldots$ from the Poisson measure of \textnormal{Definition [\ref{PoissonMeasureDefinition}]}, in the semi-classical limit $\hbar \rightarrow 0$
\begin{equation} \label{PoissonLLNStatement} \widehat{T}_2^{\uparrow}(\hbar)|_{\Upsilon_V(\cdot | \hbar)} \rightarrow T_2^{\uparrow}|_V =  |V|^2  \end{equation}

\noindent the random value $T_2^{\uparrow}(\hbar)  |_{\Upsilon_V(\cdot | \hbar)} = \hbar \sigma^2 d$ of the Wick-quantized classical harmonic oscillator Hamiltonian $\widehat{\mathcal{T}}^{\uparrow}_2(\hbar)$ in a coherent state $\Upsilon_V ( \cdot | \hbar)$ around $V$ concentrates on the non-random value $T_2^{\uparrow}|_V$ of the classical harmonic oscillator Hamiltonian $T_2^{\uparrow}$ at $V$.  Precisely, \textnormal{(\ref{PoissonLLNStatement})} is convergence in distribution to the delta measure at $T_2^{\uparrow}|_V$.
\end{theorem}
\noindent We give three proofs of Theorem [\ref{PoissonLLN}]:

\begin{itemize}
\item \textit{Computational Proof:} follows from Theorem [\ref{PoissonAOE}] by direct computation. $\square$
\item \textit{Independence Proof:} for $\tfrac{1}{\hbar} \in \N$, Theorem [\ref{PoissonAOE}] implies that our Poisson measure is infinitely-divisible as a sum of $\frac{1}{\hbar}$ independent identically distributed Poisson measures of intensity independent of $\hbar$, so the concentration of measure follows from the law of large numbers for independent averages. $\square$
\item \textit{Dynamical Proof:} by Proposition [\ref{PoissonDynamical}], use our Theorem [\ref{LLNColumn2}] for the static version of Bohr's Correspondence Principle for coherent states, and the specific form $T_2^{\uparrow}|_V = |V|^2$ of the classical harmonic oscillator Hamiltonian. $\square$
\end{itemize}

\subsubsection{Gaussian Variables as Quantum Fluctuations}

\begin{theorem} \label{PoissonCLT} For random $d=0,1,2,\ldots$ from the Poisson measure of \textnormal{Definition [\ref{PoissonMeasureDefinition}]}, quantum fluctuations of $T_2^{\uparrow}(\hbar) |_{\Upsilon_V(\cdot | \hbar)} = \hbar \sigma^2 d$ around its classical limit $T_2^{\uparrow}|_V$ occur at scale $\hbar^{1/2}$ and converge in distribution to

\begin{equation}\frac{1}{ \hbar^{1/2}} \Bigg ( \widehat{T}_2^{\uparrow} (\hbar)|_{\Upsilon_V(\cdot | \hbar)} - T_2^{\uparrow}  |_{V}  \Bigg ) \rightarrow \mathbb{G}(T_2^{\uparrow}) |_V\nonumber \end{equation}

\noindent a Gaussian random variable $\mathbb{G}(T_2^{\uparrow})|_V$ of mean $0$ and variance $|| (\nabla T_2^{\uparrow})|_V ||^2 = \sigma^2 |V|^2$ given by the square norm of the gradient of $T_2^{\uparrow}$ at $V$.
\end{theorem}
\noindent We give three proofs of Theorem [\ref{PoissonCLT}]:

\begin{itemize}
\item \textit{Computational Proof:} follows from Theorem [\ref{PoissonAOE}] by direct computation. $\square$
\item \textit{Independence Proof:} for $\tfrac{1}{\hbar} \in \N$, Theorem [\ref{PoissonAOE}] implies that our Poisson measure is infinitely-divisible as a sum of $\frac{1}{\hbar}$ independent and identically distributed Poisson measures of intensity independent of $\hbar$, so asymptotic Gaussianity follows from the central limit theorem for independent averages. $\square$
\item \textit{Dynamical Proof:} by Proposition [\ref{PoissonDynamical}], use our Theorem [\ref{CLTColumn2}] for the Gaussian correction of Bohr's Correspondence Principle for coherent states and the specific form $(\nabla T_2^{\uparrow})|_V = \sigma^2 |V|^2$ of the gradients of the classical harmonic oscillator Hamiltonian measured in the $\sigma>0$ scaling of the Euclidean metric. $\square$

\end{itemize}

\pagebreak

\subsection{\textcolor{black}{Outline}} \label{subsecOutline}

\noindent Our analysis of Poisson measures and harmonic oscillators in section \textbf{[\ref{subsecOscillations}]} proceeded in seven steps: classical, quantum, quantization, randomness, asymptotic expansions in $\hbar$, the semi-classical limit $\hbar \rightarrow 0$, and the $\hbar^{1/2}$ Gaussian quantum correction.  To do the same for Jack measures of random partitions via periodic Benjamin-Ono in section \textbf{[\ref{secPartitionsFromQuantumBenjaminOno}]}, we draw on two logically distinct sets of results: one general, one model-specific.

  \begin{figure}[htb]
\centering
\includegraphics[width=1.0\textwidth]{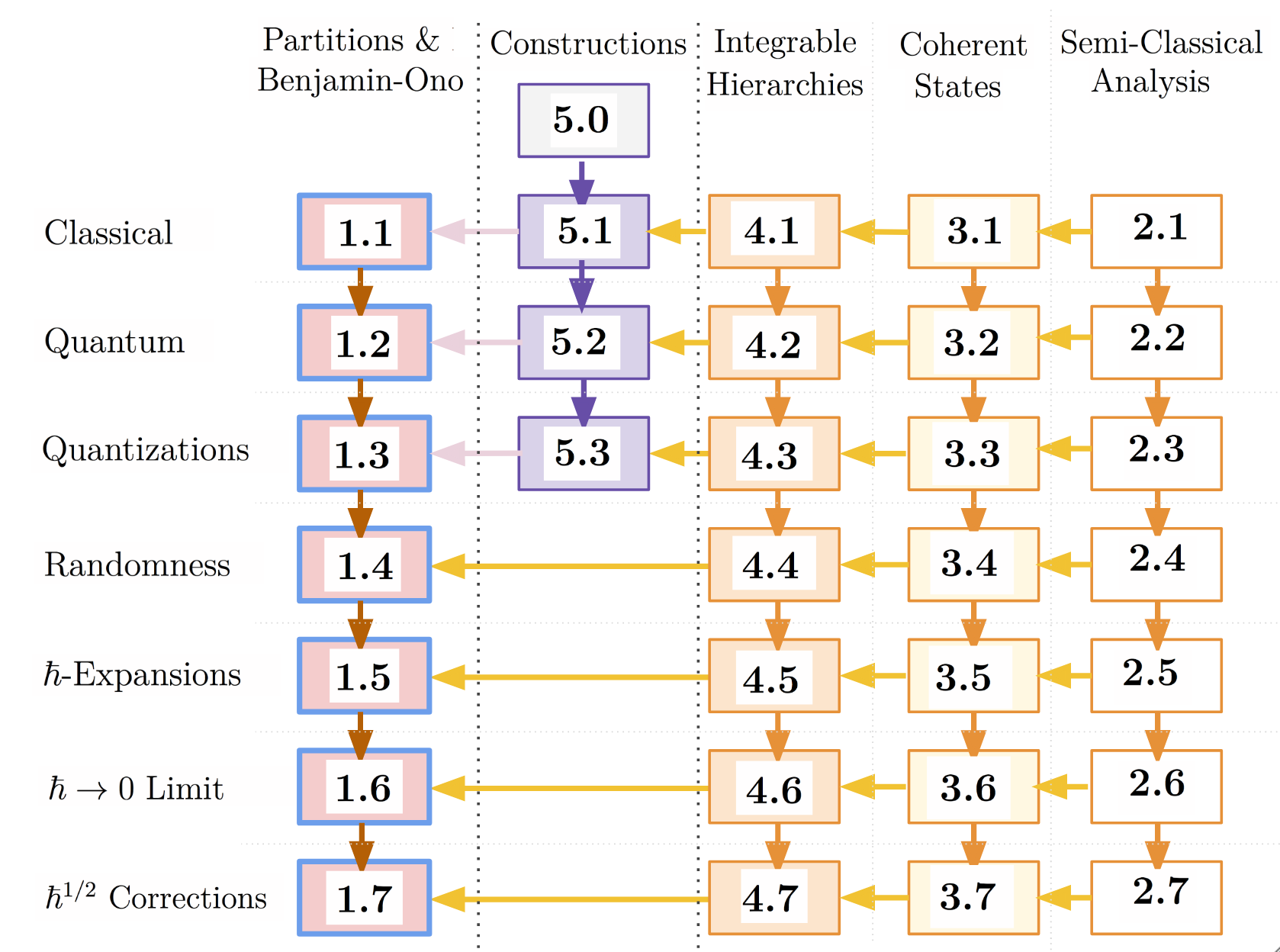}
\end{figure}

\noindent \textcolor{gray}{[General]} In section \textbf{[\ref{secColumn1}]} we give a short, self-contained review of semi-classical analysis for quantized symplectic manifolds, which we specialize in section \textbf{[\ref{secColumn2}]} to coherent states in geometrically-quantized Hermitian affine spaces.  Section \textbf{[\ref{secColumn2}]} contains the core probabilistic arguments in this paper, which we emphasize do not depend on the integrability of the Hamiltonian flow of observables in question as in Proposition [\ref{RespectTheButterflyEffect}].  That being said, we specialize section \textbf{[\ref{secColumn2}]} to integrable systems in section \textbf{[\ref{secColumn3}]}.\\
\\
\noindent \textcolor{gray}{[Model-Specific]} In section \textbf{[\ref{secConstructionsForPeriodicBenjaminOno}]}, we construct integrable hierarchies and conserved densities for classical periodic Benjamin-Ono and its integrable geometric quantization in our Theorems [\ref{CBOHConservedDensityExistence}], [\ref{QBOHConservedDensityExistence}], and [\ref{NazarovSklyaninQuantizationExistence}] via the auxiliary spectral theory of their Lax operators, namely as spectral shift functions of elliptic generalized (Fock-block) Toeplitz operators of order $1$ \cite{DeMonvelGuillemin} following Nazarov-Sklyanin \cite{NaSk2}.  We relegate these constructions to section \textbf{[\ref{secConstructionsForPeriodicBenjaminOno}]} in the same way we omitted proofs of Theorems [\ref{CHOHamiltonianSolution}] and [\ref{QHOSpectrumTheorem}] for the integrability of classical and quantum harmonic oscillators, respectively.\\
\\
\noindent Assuming as known the material in later sections, in section \textbf{[\ref{secPartitionsFromQuantumBenjaminOno}]} we derive our results for Jack measures.  While the Poisson measure enjoyed three distinct proofs of both Theorems [\ref{PoissonLLN}] and [\ref{PoissonCLT}] -- computational, independence, and dynamical proofs -- it is only the proof of dynamical type that we offer for Jack measures.  We do not pursue a computational proof via explicit formulae for Jack polynomials, nor do we find hidden independent random variables in Jack measures so as to apply the classical law of large numbers and central limit theorems.  Instead, Bohr's Correspondence Principle explains why our results occur and why they have the form they do.\\
\\
\noindent As depicted on the previous page, logical dependences between subsections is planar, so our Theorems [\ref{CBOHConservedDensityExistence}], [\ref{QBOHConservedDensityExistence}], and [\ref{NazarovSklyaninQuantizationExistence}] for periodic Benjamin-Ono hierarchies can be read independently of our probabilistic applications.  On the other hand, our results for coherent states make no reference to partitions nor to periodic Benjamin-Ono, and thus the material in sections \textbf{[\ref{secColumn1}]}, \textbf{[\ref{secColumn2}]}, \textbf{[\ref{secColumn3}]} may be read independently as well.  
 
\subsection{\textcolor{black}{Outlook}} \label{subsecOutlook}

\noindent In section \textbf{[\ref{subsecOverview}]}, we framed our study of Jack measures as that of the infinitesimal time evolution of the quantum periodic Benjamin-Ono equation in a coherent state, thus situating our work in the context of current research on quantum periodic Benjamin-Ono in which both the interaction and dispersion of phonons play a key role in accounting for the topological properties of collective excitations \cite{AbBeWi, AbWi1, LamPri2014, LamPri2015, Wieg1}.  Before we begin our analysis in earnest, let us conclude this introduction by indicating two different contexts for our results that appear only {after one takes fundamentally different interpretations of our} {parameters of quantization $\hbar$ and dispersion $\ebar$.}\\
\\
\noindent First, our path to random partitions is not without precedent.  Our reduction of the infinitesimal time evolution of the quantum periodic Benjamin-Ono hierarchy in a coherent state to Jack measures mimics the reduction by equivariant localization of Nekrasov's partition functions for SUSY gauge theories to random partitions in \cite{NekOk, Ok3} still under active investigation in geometric representation theory and enumerative geometry \cite{KimuraPestun1, MaulOk, NekYI}.  In fact, Nazarov-Sklyanin's integrable geometric quantization \cite{NaSk2} of the classical periodic Benjamin-Ono equation studied here is the same data as the quantization of a different classical system, $\mathcal{N}=2$ Yang-Mills theory on $\R^4$ in the Omega background $g_{\ee, \e}$ for abelian gauge group $U(1)$ \cite{Nek1, NekOk}.  Realizing the same data as different quantizations of different classical systems is the hallmark of \textit{duality} \cite{Aganagic0}.  Our story begins at the opposite end of the ``BPS/CFT Correspondence'' \cite{NekYI}, and our ability to work with generic $v$ and our interpretation of $\hbar = \ered$ and $\ebar = \eblue$ appear to be new.  Our doubly-graded expansion in Theorem [\ref{AOEColumn4Dispersionless}] may be regarded as a step towards a refined topological recursion for gauge theories in the Omega background.  To clarify comparison to non-abelian $U(r)$ theories hence to Liouville and Toda field theories via the ``colored Jack polynomials'' \cite{Smirnov2}, we keep careful track of the benign center $a \in \R$ in this paper, the $r=1$ case of the vacuum expectation values $\vec{a}=(a^{(1)}, \ldots, a^{(r)})$ of the Higgs field.\\
\\
\noindent A second context for random partitions is stochastic PDEs.  The Kardar-Parisi-Zhang (KPZ) equation is a model for a randomly growing interface in which tilt-dependent local growth velocity with smoothing is driven by a space-time white noise.   The subject of integrable probability \cite{BoGo0, BoPe0, Co0} revolves around a startling fact: from known spectrum of quantum integrable systems one can create discretizations of the KPZ equation which one can analyze in simultaneous limits of long time and vanishing spatial mesh.  Although quantum integrable systems play a role in the construction of such models, the meaning of $\hbar$ in the original quantum integrable system plays no role in the analysis.  While quantum integrable systems of free fermions give rise to determinantal point processes in integrable probability -- most notably, the Schur measures \cite{Ok1} generalized by the Jack measures in this paper --  fermions do not have a semi-classical limit.  Thus, in this paper we give for perhaps the first time an approach to a model of integrable probability via semi-classical analysis of an integrable system.\\
\\
\noindent The matching of our parameters of quantization $\hbar$ and dispersion $\ebar$ for quantum periodic Benjamin-Ono with parameters in gauge theories and stochastic processes is non-trivial and we hope our emphasis on the theory of coherent states can be of use in these related disciplines.  While our account of coherent states is self-contained, the literature on coherent states is vast and we do not give extensive references.  The existence of coherent states of Definition [\ref{BabyCoherentStatesDefinition}] for the quantum harmonic oscillator are at the heart of explaining why light seems to be both a stream of particles (photons) and also a continuous wave (electromagnetic radiation).  Not only do coherent states provide a phase space formulation of quantum dynamics in which Bohr's Correspondence Principle is most transparent, their Wigner quasi-probability distributions in phase space can be measured experimentally \cite{BreitenbachLASER} and are Gaussians, translations of the complex Gaussian of formula (\ref{PhaseSpaceFootprint}).  The image below is from \cite{BreitenbachLASER}.
\noindent \begin{wrapfigure}{r}{0.40 \textwidth} \begin{center}
    \includegraphics[width=0.40 \textwidth]{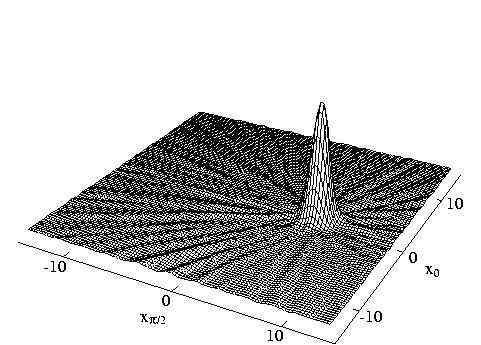}
  \end{center} 
\end{wrapfigure} 

\noindent The viewpoint of the KPZ equation suggests that our limit shape and global Gaussian fluctuation results for random partitions correspond to a hydrodynamic limit and its corrections according to Edwards-Wilkinson universality.  The viewpoint of SUSY gauge theory predicts them in Seiberg-Witten and Nekrasov-Shatashvili limits.  In this paper, our results are confirmations and corrections to the static version of Bohr's Correspondence Principle for coherent states, the correction being Gaussian because coherent states are already Gaussian as depicted.  We now turn to the semi-classical $\hbar \rightarrow 0$ and dispersionless $\ebar \rightarrow 0$ limits of Nazarov-Sklyanin's integrable geometric quantization \cite{NaSk2} of classical periodic Benjamin-Ono in a coherent state. Our story begins not with gauge theory nor with stochastic PDE but with a single principle: {integrability is not just a promise that a system is ``exactly solvable'' but} {a warning that its dynamics will differ from those of a chaotic system.}

\pagebreak

$\ldots \dotfill \ldots $ \\ 
$ \ $\\
$ \ $\\
$ \ $\\
$ \ $\\

$ \ $\\
$ \ $\\
$ \ $\\
$ \ $\\
$ \ $\\

$ \ $\\
$ \ $\\
$ \ $\\
$ \ $\\

\textsf{ \indent \indent ``The voice of truth is accompanied by a suspicious static noise\\
\indent \indent \indent \ \ \ \ \  \ \ \ \ \ \ \ \ \ to which those most closely involved turn a deaf ear."}

\textsf{\begin{center}\  \ \ \ \ \ \ \ \ \ \ \ \ \ \ \ \ \ \ Robert Musil (1880-1942)\\
 \textit{The Man Without Qualities} (1930-1943) \end{center}
}

$ \ $\\
$ \ $\\
$ \ $\\
$ \ $\\
$ \ $\\

$ \ $\\
$ \ $\\
$ \ $\\
$ \ $\\
$ \ $\\

$ \ $\\
$ \ $\\
$ \ $\\
$ \ $\\
$ \ $\\
$\ldots \dotfill \ldots $ \\ 

\pagebreak

\section{Partitions from Periodic Benjamin-Ono} \label{secPartitionsFromQuantumBenjaminOno}

\noindent In section \textbf{[\ref{subsecCBOHIntroduction}]}, we introduce the classical periodic Benjamin-Ono equation and state our first result in Theorem [\ref{CBOHConservedDensityExistence}], a construction of a new classical conserved density $dF_{\star |v} (c| \ebar)$ on the real line $c \in \R$ for a classical state $v$ with dispersion coefficient $\ebar \in \R$.  In sections \textbf{[\ref{subsecQBOHIntroduction}]} and \textbf{[\ref{subsecNazarovSklyaninIntegrableQuantizationExistence}]}, we introduce Nazarov-Sklyanin's integrable geometric quantization \cite{NaSk2} of the classical periodic Benjamin-Ono system and state our second result in Theorem [\ref{QBOHConservedDensityExistence}], a construction of a new quantum conserved density $d\widehat{F} (c | \ebar, \hbar) |_{\Psi}$ for any initial quantum state $\Psi$ of a quantization of the classical periodic Benjamin-Ono equation with quantization parameter $\hbar$.  The quantum stationary states $\Psi_{\lambda}( v | \ebar, \hbar)$ are identified with Jack polynomials in Theorem [\ref{QBOStationaryStatesAreJacks}], and in Theorem [\ref{QBOHConservedDensitiesForJacksAreAnisotropicPartitions}] we identify quantum conserved densities of quantum stationary states with the Rayleigh measures $dF_{\lambda}(c | \ee, \e)$ of anisotropic partitions after choosing Omega variables $(\ee, \e) \in \C^2$ so that \begin{eqnarray} \label{Truehbar} \hbar &=& \ered \\  \label{Trueebar} \ebar &=& \eblue \end{eqnarray}

\noindent a change of variables invariant under $\ee \longleftrightarrow \e$.  A curious feature of the periodic Benjamin-Ono system is how natural it is to use the variables $(\ee, \e)$ which treat the dimensionless parameters of quantization $\hbar$ and dispersion $\ebar$ on the same footing.  In section \textbf{[\ref{subsecRandomPartitionsIntroduction}]}, we introduce Jack measures of random partitions and state our main results in sections \textbf{[\ref{subsecAOEIntro}]}, \textbf{[\ref{subsecLLNIntro}]}, and \textbf{[\ref{subsecCLTIntro}]} in two asymptotic regimes:
{

\begin{itemize}
\item The \textit{dispersive semi-classical limit}: $\hbar \rightarrow 0$ while $\ebar >0$ is fixed.
\item The \textit{dispersionless semi-classical limit}: $\hbar, \ebar \rightarrow 0$ conflated at comparable rate 
\begin{equation} \label{Truebeta} \frac{\ebar^2}{ \hbar} \sim \Bigg ( \sqrt{ \frac{2}{ \beta}} - \sqrt{ \frac{\beta}{2}} \Bigg )^2  \end{equation}

\noindent for some parameter $\beta>0$, so that $\beta = 2$ is first taking $\ebar \rightarrow 0$ for fixed $\hbar >0$.
\end{itemize}
}

\noindent For random Rayleigh measures of Jack measures, in the dispersive and dispersionless regimes, we prove asymptotic expansions in $\hbar$ and $\ebar$ of joint cumulants of linear statistics over a new combinatorial object we call ``ribbon paths'' in Theorems [\ref{AOEColumn4Dispersive}] and [\ref{AOEColumn4Dispersionless}], concentration of measure on limit shapes in Theorems [\ref{LLNColumn4Dispersive}] and [\ref{LLNColumn4Dispersionless}], and Gaussianity of quantum corrections at scale $\hbar^{1/2}$ in Theorems [\ref{CLTColumn4Dispersive}] and [\ref{CLTColumn4Dispersionless}].\\

\noindent $$\xymatrix{\text{\textcolor{gray}{Jack Measures}} \ar@{<=>}[d]^{\small{\text{\ \ Coherent States}}}_{\small{\text{Born's Rule\ \ }}} \ \ \mathbf{M}_v(\ebar, \hbar) \ \  & \ & \ \ \mathbf{M}_v(0, \hbar) \ \ \text{\textcolor{gray}{Schur Measures}} \ar@{<=>}[d]^{\small{\text{\ \ Coherent States}}}_{\small{\text{Born's Rule \ \ }}}\\ \text{\textcolor{gray}{Quantum Benjamin-Ono}} \ \ (\ebar, \hbar) \ \ \ar@{~>} [rrdd]\ar[dd]_{\small{\text{Semi-Classical Limit}}}^{\hbar \rightarrow 0}\ar[rr]_{\ebar \rightarrow 0}^{\small{\text{Dispersionless Limit}}}& \ & \ \ \ \ ( 0, \hbar) \ \  \text{\textcolor{gray}{Quantum Dispersionless BO \ \ }} \ar[dd]^{\small{\text{Semi-Classical Limit}}}_{\hbar \rightarrow 0} \\
\ & \ & \ \\ \text{\textcolor{gray}{Classical Benjamin-Ono}} \ \ (\ebar,0) \ \ \ \ar[rr]^{\ebar \rightarrow 0}_{\small{\text{Dispersionless Limit}}}  & \ & \ \ \ \  (0,0) \ \ \text{\textcolor{gray}{Classical Dispersionless BO \ \ }} \\ }$$

\pagebreak

\subsection{Classical Periodic Benjamin-Ono Waves} \label{subsecCBOHIntroduction}

\noindent The \textit{classical periodic Benjamin-Ono equation} \cite{Benj, Ono} \textit{at criticality}  \begin{equation} \label{CBOE} \begin{cases} v_t + v v_x = \ebar J [ v_{xx}] \\ \ \ v(x,0) \in (\mathscr{M}(a), J, \mathsf{g}_{- \frac{1}{2}}, \omega_{- \frac{1}{2}})\end{cases} \end{equation} \noindent is a non-linear, non-local wave equation in (1+1)-dimensions.   $J$ is the spatial periodic Hilbert transform of Definition [\ref{PeriodicHilbertTransformDefinition}] and $\ebar \in \R$ is the coefficient of dispersion.  (\ref{CBOE}) is posed in the real $L^2$-Sobolev space of $2 \pi$-periodic distributions $v$ of mean $V_0 = a \in \R$ of regularity $s = - \frac{1}{2}$.  We construct $(\mathscr{M}(a), J, \mathsf{g}_{-1/2}, \omega_{-1/2})$ in section \textbf{[\ref{subsubsecSobolevSpaces}]} and verify that (\ref{CBOE}) is Hamiltonian in section \textbf{[\ref{subsubsecCBOHamiltonian}]}.  We discuss the problems of finding classical stationary states and proving Liouville integrability in section \textbf{[\ref{subsubsecCBOStationaryStates}]}.  In \textbf{[\ref{subsubsecCBOHConservedDensity}]} we state our first result, existence of a new conserved density for the classical periodic Benjamin-Ono equation, in Theorem [\ref{CBOHConservedDensityExistence}].  We later construct of the conserved density promised in Theorem [\ref{CBOHConservedDensityExistence}] in Theorem [\ref{CBOHConservedDensityConstruction}].  Finally, in section \textbf{[\ref{subsubsecCBOPhononSoliton}]} we conclude with a brief discussion of the scattering case to emphasize the challenges of working with periodic boundary conditions and to better appreciate the dynamics of the Benjamin-Ono system after quantization.

\subsubsection{Classical Phase Spaces as Sobolev Spaces} \label{subsubsecSobolevSpaces}

\noindent We now construct the symplectic leaves $\mathscr{M}(a)$ of real $L^2$-{Sobolev spaces} $(\mathscr{M}, J, \mathsf{g}_s, \omega_s)$ on the unit circle $\mathbb{T}$, a special case of the Hermitian affine spaces of section \textbf{[\ref{subsecHermitianAffineSpaces}]}.  Basic notions of linear algebra and K\"{a}hler geometry are invoked here with appropriate links to section \textbf{[\ref{subsecHermitianAffineSpaces}]} where all definitions appear in full.\\

\begin{wrapfigure}{r}{0.32 \textwidth} \begin{center}
    \includegraphics[width=0.32 \textwidth]{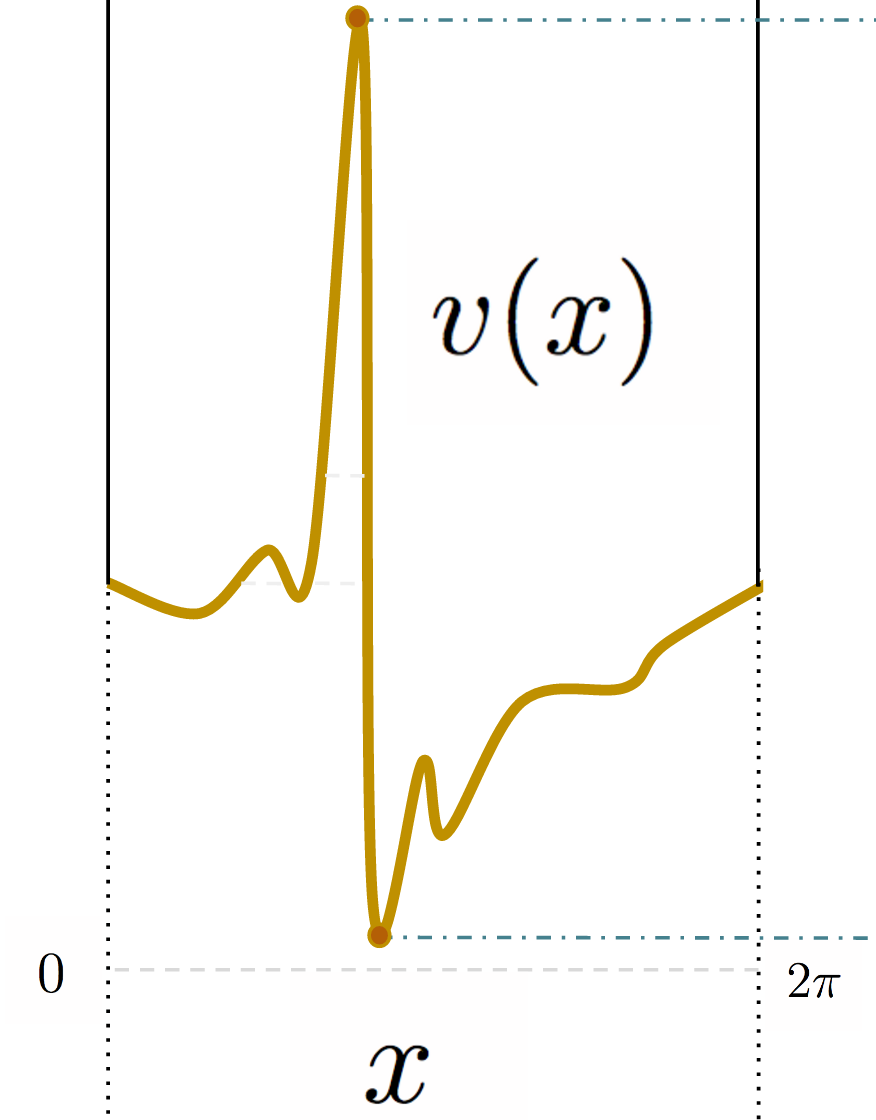}
  \end{center} 
\end{wrapfigure}

\noindent Throughout, write real $2\pi$-periodic distributions $v$ in a spatial variable $x \in \R$ with carefully chosen conventions
\begin{equation} \label{vFourierSeriesFormula} v(x) = V_0 +  \sum_{k=1}^{\infty} \big ( \overline{V_k} e^{+ \textnormal{\textbf{i}} kx} + {V_k} e^{ - \textnormal{\textbf{i}} kx}  \big ) \end{equation} \noindent for Fourier modes $V_k \in \C$.  Identify $[0, 2 \pi)$ with \begin{equation} \mathbb{T} = \{ w \in \C : |w|=1\} \end{equation} by the change of variables $w = e^{\textnormal{\textbf{i}}x}$.
 \begin{definition} Let ${M}$ be the \underline{free loop space of $\R$}, the space of smooth $v : \mathbb{T} \rightarrow \R$. \end{definition} 

\begin{definition} Let ${M}(a) \subset {M}$ be the \underline{leaf} of those $v \in {M}$ with mean $a \in \R$ 
\begin{equation} a = \oint_{\mathbb{T}} v(w) \frac{ dw}{2 \pi \textnormal{\textbf{i}}w}.\\ \end{equation}
 \end{definition}

\begin{proposition} The free loop space ${M}$ is an infinite-dimensional real vector space foliated by real affine subspaces $M(a)$, the leaves indexed by $a \in \R$.\end{proposition}

\begin{definition} \label{PeriodicHilbertTransformDefinition} The \underline{spatial periodic Hilbert transform} $J$ on $M(a)$ is the singular integral operator defined by its Fourier multiplier \begin{equation} \label{PeriodicHilbertTransformFormula} J e^{\textnormal{\textbf{i}} k x} =  \textnormal{\textbf{i}} \cdot \textnormal{sign}(k) e^{ \textnormal{\textbf{i}} k x} \end{equation} 

\noindent with the convention $\textnormal{sign}(0)=0$.

 \end{definition} 

\begin{proposition} The spatial periodic Hilbert transform $J$ provides a complex structure on each leaf ${M}(a)$ as in \textnormal{Definition [\ref{ComplexStructureDefinition}]}. \end{proposition}

\begin{proposition} \label{PropositionJCoordiantes} The complexification ${M} \otimes \C$ is the free loop space of $\C$ and thus $M(a) \otimes \C$ has a $\C$-basis $\{e_{\pm k}\}_{k=1}^{\infty}$ of $e_{\pm k} : \mathbb{T} \rightarrow \C$ given by the plane waves
\begin{equation} e_{\pm k} (x) = e^{ \textnormal{\textbf{i}} kx} \end{equation}

\noindent satisfying $J e_{\pm k} = \pm \textnormal{\textbf{i}} e_{\pm k}$ by \textnormal{Definition [\ref{PeriodicHilbertTransformFormula}]}, so the {Fourier modes} of $v$ \textnormal{(\ref{vFourierSeriesFormula})} are $J$-(anti)-holomorphic coordinates $\{\overline{V}_k, V_k\}_{k=1}^{\infty}$ on ${M}(a)$ of mean $V_0 = a \in \R$.  \end{proposition}

\begin{definition} \label{OperatorsOnCircleDefinition} Let $D = \frac{1}{ \textnormal{\textbf{i}} } \frac{ d}{dx } =  w \frac{\partial}{\partial w}$, $- \Delta = D^2$, and
\begin{equation} |D| = (- \Delta)^{1/2} \end{equation}

\noindent be the operators on $\mathbb{T}$ with Fourier multipliers $k$, $k^2$, and $|k|$, respectively. \end{definition}

\begin{definition} For $a,s \in \R$, \underline{real $L^2$-Sobolev pairing} $\mathsf{g}_s: {M}(a)\times {M}(a) \rightarrow \R$ \begin{equation} \mathsf{g}_s ( v^{\textnormal{out}}, v^{\textnormal{in}}) = \mathsf{g}_0 ( |D|^s v^{\textnormal{out}}, |D|^s v^{\textnormal{in}} ) = \sum_{k =1}^{\infty} |k|^{2s}  \Big ( \overline{V_k^{\textnormal{out}}} V_k^{\textnormal{in}} + V_k^{\textnormal{out}} \overline{V_k^{\textnormal{in}}} \Big ) \end{equation}

\noindent where $\mathsf{g}_0$ is the real $L^2$ pairing on $\mathbb{T}$ and $v^{\textnormal{out}}, v^{\textnormal{in}} \in M(a)$.

\end{definition}

\begin{proposition} \label{SobolevPairingDefinition} For $a,s \in \R$, the Sobolev pairing $\mathsf{g}_s$ is a real inner product on the leaf $M(a)$ the sense of \textnormal{Definition [\ref{RiemannianStructureDefinition}]}. \end{proposition}

\begin{definition} The \underline{real $L^2$-Sobolev space} $(\mathscr{M}(a), {\mathsf{g}}_s)$ is the completion of the leaf $M(a)$ of the free loop space $M$ of $\R$ with respect to the $L^2$-Sobolev pairing ${\mathsf{g}}_s$. \end{definition}

\begin{proposition} For $s \in \R$, the Sobolev pairing $\mathsf{g}_s$ is compatible with the spatially periodic Hilbert transform $J$ in sense of \textnormal{Definition [\ref{CompatibleDefinition}]}. \end{proposition}

\begin{corollary} \label{SobolevLeavesAreHermitianAffineSpaces} For $a,s \in \R$, the real $L^2$ Sobolev space $(\mathscr{M}(a), J, \mathsf{g}_s, \omega_s)$ is a Hermitian affine space as in \textnormal{Definition [\ref{HermitianAffineSpaceDefinition}]} with Hermitian metric \begin{equation} \mathsf{h}_s ( v^{\textnormal{out}}, v^{\textnormal{in}} ) = 2 \sum_{k=1}^{\infty} |k|^{2s} \overline{V^{\textnormal{out}}_k} V^{\textnormal{in}}_k \end{equation} independent of $a \in \R$ and symplectic structure as in \textnormal{Definition [\ref{SymplecticStructureDefinition}]} \begin{equation} \omega_s (v^{\textnormal{out}}, v^{\textnormal{in}})= \frac{1}{ \textnormal{\textbf{i}}}  \sum_{k=1}^{\infty} |k|^{2s}  \Big ( \overline{V_k^{\textnormal{out}}} V_k^{\textnormal{in}} - V_k^{\textnormal{out}} \overline{V_k^{\textnormal{in}}} \Big ) \end{equation} 
\noindent so $\mathscr{M}(a)$ is a symplectic leaf in the Poisson manifold ${\mathscr{M}}$ and its $J$-(anti)-holomorphic coordinates $\{V_k, \overline{V_k}\}_{k=1}^{\infty}$ of \textnormal{Proposition [\ref{PropositionJCoordiantes}]} are $\sigma$-coordinates of \textnormal{Definition [\ref{SigmaCoordinatesDefinition}]}.
\end{corollary}

\noindent We have described leaves $(\mathscr{M}(a), J, \mathsf{g}_s, \omega_s)$ of $L^2$-Sobolev spaces as Hermitian affine spaces.  By Corollary [\ref{GlobalKahlerPotentialDefinition}], such trivial K\"{a}hler manifolds have Poisson structure \begin{equation} \label{SobolevPoissonBracket} \{O_1, O_2 \}_s = \mathsf{g}_s (J \nabla_s O_1, \nabla_s O_2) .\end{equation}

\noindent The gradient $\nabla_s O$ of $O \in C^{\infty}(\mathscr{M}(a), \R)$ depends on $\mathsf{g}_s$ hence explicitly on $s \in \R$ by

\begin{equation} (\nabla_s O) \Big |_v = \sum_{k=1}^{\infty} \frac{1}{ |k|^{2s}} \Big ( \frac{ \partial O }{ \partial V_k} \Big |_v e^{ - \textbf{i} kx}  + \frac{\partial O }{ \partial \overline{V_k}} \Big |_{v} e^{ + \textbf{i} k x} \Big )  \end{equation}

\noindent after a choice of identification $T_v \mathscr{M} \cong \mathscr{M}$, so Definition [\ref{SobolevPairingDefinition}] implies

\begin{equation} \label{SobolevPoissonBracketFormula} \{O_1, O_2\}_s = \frac{1}{ \textbf{i}} \sum_{k=1}^{\infty} \frac{1}{|k|^{2s}} \Big (\frac{ \partial O_1}{\partial V_k} \frac{ \partial O_2}{\partial \overline{V_k}} - \frac{ \partial O_2}{ \partial V_k} \frac{ \partial O_1}{ \partial \overline{V_k}} \Big ) \end{equation}

\noindent The classical periodic Benjamin-Ono equation (\ref{CBOE}) is posed at $s = - \frac{1}{2}$.  The Poisson structure (\ref{SobolevPoissonBracketFormula}) for the leaves $(\mathscr{M}(a), J, \mathsf{g}_{- \frac{1}{2}}, \omega_{- \frac{1}{2}})$ at this special regularity $s= - \tfrac{1}{2}$ appears as the ``Important Example'' in Chapter 1.1 in \cite{DubrovinKricheverNovikov}:

\begin{definition} \label{GFZBracketDefinition} For $s = - \frac{1}{2}$, the Poisson bracket in the leaf of the real $L^2$-Sobolev space $(\mathscr{M}(a), J, {\mathsf{g}}_{- \frac{1}{2}}, \omega_{- \frac{1}{2}})$ is known as the \underline{Gardner-Faddeev-Zakharov bracket}

\begin{equation} \label{GFZBracket} \{O_1, O_2\}_{- \frac{1}{2}} = \frac{1}{ \textnormal{\textbf{i}}} \sum_{k=1}^{\infty} k \cdot \Big ( \frac{ \partial O_1}{\partial V_k} \frac{ \partial O_2}{\partial \overline{V_k}} - \frac{ \partial O_2}{ \partial V_k} \frac{ \partial O_1}{ \partial \overline{V_k}} \Big ) .\end{equation}

\end{definition}

\subsubsection{Classical Hamiltonian at Critical Regularity} \label{subsubsecCBOHamiltonian}

\noindent The {classical periodic Benjamin-Ono equation} (\ref{CBOE}) is an infinite-dimensional classical Hamiltonian system in $(\mathscr{M}(a), J, {\mathsf{g}}_s, \omega_s)$ the real $L^2$ Sobolev space on $\mathbb{T}$ for $s = - \frac{1}{2}$:

\begin{definition} \label{CBOHamiltonianDefinition} For $\ebar >0$, the \underline{classical periodic Benjamin-Ono Hamiltonian} \begin{eqnarray} \label{CBOHamiltonian} {T^{\uparrow}_3}(\ebar) \big |_v &=&  \frac{1}{4} \oint_{\mathbb{T}}   \Bigg ( v(w)^{3} +  2\ebar v(w) |D| v(w) \Bigg ) \frac{dw}{ 2 \pi \textnormal{\textbf{i}} w}   \\ &=& \frac{1}{4} \sum_{h_1 = -\infty}^{\infty} \sum_{h_2 = - \infty}^{\infty} V_{-h_2} V_{h_2 - h_1} V_{h_1} + \frac{\ebar}{2} \sum_{k=- \infty}^{\infty} |k| V_{-k} V_k \\ &=&  \label{CBOHamiltonianFinalFormula}  \sum_{h_1 = 0}^{\infty} \sum_{h_2 = 0}^{\infty} V_{-h_2} V_{h_2 - h_1} V_{h_1} + \ebar \sum_{h=0}^{\infty} h V_h V_{-h} \end{eqnarray} 

\noindent is a generalized polynomial of degree $3$ on $\mathscr{M}(a)$ as in \textnormal{Definition [\ref{GeneralizedPolynomialDefinition}]}.\end{definition}

\noindent Here we use a new non-trivial notation ${V_{-k} = \overline{V_k}}$ that we will exploit in the remainder.

\begin{proposition} \label{CBOisHamiltonian}The classical Benjamin-Ono equation \textnormal{(\ref{CBOE})} is equivalent to \begin{equation} v_t = \{  v, T^{\uparrow}_3 (\ebar) \}_{-\frac{1}{2}} \end{equation} in the leaf $\mathscr{M}(a)$ where $T_3^{\uparrow}(\ebar)$ is the Hamiltonian of \textnormal{Definition [\ref{CBOHamiltonianDefinition}]} and $\{ \cdot, \cdot \}_{- \frac{1}{2}}$ is the Gardner-Faddeev-Zakharov bracket \textnormal{(\ref{GFZBracket})} at critical regularity $s = - \frac{1}{2}$.

\end{proposition}

\noindent Proposition [\ref{CBOisHamiltonian}] follows from direct computation but is merely a formal rewriting of the equation and does not imply that the system (\ref{CBOE}) is well-posed.

\subsubsection{Classical Stationary States and Integrability} \label{subsubsecCBOStationaryStates}

\noindent One may search for classical stationary states of (\ref{CBOE}) as in Definition [\ref{ClassicalStationaryStatesDefinition}].  We do not classify them in this paper, but must mention them in passing, as their quantum analogs discussed in subsection \textbf{[\ref{subsubsecQuantumStationaryStatesareJacks}]} {are} known to be \textit{Jack polynomials}, special functions with an extensive literature which play a key role for us below.

\begin{proposition} \label{CBOIsHamiltonian} For any $\ebar $, the classical periodic Benjamin-Ono Hamiltonian \begin{equation} \{T_2^{\uparrow}, T_3^{\uparrow}(\ebar) \}_{- \frac{1}{2}} = 0 \end{equation}

\noindent Poisson commutes with the classical observable given by half the $L^2$ norm 
 \begin{equation} \label{ClassicalT2} T_2^{\uparrow} |_v = \sum_{k=1}^{\infty} V_{+k} V_{-k}  \end{equation} \noindent with respect to the Gardner-Faddeev-Zakharov bracket \textnormal{(\ref{GFZBracket})}, using again $V_{-k} = \overline{V_k}$.
 \end{proposition}
 
 \begin{corollary} Any classical mixed state $\rho$ which is stationary for the classical periodic Benjamin-Ono flow generated by $T_3^{\uparrow}(\ebar)$ must also be stationary for $T_2^{\uparrow}$. \end{corollary}
 
\noindent So far, knowledge of $T_2^{\uparrow}$ and $T_3^{\uparrow}(\ebar)$ alone does not account for the abundance of classical stationary states of finite-dimensional and compact support in phase space which exist in light of the existence of multi-phase solutions to (\ref{CBOE}), the spatially-periodic analog of multi-soliton solutions, as discovered in \cite{ChenLeePereira1979, SatsumaIshimori1979} and studied further in \cite{DobrokhotovKrichever, Matsuno2004}.  Although existence of multi-solitons or multi-phase solutions does \textit{not} imply integrability of a classical system, for integrable systems that support such stable stationary waves, exact formulas for them are often accessible through the same means as provide the exact solution of the system.  We now discuss two a priori unrelated ways of rewriting the classical periodic Benjamin-Ono equation (\ref{CBOE}), namely as a \textit{classical Hamiltonian system} and as a \textit{classical Lax system}, and how one can attempt to prove the integrability of (\ref{CBOE}) from each of these two points of view.\\
\\
\noindent First, we just saw in Proposition [\ref{CBOIsHamiltonian}] that the classical periodic Benjamin-Ono equation (\ref{CBOE}) is a classical Hamiltonian system.  For classical Hamiltonian systems with $\tealN < \infty$ degrees of freedom, the Liouville-Arnold theorem says that pairwise Poisson commutativity of $\tealN$ algebraically independent conserved quantities guarantees the integrability of the system, i.e. the existence of action-angle variables \cite{DubrovinKricheverNovikov}.  We will see in Theorem [\ref{CBOHConservedDensityExistence}] the existence of conserved quantities (\ref{ConservedQuantitiesIntro}) that pairwise Poisson commute for $\{ \cdot, \cdot \}_{-\frac{1}{2}}$.  Unfortunately, there is no Liouville-Arnold theorem in infinite dimensions for this to imply integrability of (\ref{CBOE}).\\
\\
\noindent Second, (\ref{CBOE}) is a classical Lax system, i.e. it can be written as the compatibility condition for an auxiliary linear system.  Precisely, we associate to $v(x,t) \in \mathscr{M}$ a \textit{Lax operator} $L_{\bullet}(v(x,t)| \ebar)$ and a \textit{companion operator} $P_3 ( v(x,t) | \ebar)$ on an \textit{auxiliary space} $H_{\bullet}$ so that (\ref{CBOE}) is formally equivalent to
\begin{equation} \label{LaxEquation} \frac{ \partial L_{\bullet} ( v(x,t) |\ebar)}{ \partial t} = [ P_3 ( v(x,t) | \ebar) , L_{\bullet}(v(x,t) | \ebar) ] \end{equation}
\noindent Although formally it is known \cite{BockKruskal, Nakamura1979} how to choose $L_{\bullet}(v | \ebar)$ and $P_3(v | \ebar)$ to recover (\ref{CBOE}) from (\ref{LaxEquation}), this rewriting alone does not guarantee integrability in infinite dimensions.  One needs to extract action-angle variables from the spectral theory of $L_{\bullet}(v | \ebar)$, together with an inverse spectral theory to reconstruct $L_{\bullet}(v | \ebar)$ from such variables.  Even if the Lax operator depends formally on $v$, the direct and inverse spectral problems must depend on the boundary conditions of (\ref{CBOE}).  This is because an integrable system is not an {equation} but an {equation together with boundary conditions}.  For (\ref{CBOE}), action-angle variables have been found for rapidly decaying $v$ on $\R$ by an inverse scattering transform \cite{AblCla, AblowitzFokas1983, KaupLakobaMatsuno1, KaupMatsuno}, but never for the case (\ref{CBOE}) of periodic boundary conditions.  What is missing is an analog for (\ref{CBOE}) of the finite-gap integration theory for the classical periodic Korteweg-de Vries equation where action-angles are carried by periodic and Dirichlet spectrum of Hill's operator \cite{DubrovinKricheverNovikov, HSW}.  Our Theorem [\ref{CBOHConservedDensityExistence}] stated in the next section is a step in this direction.\\
\\
\noindent Despite a lack of action-angle variables for (\ref{CBOE}), results are known for classical periodic Benjamin-Ono that would follow from having such variables: existence of rational multi-phase solutions (spatially-periodic analog of multi-soliton solutions) in \cite{ChenLeePereira1979, DobrokhotovKrichever, Matsuno2004, SatsumaIshimori1979}, construction of invariant measures in \cite{Tzvetkov2010, TzvetkovVisciglia2013}, as well as global well-posedness in the subcritical regime $s \geq 0$ in \cite{Molinet}.\\
\\
\noindent Finally, note that a large class of classical integrable systems that are both Hamiltonian integrable and Lax integrable are the classical \textit{Adler-Kostant-Symes integrable systems}, certain integrable systems whose phase spaces occur as coadjoint orbits in the dual $\mathfrak{g}^*$ of a Lie algebra ${\mathfrak{g}}$, see chapter 4.4 in \cite{AdlerEtAlBook}.  In this paper, we neither confirm nor deny that the ingredients in our study of classical periodic Benjamin-Ono are a version the Adler-Kostant-Symes construction for an infinite-dimensional Lie algebra ${\mathfrak{g}}$.
\subsubsection{Classical Conserved Densities as Interlacing Sequences} \label{subsubsecCBOHConservedDensity}

\noindent We now state our first result, the existence of a new conserved density $dF_{\star |v}(c | \ebar)$ on the real line $c \in \R$ for the classical periodic Benjamin-Ono equation (\ref{CBOE}) in Theorem \textbf{[\ref{CBOHConservedDensityExistence}]}.  We prove this result in Theorem [\ref{CBOHConservedDensityConstruction}] when we construct $dF_{\star |v} ( c | \ebar)$ as a spectral shift function in an auxiliary spectral theory of elliptic generalized Toeplitz operators of order $1$ with symbol $v$ in section \textbf{[\ref{secConstructionsForPeriodicBenjaminOno}]}.  Our Theorem [\ref{CBOHConservedDensityExistence}], as well as our Theorem [\ref{QBOHConservedDensityExistence}] for the quantum periodic Benjamin-Ono equation, are both humble corollaries of the pioneering work of Nazarov-Sklyanin \cite{NaSk2}.\\
\\
\noindent Before we introduce our $dF_{\star |v}( c | \ebar)$, let us start with $dF_{\star |v}(c)$ the conserved densities of the \textit{classical dispersionless periodic Benjamin-Ono equation} in (1+1)-dimensions \begin{equation} \label{CHBE} \begin{cases} v_t + v v_x = 0 \\ \ \ v(x,0) \in (\mathscr{M}(a), J, \mathsf{g}_{- \frac{1}{2}}, \omega_{- \frac{1}{2}})\end{cases} \end{equation}

\noindent posed at Sobolev regularity $s = - \tfrac{1}{2}$ for $v \in \mathscr{M}(a)$ of mean $a \in \R$.  While (\ref{CHBE}) is often called the ``Riemann-Hopf equation'' or ``inviscid Burgers' equation'' or the ``dispersionless KdV equation'', here it is the dispersionless limit $\ebar \rightarrow 0$ of (\ref{CBOE}).
\pagebreak

\noindent Although (\ref{CHBE}) forms shock singularities in finite time \cite{ArnoldODE}, for generic initial data $v$ it has a \textit{classical conserved density} $dF_{\star |v}(c)$ with cumulative distribution function $F_{\star |v}(c)$ and associated profile $f_{\star |v}$ by $F(c) = \frac{ 1 + f'(c)}{2}$ depicted here and defined below.

  \begin{figure}[htb]
\centering
\includegraphics[width=0.90\textwidth]{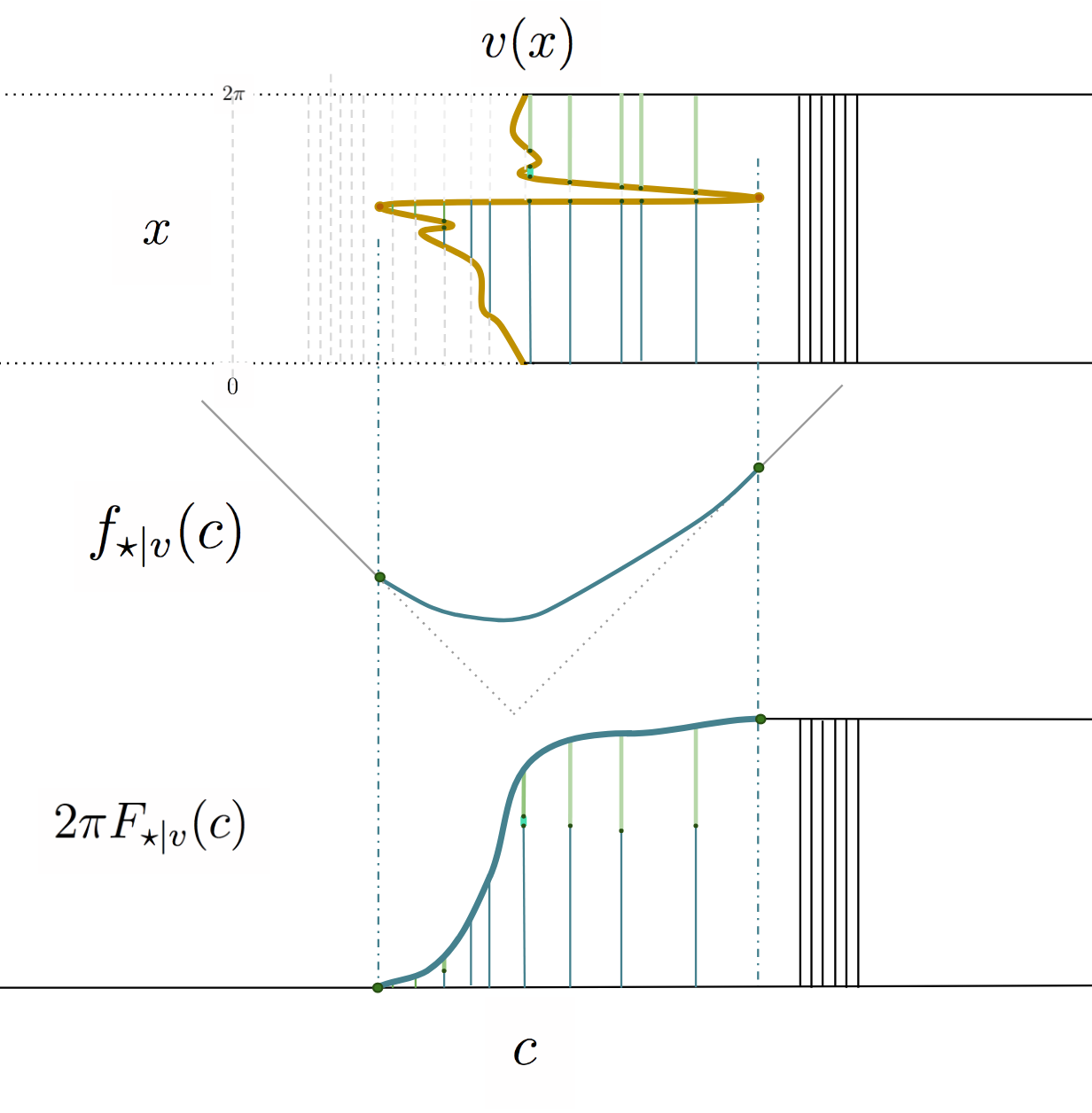}
\end{figure}

\noindent 
\begin{definition} \label{CHBHConservedDensityDefinition} For bounded real $v$, the cumulative distribution function
\begin{equation} F_{\star |v}(c) = \oint_{\mathbb{T}} \mathbbm{1}_{v(w) \leq c} \frac{ dw}{ 2 \pi \textnormal{\textbf{i}} w}.\end{equation} 
 defines $dF_{\star|v} = v_{*} \rho_{\star | \mathbb{T}}$ the \underline{push-forward of the {normalized} {uniform measure} $\rho_{\star | \mathbb{T}}$ on the} \underline{unit circle $\mathbb{T}$ {along} $v: \mathbb{T} \rightarrow \R$} \end{definition}

\begin{theorem} \label{CHBHConservedDensityIntro} Up until the gradient catastrophe, the convex profile $f_{\star |v}: \R \rightarrow \R_{ \geq 0}$ associated to $F_{\star |v}(c)$ by $F(c) = \tfrac{1 + f'(c)}{2}$ does not change if $v$ evolves by \textnormal{(\ref{CHBE})}. \end{theorem}
 \noindent \textit{Proof:} \textnormal{Integrate by parts, or recognize that the characteristics are straight lines. $\square$}

\pagebreak

\noindent Our investigation comes to life since the classical periodic Benjamin-Ono Hamiltonian $T_3^{\uparrow}(\ebar)$ is highly non-generic.  Here is our first result:

\begin{theorem}  \label{CBOHConservedDensityExistence} For bounded real $v$ and $0<\ebar$, there exists an interlacing sequence 
\begin{equation} \label{CBOHInterlacingSequence} c_1^{\uparrow}(\ebar) |_v <  c_1^{\downarrow}(\ebar)|_v <  c_2^{\uparrow}(\ebar) |_v < \cdots <  c_{e+1}^{\uparrow}(\ebar)|_v < \cdots \leq +  \infty \end{equation}

\noindent with $+ \infty$ as its only accumulation point so that

\begin{enumerate}
\item \textnormal{\textcolor{gray}{[Conserved Density]}} The following signed measure is invariant under \textnormal{(\ref{CBOE})} \begin{equation} \label{CBOHConservedDensityFormula} dF_{\star |v}(c | \ebar) = \sum_{i=1}^{\infty} \delta(c - c_i^{\uparrow}(\ebar)|_v) - \sum_{i=1}^{\infty} \delta(c - c_i^{\downarrow}(\ebar)|_v ) \end{equation} \noindent hence so is the piecewise-linear profile $f_{\star|v}(c| \ebar)$ for $F_{\star|v}(c | \ebar)$ and $F(c) = \tfrac{1+f'(c)}{2}$.  Moreover, \textnormal{(\ref{CBOHConservedDensityFormula})} is a classical conserved density $dF_{\star |v} ( c | \ebar)$ on $\mathbb{X} = \mathbb{R}$ in the sense of \textnormal{Definition [\ref{ClassicalConservedDensityDefinition}]}, namely:
\item \textnormal{\textcolor{gray}{[Integrable Hierarchy]}} The $c^l$-averages
\begin{equation} \label{ConservedQuantitiesIntro} O_l( \ebar) |_v = \int_{- \infty}^{+\infty} c^l dF_{\star |v}( c | \ebar), \end{equation}
are a classical integrable hierarchy $\{O_l(\ebar)\}_{l=1}^{\infty}$ in the sense of \textnormal{Definition [\ref{ClassicalIntegrableHierarchyDefinition}]}, i.e. in involution for the Gardner-Faddeev-Zakharov Poisson bracket \textnormal{(\ref{GFZBracket})}.
\item \textnormal{\textcolor{gray}{[Regularity of Observables]}} $O_l(\ebar)|_v$ is a generalized polynomial of degree $l \in \N$ in the $\sigma$-coordinates $\{V_k, \overline{V_k}\}$ of $v \in \mathscr{M}_0$ as in \textnormal{Definition [\ref{GeneralizedPolynomialDefinition}]}
\noindent \begin{equation} \{ O_{l_1}(\ebar)  , O_{l_2}(\ebar) \}_{- \frac{1}{2}} = 0 \end{equation}
\item \textnormal{\textcolor{gray}{[Periodic Benjamin-Ono]}} The span of $\{O_l(\ebar)\}_{l=1}^{\infty}$ includes the classical periodic Benjamin-Ono Hamiltonian $T_3^{\uparrow}(\ebar)$ \textnormal{(\ref{CBOHamiltonian})} and also $T_2^{\uparrow}$ \textnormal{(\ref{ClassicalT2})}.
\item \textnormal{\textcolor{gray}{[``Finite Gap Potentials'']}} The signed measure $dF_{\star|v}(c | \ebar)$ has compact support, hence $f_{\star|v}(c| \ebar)$) has finitely-many extrema, if $v$ is a Laurent polynomial in $e^{\pm \textnormal{\textbf{i}} kx}$.
\item \textnormal{\textcolor{gray}{[Dispersionless Limit]}} As $\ebar \rightarrow 0$ have weak convergence of signed measures to
\begin{equation}  dF_{\star |v}(c| \ebar) \rightarrow dF_{\star |v}(c) \end{equation} the classical dispersionless conserved density for \textnormal{(\ref{CHBE})} from \textnormal{Theorem [\ref{CHBHConservedDensityIntro}]}.
\end{enumerate}
\end{theorem} 

\begin{itemize}
\item \textit{Proof:} Follows from \textnormal{Theorem [\ref{CBOHConservedDensityConstruction}]}. $\square$
\end{itemize}

\noindent Our Theorem [\ref{CBOHConservedDensityExistence}] result is a dispersive version of Theorem [\ref{CHBHConservedDensityIntro}].  The non-linearity $v v_x$ in (\ref{CBOE}) is quadratic, hence the sign of the dispersion coefficient $\ebar \in \R$ does not correspond to a qualitative difference between focusing and defocusing regimes, so we could have taken $\ebar <0$ in Theorem [\ref{CBOHConservedDensityExistence}] just by changing the sign of the interlacing extrema which would then accumulate at $- \infty$.
\pagebreak

\begin{figure}[htb]
\centering
\includegraphics[width=0.85\textwidth]{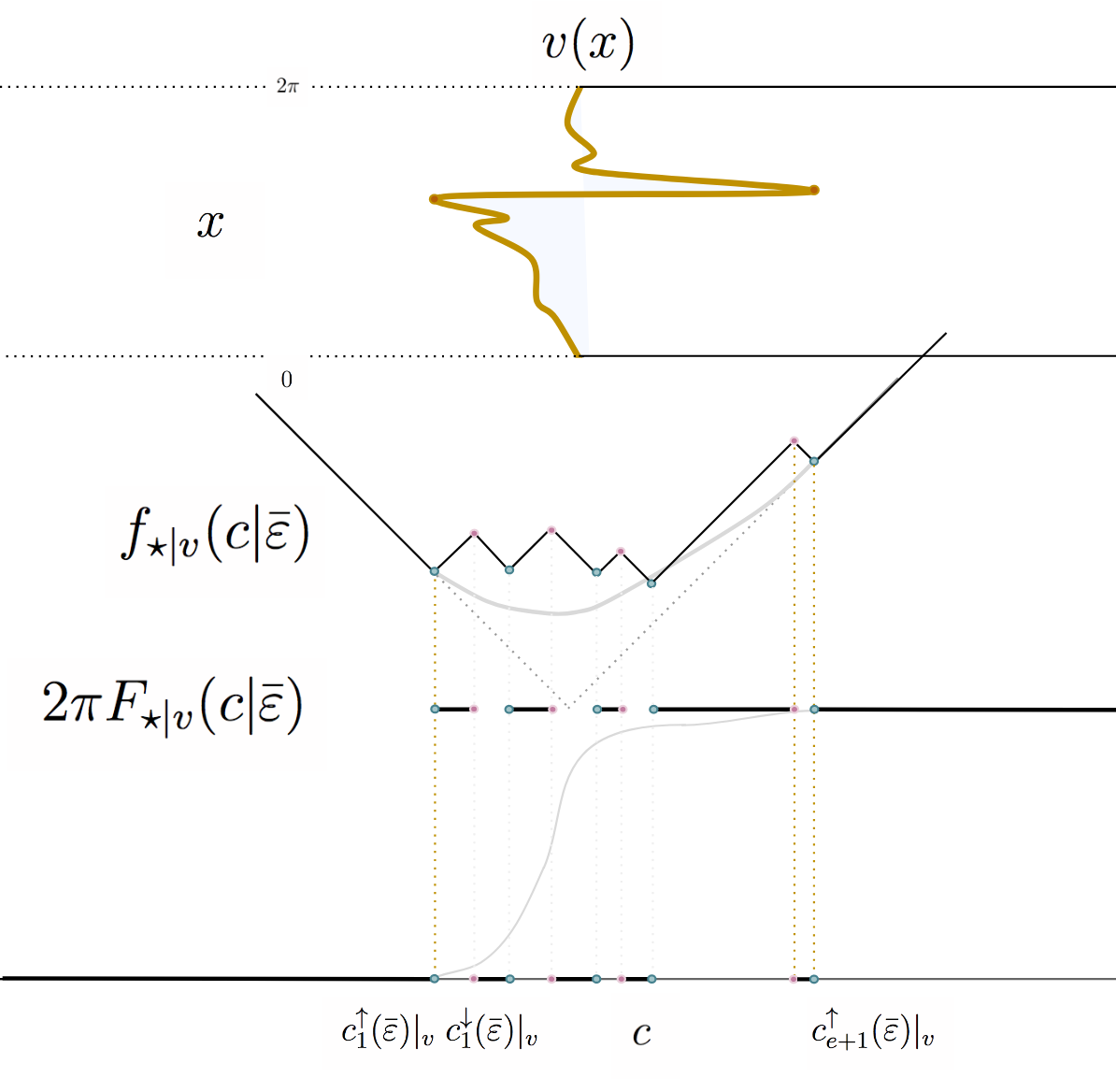}
\end{figure}

\noindent A different version of the classical periodic Benjamin-Ono hierarchy $\{T^{\downarrow}_{\ell}( \ebar)\}_{\ell=0}^{\infty}$ with a different conserved density $d\tau^{\downarrow}_{\star |v} ( c | \ebar)$ was discovered by Nazarov-Sklyanin in \cite{NaSk2}.  In \textbf{[\ref{subsubsecCBOHConservedDensityConstruction}]} our $O_l(\ebar)$ are written through their $T^{\downarrow}_{\ell}(\ebar)$, hence we do not have to prove the non-trivial Poisson commutativity of our $O_l(\ebar)$ but merely derive it from \cite{NaSk2}.  As we discussed in \textbf{[\ref{subsubsecCBOStationaryStates}]}, existence of an integrable hierarchy $\{O_l(\ebar)\}_{l=1}^{\infty}$ does not imply Liouville integrability of the classical periodic Benjamin-Ono equation (\ref{CBOE}).  Our Theorem [\ref{CBOHConservedDensityExistence}] should help, since locating a conserved density suggests distinguished coordinates on the range of the classical hierarchy for a moment map.\\
\\
\noindent Let us say why Laurent $v$ are ``finite gap potentials'' for (\ref{CBOE}).  As emphasized in \cite{DobrokhotovKrichever}, in the deep water limit $\mathsf{q} \rightarrow 0$, the smooth spectral curve $\Sigma(\ebar, \mathsf{q})|_v$ of the classical periodic intermediate long wave equation \cite{AblowitzFokasSatsumaSegur} degenerates to singular rational spectral curve $\Sigma(\ebar, 0)|_v \cong \mathbb{P}^1$ of classical periodic Benjamin-Ono.  For generic $v$, $\Sigma(\ebar, \mathsf{q})|_v$ is of infinite genus, just as our interlacing sequence (\ref{CBOHInterlacingSequence}) is infinite for generic $v$.  \\
\\
\noindent In the dispersionless limit $\ebar \rightarrow 0$, the number of interlacing extrema in an interval must grow for $v$ fixed to give weak convergence $dF_{\star |v}(c | \ebar) \rightarrow dF_{\star |v}(c)$ and of profiles $f_{\star |v}(c | \ebar) \rightarrow f_{\star|v}(c)$ as $\ebar \rightarrow 0$ at fixed $v$.  Remarkably, in the dispersionless limit, the non-convex profiles $f_{\star|v}( c | \ebar)$ of interlacing sequences concentrate on a convex $f_{\star|v}(c)$.

\pagebreak

\subsubsection{Classical Radiation and Solitons} \label{subsubsecCBOPhononSoliton}

\noindent The study of the Benjamin-Ono equation with periodic boundary conditions requires different techniques than those used for the scattering case of $v(x)$ rapidly decaying as $x \rightarrow \pm \infty$.  In this paper, we fix periodic boundary conditions of size $L= 2\pi$, a compact space in which scattering does not occur.  One can degenerate $L \rightarrow \infty$ from $L$-periodic to rapidly decaying $v$.  For the microscopic $\tealN$-body problem underlying the phonon interpretation of (\ref{CBOE}), $L \rightarrow \infty$ is the degeneration of the ``trigonometric'' Calogero-Sutherland system to the ``rational'' Calogero-Moser system \cite{Eti0}.\\
\\
\noindent Coincidentally, the Sobolev regularity $s = - \tfrac{1}{2}$ for which (\ref{CBOE}) is Hamiltonian is also the \textit{critical regularity} for (\ref{CBOE}).  If $v(x,t)$ solves (\ref{CBOE}), rescaling $v$ by $L, s \in \R$ \begin{equation} v (x,t) \longrightarrow \frac{1}{L^{2s}} \cdot v \Big ( \frac{ x}{L}, \frac{t}{L^2}\Big ) \end{equation}
\noindent the result still solves (\ref{CBOE}) if we take the very special choice of $s = s_c = - \frac{1}{2}$.  According to the \textit{Dispersive Scaling Principle}, Principle 3.1 in \cite{Tao0B}, if we take initial data for (\ref{CBOE}) with Sobolev regularity $s$, we expect the contribution of the non-linear term to the long-time behavior of the system to differ qualitatively depending on whether $s>s_c$ (subcritical), $s=s_c$ (critical), or $s<s_c$ (supercritical).  Let us now state more precisely what one expects at long time for subcritical initial data.\\
\\
\noindent According to \cite{Tao0B}, in a non-compact spatial geometry, most classical phonons of subcritical regularity  \textit{scatter}: at long time, their supremum norm vanishes despite energy conservation, and due to this, the classical phonon decays to \textit{radiation}, solutions of the underlying linear equation valid in the approximation of small amplitude.  For the classical Benjamin-Ono equation, the underlying linear equation is still non-local \begin{equation} v_t = \ebar J [v_{xx}] \end{equation} \noindent so radiation consists of real combinations of plane waves $A e^{ \textbf{i} (k x - \omega(k) t)}$ of phase velocity
\begin{equation} \label{BenjaminOnoDispersionRelation} \frac{\omega(k)}{k} = -\ebar |k|. \end{equation} \noindent While most initial data scatter into radiation, exceptional initial data may exist that maintain their form: these are the \textit{multi-soliton solutions}.  For classical Benjamin-Ono, multi-solitons were found in \cite{Matsuno1979}, see also \cite{AbBeWi, Matsuno2004}.  The \textit{soliton resolution conjecture} states that generic subcritical initial data can be ``resolved'' into a sum of a term that scatters to radiation and a multi-soliton term which separates into independent solitons of different speeds \cite{Tao0B}.  The classical Benjamin-Ono equation on the real line with rapidly-decaying initial data is integrable \cite{AblCla, AblowitzFokas1983, KaupLakobaMatsuno1, KaupMatsuno}, thus one anticipates this resolution can be verified exactly.  Indeed, while one can tell from the dispersion relation (\ref{BenjaminOnoDispersionRelation}) that all radiation moves to the \textit{left}, it turns out that all solitons move to the \textit{right}.  We emphasize the resolution of subcritical classical phonons into radiation and solitons to appreciate the analogous decomposition of quantum phonons into \textit{quasi-holes} and \textit{quasi-particles} \cite{AbWi1, AbBeWi, AbBeWi2, LamPri2014, LamPri2015, Wieg1} in section \textbf{[\ref{subsubsecQBOSolitonPhonon}]}.  Quasi-holes and quasi-particles are exchanged under a non-trivial duality no longer visible in the semi-classical limit $\hbar \rightarrow 0$.

\pagebreak

\subsection{Quantum Periodic Benjamin-Ono Waves} \label{subsecQBOHIntroduction}
\noindent The \textit{quantum periodic Benjamin-Ono equation} is

\begin{equation} \label{QBOE} \begin{cases} \textnormal{\textbf{i}} \hbar \partial_t \Psi = \widehat{\mathcal{T}}_3^{\uparrow}(\ebar, \hbar) \Psi  \\ \ \ \Psi \in (\mathscr{F}(a), \langle \cdot, \cdot \rangle_{\hbar, - \frac{1}{2}})\end{cases} \end{equation} 

\noindent with $\hbar >0$ the free parameter of quantization.  (\ref{QBOE}) is posed in the Fock space $\mathscr{F}(a)$ completion of the reproducing kernel Hilbert space of $J$-holomorphic functions on the leaf $(\mathscr{M}(a), J, \mathsf{g}_{-1/2}, \omega_{-1/2})$ of the critical Sobolev space, where $J$ is still the spatial periodic Hilbert transform.  We construct Fock space $\mathscr{F}(a) \supset L^2_{\textnormal{J-hol}} ( \mathscr{M}(a), d \rho_{\hbar, \mathsf{g}_{- 1/2}})$ in section \textbf{[\ref{subsubsecFockSobolevSpaces}]} and define the quantum Hamiltonian $\widehat{\mathcal{T}}^{\uparrow}_3(\ebar, \hbar)$ of the Schr\"{o}dinger equation (\ref{QBOE}), a dispersive generalization of Goulden-Jackson's cut-joint operator $\widehat{\mathcal{T}}_3^{\uparrow}(0, \hbar)$, in section \textbf{[\ref{subsubsecQBOHamiltonian}]}.  In section \textbf{[\ref{subsubsecQuantumStationaryStatesareJacks}]}, we identify the quantum stationary states of (\ref{QBOE}), the eigenfunctions of $\widehat{\mathcal{T}}^{\uparrow}_3(\ebar, \hbar)$, with Jack polynomials, a dispersive generalization of Schur polynomials.  In \textbf{[\ref{subsubsecQBOHConservedDensity}]} we state our second result, existence of a new conserved density for the quantum periodic Benjamin-Ono equation, in Theorem [\ref{QBOHConservedDensityExistence}].  We later construct the conserved density promised in Theorem [\ref{QBOHConservedDensityExistence}] in Theorem [\ref{QBOHConservedDensityConstruction}].  In addition, we identify of the conserved densities of quantum stationary states with the Rayleigh measures $dF_{\lambda}( c | \ee, \e)$ of profiles of anisotropic partitions in Theorem [\ref{QBOHConservedDensitiesForJacksAreAnisotropicPartitions}], which we later prove in Theorem [\ref{QBOHConservedDensitiesForJacksAreAnisotropicPartitionsRevisited}].  We conclude in section \textbf{[\ref{subsubsecQBOSolitonPhonon}]} with a brief discussion of the scattering case.

\subsubsection{\textcolor{black}{Quantum State Spaces and Fractional Brownian Motions}} \label{subsubsecFockSobolevSpaces}

\noindent For the basics of fractional Gaussian fields we refer to the recent survey \cite{FGFsurvey}.

\begin{definition} \label{FGFDefinition} Fix a compact smooth manifold $\Omega$ with boundary $\partial \Omega$ of finite dimension.  The \underline{fractional Gaussian field $d\mathbb{G}^{\Omega}_{s, b}(x| \hbar)$ on $\Omega$ of Sobolev regularity $s \in \R$,} \underline{Dirichlet boundary condition $b: \partial \Omega \rightarrow \R$, and variance $\hbar>0$} is the $\hbar^{1/2}$ multiple of the standard Gaussian in the real $L^2$-Sobolev space of distributions $v(x)$ on $\Omega$ of regularity $s$ with $v |_{\partial \Omega} = b$.  Explicitly, $d\mathbb{G}^{\Omega}_{s,b}(x | \hbar)$ is a random distribution whose average against a test function $\phi$ of dual regularity $-s$ is a Gaussian random variable

\begin{equation} \int_{\Omega} \phi(x) d\mathbb{G}^{\Omega}_{s,b}(x)  \in \mathsf{N} \Big (\int_{\Omega} \phi(x) v_{\star |b} (x) dx \  ,  \  \hbar || \phi||_{-s}^2 \Big ) \end{equation}

\noindent of mean given by the unique fractional-harmonic extension $(-\Delta)^{-s} v_{\star |b} = 0$ of $b$ and variance $\hbar || \phi||_{-s}^2$ determined by the dual Sobolev norm.
\end{definition}
\noindent Note that our $s$ refers to the regularity of the random distribution $d\mathbb{G}^{\Omega}_{s,b}(x| \hbar)$, not to the dual regularity of test functions as is also common in the literature and in \cite{FGFsurvey}.

\begin{definition} The \underline{Hurst index} of the fractional Gaussian field $d\mathbb{G}^{\Omega}_{s,b}(x | \hbar)$ is
\begin{equation} \textnormal{Hurst} \Big (d\mathbb{G}^{\Omega}_{s,b}(x | \hbar ) \Big ) = (-s) - \tfrac{ \dim \Omega}{2}. \end{equation}
\end{definition}

\noindent For $\Omega = \mathbb{T}$ and $- \tfrac{3}{2} < s < -\tfrac{1}{2}$ so $0 < \text{Hurst} < 1$, the fractional Gaussian field $d\mathbb{G}^{\mathbb{T}}_{s,a}(x| \hbar )$ is the \textit{periodic fractional Brownian motion} of Mandelbrot-van Ness \cite{MandelbrotVanNess} generalizing Brownian motion ($s = -1$).  Circles have no boundary $\partial \mathbb{T} = \emptyset$, so instead of the boundary condition $v |_{\partial \Omega} = b$ we must take an averaging condition $\oint_{\mathbb{T}} v(w) \tfrac{dw}{2 \pi \textbf{i} w} = V_0 = a \in \R$ and henceforth work in a leaf $\mathscr{M}(a)$ of the Sobolev space.\\
\\
\noindent As $s \rightarrow - \tfrac{1}{2}$ approaches the critical regularity of the classical Benjamin-Ono equation, $0 \leftarrow \text{Hurst}$ so the resulting fractional Gaussian field $d\mathbb{G}^{\mathbb{T}}_{- 1/2, a}(x| \hbar)$ is no longer defined pointwise but remains well-defined as a random distribution of Hurst index $0$, a log-correlated Gaussian field known as ``pink'' $1/k$ noise or ``$H^{1/2}$ noise''  \cite{FyodorovKeatingFreezing}.\\
\\
\noindent We now use the fractional Gaussian field $d\mathbb{G}^{\mathbb{T}}_{- 1/2, a}(x| \hbar)$ to define the Hilbert space of states for the quantum periodic Benjamin-Ono system.  First, for arbitrary $s \in \R$,

\begin{definition} \label{FockSobolevDefinition} The \underline{Fock-Sobolev space} $(\mathscr{F}(a), \langle \cdot, \cdot \rangle_{\hbar, s})$ of regularity $s \in \R$ centered at $a \in \R$ is the Hilbert space completion of the reproducing kernel pre-Hilbert space
\begin{equation} L^2_{\textnormal{J-hol}} ( \mathscr{M}(a), d \rho_{\hbar, \mathsf{g}_{s}})  \end{equation}
\noindent of $J$-holomorphic functions on the leaf $(\mathscr{M}(a), J, \mathsf{g}_s, \omega_s)$ for $\Omega = \mathbb{T}$ square-integrable against $d \rho_{\hbar, \mathsf{g}_{s}}$ the law of the fractional Gaussian field $d \mathbb{G}_{s, a}^{\mathbb{T}} ( x | \hbar)$ of \textnormal{Definition [\ref{FGFDefinition}]}. \end{definition}

\noindent Fock-Sobolev spaces are instances of a general construction of Fock spaces in \textnormal{Definition [\ref{FockSpaceDefinition}]} associated to arbitrary Hermitian affine spaces.  In this context, the law $d \rho_{\hbar, s}$ of the fractional Gaussian field $d\mathbb{G}^{\mathbb{T}}_{s, a}(x| \hbar)$ is known as the Segal-Bargmann weight.  

\begin{proposition} \label{FockSobolevBosonicBasisProposition} For any $s, a \in \R$, the Fock-Sobolev space $(\mathscr{F}(a), \langle \cdot, \cdot \rangle_{\hbar, s})$ has a dense subspace of polynomial excitations of the coherent state $\Upsilon_a(\cdot | \hbar)$ around $v \equiv a$ \begin{equation} \mathcal{F}(a) = \C[{{V}}_{+1}, {{V}}_{+2}, \ldots ] \Upsilon_a( \cdot  | \hbar) \end{equation}

\noindent where $\Upsilon_a( \cdot | \hbar) = 1$ in $\mathscr{F}(a)$ is spanned by the orthonormal basis \begin{equation} V_{\mu} = V_1^{d_1} V_2^{d_2} \cdots \end{equation}

\noindent indexed by partitions $\mu$ with occupation variables $d_k = \# \{ i : \mu_i = k \}$ and norm 

\begin{equation} || V_{\mu}||^2_{\hbar, s} = \prod_{k=1}^{\infty} (\hbar |k|^{-2s} )^{d_k } d_k! \end{equation}

\end{proposition}

\begin{itemize}
\item \textit{Proof:} follows from Corollary [\ref{OrthogonalBasisFockSpace}]. $\square$
\end{itemize}

\noindent The reader whose starting point is the theory of symmetric functions may verify that they have often seen the Fock space of Definition [\ref{FockSobolevDefinition}] at the critical regularity of the classical periodic Benjamin-Ono equation \textnormal{(\ref{CBOE})}:

\begin{proposition} \label{MacdonaldInnerProductDictionary}For Sobolev $s = -\frac{1}{2}$, the inner product on Fock space becomes the inner product VI.1.4 in \textnormal{\cite{Mac}} after the identifications in \textnormal{Definitions [\ref{OmegaVariablesDefinition}], [\ref{FractionalChargeDefinition}].} \end{proposition}

\noindent By the Sobolev Boundary Trace Theorem, the fractional Gaussian field $d\mathbb{G}^{\mathbb{T}}_{- {1}/{2}, \hbar}(x)$ is the restriction of a Gaussian free field $d \mathbb{G}^{\Omega}_{-1, b}(z, \overline{z} | \hbar)$ in any Riemann surface $\Omega$ to an infinitesimally small loop in the bulk \cite{FGFsurvey, Sheffield0}.  This analytic relation explains the ubiquity of the symmetric function inner product of Corollary [\ref{MacdonaldInnerProductDictionary}] in the two-dimensional conformal field theory.  Although this observation involves only standard material in string theory, conformal field theory, and representation theory \cite{DFMS, Kac0, KgMk}, we need to state it precisely in order to proceed.  Our goal is to prove Proposition [\ref{FGFfromVacua}] relating vertex algebras and fractional Gaussian fields, a spectral-theoretic statement that does not seem to appear readily in the literature.\\
\\
\noindent To begin, let $(\mathscr{F}, \langle \cdot, \cdot \rangle)$ be a possibly infinite-dimensional $\C$-Hilbert space and $\mathfrak{gl}(\mathscr{F})$ the space of possibly unbounded but nevertheless densely-defined operators on $\mathscr{F}$.  

\begin{definition} \label{BlockSymbolDefinition} $\widehat{\boldsymbol{\varphi}}$ is a \underline{$\mathfrak{gl}(\mathscr{F})$-valued distribution on $\mathbb{T}$} if for an appropriate class $\mathcal{C}$ of complex-valued test functions $\phi : \mathbb{T} \rightarrow \C$ the averages

\begin{equation} \label{TorusAverage} \oint_{\mathbb{T}} {\phi(w)} \widehat{\boldsymbol{\varphi}}(w) \frac{ dw}{ 2 \pi \textnormal{\textbf{i}} w} \end{equation} are well-defined elements of $\mathfrak{gl}(\mathscr{F})$.  The dual space $\mathcal{C}^{\vee}$ is the \underline{regularity class} of $\widehat{\boldsymbol{\varphi}}$. \end{definition}

\noindent One may specify a $\mathfrak{gl}(\mathscr{F})$-valued distribution on $\mathbb{T}$ by its Fourier modes \begin{equation} \widehat{{\boldsymbol{\varphi}}}(w) = \sum_{k = - \infty}^{\infty} \widehat{{\boldsymbol{\varphi}}}_{-k} w^k \end{equation} a sequence $\{\widehat{{\boldsymbol{\varphi}}}_k\}_{k=- \infty}^{+\infty}$ of possibly unbounded operators on $\mathscr{F}$.  In this basis, for test functions $\phi \in \mathcal{C}$ with Fourier expansion $\phi(w) = \sum_h \phi_h w^h$, formula (\ref{TorusAverage}) reads

\begin{equation} \oint_{\mathbb{T}} {\phi(w)} \widehat{\boldsymbol{\varphi}}(w) \frac{ dw}{ 2 \pi \textbf{i} w} = \sum_{k = - \infty}^{+\infty} \phi_k \widehat{{\boldsymbol{\varphi}}}_{k}. \end{equation} 

\noindent For the Fock-Sobolev space $(\mathscr{F}(a), \langle \cdot, \cdot \rangle_{\hbar; - \frac{1}{2}})$ centered at $a \in \R$ of Definition [\ref{FockSobolevDefinition}], we now construct a $\widehat{\mathfrak{gl}}(\mathscr{F}(a))$-valued distribution on $\mathbb{T}$ by specifying its Fourier modes.

\begin{definition} \label{CreationOperatorDefinitionIntro}The \underline{creation operator} $\widehat{\mathcal{V}}_{+k}$ is the unbounded operator on Fock space $(\mathscr{F}(a), \langle \cdot, \cdot \rangle_{- 1/2, \hbar})$ defined by multiplication by the holomorphic variable $V_k$. \end{definition}

\begin{definition} \label{AnnihilationOperatorDefinitionIntro}The \underline{annihilation operator} is the unbounded operator on Fock space $(\mathscr{F}(a), \langle \cdot, \cdot \rangle_{- 1/2, \hbar})$ defined by differentiation in the holomorphic variable $V_k$ as \begin{equation} \widehat{\mathcal{V}}_{-k} ( \hbar) = \hbar k \frac{\partial}{\partial V_k} \end{equation} \end{definition}

\noindent Creation and annihilation operators are defined for the Fock spaces of any Hermitian affine space $(\mathscr{M}, J, \mathsf{g}, \omega)$ in Definitions [\ref{CreationOperatorDefinition}] and [\ref{AnnihilationOperatorDefinition}].  The canonical commutation relation of the creation and annihilation operators defined here \begin{equation} \label{ThankGod} [ \widehat{\mathcal{V}}_{-k}, \widehat{\mathcal{V}}_{+k'}] = \hbar k  \delta(k-k') \end{equation}

\noindent reflect the Gardner-Faddeev-Zakharov Poisson bracket (\ref{GFZBracket}).  Just as $V_k, \overline{V_k}$ are Fourier modes of a classical field $v$, $\widehat{\mathcal{V}}_{\pm k}$ are Fourier modes of a quantum field $\vcurrent( \cdot | \hbar)$:

\begin{definition} \label{KacMoodyCurrentDefinition}The \underline{$\widehat{\mathfrak{gl}_1}$ affine Kac-Moody current $\vcurrent ( \cdot | \hbar)$ at level $\hbar>0$} is \begin{equation} \label{KacMoodyCurrent} \vcurrent ( w| \hbar) = \widehat{\mathcal{V}}_0 + \sum_{k= 1}^{+\infty} \big(  \widehat{\mathcal{V}}_{-k} w^k +\widehat{\mathcal{V}}_{+k} w^{-k} \big ) \end{equation} \noindent the $\mathfrak{gl}(\mathscr{F}(a))$-valued distribution as in \textnormal{Definition [\ref{BlockSymbolDefinition}]} where $\widehat{\mathcal{V}}_{\pm k}$ are the creation and annihilation operators of \textnormal{Definitions [\ref{CreationOperatorDefinitionIntro}] and [\ref{AnnihilationOperatorDefinitionIntro}]} and the zero mode $\widehat{\mathcal{V}}_0$ is central $[\widehat{\mathcal{V}}_0, \widehat{\mathcal{V}}_{\pm k}]=0$ and acts in $\mathscr{F}(a)$ by the scalar $a \in \R$.
\end{definition}

\noindent Note that $\widehat{\mathcal{V}}_{\pm k}^{\dagger} = \widehat{\mathcal{V}}_{\mp k}$ by Lemma [\ref{MutualAdjointsLemma}], so the Kac-Moody current $\vcurrent$ is actually a distribution on $\mathbb{T}$ taking values in $\mathbf{i} \mathfrak{u}( \mathscr{F}) \subset \mathfrak{gl}( \mathscr{F}(a))$ the space of possibly unbounded self-adjoint operators on $\mathscr{F}(a)$.  It is simple to check that:
\begin{proposition} The regularity class $\mathcal{C}^{\vee}$ of the Kac-Moody current $\vcurrent ( \cdot | \hbar)$ is the Sobolev space $(\mathscr{M}(a), J, \mathsf{g}_{-1/2}, \omega_{-1/2})$ underlying the Fock space $(\mathscr{F}(a), \langle  \cdot, \cdot \rangle_{- 1/2, \hbar})$.
\end{proposition}

\noindent Finally, we arrive at the desired dictionary between the infinite-dimensional Lie algebras and fractional Gaussian fields:
\begin{proposition} \label{FGFfromVacua} The random value of the $\phi$-average of the Kac-Moody current at level $\hbar$ in the vacua $\Upsilon_a( \cdot | \hbar) = 1 \in \mathscr{F}(a)$ centered at $a$ is 

\begin{equation} \oint_{\mathbb{T}} \phi(w) \vcurrent (w | \hbar) \tfrac{dw}{ 2 \pi \textnormal{\textbf{i}} w} \Bigg |_{\Upsilon_a( \cdot | \hbar)} = \oint_{\mathbb{T}} \phi(w) d \mathbb{G}^{\mathbb{T}}_{- \frac{1}{2}, a}(x | \hbar) \end{equation}

\noindent the $\phi$-average of the fractional Gaussian field $d \mathbb{G}^{\mathbb{T}}_{- \frac{1}{2}, a} ( w | \hbar)$ on $\mathbb{T}$ of \textnormal{Definition [\ref{FGFDefinition}].}

\end{proposition}

\begin{itemize}
\item \textit{Proof:} By Born's Rule and characteristic functions of Gaussians, we want: 
\begin{equation} \label{ThingThatFactorizes} \frac{\langle \Upsilon_a( \cdot | \hbar) | e^{ \textbf{i} t \oint_{\mathbb{T}} \phi (w) \vcurrent(w | \hbar) \frac{dw}{ 2 \pi \textbf{i} w} }  | \Upsilon_a ( \cdot | \hbar) \rangle_{- 1/2, \hbar} } { \langle \Upsilon_a( \cdot | \hbar) | \Upsilon_a( \cdot | \hbar) \rangle_{-1/2, \hbar} } = e^{ \textbf{i} t a \phi_0  - \frac{t^2}{2} || \phi||_{1/2}^2} \end{equation}
\noindent The $\delta(k-k')$ in formula (\ref{ThankGod}) factors (\ref{ThingThatFactorizes}) and reduces the proof to \begin{equation} 
\frac{\langle \Upsilon_a( \cdot | \hbar) | e^{ \textbf{i} t  ( \phi_k \widehat{\mathcal{V}}_{-k} + \phi_{-k} \widehat{\mathcal{V}}_{+k} )} | \Upsilon_a ( \cdot | \hbar) \rangle_{- 1/2, \hbar} } { \langle \Upsilon_a( \cdot | \hbar) | \Upsilon_a( \cdot | \hbar) \rangle_{-1/2, \hbar} } = e^{ - \tfrac{t^2}{2} |k| | \phi_k|^2 } \end{equation}

\noindent The desired result follows by the Baker-Campbell-Hausdorff formula  \begin{equation} e^{ \textbf{i} t  ( \phi_k \widehat{\mathcal{V}}_{-k} + \phi_{-k} \widehat{\mathcal{V}}_{+k} ) }= e^{ \textbf{i} t   \phi_k \widehat{\mathcal{V}}_{-k}} \cdot e^{ [ \textbf{i} t \phi_k \widehat{\mathcal{V}}_{-k}, \textbf{i} t \phi_{-k} \widehat{\mathcal{V}}_k]} \cdot e^{\textbf{i} t \phi_{-k} \widehat{\mathcal{V}}_{+k} } \end{equation} in the special case $[\widehat{\mathcal{V}}_{-k}, \widehat{\mathcal{V}}_{+k}]= \hbar k$ the commutator is central. $\square$

\end{itemize}
\noindent As we can see, the rigorous proof of Proposition [\ref{FGFfromVacua}] relies only on the standard functional analytic formulation of a single quantum harmonic oscillator.

\pagebreak

\subsubsection{\textcolor{black}{Quantum Hamiltonian as Cut-Join-Twist Operator}} \label{subsubsecQBOHamiltonian}

\noindent We have seen that the classical periodic Benjamin-Ono system (\ref{CBOE}) is formally Hamiltonian in the Sobolev space $(\mathscr{M}(a), J, \mathsf{g}_{-1/2}, \omega_{-1/2})$.  To define the quantum periodic Benjamin-Ono equation (\ref{QBOE}), we now choose a self-adjoint operator in the Fock-Sobolev space and declare it to be the quantum periodic Benjamin-Ono Hamiltonian and take (\ref{QBOE}) to be its Schr\"{o}dinger equation (\ref{SchrodingerEquation}).  Our definition follows the treatment of the dispersionless case $\ebar = 0$ by Dubrovin in \cite{Dubrovin2014}.  
\begin{definition} \label{QBODefinition} The \underline{quantum periodic Benjamin-Ono Hamiltonian} $\widehat{\mathcal{T}}^{\uparrow}_3(\ebar , \hbar)$ is the unbounded operator in the completed Fock-Sobolev space $\mathscr{F}(a)$ centered at $a \in \R$ of \textnormal{Definition [\ref{FockSobolevDefinition}]} written through the $\hbar$-dependent creation and annihilation operators of \textnormal{Definitions [\ref{CreationOperatorDefinitionIntro}], [\ref{AnnihilationOperatorDefinitionIntro}]} as 

\begin{equation} \label{QBOHamiltonian} \widehat{\mathcal{T}}^{\uparrow}_3(\ebar, \hbar) = \sum_{h_1, h_2 = 0}^{\infty} \widehat{\mathcal{V}}_{h_1} \widehat{\mathcal{V}}_{h_2 - h_1} \widehat{\mathcal{V}}_{-h_2} + \ebar \sum_{h=0}^{\infty} h \widehat{\mathcal{V}}_h \widehat{\mathcal{V}}_{-h} \end{equation}
\end{definition}

\noindent By what rule did we select $\widehat{\mathcal{T}}^{\uparrow}(\ebar, \hbar)$ to define the ``quantization'' of the classical periodic Benjamin-Ono Hamiltonian $T^{\uparrow}_3(\ebar)$?  Most curiously, it seems we chose (\ref{QBOHamiltonian}) to be the result of substituting $V_k, \overline{V_k} \rightarrow \widehat{\mathcal{V}}_k, \widehat{\mathcal{V}}_{-k}$ into the final formula (\ref{CBOHamiltonianFinalFormula}) for the classical periodic Benjamin-Ono Hamiltonian \textit{without normal ordering}.  By contrast, performing the same substitution in any of the other numerically equivalent expressions in (\ref{CBOHamiltonian}) results in a different (and typically divergent) operator.  Our choice is a shadow of Nazarov-Sklyanin's integrable geometric quantization \cite{NaSk2} we discuss in section \textbf{[\ref{subsecNazarovSklyaninIntegrableQuantizationExistence}]}.  By inspecting the formula,
\noindent \begin{proposition}$\widehat{\mathcal{T}}_3^{\uparrow}(\ebar, \hbar)$ is self-adjoint in $\mathscr{F}(a)$. \end{proposition}
\noindent By inspecting the literature,
\begin{proposition} \label{LehnFormulaAppears} $\widehat{\mathcal{T}}^{\uparrow}_3(\ebar, \hbar)$ coincides with Lehn's formula for the cup product by $c_1(\mathcal{O}(1))$ in the $(\C^{\times})^2$-equivariant cohomology of the Hilbert scheme of points in $\C^2$, where creation and annihilation operators are Nakajima's operators \textnormal{\cite{MaulOk, Nak0}} and $(\ee, \e)$ of \textnormal{Definition [\ref{OmegaVariablesDefinition}]} are torus weights \begin{equation} H^*_{(\C^{\times})^2} ( \bullet ) = H^* ( B( \C\nolimits^{\times})^2) = H^*(\mathbb{CP}^{\infty} \times \mathbb{CP}^{\infty}) = \Z [ \ee, \e]. \end{equation}
\end{proposition}
\begin{proposition} $\widehat{\mathcal{T}}^{\uparrow}_3(0, \hbar)$ at zero dispersion $\ebar = 0$ coincides with Goulden-Jackson's cut and join operator \textnormal{\cite{GouldenJackson1997}} using the conventions of \textnormal{Definitions [\ref{OmegaVariablesDefinition}] and [\ref{FractionalChargeDefinition}]}.
\end{proposition}

\noindent The cut and join operator earns its name from the cubic term $\widehat{\mathcal{T}}_3^{\uparrow}(0, \hbar)$ in (\ref{QBOHamiltonian}): \begin{equation}  \sum_{h_1, h_2 = 0}^{\infty} \widehat{\mathcal{V}}_{h_1} \widehat{\mathcal{V}}_{h_2 - h_1} \widehat{\mathcal{V}}_{-h_2} =  \sum_{h_1 > h_2 \geq 0}^{\infty} \widehat{\mathcal{V}}_{h_1} \widehat{\mathcal{V}}_{h_2 - h_1} \widehat{\mathcal{V}}_{-h_2} +  \sum_{0 \leq h_1 < h_2 }^{\infty} \widehat{\mathcal{V}}_{h_1} \widehat{\mathcal{V}}_{h_2 - h_1} \widehat{\mathcal{V}}_{-h_2}  \end{equation}

\noindent Considering the sign of $h_1 - h_2$ means there are two types of cubic terms: one with two annihilation operators and one with two creation operators.  The enumerative combinatorics and algebraic geometry of $\widehat{\mathcal{T}}^{\uparrow}_3(\ebar, \hbar)$ carry dynamical meaning: \begin{quote} ``The cubic terms in [$\widehat{\mathcal{T}}_3^{\uparrow}(\ebar, \hbar)$] describe the disintegration of one phonon to two and the merging of two to one'' \cite{LamPri2014}. \end{quote}

\pagebreak

\subsubsection{\textcolor{black}{Quantum Stationary States as Jack Polynomials}} \label{subsubsecQuantumStationaryStatesareJacks}

\noindent Now that we have introduced the quantum periodic Benjamin-Ono Hamiltonian $\widehat{\mathcal{T}}^{\uparrow}_3(\ebar, \hbar)$, we now turn to the classification of its quantum stationary states $\Psi_{\lambda,a}( \cdot | \ebar, \hbar)$ where $\lambda \in \mathbb{Y}(\mathscr{M}(a))$ lives in some index set that must reflect the underlying geometry of the classical phase space $\mathscr{M}(a)$.  For our Sobolev leaves $\mathscr{M}(a)$ on the circle $\mathbb{T}$, the resulting index set are ``partitions'' in the title of this paper:

\begin{definition} \label{PartitionsDefinition} A \underline{partition} $\lambda$ is a weakly-decreasing sequence $0 \leq \cdots \leq \lambda_2 \leq \lambda_1$ of non-negative integer ``row lengths'' $\lambda_i \in \N$ so that $\textnormal{deg}(\lambda)=\sum_i \lambda_i < \infty$.  Let $\mathbb{Y}$ be the set of all partitions.\end{definition} 

\noindent A single computation below $[\widehat{\mathcal{T}}_2^{\uparrow}, \widehat{\mathcal{T}}_3^{\uparrow}(\ebar, \hbar)] = 0$ has a big consequence: 

\begin{proposition} \label{JacksExist} The quantum periodic Benjamin-Ono Hamiltonian $\widehat{\mathcal{T}}_3^{\uparrow}(\ebar, \hbar)$ has discrete spectrum in $\mathscr{F}(a)$ with corresponding quantum stationary states \begin{equation} \Psi_{\lambda, a}( V_1, \ldots | \hbar, \ebar) = P_{\lambda}( V_1, \ldots | \ebar, \hbar) \Upsilon_{a}( \cdot | \hbar) \end{equation} for $\Upsilon_a( \cdot | \hbar) = 1 \in \mathscr{F}(a)$ and $P_{\lambda}$ polynomials in $V_1, V_2, \ldots$ indexed by partitions $\lambda \in \mathbb{Y}$ and independent of $a$.\end{proposition} \begin{itemize}
\item  \textit{Proof:} One can check that $\widehat{\mathcal{T}}_3^{\uparrow}(\ebar, \hbar)$ commutes with the total Hamiltonian \begin{equation} \label{MomentumOperatorFormula} \widehat{\mathcal{T}}_2^{\uparrow}(\hbar) = \sum_{k=1}^{\infty} \widehat{\mathcal{V}}_{+k} \widehat{\mathcal{V}}_{-k} \end{equation} \noindent of the system of $k=1,2,3,\ldots$ independent quantum harmonic oscillators with angular frequency $\sigma_k^2 = |k|^{-2s} =k $ for $s =- 1/2$.  For any $d \in \N$, this implies $\widehat{\mathcal{T}}_3(\ebar, \hbar)$ preserves $\mathcal{F}(a)[d]$, the finite-dimensional space spanned by $V_{\mu}$ with $\deg \mu = \sum_i \mu_i = d$.  Second, $\widehat{\mathcal{T}}_3^{\uparrow}(\hbar, \ebar)$ is a $\C$-symmetric operator on $(\mathcal{F}(a), \langle \cdot, \cdot \rangle_{\hbar ; - \frac{1}{2}})$, and hence its restriction to $\mathcal{F}(a)[d]$ is self-adjoint, so by the spectral theorem in finite-dimensions there are $\Psi_{\lambda,a}( \cdot | \hbar, \ebar) \in \mathcal{F}(a)$ eigenfunctions of $\widehat{\mathcal{T}}_3^{\uparrow}(\ebar, \hbar)$ that are ordinary polynomials in $V_1, V_2, \ldots$.  By dimension count, they must be indexed by partitions $\lambda$ as in \textnormal{Definition [\ref{PartitionsDefinition}]}, as partitions $\mu$ index the quantum stationary states of the quantum oscillator flow.  Independence of $a$ follows from role of $a$ in $\widehat{\mathcal{T}}_3^{\uparrow}(\ebar, \hbar)|_{\mathscr{F}(a)}$ as just a scalar multiple of $\widehat{\mathcal{T}}_2^{\uparrow}(\hbar)$. $\square$
\end{itemize}

\begin{definition} The \underline{quantum periodic Benjamin-Ono characters} are

\begin{equation} \chi^{\lambda}_{\mu}( \ebar, \hbar) = \langle P_{\lambda}( \cdot | \ebar, \hbar)  , V_{\mu} \rangle_{- \frac{1}{2} , \hbar} \end{equation}
\noindent the coefficients of the change of basis in Fock space $\mathscr{F}(a)$. \end{definition}

\noindent We next show that quantum periodic Benjamin-Ono characters generalize Lasalle's ``Jack characters'' \cite{Las1} due to the extra grading provided by the dimensionless quantum scale $\hbar>0$.  To identify the polynomials $P_{\lambda}( V_1, \ldots | \ebar, \hbar)$ of Proposition [\ref{JacksExist}] with known polynomials, the \textit{Jack polynomials}, we need an important change of variable:

\begin{definition} \label{OmegaVariablesDefinition} The \underline{Omega variables} $(\ee, \e) \in \C^2$ in the regime 
\begin{equation} \ee < 0 < \e \end{equation} 

\noindent parametrize our coefficients of quantization and dispersion as \begin{eqnarray} \hbar &=& \ered \\ \ebar &=& \eblue \end{eqnarray} \noindent with $\hbar > 0, \ebar \in \R$.  The original ``Jack parameter'' $\alpha = \frac{2}{ \beta} =  \frac{\e}{- \ee}$ in \textnormal{\cite{Mac}}, where $\beta$ is the traditional parameter of circular $\beta$-ensembles \textnormal{\cite{For}}.\end{definition}

\noindent The name ``Omega variables'' is in deference to Nekrasov's Omega background in supersymmetric gauge theory \cite{Nek1, NekYI, NekOk, NekPes, NekPesSha}, to which the Omega variables in our paper enjoy a structural connection: we begin at the ``CFT'' end of the BPS/CFT Correspondence \cite{NekYI} for $\mathscr{N}=2$ SUSY gauge theory on $\R^4$ in the Omega background with gauge group $U(1)$ and matter content determined by $v$, though such a correspondence for generic $v$ has not yet appeared in full in the literature.  Whereas $\ee \longleftrightarrow \e$ is typically the indication of non-trivial dualities on the gauge theory side,

\begin{proposition} \label{GiveMeAJob} Permuting $\ee \longleftrightarrow \e$ doesn't change $\hbar$ or $\ebar$ of \textnormal{Definition [\ref{OmegaVariablesDefinition}]}. \end{proposition}

\noindent A second change of variables is necessary to arrive at Jack polynomials:
\begin{definition} \label{FractionalChargeDefinition} The \underline{power sums} $p_k$ are rescalings by an Omega variable \begin{equation} V_k = (- \ee) p_k \end{equation}
\noindent of the distinguished $J$-holomorphic coordinates $V_k$ on the Sobolev space \end{definition}

\begin{theorem} \label{QBOStationaryStatesAreJacks} With conventions of \textnormal{Definitions [\ref{OmegaVariablesDefinition}] and [\ref{FractionalChargeDefinition}]}, the polynomial quantum stationary states of \textnormal{Proposition [\ref{JacksExist}]} may be identified 
\begin{equation} \Psi_{\lambda,a}( v | \hbar, \ebar) = P_{\lambda}( \tfrac{V_1}{-\ee}, \tfrac{V_2}{ -\ee} , \ldots | \tfrac{\e}{-\ee}) \Upsilon_a( \cdot | \hbar) \end{equation}

\noindent with \textit{Jack polynomials} $P_{\lambda} ( p_1, p_2, \ldots | \alpha)$ in the power sums \textnormal{\cite{Mac}}.
\end{theorem}

\noindent In this paper, we do not require a closed formula for Jack polynomials, just as we do not require exact description of classical stationary states of periodic Benjamin-Ono (\ref{CBOE}) which would follow from knowledge of the action-angle variables.  From the starting point of the periodic Benjamin-Ono system, the invariance $\ee \longleftrightarrow \e$ of Proposition [\ref{GiveMeAJob}] must be visible in the Jack polynomials themselves, as after the changes of variable in Definitions [\ref{OmegaVariablesDefinition}] and [\ref{FractionalChargeDefinition}] we must be giving two equivalent descriptions of the same state $\Psi_{\lambda, a}( \cdot | \ebar, \hbar)$:
\begin{corollary} \label{SymmetryJackPolynomials} Jack polynomials obey the $\ee \leftrightarrow \e$ symmetry \textnormal{(}that is, $\alpha \longleftrightarrow 1/\alpha$\textnormal{)}

\begin{equation} P_{\lambda} ( p_1, p_2, \ldots | \tfrac{\e}{- \ee}) = P_{\lambda'} ( \tfrac{ \ee}{\e} p_1, \tfrac{\ee}{\e} p_2, \ldots | \tfrac{\ee}{-\e}) \end{equation}

\noindent where $\lambda'$ is the transpose of the partition $\lambda$ {\textnormal{\cite{Mac}}}.
 \end{corollary}

\pagebreak

\subsubsection{\textcolor{black}{Quantum Conserved Densities and Anisotropic Partitions}} \label{subsubsecQBOHConservedDensity}

\noindent In Theorem [\ref{CBOHConservedDensityExistence}], for any bounded real initial data $v \in \mathscr{M}(a)$ we asserted existence of a non-random signed measure $dF_{\star |v}(c  | \ebar)$ on the real line $\mathbb{X} = \R$ which does not change if $v$ evolves according to the classical periodic Benjamin-Ono equation (\ref{CBOE}).  Here is its quantum analog:

\begin{theorem} \label{QBOHConservedDensityExistence} For any $\Psi \in \mathscr{F}(a)$, there exists a random signed measure \begin{equation} d \widehat{F}(c | \ebar, \hbar) |_{\Psi} \end{equation} \noindent on the real line $\mathbb{X} = \R$ so that

\begin{enumerate}
\item \textnormal{\textcolor{gray}{[Conserved Density}]} The law of $d\widehat{F}(c| \ebar, \hbar)|_{\Psi}$ does not change if $\Psi$ evolves by the quantum periodic Benjamin-Ono equation \textnormal{(\ref{QBOE})}, i.e. it is a (realized) quantum conserved density in the sense of \textnormal{Definitions [\ref{QuantumConservedDensityDefinition}] and [\ref{TheModelColumn3PreDefinition}]}.
\item \textnormal{\textcolor{gray}{[Integrable Hierarchy]}} There exist quantum observables $\{\widehat{\mathcal{O}}_{l}( \ebar, \hbar)\}_{l=1}^{\infty}$ which \noindent \begin{equation} [ \widehat{\mathcal{O}}_{l_1}(\ebar, \hbar)  , \widehat{\mathcal{O}}_{l_2}(\ebar, \hbar) ]= 0  \end{equation}\noindent pairwise commute and whose random values in a state $\Psi$ are identically \begin{equation} \widehat{\mathcal{O}}_l (\ebar, \hbar) \big |_{\Psi} = \int_{- \infty}^{\infty} c^l d\widehat{F} ( c | \ebar, \hbar)|_{\Psi}. \end{equation}

\item \textnormal{\textcolor{gray}{[Regularity of Observables]}} $\widehat{\mathcal{O}}_l(\ebar, \hbar)$ is a generalized non-commutative polynomial of degree $l \in \N$ in creation and annihilation operators $\widehat{\mathcal{V}}_{\pm k}$ as \textnormal{Definition [\ref{GeneralizedPolynomialDefinition}]}

\item \textnormal{\textcolor{gray}{[Periodic Benjamin-Ono]}} The span of $\{\widehat{\mathcal{O}}_l(\ebar, \hbar)\}_{l=1}^{\infty}$ includes the quantum periodic Benjamin-Ono Hamiltonian $\widehat{\mathcal{T}}_3^{\uparrow}(\ebar, \hbar)$ \textnormal{(\ref{QBOHamiltonian})} and also $\widehat{\mathcal{T}}_2^{\uparrow}(\hbar)$ \textnormal{(\ref{MomentumOperatorFormula})}
\end{enumerate}

\end{theorem}

\noindent We construct $d \widehat{F}( c | \ebar, \hbar)|_{\Psi}$ in Theorem [\ref{QBOHConservedDensityConstruction}] and discuss the proof in section \textbf{[\ref{subsubsecNekrasovQQCharactersSturmLiouvilleTheory}]}.\\
\\
\noindent In our analysis of the classical periodic Benjamin-Ono equation, we did not identify the classical stationary states, but mentioned they would be controlled by the Liouville tori in an action-angle variable decomposition of phase space.  Just as the quantum Calogero systems were solved first before their classical counterparts, it is a historical accident that the quantum stationary states of quantum periodic Benjamin-Ono have been first identified before their classical counterparts.  We thus specialize Theorem [\ref{QBOHConservedDensityExistence}] to the quantum stationary states:

\begin{theorem} \label{QBOHConservedDensitiesForJacksAreAnisotropicPartitions} Under the identifications in \textnormal{Definitions [\ref{OmegaVariablesDefinition}] and [\ref{FractionalChargeDefinition}]}, in the case of quantum stationary states $\Psi = \Psi_{\lambda, a} ( \cdot | \ebar, \hbar)$ which we identified with Jack polynomials in \textnormal{Theorem [\ref{QBOStationaryStatesAreJacks}]}, quantum conserved densities of \textnormal{Theorem [\ref{QBOHConservedDensityExistence}]} are \begin{equation} d \widehat{F}_{\Psi_{\lambda, a}( \cdot | \ebar, \hbar)} ( c | \ebar, \hbar) = d F_{\lambda} ( c - a | \ee, \e) \end{equation}

\noindent non-random Rayleigh measures of profiles $f_{\lambda} (c - a | \ee, \e)$ of anisotropic partitions.
\end{theorem}

\noindent We later verify this result in Theorem [\ref{QBOHConservedDensitiesForJacksAreAnisotropicPartitionsRevisited}].  Just like Theorem [\ref{CBOHConservedDensityExistence}], both Theorems [\ref{QBOHConservedDensityExistence}] [\ref{QBOHConservedDensitiesForJacksAreAnisotropicPartitions}] are merely extensions of the pioneering work of Nazarov-Sklyanin \cite{NaSk2}.  We now draw and define anisotropic partitions following \cite{Ke2}.

 \begin{wrapfigure}{r}{0.30 \textwidth}
  \begin{center}
    \includegraphics[width=0.30 \textwidth]{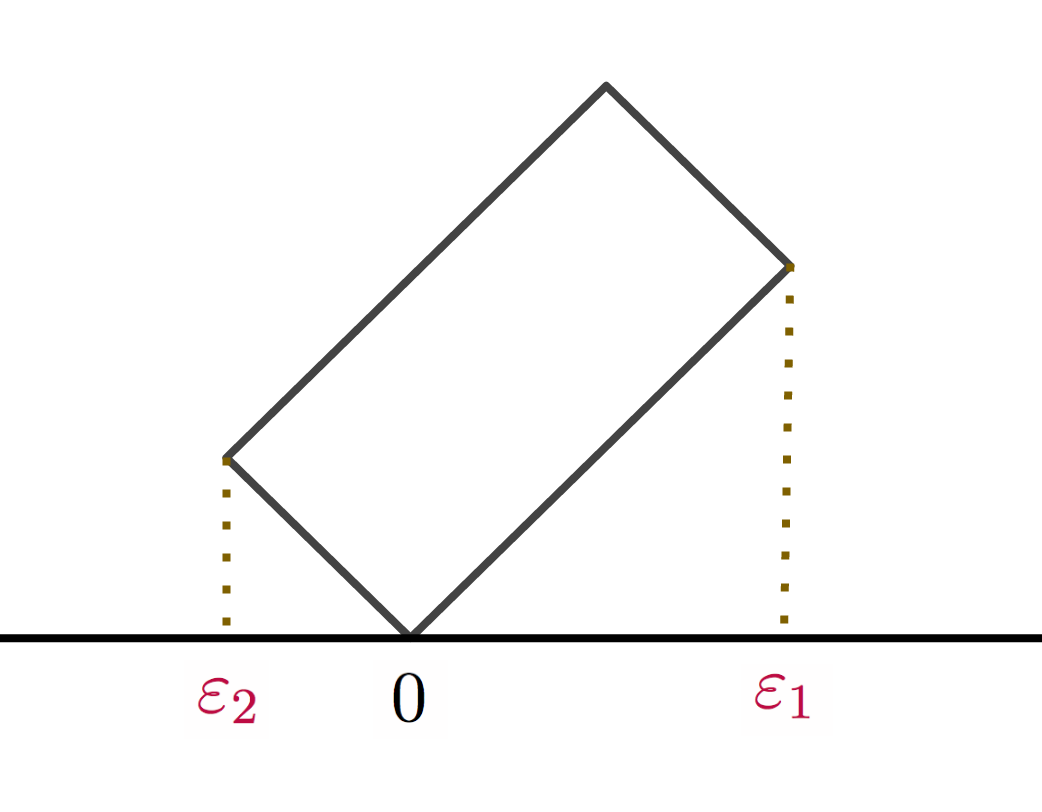}
  \end{center}

\end{wrapfigure}
  
\noindent The Omega variables of Definition [\ref{OmegaVariablesDefinition}] in the regime  \begin{equation} \label{EpsilonParameters} \ee < 0 < \e \end{equation} 

\noindent define rectangular (``anisotropic'') boxes $\square_{\ee, \e}$ of area twice $\hbar = \ered$ determined by the two vectors $(\e, \e)$ and $(\ee,-\ee) \in \R^2$.

\begin{definition} An \underline{anisotropic partition centered at $a$} is the data of a partition $\lambda$, $a \in \R$, and $\ee < 0 <  \e$.  \end{definition}

\noindent Draw an anisotropic partition as a pile of $\text{deg}(\lambda )$ identical anisotropic boxes $\square_{\ee, \e}$ in the corner $|c-a|$.  The $i$th row has $\lambda_i$ boxes stacked in the direction of positive slope.  Below is $\lambda =\{ 0 \leq 1 \leq 1 \leq 1 \leq 2 \leq 5 \leq 7\}$ with $\text{deg}( \lambda )= 17$ and $2 \e + \ee =0$.
 \begin{figure}[htb]
\centering
\includegraphics[width=0.9 \textwidth]{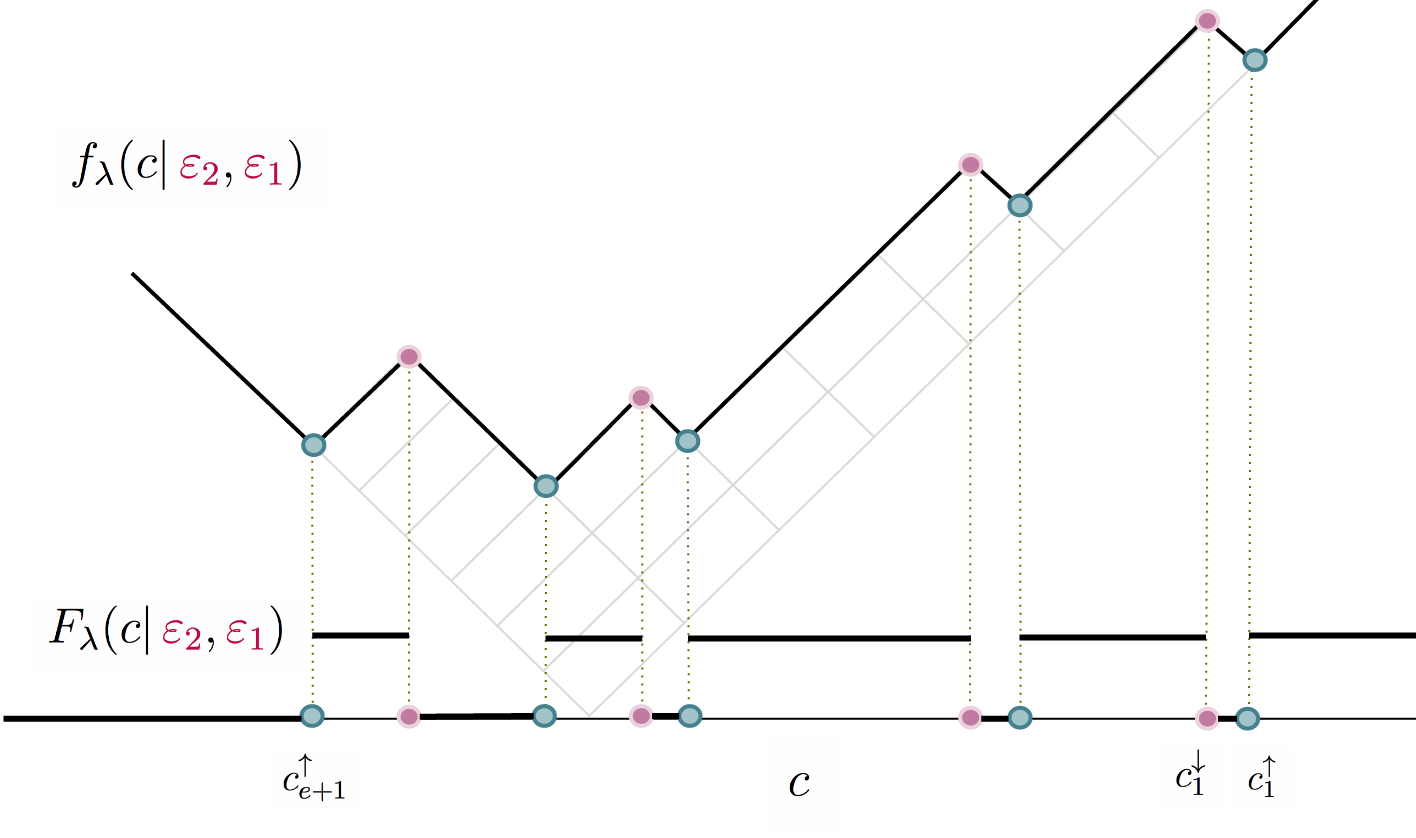}
\end{figure}

\begin{definition} \label{AnisotropicProfileDef}  The \underline{anisotropic profile} $f_{\lambda}(c-a|   \ee, \e)$ of an anisotropic partition is the piecewise-linear function outlined by the outermost boxes of the anisotropic partition.  The interlacing extrema $c_{e+1}^{\uparrow} < c_e^{\downarrow} < c_e^{\uparrow} < \cdots < c_2^{\uparrow} < c_1^{\downarrow} < c_1^{\uparrow}$ parametrize supports of the \underline{Rayleigh measure} $ d F_{\lambda}( c-a | \ee, \e)  = \sum_{i=1}^{e+1} \delta( c  - c_{i}^{\uparrow})  dc - \sum_{j=1}^e \delta( c - c_j^{\downarrow} ) dc$.
\end{definition}

\noindent The isotropic case $(\ee, \e) \rightarrow ( \eebrown, \ebrown)$ gives the usual drawing of $\lambda$ as a \textit{Young diagram}.\\
\\
\noindent Compared to the piecewise-linear classical conserved densities of Theorem [\ref{CBOHConservedDensityExistence}], for the piecewise-linear profiles of anisotropic partitions the footprint in complex plane is ``quantized'': the interface must be the result of stacking rectangles of fixed anisotropy $\ebar$ each of the same area $2 \hbar$.  We later derive this result in Theorem [\ref{QBOHConservedDensitiesForJacksAreAnisotropicPartitionsRevisited}] by writing the quantum Benjamin-Ono equation (\ref{QBOE}) as a quantum Lax system, though one expects to also be able to see these profiles of anisotropic partitions emerge directly from the Bohr-Sommerfeld quantization conditions valid only for integrable systems.

\subsubsection{\textcolor{black}{Quantum Quasi-Holes and Quasi-Particles}} \label{subsubsecQBOSolitonPhonon}

\noindent The classical periodic Benjamin-Ono equation (\ref{CBOE}) is formally Hamiltonian by Proposition [\ref{CBOisHamiltonian}] and globally well-posed for initial data that are subcritical and periodic \cite{Molinet} or rapidly-decaying \cite{TaoBenjaminOno}.  If one could show that the flow map preserves volumes in the classical phase space $(\mathscr{M}(a), J, \mathsf{g}_{- 1/2}, \omega_{-1/2})$, one can conclude that the flow cannot have attractive points, so phonons $v(x,t)$ cannot dissipate irreversibly to fixed equilibrium.  In \cite{Wieg1}, Wiegmann emphasizes that the $\ebar J [v_{xx}]$ term in (\ref{CBOE}) ``has dimensions of viscosity, but contrary to real viscosity does not produce dissipation'' because after quantization it is this ``dissipationless viscosity'' which accounts for the fractional charges and braid statistics of \textit{quasi-particles} and \textit{quasi-holes} in the quantum Benjamin-Ono equation \cite{AbBeWi, AbBeWi2, AbWi1, LamPri2015, LamPri2014, Wieg1}.  Dispersion $\ebar \neq 0$ is crucial for the existence of classical solitons, their quantum analogs being the right-moving collective excitations of phonons known as \textit{quasi-particles}. Moreover, only after quantization does one see exact left-moving collective excitations of phonons: these are the \textit{quasi-holes}.  We conclude this section with three remarks about such solutions.\\
\\
\noindent First, quasi-holes and quasi-particles are exchanged under a non-trivial duality with no semi-classical limit: ``in the classical limit the fractional (``hole'') branch disappears, but the quantized ``particle'' branch remains'' \cite{Wieg1}.  This disappearance is a footprint of the scattering of classical phonons into radiation discussed in section \textbf{[\ref{subsubsecQBOSolitonPhonon}]}.\\
\\
\noindent Second, in Corollary [\ref{SymmetryJackPolynomials}] we saw the same duality stated not for quantum phonons or quantum solitons but for quantum stationary states, namely the Jack polynomials.  We emphasize that Jacks are not the quantum analog of the classical multi-phase solutions corresponding to the ``finite gap'' Laurent $v$ in Theorem [\ref{CBOHConservedDensityExistence}].  However, they remarkably coincide in the absence of dispersion $\ebar = 0$ (Schur polynomials).  At $\ebar = 0$ but $\hbar > 0$, one has a non-trivial identification of quantum dispersionless periodic Benjamin-Ono with a chiral conformal field theory of free fermions \cite{Dubrovin2014}.  This is a salient difference, as fermions have no semi-classical limit.\\
\\
\noindent Third, the aforementioned duality between quasi-holes and quasi-particles is also visible and better studied for the quantum Calogero-Sutherland $\tealN$-body system \cite{Ha1995Fractional, Pasq0, Poly0}, which degenerates to our quantum periodic Benjamin-Ono in the \textit{chiral sector} in which the density field is \textit{approximately uniform} \cite{AbBeWi, Poly1995, StoneAnduagaXing, StoneGutman}.  In light of this hydrodynamic degeneration to quantum periodic Benjamin-Ono, along with the intricacies of the bulk-boundary correspondence, in \cite{Wieg1} Wiegmann proposes (\ref{QBOE}) as an effective description of edge excitations in the fractional quantum Hall effect.  In the simpler dispersionless case $\ebar = 0$, the bulk-boundary correspondence is the abelian $r=1$ case of a fascinating chapter in the pure mathematics of quantum groups, modular tensor categories, and knot invariants: Witten's holographic duality between $U(r)$ Chern-Simons theory in the bulk and the chiral Wess-Zumino-Novikov-Witten conformal field theory at the edge \cite{Witten1989}.  To refine this story accounting for dispersion $\ebar \neq 0$ as advocated in \cite{AbWi1, LamPri2014, LamPri2015, Wieg1}, one must confront quantum integrability, which means having not just a choice of Schr\"{o}dinger operator but a choice of quantization.

\pagebreak

\subsection{\textcolor{black}{Quantized Periodic Benjamin-Ono Waves}}\label{subsecNazarovSklyaninIntegrableQuantizationExistence}

\noindent In section \textbf{[\ref{subsubsecNSIntegrableQuantizationExistenceAppears}]} we state the existence of Nazarov-Sklyanin's integrable geometric quantization in Theorem [\ref{NazarovSklyaninQuantizationExistence}] later constructed in Theorem [\ref{NazarovSklyaninQuantizationConstruction}].  In section \textbf{[\ref{subsubsecNekrasovQQCharactersSturmLiouvilleTheory}]}, we discuss our proof of Theorem [\ref{NazarovSklyaninQuantizationExistence}] in light of appearances of the quantum periodic Benjamin-Ono hierarchy in geometric representation theory and gauge theory.
\subsubsection{\textcolor{black}{Nazarov-Sklyanin's Integrable Geometric Quantization}} \label{subsubsecNSIntegrableQuantizationExistenceAppears}

\noindent A quantization is a map from classical observables to quantum observables defined precisely in \textnormal{Definition [\ref{QuantizationDefinition}].}  The following theorem is implicit in the results of \cite{NaSk2}.  Informally, it says that our choice to associate the quantum periodic Benjamin-Ono hierarchy to the classical periodic Benjamin-Ono hierarchy comes from a quantization rule satisfying certain structural and regularity assumptions.

\begin{theorem} \label{NazarovSklyaninQuantizationExistence} \textnormal{(Nazarov-Sklyanin \cite{NaSk2})} There exists a quantization $Q_{NS}$ of the Poisson algebra $\mathsf{A}^{\textnormal{genpoly}}$ of generalized polynomials on the leaf $(\mathscr{M}(a), J, \mathsf{g}_{-1/2}, \omega_{-1/2})$ in the sense of \textnormal{Definition [\ref{GeneralizedPolynomialDefinition}]} so that \begin{itemize}
\item \textcolor{gray}{\textnormal{[Structure of Quantization]}} $Q_{NS}$ is an integrable quantization of $\mathsf{T} \subset \mathsf{A}^{\textnormal{genpoly}}$ as in \textnormal{Definition [\ref{IntegrableQuantizationDefinition}]}, where $\mathsf{T}$ is the Poisson-commutative subalgebra generated by the classical periodic Benjamin-Ono hierarchy $\{O_l(\ebar)\}_{l=1}^{\infty}$ of \textnormal{Theorem [\ref{CBOHConservedDensityExistence}]}.
\item \textcolor{gray}{\textnormal{[Regularity of Quantization]}} $Q_{NS}$ is an $\eta$-quantization of the pair $\mathsf{T} \subset \mathsf{A}^{\textnormal{genpoly}}$ of generalized polynomials for an ordering $\eta$ that is close to Wick quantization of the type constructed in \textnormal{Theorem [\ref{EtaQuantizationsAreCanonicalQuantizations}]}.
\end{itemize}

\noindent hence $Q_{NS}$ satisfies the structural assumptions of \textnormal{Definition [\ref{TheModelColumn3Definition}]} and the regularity assumptions of \textnormal{Proposition [\ref{TheModelColumn3RegularityAssumptions}]}.
\end{theorem}

\begin{itemize}
\item \textit{Proof:} Follows as Theorem [\ref{NazarovSklyaninQuantizationConstruction}] below. $\square$
\end{itemize}

\noindent We later construct this quantization $Q_{NS}$ using the auxiliary spectral theory of elliptic generalized Fock-block Toeplitz operators of order $1$ \cite{DeMonvelGuillemin} in Theorem [\ref{NazarovSklyaninQuantizationConstruction}].

\subsubsection{\textcolor{black}{Nekrasov qq-Characters and Sturm-Liouville Theory}} \label{subsubsecNekrasovQQCharactersSturmLiouvilleTheory}

\noindent We now discuss the context and proof of Theorem [\ref{NazarovSklyaninQuantizationExistence}].  We saw in Proposition [\ref{LehnFormulaAppears}] that Jack polynomials $\Psi_{\lambda}( \cdot | \ee, \e)$ appear in the $(\C^{\times})^2$ equivariant cohomology of the Hilbert schemes $\overline{\mathcal{M}}(1)= \coprod_{d=0}^{\infty} \overline{\mathcal{M}}(1, d)$ of $d$ points in $\C^2$.  This cohomology ring is chosen to be the quantum state space of $\mathcal{N}=2$ SUSY Yang-Mills on $\R^4 \cong \C^2$ in the Omega background $g_{\ee, \e}$ with abelian gauge group $U(1)$.  Our Theorem [\ref{QBOHConservedDensityExistence}] constructs the quantum periodic Benjamin-Ono hierarchy of commuting operators $\widehat{\mathcal{O}}_{\ell}(\ee, \e)$ of cup product by Chern classes of the tautological bundle explicitly through the Nakajima operators $\widehat{\mathcal{V}}_{\pm k}$, our creation and annihilation operators on Fock space.  We extend the work of Nazarov-Sklyanin \cite{NaSk2} to enable generalizations to other Nakajima quiver varieties, e.g. $\overline{\mathcal{M}}(r) = \coprod_{d=0}^{\infty} \overline{\mathcal{M}}(r,d)$ for the same SUSY theory with non-abelian gauge group $U(r)$ for $d$ the instanton charge \cite{MarshakovNekrasov, Nek1, NekOk, Smirnov2}.\\
\\
\noindent To construct the commuting Chern operators $\widehat{\mathcal{O}}_l(\ee, \e)$ in the Hilbert scheme case, we will realize them in Theorem [\ref{QBOHConservedDensityConstruction}] as coefficients of $u^{-l}$ of the logarithmic derivative \begin{equation} \label{LogDerivBebe} \widehat{\mathcal{O}}(u | \ee, \e) = \frac{\partial}{\partial u} \log \widehat{\mathcal{T}}^{\uparrow}( u | \ee, \e) \end{equation} of Nazarov-Sklyanin's quantum transfer operator $\widehat{\mathcal{T}}^{\uparrow}(u|\ee, \e)$.  The relation (\ref{LogDerivBebe}) which gives local observables by the logarithmic derivative of a generating function of non-local observables is a form frequently encountered in both classical and quantum inverse scattering method, e.g. for classical KdV \cite{DubrovinKricheverNovikov, HSW} or quantum XXZ \cite{Faddeev0}.  Nazarov-Sklyanin's quantum transfer operator is the same chiral and approximately uniform density hydrodynamic limit $\tealN \rightarrow \infty$ of the quantum Calogero-Sutherland $\tealN$-body problem discussed in section \textbf{[\ref{subsecOverview}]} applied to the transfer matrix of the Yangian in \cite{BernardGaudinHaldanePasquier}.  For more related constructions, see \cite{Dubrovin2014, MaulOk, SergVes, SergVes2}.\\
\\
\noindent What makes non-local transfer operators so special is the role they play in the definition of an integrable quantization.  As we will see in Theorem [\ref{NazarovSklyaninQuantizationConstruction}], the proof of Theorem [\ref{NazarovSklyaninQuantizationExistence}] just stated above, Nazarov-Sklyanin's integrable geometric quantization comes simply from exchanging the classical real $2 \pi$-periodic $v(x)$ from our Sobolev space in the classical transfer observable $T^{\uparrow}(u | \ebar)|_v$  for the the affine $\widehat{\mathfrak{gl}_1}$ Kac-Moody current $\vcurrent ( x | \hbar)$ at level $\hbar$ after the change of variables in Definition [\ref{OmegaVariablesDefinition}].  What is remarkable is that after this substitution, the classical transfer observable for periodic Benjamin-Ono becomes a quantum transfer operator, i.e. a family of commuting operators indexed by the spectral parameter $u \in \C \setminus \R$.  A closely related family of operators are Nekrasov's ``bulk $\widehat{\mathcal{Y}}$ observables'' $\widehat{\mathcal{Y}}(u | \ee, \e)$, introduced in \cite{NekYI}, further explored in \cite{KimuraPestun1}, quantizing the observables from \cite{NekPes} and the basis for Nekrasov's theory of $qq$-characters in \cite{NekYI}.  Although bulk $\widehat{\mathcal{Y}}$ observables are defined for much more general gauge theories than those associated to Jacks and Hilbert schemes, unlike $\widehat{\mathcal{T}}^{\uparrow}(u | \ee, \e)$ the $\widehat{\mathcal{Y}}(u | \ee,\e)$ do not commute.  Our study of Nazarov-Sklyanin's quantum transfer operator culminates in section \textbf{[\ref{subsubsecAnisotropicSSF}]} with a proof of Theorem [\ref{QBOHConservedDensitiesForJacksAreAnisotropicPartitions}] stated above, realizing the interlacing extrema of the profile of an anisotropic partition as interlacing eigenvalues of a self-adjoint operator $\widehat{\mathcal{L}}_{\bullet} (\ee, \e) |_{\lambda}$ and its minor $\widehat{\mathcal{L}}_+ (\ee, \e)|_{\lambda}$ in finite-dimensional spaces $\mathscr{H}_{\bullet}[ \lambda; \ee, \e]$ we call \textit{Jack-Lax orbits}.  We expect Jack-Lax orbits to be identified with factorizations in \cite{KimuraPestun1, MaulOk, NekYI}.\\
\\
\noindent Finally, how does one prove that quantum transfer operators commute, i.e. Theorem [\ref{NSQuantumCore}] hence Theorem [\ref{NazarovSklyaninQuantizationExistence}]?  Nazarov-Sklyanin's method is to first stabilize $\tealN \rightarrow \infty$ the Sekiguchi-Debiard determinants in \cite{NaSk1} then make sense of their shifted ratio in \cite{NaSk2}, but their discovery of the quantum Lax operator $\widehat{\mathcal{L}}_{\bullet}(\ebar, \hbar)$ suggests a proof ought to exist without truncation by $\tealN < \infty$, perhaps by a Yang-Baxter relation for the quantum Lax operator resolvent like one has for the instanton $\mathbf{R}$-matrix \cite{MaulOk}.  Our work in section \textbf{[\ref{secConstructionsForPeriodicBenjaminOno}]} suggests that a proof of Theorem [\ref{NSQuantumCore}] could begin not with equivariant cohomology (where the explicit spectrum via Young diagrams is transparent) but from the Sturm-Liouville oscillation theory of Fock-block Toeplitz operators (where Nakajima's creation and annihilation operators are transparent).

\pagebreak

\subsection{\textcolor{black}{Random Partitions from Coherent States}} \label{subsecRandomPartitionsIntroduction}

\noindent We introduce Jack measures in \textbf{[\ref{subsubsecJackMeasures}]} and our regularity assumptions in \textbf{[\ref{subsubsecJackMeasuresRegularityAssumptions}]}. \subsubsection{\textcolor{black}{Born's Rule: Jack Measures as Dispersive Schur Measures}} \label{subsubsecJackMeasures}

\noindent We first give a non-dynamical definition of Jack measures, then use Born's Rule to derive them from the quantization of the periodic Benjamin-Ono equation.

\begin{definition} \label{TheModelColumn4Definition} Under the conventions of \textnormal{Definitions [\ref{OmegaVariablesDefinition}] and [\ref{FractionalChargeDefinition}]}, the data of $\hbar >0$, $\ebar \in \R$, and a real-valued $2\pi$-periodic distribution $v$ of Sobolev regularity $s = - \frac{1}{2}$ and mean $a \in \R$ define a \underline{Jack measure} $\mathbf{M}_v ( \ebar, \hbar)$ on partitions $\lambda$

\begin{equation} \label{JackLaw} \textnormal{Prob}( \lambda) \propto \Big | P_{\lambda}^{\textnormal{norm}} ( V_1, V_2, \ldots | \ebar, \hbar) \Big |^2 \end{equation}

\noindent whose law is proportional to the square modulus of the normalized Jack polynomial $P^{\textnormal{norm}}_{\lambda}( V_1, V_2, \ldots | \hbar, \ebar)$ of \textnormal{Theorem [\ref{QBOStationaryStatesAreJacks}]} evaluated in the Fourier modes $V_k \in \C$ of $v$.\end{definition}

\noindent The fact that Jack measures are well-defined at the regularity $s = - \frac{1}{2}$ is identical to Stanley's Cauchy identity \cite{Stanley} in which the term $k^{-1}$ is truly $k^{2s}$: \begin{equation} \text{exp} \Bigg ( \frac{1}{\hbar} \sum_{k=1}^{\infty} \frac{\overline{V_k^{\text{out}}} V_k^{\text{in}}}{k} \Bigg ) = \sum_{\lambda} \overline{P^{\text{norm}}_{\lambda} ( V_k^{\text{out}} | \ebar, \hbar)} P_{\lambda}^{\text{norm}} ( V_k^{\text{in}} | \ebar, \hbar) \end{equation}

\noindent where $\{V_k^{\text{out}}\}_{k=1}^{\infty}$, $\{V_k^{\text{in}}\}_{k=1}^{\infty}$ are two infinite collections of indeterminates.\\
\\
\noindent Our original motivation to introduce Jack measures was to unify $\ebar$ and $v$-deformations of \textit{Poissonized Plancherel measures} $M_{\text{PL}}(0, \hbar)$ in asymptotic representation theory of symmetric groups \cite{Ke0, Ol0}, again using conventions of Definitions [\ref{OmegaVariablesDefinition}] and [\ref{FractionalChargeDefinition}].  In the dispersionless limit $\ebar \rightarrow 0$, Jack measures degenerate to \textit{Schur measures} $\mathbf{M}_v( 0,\hbar)$ originally introduced by Okounkov in \cite{Ok1}.  Alternatively, the Jack measure $M_{\text{PL}}(\ebar, \hbar)$ with symbol \begin{equation} v_{\text{PL}}(x) =2 \cos x \end{equation} which is $v_{\text{PL}}(w) = w + \tfrac{1}{w}$ are the \textit{Poissonized Jack-Plancherel measures}, a mixture of {Jack-Plancherel measures} \cite{Ke4} by a Poisson measure of intensity $\tfrac{1}{\textcolor{black}{ \hbar}}$.  By the relationship of Jacks to Hilbert schemes, Jack-Plancherel measures $M_{PL}( \ebar, \hbar)$ are the random partitions determined by \textit{Nekrasov's partition function} for pure $\mathcal{N}=2$ Yang-Mills on $\R^4$ with gauge group $U(1)$ in the Omega background \cite{Nek1, NekOk, Ok2, Ok3}.  While other $v$ may correspond to different choices of matter fields, the case of generic $v$ appears for the first time in this paper in the simpler case of abelian gauge group.\\
\\
\noindent We now derive Jack measures from the theory of coherent states.  In Theorem [\ref{QBOHConservedDensityExistence}], we used Born's Rule to realize a quantum conserved density of the quantum periodic Benjamin-Ono hierarchy in any state $\Psi \in \mathscr{F}(a)$ as a random signed measure $d \widehat{F}( c | \ebar, \hbar)|_{\Psi}$ on $\mathbb{X} = \R$. We also saw in Theorem [\ref{QBOHConservedDensitiesForJacksAreAnisotropicPartitions}] that in the case of quantum stationary states (Jack polynomials), the quantum conserved densities are non-random and given by $dF_{\lambda}( c | \ebar, \hbar)$ the Rayleigh measure of the profile of an anisotropic partition.  Combining these two results, get quantum conserved densities from random partitions:
\begin{definition} For any $\Upsilon(\hbar) \in \mathscr{F}(a)$, define $\mathbf{M}_{\Upsilon(\hbar)}(\ebar, \hbar)$ random partitions $\lambda \in \mathbb{Y}$
\begin{equation} \label{RespectTheLaw} \textnormal{Prob}(\lambda) = \frac{1}{ || \Upsilon(\hbar)||^2} \Big |\Big \langle \Upsilon(\hbar), P_{\lambda}^{\textnormal{norm}}(  \cdot | \ebar, \hbar) \Big \rangle_{- \frac{1}{2}, \hbar} \Big |^2
 \end{equation}
 \noindent determined by expressing $\Upsilon(\hbar)$ as a superposition of normalized Jack polynomials.
\end{definition}

\begin{proposition} \label{TheSecretLifeOfPlants} For any $\Upsilon(\hbar) \in \mathscr{F}(a)$, its quantum conserved density \begin{equation} d \widehat{F}(c | \ebar, \hbar)|_{\Upsilon(\hbar)} = d F_{\lambda}( c | \ebar, \hbar)|_{\Upsilon(\hbar)} \end{equation}

\noindent of \textnormal{Theorem [\ref{QBOHConservedDensityExistence}]} is equal in law to the random Rayleigh measure of the profile of a random anisotropic partition $\lambda$ sampled according to the law \textnormal{(\ref{RespectTheLaw})} of $\mathbf{M}_{\Upsilon(\hbar)}(\ebar, \hbar)$. 
\end{proposition}
\begin{itemize}
\item \textit{Proof:} Follows from Theorem [\ref{JacksExist}] and Proposition [\ref{QuantumConservedDensitiesIfDiscreteSpectrum}]. $\square$
\end{itemize}

\noindent We now have a model $\mathbf{M}_{\Upsilon(\hbar)}(\ebar, \hbar)$ of random partitions for every $\hbar>0$, $\ebar \in \R$, and $\hbar$-dependent quantum state $\Upsilon(\hbar)$.  However, not every $\Upsilon(\hbar)$ has a meaningful semi-classical limit in which we expect to see correspondence to the classical periodic Benjamin-Ono system.  Such states are \textit{quasi-classical states}, discussed in sections \textbf{[\ref{subsecStaticBohrCP}], [\ref{subsecDynamicalBohrCP}]}.  A special class of quasi-classical states are the \textit{coherent states}, analyzed in depth for Hermitian affine spaces in \textbf{[\ref{secColumn2}]} and presented here for $\mathbb{T}$ Sobolev leaves.

\begin{definition} For any $2\pi$-periodic distributions $v^{\textnormal{out}}, v^{\textnormal{in}} \in (\mathscr{M}(a), J, \mathsf{g}_{-1/2}, \omega_{-1/2})$ of Sobolev regularity $s = - \frac{1}{2}$ and mean $a \in \R$, the \underline{coherent state $\Upsilon_{v^{\textnormal{out}}} ( \cdot | \hbar)$ around $v^{\textnormal{out}}$} is the quantum state in the Fock-Sobolev space $(\mathscr{F}(a), \langle \cdot, \cdot \rangle_{- \frac{1}{2}, \hbar})$ defined by
\begin{equation} \Upsilon_{v^{\textnormal{out}}} ( v^{\textnormal{in}} | \hbar) = \textnormal{exp} \Bigg (\frac{1}{\hbar} \sum_{k=1}^{\infty} \frac{\overline{V_k^{\textnormal{out}}} V_k^{\textnormal{in}}}{k} \Bigg )  \end{equation}
\noindent for $V_k^{\textnormal{in}}$ are $J$-holomorphic coordinates on $\mathscr{M}(a)$ in which the metric is diagonal.
\end{definition}
\begin{corollary} \label{PlantCorollary} By \textnormal{Proposition [\ref{TheSecretLifeOfPlants}]}, the Rayleigh measure $dF_{\lambda}( c | \ebar, \hbar)$ of the anisotropic partition $\lambda$ sampled from a Jack measure $\mathbf{M}_v ( \ebar, \hbar)$ is equal in law to the quantum conserved density of a coherent state $\Upsilon_v ( \cdot | \hbar)$ around $v$. \end{corollary}

\noindent The next result is one of the two most crucial links in our chain of arguments in this paper, as it is what our results in below in the bulk of the paper to be specialized in this section to derive results for Jack measures.

\begin{proposition} \label{JackMeasuresCrucialLinkStructural} Jack measures satisfy structural assumptions of \textnormal{Definition [\ref{TheModelColumn3Definition}]}.
 \end{proposition}
\begin{itemize}
\item \textit{Proof:} By taking a coherent state we satisfy the structure of state assumption of Definition [\ref{TheModelColumn3Definition}], while the remaining structural assumptions are satisfied by previous results, namely the structure of observables by Theorem [\ref{CBOHConservedDensityExistence}] and the structure of quantization by Theorem [\ref{NazarovSklyaninQuantizationExistence}]. $\square$
\end{itemize}

\subsubsection{\textcolor{black}{Finite Joint Moments from Regularity Assumptions}} \label{subsubsecJackMeasuresRegularityAssumptions}

\noindent Jack measures $\mathbf{M}_v( \ebar, \hbar)$ determine a random partition $\lambda \in \mathbb{Y}$, hence an infinite collection $0 \leq \cdots \leq \lambda_2 \leq \lambda_1$ of correlated discrete random variables $\lambda_i \in \N$ with $\deg \lambda := \sum_{i} \lambda_i < \infty$ sampled randomly by the law (\ref{JackLaw}).  Given the complexity of Jack special functions, we cannot regard this law as very explicit rule for the correlation of the variables $\lambda_i$, let alone a local rule for their interaction.  In this paper, we do not study arbitrary observables of the random $\lambda$ but only those observables that depend only on the profile $f_{\lambda} ( c | \ee, \e)$ of the anisotropic partition $\lambda$ in Omega variables of Definition [\ref{OmegaVariablesDefinition}].  

\begin{definition} \label{PhiAveragesDefinition} For any test function $\phi : \R \rightarrow \R$, the \underline{$\phi$-average} or $\phi$-linear statistic 

\begin{equation} \widehat{O}_{\phi}^{\eta_{NS}} ( \ebar, \hbar)|_{\Upsilon_v ( \cdot | \hbar)} = \int_{- \infty}^{+\infty} \phi(c) d F_{\lambda}( c | \ebar, \hbar)|_{\Upsilon_v ( \cdot | \hbar)} \end{equation}

\noindent is a random variable if $dF_{\lambda}( c | \ebar, \hbar)|_{\Upsilon_v ( \cdot | \hbar)}$ is the random Rayleigh measure of the profile of an anisotropic partition $\lambda \in \mathbb{Y}(a; \ee, \e)$ sampled from Jack measure.  Our notation reflects the realization of Jack measures as quantum conserved densities of the quantum periodic Benjamin-Ono hierarchy under the Nazarov-Sklyanin integrable quantization $\eta_{NS}$.  We abbreviate the polynomial $\phi(c)=c^l$ averages by $\widehat{O_l}^{\eta_{NS}} ( \ebar, \hbar)|_{\Upsilon_v ( \cdot | \hbar)}$. \end{definition}

\noindent We now discuss the regularity assumptions for Jack measures, the companion of the structural assumptions met by Jack measures in Proposition [\ref{JackMeasuresCrucialLinkStructural}].
\begin{proposition} \label{TheModelColumn4RegularityAssumptions} If $v$ is a Laurent polynomial in $e^{ \pm \textnormal{\textbf{i}} k x}$
\begin{equation} v(x) = V_0 + \sum_{k=1}^{K(v)}  \big ( \overline{V_k} e^{ + \textnormal{\textbf{i}} k x} + V_k e^{ - \textnormal{\textbf{i}} k x} \big ) \end{equation} for some $K(v) < \infty$ so $V_k \equiv 0$ for almost every $1 \leq k \leq \infty$, then the corresponding Jack measure satisfies the regularity assumptions of \textnormal{Proposition [\ref{TheModelColumn3RegularityAssumptions}]}, hence for all $l_1, \ldots, l_n \in \N$, all joint moments

\begin{equation} \mathbb{E} \Bigg [ \widehat{O}_{l_1}^{\eta_{NS}}(\hbar)|_{\Upsilon_v ( \cdot | \hbar)} \cdots \widehat{O}_{l_n}^{\eta_{NS}}(\hbar)|_{\Upsilon_v ( \cdot | \hbar)} \Bigg ]  < \infty\end{equation}
\noindent of all $c^l$-averages $\widehat{O}_l^{\eta_{NS}}(\hbar)|_{\Upsilon_v ( \cdot | \hbar)}$ as in \textnormal{Definition [\ref{PhiAveragesDefinition}]} of the random Rayleigh measure $dF_{\lambda}( c | \hbar)|_{\Upsilon_v ( \cdot | \hbar)}$ sampled from Jack measures in \textnormal{Definition [\ref{TheModelColumn4Definition}]} are finite.
\end{proposition}

\begin{itemize} \item \textit{Proof:} use Proposition [\ref{TheModelColumn3RegularityAssumptions}], the regularity of observables in Theorem [\ref{CBOHConservedDensityExistence}], and the regularity of Nazarov-Sklyanin's quantization in Theorem [\ref{subsecNazarovSklyaninIntegrableQuantizationExistence}]. $\square$ \end{itemize}

\noindent Proposition [\ref{TheModelColumn4RegularityAssumptions}] is the second crucial link after Proposition [\ref{JackMeasuresCrucialLinkStructural}] that ensures the general arguments in the bulk of the paper to come imply the results for Jack measures we state in the next three sections.

\pagebreak

\subsection{\textcolor{black}{Asymptotic Expansion of Joint Cumulants}} \label{subsecAOEIntro}

\noindent In this section, we give a polynomial asymptotic expansion in $\hbar$ for fixed $\ebar \in \R$ of joint cumulants of $c^l$-averages for random partitions sampled from Jack measures in Theorem [\ref{AOEColumn4Dispersive}].  We strengthen Theorem [\ref{AOEColumn4Dispersive}] in Theorem [\ref{AOEColumn4Dispersionless}], a doubly-graded polynomial asymptotic expansion in both $\hbar$ and $\ebar$ of the same joint cumulants, together with an enumerative interpretation of the coefficients via ``ribbon paths.'' Ribbon paths are new combinatorial objects that emerge from the auxiliary spectral theory of the quantum Benjamin-Ono Lax operator $\widehat{\mathcal{L}}_{\bullet}(\ebar, \hbar)$, an elliptic Fock-block generalized Toeplitz operator of order $1$ we study extensively in section \textbf{[\ref{subsecQBOHConstruction}]}.
\subsubsection{\textcolor{black}{Ribbon Paths in Dispersive and Dispersionless Regimes}}

\noindent Let $W_n(\{\mathbb{O}_i\}_{i=1}^n)$ be the joint cumulant of $n$ random variables $\mathbb{O}_1, \ldots, \mathbb{O}_n$.

\begin{theorem} \label{AOEColumn4Dispersive} For the random Rayleigh measure $dF_{\lambda}( c | \ebar, \hbar)|_{\Upsilon_v ( \cdot | \hbar)}$ in $\mathbb{X} = \R$ sampled from a Jack measure $\mathbf{M}_v ( \ebar, \hbar)$ of \textnormal{Definition [\ref{TheModelColumn4Definition}]} satisfying the regularity assumptions of \textnormal{Proposition [\ref{TheModelColumn4RegularityAssumptions}]}, all joint cumulants of all random $c^l$-averages \begin{equation} \widehat{O_{l_i}(\ebar)}^{\eta_{NS}} ( \hbar) \big |_{\Upsilon_v ( \cdot | \hbar)}= \int_{- \infty}^{+\infty} c^l d F_{\lambda}( c | \ebar, \hbar) |_{\Upsilon_v ( \cdot | \hbar)} \end{equation} \noindent are polynomials in $\hbar$ \begin{equation} \label{AOEColumn4Formula} W_n \Big ( \Big \{ \widehat{O_{l_i}(\ebar)}^{\eta_{NS}} ( \hbar) \big |_{\Upsilon_v ( \cdot | \hbar) } \Big \}_{i=1}^n \Big ) = \sum_{g =0}^{ \lfloor \frac{ l_1 + \cdots + l_n}{2} \rfloor} \hbar^{n-1+g} W_{\eta_{NS}, n,g}(O_{l_1}(\ebar), \ldots, O_{l_n}(\ebar))|_v \end{equation}

\noindent with order of vanishing at least $n-1$ at $\hbar = 0$ and whose coefficients are polynomials $W_{\eta_{NS}, n,g}(O_{l_1}(\ebar), \ldots, O_{l_n} (\ebar))|_v$ in the $\sigma$-coordinates $\{V_k , \overline{V_k}\}_{k=1}^{K}$ of $v \in \mathscr{M}(a)$.
\end{theorem}

\begin{itemize}
\item \textit{Proof:} Follows from Theorem [\ref{AOEColumn3}] after Propositions [\ref{JackMeasuresCrucialLinkStructural}] and [\ref{TheModelColumn4RegularityAssumptions}]. $\square$
\end{itemize}

\noindent Theorem [\ref{AOEColumn4Dispersive}] is a purely semi-classical result $\hbar \rightarrow 0$ that follows from our much more general Theorem [\ref{AOEColumn3}] for \textit{arbitrary quantum integrable systems}.  Although the key {arguments} for Theorem [\ref{AOEColumn3}] are identical to those for our even more general Theorem [\ref{AOEColumn2}] in which we do not assume integrability, knowledge of \textit{all} higher invariants $W_{\eta_{NS}, n, g}(O_{l_1}(\ebar), \ldots, O_{l_n}(\ebar))|_v$ {should} reflect the integrability of the underlying classical system, since high moments $l_1, \ldots, l_n \rightarrow \infty$ probed at rate comparable to $\hbar \rightarrow 0$ are non-local quantities unlike those in Proposition [\ref{RespectTheButterflyEffect}].\\
\\
\noindent Taking a closer look at the particular integrable system at hand, we can consider not just the coefficient of quantization $\hbar$ but also the coefficient of dispersion $\ebar$.  We already gave a non-trivial result in the disperisonless limit $\ebar \rightarrow 0$ at the classical scale $\hbar = 0$ in our Theorem [\ref{CBOHConservedDensityExistence}].  Our next result is a stronger version of Theorem [\ref{AOEColumn4Dispersive}] which follows not only from the general arguments of sections \textbf{[\ref{secColumn2}]}, \textbf{[\ref{secColumn3}]} but also from the auxiliary spectral theory for periodic Benjamin-Ono of section \textbf{[\ref{secConstructionsForPeriodicBenjaminOno}]}.  By Theorem [\ref{SzegoFirstTheorem}] below, our next Theorem [\ref{AOEColumn4Dispersionless}] can be seen as a distinguished multi-point, dispersive, and quantum generalization of Szeg\H{o}'s First Theorem for Toeplitz determinants.

\begin{theorem} \label{AOEColumn4Dispersionless} The coefficients  $W_{\eta_{NS}, n,g}(O_{l_1}(\ebar), \ldots, O_{l_n} (\ebar))|_v$ in \textnormal{Theorem [\ref{AOEColumn4Dispersive}]} are also polynomials in the coefficient of dispersion $\ebar \in \R$ \begin{equation} W_{\eta_{NS}, n,g}(O_{l_1}(\ebar), \ldots, O_{l_n} (\ebar))|_v  = \sum_{m=0}^{l_1 + \cdots + l_n - 2(n-1+g)} \ebar^m W_{\eta_{NS}, n, g, m}(O_{l_1}, \ldots, O_{l_n}) |_v \end{equation}  \noindent so that joint cumulants of Jack measure $c^l$-averages \textnormal{(\ref{AOEColumn4Formula})} are \begin{equation} \label{AOEColumn4SecondFormula} W_n \Big ( \Big \{ \widehat{O_{l_i}(\ebar)}^{\eta_{NS}} ( \hbar) \big |_{\Upsilon_v ( \cdot | \hbar) } \Big \}_{i=1}^n \Big ) = \sum_{g=0}^{\infty} \sum_{m = 0}^{\infty} \sum_{\mu \in \mathbb{Y}}\sum_{\overline{\mu} \in \mathbb{Y}} W_{\eta_{NS}, n, g,m} \{ \mu, \overline{\mu}\} \hbar^{n-1+g} \ebar^m V_{\mu} \overline{V_{\overline{\mu}}} \end{equation}

\noindent polynomials in $\hbar$, $\ebar$, $V_{\mu} = V_1^{d_1} V_2^{d_2} \cdots$, and $\overline{V_{\overline{\mu}}} = \overline{V_1}^{\overline{d_1}} \overline{V_2}^{\overline{d_2}} \cdots$ with {non-negative integer coefficients} $W_{n, g,m} \{ \mu, \overline{\mu} \} \in \N$ determined by the explicit formula as a sum \begin{equation} W_{\eta_{NS}, n, g, m} ( O_{l_1}, \ldots, O_{l_n})|_v = \sum_{\Gamma \in \mathscr{P}^{\circ}_{n,g,m}(l_1, \ldots, l_n)} \mathscr{W} (\Gamma)|_v \end{equation}
\noindent over connected ribbon paths $\Gamma$ on $n$ sites of lengths $l_1, \ldots, l_n$ with $n-1+g$ pairings as in \textnormal{Definition [\ref{RibbonPathWithPairingsDefinition}]} weighted by the quantum periodic Benjamin-Ono ribbon path weight $\mathscr{W}(\Gamma)|_v$ of \textnormal{Definition [\ref{QBORibbonPathWeightDefinition}]}. \end{theorem}
 \begin{itemize}
 \item \textit{Proof:} Follows from Theorem [\ref{AOEColumn4Dispersive}], Theorem [\ref{NSQuantumCore}], Theorem [\ref{QBOHConservedDensityConstruction}], and the logarithmic relation between generating functions of $T_{\ell}^{\uparrow}$ and $O_l$ in Corollary [\ref{KMKsupport}] of Kerov's Markov-Kre\u{\i}n Correspondence. $\square$
 \end{itemize}
   
  \begin{figure}[htb]
\centering
\includegraphics[width=0.95\textwidth]{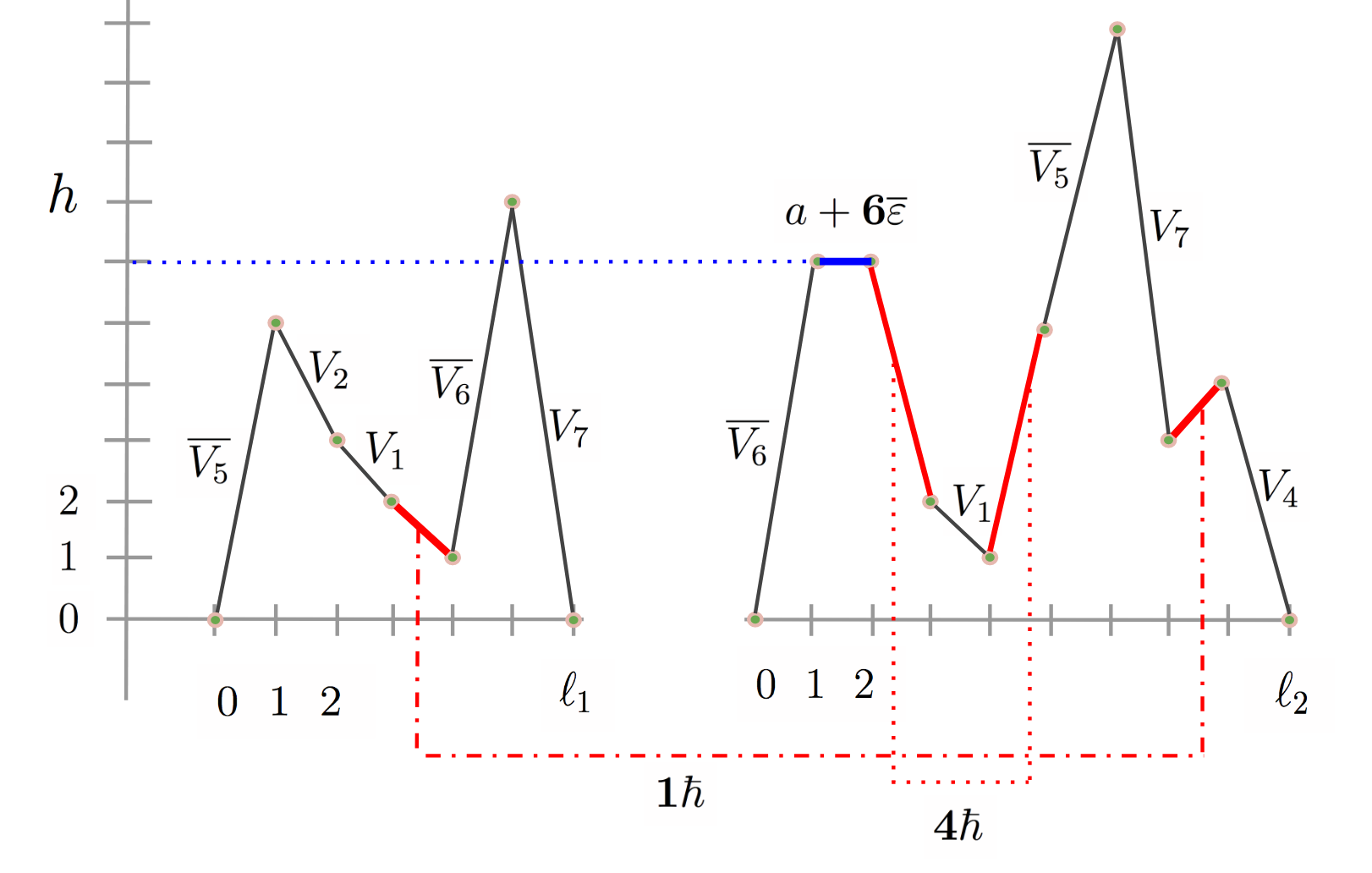}
\end{figure}

   \noindent We now define ribbon paths with pairings and the weight $\mathscr{W}(\Gamma)|_v$ of Theorem [\ref{AOEColumn4Dispersionless}].
   
\begin{definition} \label{RibbonPathDefinition} A \underline{ribbon path on $n$ sites} $\Gamma$ is a choice of $\ell_1, \ldots, \ell_n \in \N$ and maps \begin{equation} h^{(i)} : \{0, 1, 2, \ldots, \ell_i\} \longrightarrow \{0,1,2,3,\ldots \} \end{equation} \noindent $j \mapsto h^{(i)}_j$ with boundary conditions $h^{(i)}_0 = h^{(i)}_{\ell_i}=0$.  $\ell_i$ is the ``length of the $i$th site.''

\end{definition}

\begin{definition} \label{GraphOfRibbonPathDefinition} The \underline{graph} of a ribbon path on $n$ sites with lengths $\ell_1, \ldots, \ell_n$ is the disjoint union of the graphs of each map $h^{(i)}$ whose vertices are $(j, h_j^{(i)}) \in \N^2$ and whose edges $e_j (h^{(i)})$ join $(j, h^{(i)}_j) \sim (j+1, h^{(i)}_{j+1})$ for $0 \leq j \leq \ell_i - 1$. \end{definition}

\begin{definition} The \underline{jump size} of an edge $e_j(h^{(i)})$ in the graph of a ribbon path is 
\begin{equation} \textnormal{jump}(e_j(h^{(i)})) = h_{j+1} - h_j \end{equation}

\noindent and takes values in $\Z$.
 \end{definition}

\begin{definition} \label{RibbonPathWithPairingsDefinition} A \underline{ribbon path on $n$ sites with $g$ pairings} $(\Gamma, P)$ is a ribbon path $\Gamma$ on $n$ sites of lengths $\ell_1, \ldots, \ell_n$ together with a choice $P$ of $g$ distinct ordered pairs of edges $e_j(h^{(i)}), e_{j'}(h^{(i')})$ so that $i \leq i'$, $j < j'$ if $i = i'$, and \begin{eqnarray} \textnormal{jump} ( e_j(h^{(i)})) &=& - k  \\  \textnormal{jump} ( e_{j'}(h^{(i')})) &=& +k \end{eqnarray} \noindent for some $k=1,2,3,\ldots$.  We say $k_1, \ldots, k_g$ are the sizes of the pairings.
\end{definition}

\begin{definition} The \underline{underlying graph} of a ribbon path with $n$ sites and $g$ pairings is the graph with $n$ vertices $\textnormal{site}^{(1)}, \ldots, \textnormal{site}^{(n)}$ and a directed edge joining $\textnormal{site}^{(i)}$ and $\textnormal{site}^{(i')}$ for $i \leq i$ for every pairing between site $i$ and $i'$.  \end{definition}

\begin{definition} A ribbon path $(\Gamma, P)$ on $n$ sites with $g$ pairings is \underline{connected} if its underlying graph is connected. \end{definition}

\begin{definition} \label{QBORibbonPathWeightDefinition} The \underline{quantum periodic Benjamin-Ono weight} $\mathscr{W}(\Gamma)|_v$ of a ribbon path on $n$ sites with $g$ pairings is the multiplicative weight with three contributions:
\begin{enumerate}
\item Pairings of size $k$ contribute the weight $k$
\item Jumps of size $\pm k$ not involved in a pairing contribute $V_{\pm k}$ where $V_{-k} = \overline{V_k}$
\item Jumps of size $0$ at height $h$ are weighted by $a + \ebar h$.
\end{enumerate}
 \end{definition}
 \begin{definition} In the dispersive regime $\ebar \neq 0$, a jump of size $0$ is called a \underline{twist} to emphasize the non-trivial height-dependent contribution to the weight.
 \end{definition}
 
 \noindent  On the previous page, we depict the graph of a connected ribbon path $\Gamma$ on $n=2$ sites of lengths $\ell_1 = 6$, $\ell_2 = 9$ with $m=1$ twist at height $h = 6$ and $2$ pairings of sizes $k=1$ and $k=4$.  In this figure we also depict the multiplicative contributions of jumps, pairings, and edges to the weight $\mathscr{W}(\Gamma)|_v$ in black, red, and blue, respectively.

\subsubsection{{\textcolor{black}{Hurwitz Numbers, Expander Graphs, and Hilbert Schemes}}} \label{EnumerativeNoobs}

\noindent Given previous work on random partitions generalized by our Jack measures, logically our ribbon paths must offer a new description of at least three objects in enumerative combinatorics and enumerative algebraic geometry: the double Hurwitz numbers \cite{ACEH2017, Ok4}, the counts of expander graphs in \cite{DoFe, DoSni2017}, and the equivariant intersection numbers of Hilbert schemes \cite{MaulOk, Nak0, Ok2}.  Although we do not exhibit bijections between these objects and our ribbon paths, we emphasize that logically they must exist simply by matching coefficients in identical expressions.\\
\\
\noindent The simpler dispersionless case $\ebar =0$ of our Theorem [\ref{AOEColumn4Dispersionless}] for Schur measures is still new, coinciding for generic $v$ with the expansion of hypergeometric 2D Toda lattice tau functions into weighted double Hurwitz numbers in \cite{ACEH2017} and at $v_{\text{PL}}(x) = 2 \cos x$ the expansion in \cite{Ey2008}.  The main challenge overcome in this paper is to find a way work with dispersion $\ebar \neq 0$, i.e $\beta \neq 2$ in the conventions of Definition [\ref{OmegaVariablesDefinition}].\\
\\
\noindent We choose the name ``ribbon paths'' to signal a comparison to the \textit{ribbon graphs} in random matrix theory and enumerative geometry \cite{LaZv}.  After the change of variables \begin{equation} \label{ChangeOV} \hbar=  \frac{2}{ \beta} \cdot \frac{1}{ \textcolor{teal}{N}^2} \ \ \ \ \ \ \ \ \ \ \ \ \ \ \  \ebar = \Big ( \frac{2}{ \beta} - 1 \Big )  \frac{1}{\textcolor{teal}{N}} \end{equation} \noindent our Theorem [\ref{AOEColumn4Dispersionless}] is identical in form to the $1/\textcolor{teal}{N}$ refined topological expansion of joint cumulants of linear statistics for the \textcolor{black}{one-cut} $\beta$-ensembles on $\R$.  Assuming existence of a $1/\textcolor{teal}{N}$ expansion, for one-cut polynomial potentials $V$, Chekhov-Eynard proved \cite{ChEy1, ChEy2} that $ \widehat{W}_{n,g,m}^V (\ell_1, \ldots, \ell_{{n}}) $  enumerate the number of \textit{ribbon graphs} of genus ${g}$ with ${m}$ M\"{o}bius strips built from ${n}$ vertices of degree $\ell_i$ with vertex weights depending on $V$.  For details, see chapter 10 in \cite{ChEyMrch}.  The existence of this $1/ \textcolor{teal}{N}$ expansion was verified by Borot-Guionnet in \cite{BtGu1}.  A modification of \cite{ChEy2} to accommodate the case of $V$ with $V'$ a rational function is given independently in \cite{BrMaSt, Ch}.\\
\\
\noindent Although it is with an eye for random partitions that researchers have found new ways to handle problems of enumeration, the ribbon paths discovered here are new combinatorial objects of interest in their own right and are defined without reference to the Jack measures in which we discovered them.  Namely, the periodic initial data $v$ around which we take a coherent state $\Upsilon_v ( \cdot | \hbar)$ appears only as a choice of \textit{weights} of the jumps in a ribbon path.  We hope that our ribbon paths can gain a suitable geometric description in the future.  For this, we single out the work of \cite{MaulOk} in which \begin{eqnarray} \ebar &=& \eblue\\
\hbar &=& \ered
\end{eqnarray}
\noindent after the passage to Omega variables of Definition [\ref{OmegaVariablesDefinition}], our coefficients of dispersion and quantization are respectively the \textit{weight of the symplectic form} on the Hilbert scheme $\overline{\mathcal{M}}(1,d)$ and the \textit{handle-gluing element} in the Frobenius algebra $H^{*}_{(\C^{\times})^2} (\C^2)[\tfrac{1}{\hbar}]$.

\subsubsection{\textcolor{black}{Refined Topological Recursions and Spectral Curves}} \label{subsubsecWeldingOperatorRefinedTopologicalRecursion}

\noindent What makes the enumeration problems just discussed in section \textbf{[\ref{EnumerativeNoobs}]} tractable are the existence of non-trivial recursions between counts which admit exact solutions.  In the dispersionless $\ebar = 0$ or dispersive $\ebar \neq 0$ cases, these recursions are known as \textit{topological recursions} \cite{Ey0} and \textit{refined topological recursions} \cite{ChEy1, ChEy2, ChEyMrch}, respectively.  We do not verify that the constructions in this paper are an instance of one of the many existing definitions of the topological recursion, but the applications of the refined topological recursion to continuous $\beta$-ensembles was a main inspiration for this paper.  We now discuss our treatment of dispersion and quantization in our Theorem [\ref{AOEColumn4Dispersionless}] and the sense in which we have discovered a refined topological recursion.\\
\\
\noindent Our goal in the next two sections is to analyze Jack measures in two semi-classical regimes: the $\hbar \rightarrow 0$ limit with fixed dispersion $\ebar \in \R$, and the $\hbar \rightarrow 0$ limit taken simultaneously with the dispersionless limit $\ebar \rightarrow 0$ at comparable rate
\begin{equation} \label{Frog} \frac{\ebar^2}{ \hbar} \sim \Bigg ( \sqrt{ \frac{2}{ \beta}} - \sqrt{ \frac{\beta}{2}} \Bigg )^2  \end{equation}
\noindent where $\beta$ is that of Definition [\ref{OmegaVariablesDefinition}].  The confusion of \textbf{two} distinct dynamical concepts of quantization and dispersion due to standard use of the $\beta$ parameter appears in more than just the theory of random partitions.  As $\frac{2}{ \beta} = \alpha$ is the original Jack parameter in \cite{Mac}, $\beta$ in the rate (\ref{Frog}) is identical to the $\beta$ in $\tealN$-point circular $\beta$-ensembles \cite{For}.  Using Born's Rule, one may realize circular $\beta$-ensembles as the ground state distribution of the quantum Calogero-Sutherland $\tealN$-body problem on $\mathbb{T}$, of which quantum periodic Benjamin-Ono is a hydrodynamic limit $\tealN \rightarrow \infty$ as discussed in section \textbf{[\ref{subsecOverview}]}.  From this point of view, in the standard conventions of formula (\ref{ChangeOV}), the ``$\tealN \rightarrow \infty$ limit'' so routinely investigated in random matrix theory is a \textbf{triple} conflation of hydrodynamic, classical, and dispersionless limits.  Our Theorem [\ref{AOEColumn4Dispersive}] is a specialization of our Theorem [\ref{AOEColumn3}] for arbitrary integrable systems, which could be specialized e.g. to Calogero systems with any boundary conditions hence applied to continuous $\beta$-ensembles on $\mathbb{T}$ or $\mathbb{R}$ at fixed $\tealN < \infty$.\\
\\
\noindent After clarifying that quantum Calogero systems have not 1 but at least 3 important parameters $\hbar, \ebar, \tealN$, our next step is to bring front and center our choice of quantization.  If one identifies the enumeration of more and more complicated objects with higher and higher $\hbar$ corrections, these corrections become more and more sensitive to the choice of quantization.  In this paper, ribbon paths and hence Hurwitz numbers, expander graphs, and Hilbert scheme invariants of section \textbf{[\ref{EnumerativeNoobs}]} rely on the choice of Nazarov-Sklyanin's integrable geometric quantization in Theorem [\ref{NazarovSklyaninQuantizationExistence}] of the Sobolev leaves $(\mathscr{M}(a), J, \mathsf{g}_{-1/2}, \omega_{-1/2})$.  The starting point of topological recursions is a ``spectral curve'' that is not necessarily a spectral curve of a classical integrable system.  In our case, we do start with spectral curves $\Sigma(\ebar)|_v$ of classical periodic Benjamin-Ono as discussed in section \textbf{[\ref{subsubsecCBOHConservedDensity}]}, equivalent to the data of the conserved density of the classical integrable hierarchy of our Theorem [\ref{CBOHConservedDensityExistence}], the leading order term $\hbar^0$ precisely of interest in the semi-classical limit $\hbar \rightarrow 0$ and in the next section.

\pagebreak

\subsection{\textcolor{black}{Limit Shapes in Static Semi-Classical Limits}} \label{subsecLLNIntro}

\noindent In section \textbf{[\ref{subsubsecConcentrationColumn4}]}, we prove Theorems [\ref{LLNColumn4Dispersive}] and [\ref{LLNColumn4Dispersionless}] in dispersive and dispersionless semi-classical limits, respectively, which assert concentrations of measure as $\hbar \rightarrow 0$ that can be described in two ways.  Computationally, we describe how the random profiles of random partitions sampled from Jack measures concentrate on non-random limit shapes.  Conceptually, we verify the static version of Bohr's Correspondence Principle of Definition [\ref{StaticBohr}] for coherent states with respect to the quantum periodic Benjamin-Ono hierarchy.    We discuss the context for these results in section \textbf{[\ref{subsubsecSCLDP}]}.

\subsubsection{\textcolor{black}{Concentration in Dispersive and Dispersionless Regimes}} \label{subsubsecConcentrationColumn4}

\noindent Random partitions from Jack measures form limit shapes in the semi-classical limit.
\begin{theorem} \label{LLNColumn4Dispersive} For the random Rayleigh measure $d{F}_{\lambda}( c | \ee, \e)|_{\Upsilon_v ( \cdot | \hbar)}$ in $\mathbb{X} = \R$ sampled from the Jack measure $\mathbf{M}_v ( \ebar, \hbar)$ of random partitions $\lambda$ of \textnormal{Definition [\ref{TheModelColumn3Definition}]} satisfying the regularity assumptions of \textnormal{Proposition [\ref{TheModelColumn4RegularityAssumptions}]} with the conventions of \textnormal{Definition [\ref{OmegaVariablesDefinition}]}, in the semi-classical limit $\hbar \rightarrow 0$ this quantum conserved density of the $\eta_{NS}$-quantized periodic Benjamin-Ono integrable hierarchy in a coherent state $\Upsilon_v( \cdot | \hbar)$ around $v$ concentrates on a limit shape \begin{equation} \label{LLNColumn4Formula} d\widehat{F}^{\eta_{NS}}_{\lambda}( c |\ebar, \hbar)|_{\Upsilon_v ( \cdot | \hbar)} \rightarrow dF_{\star |v}(c)  \end{equation}

\noindent the non-random classical conserved density $dF_{\star |v}(c | \ebar)$ of the classical periodic Benjamin Ono hierarchy at $v$ from \textnormal{Theorem [\ref{CBOHConservedDensityExistence}]} for \textnormal{(\ref{LLNColumn4Formula})} convergence in distribution of $c^l$-averages.

\end{theorem}

\begin{itemize}
\item \textit{Proof:} follows from Theorem [\ref{LLNColumn3}] and Propositions [\ref{JackMeasuresCrucialLinkStructural}] and [\ref{TheModelColumn4RegularityAssumptions}]. $\square$
\end{itemize}

\noindent In this regime $\hbar \rightarrow 0$ with $\ebar \neq 0$ fixed, the random profile $f_{\lambda}( c -a| \ee, \e)$ sampled from a Jack measure, a random piecewise-linear profile whose extrema must arise from a pile of identically anisotropic boxes in the corner $|c-a|$, concentrates on a limiting profile $f_{\star |v}(c | \ebar)$ which is also piecewise-linear if $\ebar \neq 0$ by Theorem [\ref{CBOHConservedDensityExistence}].  See figures for $f_{\star |v}(c | \ebar)$ in section \textbf{[\ref{subsubsecCBOHConservedDensity}]}.  Limit shapes in this regime were discovered in \cite{DoSni2017} for different models of random partitions parametrized not explicitly by a symbol $v$ but implicitly by a factorization condition, the overlap being the case of Jack-Plancherel measures which is the simplest $2\pi$-periodic initial data $v_{\textnormal{PL}}(x) = 2 \cos x$.\\
\\
\noindent Due to the ``finite-gap potential'' portion of Theorem [\ref{CBOHConservedDensityExistence}], the limit shape $f_{\star | v}(c | \ebar)$ in Theorem [\ref{LLNColumn4Dispersive}] has finitely-many extrema because we have assumed $v$ to be a Laurent polynomial in our regularity assumptions of Proposition [\ref{TheModelColumn4RegularityAssumptions}].  We have not pursed the greatest analytic generality in our semi-classical analysis, but we do emphasize that in our study of the auxiliary spectral theory of the classical Lax operator for Benjamin-Ono we kept careful track of the regularity of $v$ to ensure the realization of $f_{\star|v}(c | \ebar)$ as a spectral shift function for symbols $v$ of weaker regularity.\\
\\
\noindent We now prove a version of Theorem [\ref{LLNColumn4Dispersive}] in a second asymptotic regime: not in the semi-classical limit $\hbar \rightarrow 0$ at fixed dispersion but in a conflation $\hbar, \ebar \rightarrow 0$ of the semi-classical and dispersionless limits taken at comparable rate depending on $\beta$.

\pagebreak

  \begin{figure}[htb]
\centering
\includegraphics[width=1.0 \textwidth]{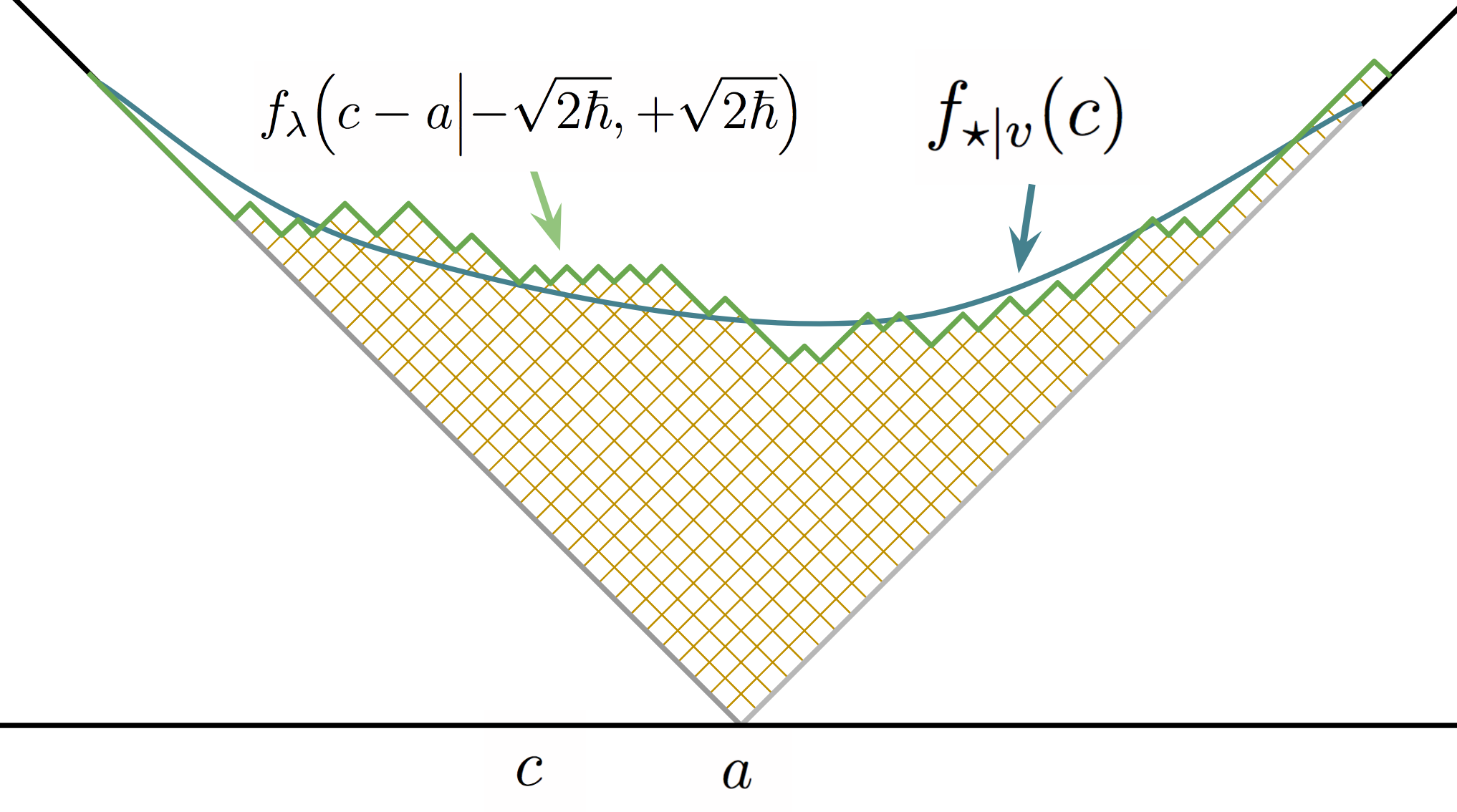}
\end{figure}

\begin{theorem} \label{LLNColumn4Dispersionless} For the random Rayleigh measure $d{F}_{\lambda}( c | \ee, \e)|_{\Upsilon_v ( \cdot | \hbar)}$ in $\mathbb{X} = \R$ sampled from the Jack measure $\mathbf{M}_v ( \ebar, \hbar)$ of random partitions $\lambda$ of \textnormal{Definition [\ref{TheModelColumn3Definition}]} satisfying the regularity assumptions of \textnormal{Proposition [\ref{TheModelColumn4RegularityAssumptions}]} with the conventions of \textnormal{Definition [\ref{OmegaVariablesDefinition}]}, in the simultaneous semi-classical $\hbar \rightarrow 0$ and dispersionless $\ebar \rightarrow 0$ limits taken at comparable rate \begin{equation} \frac{\ebar^2}{ \hbar} \sim \Bigg ( \sqrt{ \frac{2}{ \beta}} - \sqrt{ \frac{\beta}{2}} \Bigg )^2 \end{equation} \noindent this quantum conserved density of the $\eta_{NS}$-quantized periodic Benjamin-Ono integrable hierarchy in a coherent state $\Upsilon_v( \cdot | \hbar)$ around $v$ concentrates on a limit shape \begin{equation} \label{LLNColumn4DispersiveFormula} d\widehat{F}^{\eta_{NS}}_{\lambda}( c |\ebar, \hbar)|_{\Upsilon_v ( \cdot | \hbar)} \rightarrow dF_{\star |v}(c)  = v_* \rho_{\star | \mathbb{T}} \end{equation} \noindent the non-random classical conserved density $dF_{\star |v}(c )$ of the classical \underline{dispersionless} periodic Benjamin Ono hierarchy at $v$ from \textnormal{Theorem [\ref{CHBHConservedDensityIntro}]} for \textnormal{(\ref{LLNColumn4DispersiveFormula})} convergence in distribution of $c^l$-averages.  Explicitly, the limiting profile $f_{\star |v}(c)$ is {independent of $\beta$} has Rayleigh measure $dF_{\star |v} =v_*\rho_{\star| \mathbb{T}}$ the {push-forward of the uniform measure on the} {circle $\mathbb{T}$ along $v: \mathbb{T} \rightarrow \R$}. \end{theorem}

\begin{itemize}
\item \textit{Proof:} Follows from Theorem [\ref{LLNColumn3}], Propositions [\ref{JackMeasuresCrucialLinkStructural}] and [\ref{TheModelColumn4RegularityAssumptions}], the classical dispersionless limit in Theorem [\ref{CBOHConservedDensityExistence}], and Theorem [\ref{CHBHConservedDensityIntro}]. $\square$
\end{itemize}

\noindent Theorem [\ref{LLNColumn4Dispersionless}] generalizes the limit shape result for Schur measures of Okounkov in \cite{Ok2} in the case $\beta=2$ in which one first sends $\ebar \rightarrow 0$ and only then takes the semi-classical limit $\hbar \rightarrow 0$.  Above we depict this $\hbar \rightarrow 0$ concentration of Young diagrams built from squares of anisotropy $\ebar = 0$ on some convex $f_{\star |v}(c)$ without drawing $v$.  See the figures for $f_{\star |v}(c)$ in section \textbf{[\ref{subsubsecCBOHConservedDensity}]}.
\pagebreak

\noindent Theorem [\ref{LLNColumn4Dispersionless}] also generalizes the famous case $v_{\text{PL}}(x) = 2 \cos x$ of Vershik-Kerov \cite{KeVe} and Logan-Shepp \cite{LoSh} for the Plancherel measures at $\beta=2$ and \cite{DoFe} at $\beta \neq 2$ after a simple dePoissonization argument.  Our approach gives both a new proof of and a new perspective on Okounkov's result, stemming from the non-trivial identification of quantum dispersionless periodic Benjamin-Ono equation with the chiral conformal field theory of free fermions as discussed in section \textbf{[\ref{subsubsecQBOSolitonPhonon}]} and by Dubrovin \cite{Dubrovin2014}.\\
\\
\noindent Our conclusion in the second regime is not ``more explicit'' than the semi-classical limit at fixed dispersion in Theorem [\ref{LLNColumn4Dispersive}], since the classical conserved density $dF_{\star |v}(c |0)$ is obviously simpler at zero dispersion because the non-linear wave equation itself is simpler.  In particular, in the absence of the periodic Hilbert transform $J$, the wave equation is \textit{local}.  We do give an explicit realization of the limit shapes $dF_{ \star|v}(c | \ebar)$ in terms of $v$ and $\ebar$ as spectral shift functions in Theorem [\ref{CBOHConservedDensityConstruction}].\\
\\
\noindent On the other hand, the limit shapes are qualitatively different: while the dispersive limit shape $f_{\star |v}( c | \ebar)$ is \textit{piecewise-linear} for $\ebar \neq 0$, the dispersionless limit shape $f_{\star |v}(c )$ is \textit{convex}.  Moreover, in Theorem [\ref{LLNColumn4Dispersionless}] we see that we are in the ``one-cut regime'' as soon as $v$ is continuous.  Remarkably, this is a local condition on $v$, not a global condition as one sees for continuous $\beta$-ensembles and invariant matrix models.\\
\\
\noindent Finally, we emphasize that the core argument for both limit shape results appears in Theorem [\ref{LLNColumn2}] for the random values of quantized observables in coherent states, which {does not in any fundamental way assume integrability}.  Under the additional assumption of integrability, we derive from Theorem [\ref{LLNColumn2}] a limit shape results for {arbitrary integrable systems} in Theorem [\ref{LLNColumn3}].  To truly engage with the integrability or chaotic nature of the underlying classical dynamics, one must go beyond the static regime of Bohr's Correspondence Principle, see discussions in section \textbf{[\ref{subsecStaticBohrCP}]} and \textbf{[\ref{subsecDynamicalBohrCP}]}.

\subsubsection{\textcolor{black}{Seiberg-Witten and Nekrasov-Shatashvili Limits}} \label{subsubsecSCLDP}

\noindent We have mentioned that Jack measures are Nekrasov partition functions for certain $\mathcal{N}=2$ SUSY gauge theories on $\R^4$ with gauge group $U(1)$.  For random partition models associated to non-abelian gauge theories, in \cite{NekOk} the limit shapes found when\begin{equation} \label{Smile} \ee \rightarrow 0 \leftarrow \e \end{equation} \noindent both Omega variables vanish provides the first rigorous derivation of the predictions of the \textit{Seiberg-Witten limit}.  By Definition [\ref{OmegaVariablesDefinition}], the limit (\ref{Smile}) is the conflation $\hbar, \ebar \rightarrow 0$ of semi-classical and dispersionless limits, and thus the limit shapes in \cite{NekOk} are analogs of those in our Theorem [\ref{LLNColumn4Dispersionless}], not our Theorem [\ref{LLNColumn4Dispersive}].  Instead, our Theorem [\ref{LLNColumn4Dispersive}] for limit shapes $dF_{\star |v}(c | \ebar)$ in the semi-classical limit $\hbar \rightarrow 0$ with fixed dispersion $\ebar$ corresponds to $\ee \rightarrow 0$ with $0< \e$ fixed \textbf{\textit{or}} $\ee < 0$ fixed with $0 \leftarrow \e$, the \textit{Nekrasov-Shatashvili limits} \cite{NekShat}.  For our $\hbar = \ered$, the Nekrasov-Shatashvili is a semi-classical limit with remaining Omega variable $\ebar = \eblue$ is a parameter of dispersion, while \cite{NekShat} interpret the remaining Omega variable as a parameter of quantization.  The validity of both interpretations is the mark of \textit{duality} \cite{Aganagic0}.\\
\\
\noindent Now that we have matched our two asymptotic regimes to the Seiberg-Witten and Nekrasov-Shatashvili limits in SUSY gauge theory, let us compare the method of proof of the limit shape results in \cite{NekOk} with our proofs of Theorems [\ref{LLNColumn4Dispersive}] and [\ref{LLNColumn4Dispersionless}].  The paper \cite{NekOk} does not develop techniques for the full Omega deformation -- these came later in \cite{KimuraPestun1, MaulOk, NekYI, NekPes} -- but rather exploits the fact that in the Seiberg-Witten limit one may take without loss of generality $\ebar = \eblue = 0$ and use free fermions.  Moreover, for particular choices of matter fields (which here are choices of $v$), the authors in \cite{NekOk} are able to find large deviations principles for their models of random partitions generalizing the hook integral for Plancherel measures \cite{KeVe, LoSh}.\\
\\
\noindent Large deviations principles for arbitrary Schur measures are still not known - let alone for Jack measures - but we can say that our work in section \textbf{[\ref{secConstructionsForPeriodicBenjaminOno}]} reduces the problem of finding a large deviation principle to the problem of relating the \textit{eigenfunctions} of the Nazarov-Sklyanin quantum transfer operator to its \textit{eigenvalues at those eigenfunctions}, i.e. to a problem in Sturm-Liouville oscillation theory.  Note again that we have kept careful track of the center $a \in \R$ of our random anisotropic Young diagram, the mean $a = \oint_{\mathbb{T}} v(w) \tfrac{ dw}{ 2 \pi \textbf{i} w}$ of the classical phonon field $v$, as the Higgs VEVs $a^{(1)}, \ldots, a^{(r)}$ for non-abelian theories index the centers of the $r$ Young diagrams $\lambda^{(1)}, \ldots, \lambda^{(r)}$ and play a non-trivial role in the description of the limit shapes in \cite{NekOk}.\\
\\
\noindent In this paper, it is not steepest descent and large deviations principles which we use to prove a concentration of measure at limit shapes but instead a method of moments made available by an explicit description of the quantum periodic Benjamin-Ono hierarchy via Nazarov-Sklyanin's integrable geometric quantization of Theorem [\ref{NazarovSklyaninQuantizationExistence}].  Our approach ushers the problem of limit shapes into the much more general context of geometric quantization and semi-classical analysis, so that the core of our argument for the limit shape results is our Theorem [\ref{LLNColumn2}] which takes place in the general setting of coherent states in arbitrary Hermitian affine spaces \textit{and does not assume integrability}.  We specialize Theorem [\ref{LLNColumn2}] to arbitrary integrable systems in Theorem [\ref{LLNColumn3}] and then later to the limit shape results above in the Seiberg-Witten and Nekrasov-Shatashvili limits, though the arguments and results do not truly engage with the property of integrability in the precise sense of Proposition [\ref{RespectTheButterflyEffect}].\\
\\
\noindent In conclusion, not only do our arguments not rely on pre-existing \textit{techniques} used to analyze similar models of random partitions, our proofs also do not rely on any \textit{details of the model itself} but simply on the basics of the theory of coherent states.  The static version of Bohr's Correspondence Principle plays the role of a large deviations principle for our Jack measures of random partitions: it allows us to anticipate that Theorems [\ref{LLNColumn4Dispersive}] and [\ref{LLNColumn4Dispersionless}] must occur and must have the form they do before doing any computation.  Moreover, the theory of coherent states also explains why the next-order corrections to the semi-classical limit ought to be a Gaussian field on the support of the limit shape and why their covariance ought to reflect the inverse metric on the Sobolev leaf $(\mathscr{M}(a), J, \mathsf{g}_{-1/2}, \omega_{-1/2})$, our final results to which we now turn.

\pagebreak

\subsection{\textcolor{black}{Gaussian Fields as Quantum Fluctuations}} \label{subsecCLTIntro}

\noindent In section \textbf{[\ref{subsubsecGaussianityColumn4}]}, we prove Theorems [\ref{CLTColumn4Dispersive}], [\ref{CLTColumn4Dispersionless}] in dispersive and dispersionless semi-classical limits, respectively, which assert emergence of a Gaussian field that can be described in two ways.  Computationally, we describe how the global height fluctuations of random profiles of random partitions sampled from Jack measures around their non-random limit shapes converge to a Gaussian field.  Conceptually, we verify that quantum corrections to the static version of Bohr's Correspondence Principle of Definition [\ref{StaticBohr}] are asymptotically Gaussian as predicted by the intrinsic Gaussian form of the coherent states themselves.    We discuss model-dependent and model-independent aspects of our results in sections \textbf{[\ref{subsubsecCLTDiscussion2}]} and \textbf{[\ref{subsubsecCLTDiscussion1}]} , respectively.  Notice that we never call these results ``central limit theorems'' since we do not rely on the central limit theorem for sums of independent averages.
\subsubsection{\textcolor{black}{Gaussianity in Dispersive and Dispersionless Regimes}} \label{subsubsecGaussianityColumn4}

\noindent For random partitions sampled from Jack measures in the semi-classical limit, global height fluctuations of their profiles around the limit shape are asymptotically Gaussian.

\begin{theorem} \label{CLTColumn4Dispersive} For the random Rayleigh measure $d{F}_{\lambda}( c | \ee, \e)|_{\Upsilon_v ( \cdot | \hbar)}$ in $\mathbb{X} = \R$ sampled from the Jack measure $\mathbf{M}_v ( \ebar, \hbar)$ of random partitions $\lambda$ of \textnormal{Definition [\ref{TheModelColumn3Definition}]} satisfying the regularity assumptions of \textnormal{Proposition [\ref{TheModelColumn4RegularityAssumptions}]} with the conventions of \textnormal{Definition [\ref{OmegaVariablesDefinition}]}, subtracting the non-random limit shape $dF_{\star |v}( c | \ebar)$ of \textnormal{Theorem [\ref{LLNColumn4Dispersive}]} which is the classical conserved density of \textnormal{Theorem [\ref{CBOHConservedDensityExistence}]}, in the semi-classical limit $\hbar \rightarrow 0$, fluctuations of this quantum conserved density of the $\eta_{NS}$-quantized periodic Benjamin-Ono integrable hierarchy in a coherent state $\Upsilon_v( \cdot | \hbar)$ around $v$ occur at scale $\hbar^{1/2}$ and converge to\begin{equation} \label{CLTColumn4DispersiveFormula} \frac{1}{\hbar^{1/2}} \Bigg ( dF_{\lambda}( c | \ebar, \hbar)|_{\Upsilon_v ( \cdot | \hbar)} - dF_{\star | v}(c | \ebar) \Bigg ) \rightarrow d\mathbb{G}(O(\ebar))|_v (c) \end{equation} \noindent a Gaussian field $d \mathbb{G}(O(\ebar))|_v$ on $\mathbb{X} = \R$ of mean $0$ and covariance kernel abbreviated $d \mathbb{W}(c_1, c_2 | \ebar)|_v = W_2  (  d \mathbb{G}(O(\ebar))|_v (c_1) , d\mathbb{G}(O(\ebar))|_v ( c_2) ) $ given explicitly by \begin{equation} \label{MegaBoomBoomCovariance} \iint_{\mathbb{R} \times \mathbb{R}} c_1^{l_1} c_2^{l_2} d\mathbb{W}(c_1, c_2 | \ebar)|_v =  \mathsf{g}_{-1/2} ( (\nabla O_{l_1}(\ebar))|_v , (\nabla O_{l_2}(\ebar)) |_v )  \end{equation} where \textnormal{(\ref{CLTColumn4DispersiveFormula})} is convergence in distribution of $c^l$-averages, $O_{l}(\ebar)$ are the classical periodic Benjamin-Ono hierarchy of \textnormal{Theorem [\ref{CBOHConservedDensityExistence}]}, and $\nabla$ is the gradient as defined in the leaf $(\mathscr{M}(a), J, \mathsf{g}_{-1/2}, \omega_{-1/2})$ of the Sobolev space on $\mathbb{T}$ at regularity $s = -1/2$. \end{theorem}

\begin{itemize}

\item \textit{Proof:} follows from Theorem [\ref{CLTColumn3}] and Propositions [\ref{JackMeasuresCrucialLinkStructural}] and [\ref{TheModelColumn4RegularityAssumptions}]. $\square$
\end{itemize}

\noindent The Gaussian process here given explicitly by the gradients of the classical observables $O_l(\ebar)|_v$ from Theorem [\ref{CBOHConservedDensityExistence}] later constructed via spectral shift functions of elliptic generalized Toeplitz operators of order $1$ in Theorem [\ref{CBOHConservedDensityConstruction}].\\
\\
\noindent We now prove a version of Theorem [\ref{CLTColumn4Dispersive}] in a second asymptotic regime: not in the semi-classical limit $\hbar \rightarrow 0$ at fixed dispersion but in a conflation $\hbar, \ebar \rightarrow 0$ of the semi-classical and dispersionless limits taken at comparable rate depending on $\beta$.

\pagebreak

\noindent Recall the push-forward $v_{*} \boldsymbol{\rho}$ of any distribution $\boldsymbol{\rho}$ along any $v:\mathbb{T} \rightarrow \R$ is defined by \begin{equation} \int_{- \infty}^{+\infty} \phi(c) d(v_{*} \boldsymbol{\rho})(c) = \oint_{\mathbb{T}} \phi(v(w)) d  \boldsymbol{\rho}(w) \end{equation}

\begin{theorem} \label{CLTColumn4Dispersionless} For the random Rayleigh measure $d{F}_{\lambda}( c | \ee, \e)|_{\Upsilon_v ( \cdot | \hbar)}$ in $\mathbb{X} = \R$ sampled from the Jack measure $\mathbf{M}_v ( \ebar, \hbar)$ of random partitions $\lambda$ of \textnormal{Definition [\ref{TheModelColumn3Definition}]} satisfying the regularity assumptions of \textnormal{Proposition [\ref{TheModelColumn4RegularityAssumptions}]} with the conventions of \textnormal{Definition [\ref{OmegaVariablesDefinition}]}, subtracting the non-random dispersionless limit shape $dF_{\star |v}( c )$ of \textnormal{Theorem [\ref{LLNColumn4Dispersionless}]} which is the classical conserved density of \textnormal{Theorem [\ref{CHBHConservedDensityIntro}]}, in the simultaneous semi-classical $\hbar \rightarrow 0$ and dispersionless $\ebar \rightarrow 0$ limits taken at rate \begin{equation} \frac{\ebar^2}{ \hbar} \sim \Bigg ( \sqrt{ \frac{2}{ \beta}} - \sqrt{ \frac{\beta}{2}} \Bigg )^2 \end{equation} fluctuations of this quantum conserved density of $\eta_{NS}$-quantized periodic Benjamin-Ono integrable hierarchy in a coherent state $\Upsilon_v( \cdot | \hbar)$ around $v$ occur at scale $\hbar^{1/2}$ and \begin{equation} \label{CLTColumn4DispersiveFormula} \frac{1}{\hbar^{1/2}} \Bigg ( dF_{\lambda}( c | \ebar, \hbar)|_{\Upsilon_v ( \cdot | \hbar)} - dF_{\star | v}(c | \ebar) \Bigg ) \rightarrow d\mathbb{G}(O(0))|_v (c) + \Big ( \sqrt{\frac{2}{\beta}} - \sqrt{\frac{\beta}{2}} \Big ) dG_{1,0,1}(c)|_v \end{equation} \noindent converge in distribution of $c^l$-averages to a sum of two pieces:

\begin{enumerate}
\item The random Gaussian field $d \mathbb{G}(O(0))|_v$ on $\mathbb{X} = \R$ independent of $\beta$ given by the push-forward of the fractional Gaussian field $d\mathbb{G}^{\mathbb{T}}_{- 1/2, a}(x| 1)$ of variance $1$ along $v: \mathbb{T} \rightarrow \R$, for $d \mathbb{G}^{\mathbb{T}}_{-1/2, a} ( x | \hbar)$ defining Fock space.
\item A non-random mean-shift $G_{1,1}(c)$ for $\beta \neq 2$ determined by its moments \begin{equation} \int_{- \infty}^{+\infty} c^l dG_{1,0,1}(c)|_v = W_{\eta_{NS}, 1, 0,1}(O_l)|_v \end{equation} where $W_{\eta_{NS}, n ,g,m}(O_{l_1}, \ldots, O_{l_n})$ are the invariants of our \textnormal{Theorem [\ref{AOEColumn4Dispersionless}]}.
\end{enumerate}
\end{theorem}

\noindent $d\mathbb{G}(O(0))|_v$ is the $\ebar =0$ case of the previous Theorem [\ref{CLTColumn4Dispersive}].  There is nothing mysterious about $dG_{1,0,1}(c)|_v$: it only appears because the quantum fluctuations occur at scale $\hbar^{1/2}$ but now the variable of dispersion $\ebar$ is also of order $\hbar^{1/2}$.

\begin{itemize}
\item \textit{Proof:} Follows from Theorem [\ref{CLTColumn3}], Propositions [\ref{JackMeasuresCrucialLinkStructural}] and [\ref{TheModelColumn4RegularityAssumptions}], the classical dispersionless limit in Theorem [\ref{CBOHConservedDensityExistence}], and Theorem [\ref{CHBHConservedDensityIntro}].  A key step is the explicit evaluation of formula (\ref{MegaBoomBoomCovariance}) in the dispersionless case $\ebar = 0$, so as to give an explicit description of $d\mathbb{G}(O(0))|_v = v_* (d \mathbb{G}^{\mathbb{T}}_{-1/2, a}(\cdot | 1))$ as a push-forward along $v : \mathbb{T} \rightarrow \R$.  For this one uses the basic identification of the Sobolev leaf $(\mathscr{M}(a), J, \mathsf{g}_{-1/2}, \omega_{-1/2})$ as a Hermitian affine space, the description in coordinates of the inverse Hermitian metric in Proposition [\ref{WeldingKeyProposition}] which we call the ``welding operator'', and finally the explicit form from Theorem [\ref{CHBHConservedDensityIntro}]
\begin{equation} O_l (0)|_v = \oint_{\mathbb{T}} v(w)^{\ell} \frac{dw}{ 2 \pi \textbf{i} w} \end{equation}

\noindent of the classical dispersionless periodic Benjamin-Ono hierarchy. $\square$
\end{itemize}

\noindent Global Gaussian fluctuations in both regimes were discovered in \cite{DoSni2017} for different models of random partitions parametrized not explicitly by a symbol $v$ but implicitly by a factorization condition, overlapping at Poissonized Jack-Plancherel measures $v_{\textnormal{PL}}(x) = 2 \cos x$.  At $\beta=2$, our Theorem [\ref{CLTColumn4Dispersionless}] gives a new proof of Kerov's global Gaussian fluctuation result for Plancherel measures after a simple dePoissonization argument, while at $\beta \neq 2$ we recover \cite{DoFe} via the mean shift $dG_{1,0,1}(c)|_{2\cos x}$.  Our $d\mathbb{G}(O(0))(c)|_v$ at generic $v$ coincides with the fluctuations of Borodin's biorthogonal ensembles found in \cite{BreuerDuits}.  For more such results for discrete $\beta$-ensembles, see \cite{BoCMI, BoGoGu}.\\
\\
\noindent We derive both Theorem [\ref{CLTColumn4Dispersive}] and Theorem [\ref{CLTColumn4Dispersionless}] from our general Theorem [\ref{CLTColumn3}] which proves the emergence of global Gaussian fluctuations around a limit shape for \textit{any integrable system}.  Let us conclude with two remarks on our Theorem [\ref{CLTColumn3}].

\subsubsection{\textcolor{black}{Vertex Algebras and Gaussian Multiplicative Chaos}} \label{subsubsecCLTDiscussion2}

\noindent On the one hand, by using Theorem [\ref{CLTColumn3}] below for any integrable hierarchy, our proof of Theorems [\ref{CLTColumn4Dispersive}], [\ref{CLTColumn4Dispersionless}] do not use the specific model-dependent form of the auxiliary spectral theory we develop in section \textbf{[\ref{secConstructionsForPeriodicBenjaminOno}]} for the classical and quantum periodic Benjamin-Ono hierarchies.  However, the material in section \textbf{[\ref{secConstructionsForPeriodicBenjaminOno}]} shows that linear statistics of Jack measures are determined by matrix elements of \textit{exponentials} of the affine Kac-Moody current $\vcurrent ( x | \hbar)$ of Definition [\ref{KacMoodyCurrentDefinition}].  In Proposition [\ref{FGFfromVacua}], we gave an exact relation between $\vcurrent (x | \hbar)$ and the periodic fractional Brownian motion of Hurst index $0$ and mean $a \in \R$ that is implicit but never stated in the literature, e.g. in \cite{DFMS, Kac0, KgMk, FGFsurvey}.  The same Born's Rule applied to the \textit{exponentials} of $\vcurrent(x | \hbar)$ which construct the quantum periodic Benjamin-Ono hierarchy must match with some variant of the Gaussian Multiplicative Chaos \cite{RhodesVargas0}, a distinguished choice of exponential of the fractional Gaussian field $d \mathbb{G}_{-1/2, a}^{\mathbb{T}}(x | \hbar)$ defining our Fock-Sobolev space.  The choices of regularization necessary to define Gaussian Multiplicative Chaos are related by Born's Rule to the choice of Nazarov-Sklyanin's integrable geometric quantization $Q_{NS}$ of Theorem [\ref{NazarovSklyaninQuantizationExistence}], which we have already seen is the same choice which controls the enumeration problems to all orders in section \textbf{[\ref{EnumerativeNoobs}]}.
\subsubsection{\textcolor{black}{Geometric Quantization of K\"{a}hler Manifolds}} \label{subsubsecCLTDiscussion1}

\noindent On the other hand, our Theorem [\ref{CLTColumn3}] follows from our Theorem [\ref{CLTColumn2}] which finds a Gaussian variable $\mathbb{G}(O)|_v$ as $\hbar^{1/2}$ quantum correction to the static version of Bohr's Correspondence Principle, the variance $|| (\nabla O)|_v ||_{\mathsf{g}}^2$ given by the inverse metric of a Hermitian affine space $(\mathscr{M}, J, \mathsf{g}, \omega)$.  From this point of view, our Theorems [\ref{CLTColumn4Dispersive}], [\ref{CLTColumn4Dispersionless}] are quantities of the type in Proposition [\ref{RespectTheButterflyEffect}] and hence do not fundamentally rely on the property of integrability.  What they do rely on is that the first bidifferential $B_{1}^{Q_{NS}}$ of Nazarov-Sklyanin's quantization $Q_{NS}$ reflects the choice of K\"{a}hler structure on the Sobolev leaf $(\mathscr{M}(a), J, \mathsf{g}_{-1/2}, \omega_{-1/2})$.  Geometric quantization gives a reason for our results: if one defines a random point process by special functions in Fock space, one should not be surprised to see a fractional Gaussian field, {since a fractional Gaussian field is part of Fock space by the very definition}.

\pagebreak

\section{Semi-Classical Analysis} \label{secColumn1}

\noindent In this section, we set notation and terminology in a self-contained review of classical \textbf{[\ref{subsecClassicalSystems}]} and quantum \textbf{[\ref{subsecQuantumSystems}]} Hamiltonian systems and the problems of quantizing them \textbf{[\ref{subsecQuantizations}]}, of assigning probabilistic meaning to them \textbf{[\ref{subsecTheModelAppearsColumn1}]}, and of deriving asymptotic expansions of correlation functions in the quantum scale $\hbar$ \textbf{[\ref{subsecAsymptoticExpansionsMicrolocalAnalysis}]} so as to verify the static  \textbf{[\ref{subsecStaticBohrCP}]} and dynamic \textbf{[\ref{subsecDynamicalBohrCP}]} versions of Bohr's Correspondence Principle.  Before we begin, let us highlight three key points which are essential for situating our results in relation to the traditions of deformation quantization and semi-classical analysis:

\begin{itemize}
\item We distinguish between ``quasi-classical states'' and ``semi-classical states'' in \textbf{[\ref{subsecStaticBohrCP}]}, \textbf{[\ref{subsecDynamicalBohrCP}]} to clarify that we do not study finite-time evolution in this paper.
\item We use Born's Rule to define quantum conserved quantities as random variables whose law is invariant under the Schr\"{o}dinger equation.  As a consequence, the probabilistic language usually necessary to state the concentration of measure in the static semi-classical limit $\hbar \rightarrow 0$ is packaged naturally into the phrase:  ``the quantum conserved quantities concentrate on classical conserved quantities.''
\item We do not gauge away the symmetric part $(B_1^Q)^+$ of the first bidifferential defining a quantization $Q$.  In the quantizations in section \textbf{[\ref{secColumn2}]}, this term will reflect the K\"{a}hler geometry of the classical phase space and will appear in the covariance of our emergent Gaussian fluctuations in Theorem [\ref{CLTColumn2}].
\end{itemize}
\subsection{Classical Hamiltonian Systems} \label{subsecClassicalSystems}
\noindent We recall the basic ingredients of classical Hamiltonian systems. \subsubsection{Classical Phase Spaces and Observables} \label{subsubsecClassicalPhaseSpace}

\begin{definition} A \underline{classical phase space} is a real smooth symplectic manifold $(\mathscr{M}, \omega)$. \end{definition}

\begin{definition} A \underline{classical state} \label{DefinitionClassicalState} $v$ is an element $v \in \mathscr{M}$. \end{definition}
\begin{definition} \label{MixedStatesDefinition} A \underline{classical mixed state} $\rho$ is a probability measure on $\mathscr{M}$. \end{definition}

\begin{definition} A \underline{classical observable} \label{DefinitionClassicalObservable} $O \in C^{\infty}(\mathscr{M}, \R)$ is a smooth real-valued function possibly only densely-defined on $\mathscr{M}$.  Write $O|_v$ for the value of $O$ at $v \in \mathscr{M}$.\end{definition} 

\begin{definition} \label{HamiltonianVectorFieldDefinition} A \underline{Hamiltonian vector field} $X_O$ is the vector field on $\mathscr{M}$ defined so \begin{equation} \omega ( X_O, Y) = \partial_Y O \end{equation}
\noindent is the directional derivative of a classical observable $O$ for any vector fields $Y$. \end{definition}

\begin{definition} \label{PoissonBracketDefinition} The \underline{Poisson bracket} of two classical observables $O_1, O_2$ is
\begin{equation} \{ O_1, O_2\} = \omega (X_{O_1}, X_{O_2})\end{equation}

\noindent a new classical observable defined by the Hamiltonian vector fields $X_{O_1}, X_{O_2}$. 

\end{definition}

\subsubsection{Classical Hamiltonians and Conserved Quantities}
\noindent A special class of classical systems are the classical Hamiltonian systems.

\begin{definition} \label{ClassicalHamiltonianSystemDefinition} A \underline{classical Hamiltonian system} is the data of a classical phase space $(\mathscr{M}, \omega)$ and a classical observable $T: \mathscr{M} \rightarrow \R$ called the \underline{classical Hamiltonian}. \end{definition}

\begin{definition} For a classical Hamiltonian system, the \underline{Liouville equation} \begin{equation} \label{LiouvilleEquation} \frac{\partial \rho}{\partial t} = - \{ \rho, T\}. \end{equation}

\noindent for the time evolution of mixed states $\rho$ is the dual of the equation 

\begin{equation} \label{HamiltonianFlow} \frac{\partial O}{\partial t} = \{ O, T\} \end{equation} \noindent for the time evolution of classical observables.
\end{definition}

\noindent A priori, we do not know existence or continuity of a flow map $t \times O(\cdot, 0) \rightarrow O(\cdot, t)$.

\begin{definition} \label{ClassicalConservedQuantityDefinition} A \underline{classical conserved quantity} for the Liouville equation generated by $T$ with initial condition $\rho = \delta_v$ at $v \in \mathscr{M}$ is the non-random value $O|_v$ of 
\begin{equation} \{O, T\}=0 \end{equation} a classical observable $O$ at $v$ which Poisson commutes with $T$.

 \end{definition}

\subsubsection{Classical Stationary States}

\begin{definition} \label{ClassicalStationaryStatesDefinition} A classical mixed state $\rho$ on $\mathscr{M}$ is a \underline{$T$-stationary state} if it is an invariant measure for the Liouville equation \textnormal{(\ref{LiouvilleEquation})} with Hamiltonian $T$. \end{definition}

\noindent $T$-stationary states are candidates for a classical system's long-time behavior for arbitrary initial conditions.  Notable stationary states are those supported on \textit{$T$-fixed points}, on \textit{closed $T$-periodic orbits}, or on the \textit{energy surface} $T^{-1}(E)$ for fixed $E \in \R$.



\subsection{Quantum Hamiltonian Systems} \label{subsecQuantumSystems}

\noindent We recall the basic ingredients of quantum Hamiltonian systems.  A priori, this material is parallel but unrelated to the material of subsection \textbf{[\ref{subsecClassicalSystems}]}.
 \subsubsection{Quantum State Spaces and Observables}

\begin{definition} \underline{Planck's constant} $\hbar$ is an experimentally determined fundamental physical constant with dimensions $\text{mass}^1 \cdot \text{length}^2  \cdot \text{time}^{-1}$. \end{definition}

\begin{definition} By standard abuse of notation, fix and ignore Planck's constant and write $\hbar$ instead for a dimensionless parameter $\hbar >0$ called the \underline{quantum scale}. \end{definition}

\begin{definition} A \underline{quantum state space} is a separable $\C$-Hilbert space $(\mathscr{H}, \langle \cdot, \cdot \rangle_{\hbar})$ whose Hermitian inner product depends on the quantum scale $\hbar >0$. \end{definition}

\begin{definition} A \underline{quantum state} $\Psi$ is an element $\Psi \in \mathscr{H}$. \end{definition}

\begin{definition} A \underline{quantum observable} $\widehat{\mathcal{O}}(\hbar)$ is a self-adjoint operator on $\mathscr{H}$. \end{definition}

\noindent Quantum observables $\widehat{\mathcal{O}}(\hbar)$ may be unbounded, hence densely-defined on $\mathscr{H}$, just like classical observables $O$ may be densely-defined on the classical phase space $\mathscr{M}$.  Both $\langle \cdot, \cdot \rangle_{\hbar}$ and $\widehat{\mathcal{O}}(\hbar)$ depend on $\hbar >0$ but one may often write just $\langle \cdot, \cdot \rangle$, $\widehat{\mathcal{O}}$.

\subsubsection{Quantum Hamiltonians and Conserved Quantities} \label{subsecQuantumHamiltonianDynamics}

\begin{definition} \label{QuantumSystemDefinition} A \underline{quantum Hamiltonian system} is the data of a quantum {state space} $(\mathscr{H}, \langle \cdot, \cdot \rangle_{\hbar})$ and a quantum observable $\widehat{\mathcal{T}}(\hbar)$ called the \underline{quantum Hamiltonian}. \end{definition}

\begin{definition} For a quantum Hamiltonian system, the \underline{Schr\"{o}dinger equation} \begin{equation} \label{SchrodingerEquation} \textnormal{\textbf{i}} \hbar \frac{ \partial \Psi}{ \partial t} = \widehat{\mathcal{T}} (\hbar) \Psi \end{equation} 
\noindent describes the time evolution $t \mapsto \Psi(t)$ of a quantum state $\Psi \in \mathscr{H}$.
\end{definition} 

\begin{definition} \label{QuantumConservedQuantityFirstDefinition} A \underline{quantum conserved quantity} for the Schr\"{o}dinger equation with Hamiltonian $\widehat{\mathcal{T}}(\hbar)$ and initial condition $\Psi \in \mathscr{H}$ is $d\mathsf{\mu}_{\Psi, \Psi} ( E | \widehat{\mathcal{O}}(\hbar))$ the spectral measure of a quantum observable $\widehat{\mathcal{O}}(\hbar)$ at $\Psi$ from \textnormal{Theorem [\ref{SpectralTheorem}]} provided they commute 
\begin{equation} [ \widehat{\mathcal{O}}, \widehat{\mathcal{T}}]=0. \end{equation} \end{definition}

\subsubsection{Quantum Stationary States}  \label{subsecQuantumStationaryStates}

\noindent All Schr\"{o}dinger equations (\ref{SchrodingerEquation}) are deterministic and linear, so the initial value problem is solved by writing initial states as superpositions of $\widehat{\mathcal{T}}(\hbar)$-stationary states: \begin{definition} \label{QuantumStationaryStateDefinition} A quantum state $\Psi_{\lambda}(\cdot | \hbar) \in \mathscr{H}$ is a \underline{$\widehat{\mathcal{T}}(\hbar)$-stationary state} \begin{equation} \label{EigenvalueProblem} \widehat{\mathcal{T}}(\hbar)  \Psi_{\lambda} ( \cdot | \hbar) = \widehat{T}(\hbar) |_{\lambda} \Psi_{\lambda}( \cdot | \hbar) \end{equation}

\noindent if it is an eigenstate of the quantum Hamiltonian $\widehat{\mathcal{T}}(\hbar)$ with eigenvalue $\widehat{T} (\hbar) |_{\lambda}  \in \R$.  Let $\mathbb{Y}$ be the set of $\lambda$ indexing the $\widehat{\mathcal{T}}(\hbar)$-stationary states.\end{definition}

\subsection{Quantizations} \label{subsecQuantizations}

\subsubsection{Quantizations and Phase Space Star Products}
\noindent Informally, a quantization is a rule which associates to the data of classical system from section \textbf{[\ref{subsecClassicalSystems}]} the data of a quantum system from section \textbf{[\ref{subsecQuantumSystems}]}.  Many different axiomatic definitions of quantizations exist, each with slight variations.  In this paper,

\begin{definition} \label{QuantizationDefinition}A \underline{quantization} ${Q}(\hbar)$ of a unital Poisson subalgebra $\mathsf{A} \subset C^{\infty}(\mathscr{M}, \R)$ is a choice of quantum state space $(\mathscr{H}, \langle \cdot, \cdot \rangle_{\hbar})$ at scale $\hbar>0$ and a well-defined map \begin{equation} Q(\hbar): \mathsf{A} \longrightarrow \textnormal{\textbf{i}} \mathfrak{u} ( \widehat{\mathscr{H}}, \langle \cdot, \cdot \rangle_{\hbar}) \end{equation} \noindent which associates to every classical observable $O \in A$ a quantum observable \begin{equation} Q(\hbar): O \longmapsto \widehat{O}^Q(\hbar) \end{equation} \noindent so that its \underline{phase space star product}, the pullback $\star_{\hbar}^Q$ of multiplication in $\textnormal{\textbf{i}} \mathfrak{u}(\mathscr{H}, \langle \cdot, \cdot \rangle_{\hbar})$ along $Q(\hbar)$ defined for all $O_1, O_2 \in \mathsf{A}$ by \begin{equation} Q(\hbar)({O_1 \star_{\hbar}^Q O_2}) = Q(\hbar)({O}_1)  \cdot Q(\hbar)({O}_2) \end{equation} satisfies the following axioms:
\begin{enumerate}
\item \textnormal{\textcolor{gray}{[Asymptotic Expansions]}} For $g=0,1,2,\ldots$ there exist bilinear operators 
\begin{equation} B_g^Q: \mathsf{A} \times \mathsf{A} \rightarrow \mathsf{A} \otimes \C \end{equation} \noindent so that $\star_{\hbar}^Q$ has an asymptotic expansion in the pullback of the strong topology \begin{equation} \label{StarProductFormula} {O}_1 \star_{\hbar}^Q {O}_2 \sim \sum_{g=0}^{\infty} \hbar^g {B_g^Q(O_1, O_2) } \end{equation}
\item \textnormal{\textcolor{gray}{[Identity]}} $1 \in \mathsf{A}$ is a two-sided identity: $1 \star_{\hbar}^Q O = O \star_{\hbar}^Q 1 = O$ for all $O \in \mathsf{A}$

\item \textnormal{\textcolor{gray}{[Naturality]}} $B_g^Q$ are bidifferential operators of order $(g,g)$
\item \textnormal{\textcolor{gray}{[Static Version of Bohr's Correspondence Principle for Observables]}} to leading order in $\hbar$, the $\star_{\hbar}^Q$ product recovers the commutative multiplication in $\mathsf{A}$, i.e. \begin{equation} B_0^Q ( O_1, O_2) = O_1 O_2 \end{equation} is independent of $Q$ and depends pointwise on classical observables in $\mathscr{M}$.

\item \textnormal{\textcolor{gray}{[Dynamic Version of Bohr's Correspondence Principle for Observables]}} to leading order in $\hbar$, the $\star_{\hbar}^Q$ commutator recovers the Poisson bracket on $\mathsf{A}$, i.e. if we write $B_1^Q = (B_1^Q)^+ + (B_1^Q)^-$ as a sum of its symmetric and anti-symmetric parts \begin{equation} B_1^Q( O_1, O_2)^{\pm} = \tfrac{1}{2} \big (  B_1^Q(O_1, O_2) + B_1^{Q}(O_2, O_1 ) \big )  \end{equation}
 
\noindent then we have \begin{equation} B_1^Q(\cdot, \cdot)^- = \textnormal{\textbf{i}} \{ O_1, O_2\} \end{equation} is independent of $Q$ and depends locally on classical observables in $\mathscr{M}$.

\end{enumerate}
\end{definition}

\subsubsection{Smooth Quantizations via Microlocal Analysis} \label{subsubsecSmoothQuantizations}

\noindent The larger the size of the Poisson subalgebra $\mathsf{A} \subset C^{\infty}( \mathscr{M}, \R)$, whose size will grow as regularity hypothesis defining $\mathsf{A}$ are weakened, the harder one must work analytically to derive estimates implying the asymptotic expansions required by a quantization in the sense of Definition [\ref{QuantizationDefinition}].  These difficulties are present even if $\dim \mathscr{M} < \infty$.  For Euclidean space $\mathscr{M} \cong \R^{2K}$ with canonical coordinates $\{q_k, p_k\}_{k=1}^K$,  there are two examples of quantizations of the algebra $\mathsf{A} = \mathbb{R}[ p_1, \ldots, p_K, q_1, \ldots, q_K]$ of polynomials which admit extensions to $\mathsf{A} = C^{\infty} (\mathscr{M}, \R)$ both infamous in \textit{microlocal analysis}. 
\begin{enumerate}
\item The \underline{Wigner-Weyl quantization} of $C^{\infty}(\mathscr{M}, \R)$, whose phase space star product $\star_{\hbar}^{Q}$ is known as the Moyal star product, is defined using H\"{o}rmander's theory of Fourier integral operators or Maslov's canonical operator \cite{FollandBook, GuilleminSternbergSemiClassicalAnalysis, HallBook, MartinezBook, Maslov0, Zwor0}.

\item The \underline{anti-Wick quantization} of $C^{\infty}(\mathscr{M}, \R)$, a special case of Berezin-Toeplitz quantization, is defined using general Toeplitz operators \cite{DeMonvelGuillemin, Englis, HallBook}.
\end{enumerate}

\noindent Both Wigner-Weyl and anti-Wick quantizations are \textit{canonical quantizations} which we study in section \textbf{[\ref{subsecCanonicalQuantizations}]}.  In section \textbf{[\ref{subsecCanonicalQuantizations}]} we study the opposite of anti-Wick quantization, the \textit{Wick quantization} of polynomials $\mathsf{A} =  \mathbb{R}[ p_1, \ldots, p_K, q_1, \ldots, q_K]$ on Euclidean $\mathscr{M} \cong \R^{2K}$, depending on a chosen complex structure $J$ on $\mathscr{M}$.  We then construct a large family of polynomial quantizations in a certain sense equivalent to Wick quantization in Theorem [\ref{EtaQuantizationsAreCanonicalQuantizations}], each determined by an ``ordering'' $\eta$.  In Theorem [\ref{NazarovSklyaninQuantizationExistence}] we state the existence of Nazarov-Sklyanin's integrable geometric quantization of the periodic Benjamin-Ono system \cite{NaSk2} via an exotic ordering $\eta_{NS}$, which we construct in Theorem [\ref{NazarovSklyaninQuantizationConstruction}].
 We do not develop a symbol calculus for $Q_{NS}$ and content ourselves with the modest choice of polynomial $\mathsf{A}$.  Note that a more detailed analytic definition of quantizations than our Definition [\ref{QuantizationDefinition}] is provided by Rieffel's \textit{strict deformation quantizations} in the setting of $C^{*}$-algebras \cite{Landsman, Rieffel1994, Varilly}.
 
 \subsubsection{Deformation Quantizations and K\"{a}hler Metrics} \label{subsubsecDeformationQuantizationAndComplexStructures}
 \noindent An important condition in Definition [\ref{QuantizationDefinition}] is that $Q(\hbar)$ is a well-defined map.  To the contrary, in the setting of \textit{deformation quantization} \cite{FedosovBook}, there is no well-defined map $Q$ in which $\hbar>0$ is a numerical constant, but rather one begins with the phase space star product $\star_{\hbar}^Q$ in formula (\ref{StarProductFormula}) and treats $\hbar$ as a formal parameter.  Choosing a notion of \textit{gauge equivalence} of two deformation quantizations, one checks that 
 \begin{proposition} The symmetric part $(B_1^Q)^+$ of $B_1^Q$ ``can be killed by a gauge transformation and is a coboundary in the Hochschild complex'' \textnormal{\cite{Kontsevich}}. \end{proposition}
 
 \noindent For example, Wigner-Weyl quantization has $(B_1^Q)^+ \equiv 0$ while anti-Wick quantization does not, yet they are related by special well-known explicit gauge equivalence called the \textit{Segal-Bargmann transform} \cite{HallBook}.  However, in what follows, it is absolutely crucial that we do not quotient by gauge equivalence, since while any quantization requires $(B_1^Q)^-$ to reflect the \textit{Poisson structure} of $(\mathscr{M}, \omega)$, we will work with quantizations $Q$ of certain K\"{a}hler manifolds whose $(B_1^Q)^+$ reflect the \textit{K\"{a}hler metric} of $(\mathscr{M}, J, \mathsf{g}, \omega)$.  Importantly, $(B_1^{\textnormal{Wick}})^+(O,O)|_v$ is the variance of the Gaussian in our Theorem [\ref{CLTColumn2}].

\subsection{Random Values of Observables in Quantum States}  \label{subsecTheModelAppearsColumn1}
\noindent We continue our discussion of Born's Rule from section \textbf{[\ref{subsecObservables}]}, switching from a normalized quantum state $\Psi$ to an unnormalized $\hbar$-dependent quantum state $\Upsilon( \hbar)$.
\subsubsection{Born's Rule: Randomness from Non-Random Matrices}

\noindent Specialize Definition [\ref{TheModelColumn1PreDefinition}] to quantum observables $\widehat{\mathcal{O}}(\hbar) = \widehat{O}^Q(\hbar)$ realized by a quantization $Q$ of a classical observable $O$:

\begin{definition} \label{TheModelColumn1Definition}The \underline{random value} $\widehat{O}^Q(\hbar)|_{\Upsilon( \hbar)}$ of a quantized observable $\widehat{\mathcal{O}}^Q(\hbar)$ in a quantum state $\Upsilon(\hbar) \in \mathscr{H}$ is the random variable $\widehat{O}^Q(\hbar)|_{\Upsilon(\hbar)}$ whose law is $d \mu_{\Upsilon(\hbar), \Upsilon(\hbar)}( \cdot | \widehat{\mathcal{O}}^Q(\hbar))$ the spectral measure of $\widehat{\mathcal{O}}^Q(\hbar)$ at $\Upsilon(\hbar)$.  For $\phi : \R \rightarrow \C$ bounded continuous, have two formulas for the expectation \begin{equation} { \langle \Upsilon(\hbar) | \phi(\widehat{{O}}^Q(\hbar)) | \Upsilon(\hbar) \rangle } = \mathbb{E} [ \phi( \widehat{O}^Q |_{\Psi} )]=  \int_{- \infty}^{+\infty} \phi(E) d\upmu_{\Psi, \Psi}( E | \widehat{{O}}^Q(\hbar)) \end{equation}

\end{definition}

\noindent With Born's Rule in hand, we gain a more intuitive characterization of ``quantum conserved quantities'' of Definition [\ref{QuantumConservedQuantityFirstDefinition}] that better matches our Definition [\ref{ClassicalConservedQuantityDefinition}] of classical conserved quantities:
\begin{definition} \label{RealizedQuantumConservedQuantityDefinition}A \underline{realized quantum conserved quantity} for the Schr\"{o}dinger equation generated by $\widehat{\mathcal{T}}$ with initial condition $\Upsilon(\hbar)$ is the random value $\widehat{\mathcal{O}}|_{\Upsilon(\hbar)}$ of 
\begin{equation} [ \widehat{\mathcal{O}}, \widehat{\mathcal{T}}] = 0 \end{equation}

\noindent a quantum observable $\widehat{\mathcal{O}}$ in $\Upsilon(\hbar)$ which commutes with $\widehat{\mathcal{T}}$.
\end{definition}

\noindent ``Realizations'' are the difference between a random variable and its underlying law.\\
\\ 
\noindent Our notation for the random variable $\widehat{O}^Q(\hbar)|_{\Upsilon(\hbar)}$ is consistent with our use of the same notation for eigenvalues an operator at an eigenstate in Definition [\ref{QuantumStationaryStateDefinition}]:

\begin{proposition} If $\Psi_{\lambda}(\cdot | \hbar)$ is a $\widehat{\mathcal{T}}(\hbar)$-eigenstate with eigenvalue $\widehat{T}(\hbar)|_{{\lambda}} \in \R$
\begin{equation} \widehat{\mathcal{T}}(\hbar) \ \Psi_{\lambda}( \cdot | \hbar) = \widehat{T}(\hbar)|_{{\lambda}}  \ \Psi_{\lambda}( \cdot | \hbar), \end{equation}

\noindent the random value $\widehat{T}(\hbar)|_{\Psi_{\lambda}( \cdot | \hbar)}$ of $\widehat{\mathcal{T}}(\hbar)$ in $\Psi_{\lambda}( \cdot | \hbar)$ is non-random and must be $\widehat{T}(\hbar)|_{{\lambda}}$. \end{proposition}

\begin{definition} \label{RandomQuantumStationaryStateIndexDefinition} If a quantized Hamiltonian $\widehat{{T}}^Q(\hbar)$ has discrete spectrum so that the quantum state space $(\mathscr{H}, \langle \cdot, \cdot \rangle_{\hbar})$ has a countable orthonormal basis $\{\Psi_{\lambda}^{\textnormal{norm}}( \cdot | \hbar)\}_{\lambda \in \mathbb{Y}(\mathscr{M})}$ of quantum stationary states indexed by some set of $\lambda \in \mathbb{Y}(\mathscr{M})$ reflecting the geometry of the classical phase space, the \underline{random quantum stationary state distribution of $\Upsilon(\hbar)$}

\begin{equation} \textnormal{Prob} ( \lambda) = \frac{1}{ || \Upsilon(\hbar)||_{\hbar}^2} \cdot \Bigg | \Big  \langle \Upsilon(\hbar) \big | \Psi_{\lambda}^{\textnormal{norm}}( \cdot | \hbar) \Big \rangle_{\hbar} \Bigg |^2 \end{equation}

\noindent is a probability measure on $\mathbb{Y}(\mathscr{M})$.

\end{definition}

\begin{proposition} \label{DiscreteSpectrumImpliesRandomLambda} If a quantized Hamiltonian $\widehat{{T}}^Q(\hbar)$ has discrete spectrum,  then for any quantum conserved quantity $\widehat{\mathcal{O}}|_{\Upsilon(\hbar)}$ of the flow generated by $\widehat{{T}}^Q(\hbar)$ we have \begin{equation} \widehat{\mathcal{O}}|_{\Upsilon(\hbar)} = \widehat{\mathcal{O}}|_{\lambda} \end{equation}

\noindent equality in law for $\lambda$ sampled from the random quantum stationary state distribution of $\Upsilon(\hbar)$ of \textnormal{Definition [\ref{RandomQuantumStationaryStateIndexDefinition}]}. \end{proposition}

\begin{itemize}
\item \textit{Proof:} use \textnormal{Definition [\ref{RandomQuantumStationaryStateIndexDefinition}]} and the formula for spectral measures  \begin{equation} d \mathsf{\mu}_{\Upsilon(\hbar), \Upsilon(\hbar)} ( E | \widehat{\mathcal{O}}) = \frac{1}{ || \Upsilon(\hbar)||_{\hbar}^2}  \cdot \sum_{ \lambda \in \mathbb{Y} ( \mathscr{M} ) } \delta \big ( E - \widehat{\mathcal{O}}|_{\lambda} \big ) \cdot  \Bigg | \Big  \langle \Upsilon(\hbar) \big | \Psi_{\lambda}^{\textnormal{norm}}( \cdot | \hbar) \Big \rangle_{\hbar} \Bigg |^2 \end{equation}

\noindent in absence of continuous spectrum. $\square$

\end{itemize}

\subsubsection{Finite Moments from Regularity Assumptions} \label{FiniteMomentsFromRegularityAssumptionsColumn1}

\noindent Let $\widehat{O}^Q(\hbar)|_{\Upsilon(\hbar)}$ be the random value of a quantized observable $\widehat{\mathcal{O}}^Q(\hbar)$ in a $\hbar$-dependent quantum state $\Psi = \Upsilon(\hbar)$.  By Definition [\ref{TheModelColumn1Definition}], any bounded function ${\phi} : \R \rightarrow \C$ defines a new random variable $\phi(\widehat{O}^Q(\hbar)|_{\Upsilon(\hbar)})$ with finite expectation 
\begin{equation} \label{SingleVariableBoundedExpectation} \mathbb{E} \big [  {\phi} \big (\widehat{O}^Q(\hbar)|_{\Upsilon(\hbar)}  \big )  \big ] =   \frac{ \big \langle \Upsilon(\hbar) \big | {\phi} \big ( \widehat{\mathcal{O}}^Q ( \hbar)  \big )  \big | \Upsilon(\hbar) \big \rangle_{\hbar} } {\big \langle \Upsilon(\hbar) | \Upsilon(\hbar) \big \rangle_{\hbar}  } .\end{equation}

\noindent If one formally Taylor expands a bounded ${\phi}$ around $0 \in \R$, the result is an expression ${\phi} \sim \sum_{n} \tfrac{1}{n!}{\phi}^{(n)}(0) E^n$ as an infinite linear combination of powers $E^n$ for $l=0,1,2,\ldots$ which themselves are unbounded on $\R$ unlike ${\phi}$.  This formal expression might tempt one to rewrite the correlator (\ref{SingleVariableBoundedExpectation}) as a formal infinite linear combination of moments \begin{equation} \label{SingleVariableMoment} \mathbb{E} \big [  \big ( \widehat{O}^Q(\hbar)|_{\Upsilon ( \hbar)}   \big )^n \big ] = \frac{ \big \langle \Upsilon (\hbar) \big | \ \big ( \widehat{\mathcal{O}}^Q(\hbar) \big )^n      \ \big | \Upsilon ( \hbar)  \big \rangle_{\hbar} }{ \big \langle \Upsilon( \hbar) \ \big | \ \Upsilon (\hbar) \big \rangle_{\hbar} }.\end{equation}

\noindent The $n$th moment \textnormal{(\ref{SingleVariableMoment})} is finite if $\Upsilon(\hbar)$ is in the domain of $ \big (\widehat{\mathcal{O}}^Q( \hbar) \big ) ^{\lceil n/2 \rceil}$.

\subsection{Asymptotic Expansions of Moments} \label{subsecAsymptoticExpansionsMicrolocalAnalysis}

\noindent  When well-defined, the expressions (\ref{SingleVariableBoundedExpectation}) and (\ref{SingleVariableMoment}) are functions $W(\hbar)$ of $\hbar >0$ which depend on $\hbar$ in three ways: in the observable $\widehat{\mathcal{O}}^Q(\hbar)$, in the state $\Upsilon(\hbar)$, and in the inner product $\langle \cdot, \cdot \rangle_{\hbar}$ of the quantum state space.  But if all we know is that $W(\hbar)$ is well-defined at $\hbar >0$, we do not yet know that $W(\hbar)$ is continuous at $\hbar = 0$, let alone differentiable or analytic.  Determining the regularity of quantum correlation functions $W(\hbar)$ of the type (\ref{SingleVariableBoundedExpectation}) or (\ref{SingleVariableMoment}) as $\hbar \rightarrow 0$ is a major analytic problem in the rigorous study of quantum systems, together with the problem of interpreting the coefficients $W_g$ in the resulting \textit{asymptotic expansions} $W(\hbar) \sim \sum_{g=0}^{\infty} \hbar^g W_{g}$ in $\hbar$.\\
\\
\noindent As we discussed in section \textbf{[\ref{subsubsecSmoothQuantizations}]}, to even begin to derive asymptotic expansions in $\hbar$ of correlation functions of the type (\ref{SingleVariableMoment}) for bounded ${\phi}$ without decomposing $\phi$ into infinite linear combinations of simpler yet unbounded functions to reduce to correlation functions of the latter type (\ref{SingleVariableMoment}), one requires a quantization of smooth functions $\mathsf{A} = C^{\infty}(\mathscr{M}, \R)$ which is typically a challenge to construct.  That being said, in the case of such a smooth quantization defined by a symbol calculus, one can use the dictionary between Heisenberg and Schr\"{o}dinger pictures to derive asymptotic expansions in $\hbar$ of correlation functions of bounded type (\ref{SingleVariableMoment}). Again as we discussed in section \textbf{[\ref{subsubsecSmoothQuantizations}]}, in our study of the periodic Benjamin-Ono system to come we work with a particular quantization $Q_{NS}$ for which no symbol calculus is a priori readily available, hence for simplicity we will take humble regularity assumptions so as to derive expansions of moments of type (\ref{SingleVariableMoment}).

\subsection{Concentration of Measure in Static Semi-Classical Limits} \label{subsecStaticBohrCP}

\noindent The principle that classical systems as defined in section \textbf{[\ref{subsecClassicalSystems}]} emerge as the leading order $\hbar^0$ behavior in the semi-classical limit $\hbar \rightarrow 0$ of quantum systems as defined in section \textbf{[\ref{subsecQuantumSystems}]} is known as \textit{Bohr's Correspondence Principle}.  In subsection \textbf{[\ref{subsecAsymptoticExpansionsMicrolocalAnalysis}]}, we discussed the problem of determining $\hbar \rightarrow 0$ asymptotic expansions of correlation functions of the random variable $\widehat{O}^Q(\hbar)|_{\Upsilon(\hbar)}$ of Definition [\ref{TheModelColumn1Definition}], the random value of an arbitrary quantized observable $\widehat{\mathcal{O}}^Q(\hbar)$ in arbitrary $\hbar$-dependent quantum state $\Psi = \Upsilon(\hbar)$.  However, Bohr's Correspondence Principle is only expected to hold for a special class of $\hbar$-dependent quantum states $\Upsilon(\hbar) $ known as \textit{quasi-classical states}.

\begin{definition} \label{StaticBohr} \textnormal{\textcolor{gray}{[Static Version of Bohr's Correspondence Principle for States]}} $\Upsilon_{v}( \hbar) \in \mathscr{H}$ is \underline{quasi-classical around a classical state $v$} with respect to a quantization $Q$ of a Poisson algebra $\mathsf{A} \subset C^{\infty}(\mathscr{M}, \R)$ of classical observables if for every $O \in \mathsf{A}$ as $\hbar \rightarrow 0$ we have the following concentration of measure: \begin{equation} \label{StaticBohrLimitStatement} \widehat{O}^Q(\hbar) |_{\Upsilon_{v}(\hbar)} \longrightarrow O|_v \end{equation}

\noindent the random value $\widehat{O}^Q(\hbar) |_{\Upsilon_v(\hbar)}$ of the quantized observable $\widehat{\mathcal{O}}^Q(\hbar)$ in the quantum state $\Upsilon_{v}(\hbar) \in \mathscr{H}$ concentrates as $\hbar \rightarrow 0$ on the non-random value $O|_v$ of the classical observable $O$ in the classical state $v$.  Precisely, \textnormal{(\ref{StaticBohrLimitStatement})} is convergence in distribution to the delta measure at $O|_v$.\end{definition}

\noindent The relation to the Static Version of Bohr's Correspondence Principle for Observables, the axiom $B_0^Q(O_1, O_2) = O_1 O_2$ in our Definition [\ref{QuantizationDefinition}], uses the dictionary between the ``Heisenberg picture'' and the ``Schr\"{o}dinger picture'' of quantum systems.\\
\\
\noindent We conclude this section with two remarks.  First, our terminology of quantum states being ``quasi-classical around a classical state $v$'' in Definition [\ref{StaticBohr}] is related to but distinct from the term ``semi-classical states'' in semi-classical analysis of the WKB approximation \cite{GuilleminSternbergSemiClassicalAnalysis, MartinezBook, Maslov0}.  We clarify this distinction briefly in the next subsection by using a generalization of Definition [\ref{StaticBohr}] for mixed states $\rho$ of Definition [\ref{MixedStatesDefinition}]:

\begin{definition} \label{MixedStaticBohr} \textnormal{\textcolor{gray}{[Static Version Bohr's Correspondence Principle for Mixed States]}} $\Upsilon_{\rho}(\hbar) \in \mathscr{H}$ is \underline{quasi-classical around a classical mixed state $\rho$} with respect to a given quantization $Q$ of a Poisson algebra $\mathsf{A} \subset C^{\infty}(\mathscr{M}, \R)$ of classical observables if for every $O \in \mathsf{A}$ as $\hbar \rightarrow 0$ we have concentration of measure \begin{equation} \label{StaticBohrLimitStatement2} \widehat{O}^Q(\hbar) |_{\Upsilon_{\rho}(\hbar)} \longrightarrow O|_{\rho} \end{equation}

\noindent on the non-random average $O|_{\rho} = \int_{\mathscr{M}} O|_v d \rho (v)$ of the classical observable $O$ in the classical mixed state $\rho$. \end{definition}

\noindent As a second remark, we say that Definitions [\ref{StaticBohr}] and [\ref{MixedStaticBohr}] are ``static'' because they do not involve the behavior of observables and states under classical and quantum time evolution by Hamiltonians $T$ and $\widehat{\mathcal{T}}(\hbar)$.  In the next section, we briefly clarify how our investigation compares to the usual investigations in semi-classical analysis by defining two ``dynamic'' versions of Bohr's Correspondence Principle.  

\subsection{Integrability and Chaos in Quantum Fluctuations} \label{subsecDynamicalBohrCP}

\begin{definition} \label{StrongBohr} \textnormal{\textcolor{gray}{[Strong Version of Bohr's Correspondence Principle for States]}} Suppose a quantum Hamiltonian $\widehat{\mathcal{T}}^Q(\hbar)$ on $(\mathscr{H}, \langle \cdot, \cdot \rangle_{\hbar})$ is the quantization by $Q$ of a classical Hamiltonian $T$ on $(\mathscr{M}, \omega)$.  An $\hbar$-dependent family of quantum states $\Upsilon_{v}(\hbar) \in \mathscr{H}$ is \underline{dynamically coherent around a classical $v \in \mathscr{M}$} if 
\begin{enumerate}
\item The quantum time evolution of $\Upsilon_{v( \cdot, 0)}(\hbar)$ by $\widehat{\mathcal{T}}^Q(\hbar)$ for time $t$ results in $\Upsilon_{v(\cdot, t)}(\hbar)$ where $v(\cdot, t)$ is defined by the classical flow by $T$ with initial data $v( \cdot, 0)$
\item For all $t \geq 0$, $\Upsilon_{v ( \cdot, t)}(\hbar)$ is quasi-classical around $v(\cdot, t)$ with respect to the quantization $Q$ as defined in \textnormal{Definition [\ref{StaticBohr}]}.
\end{enumerate} \end{definition}

\noindent Dynamical coherence is a strong condition since $\{\Upsilon_v(\hbar)\}_{v \in \mathscr{M}}$ is time-independent.  In fact, one expects that dynamical coherence is only possible for quantizations of classical integrable systems.  In the search for ``{quantum chaos}'', i.e. for universal features of quantum systems that reflect the integrable or chaotic nature of the corresponding classical system \cite{Berry1989, Gutz0}, one studies the double scaling limit $t \rightarrow \infty$, $\hbar \rightarrow 0$ of the Schr\"{o}dinger equation (\ref{SchrodingerEquation}) in four notable regimes:

\begin{enumerate}
\item \textcolor{gray}{[Underlying Regime]} $\hbar \rightarrow 0$ first, i.e. $t \rightarrow \infty$ for the underlying classical system.
\item \textcolor{gray}{[Ehrenfest Regime]} $\hbar \rightarrow 0, t \rightarrow \infty$ at rate 
\begin{equation} \frac{t}{\hbar} \sim \frac{1}{\hbar} \cdot \log \frac{1}{\hbar} \end{equation}
\item \textcolor{gray}{[Heisenberg Regime]} $\hbar \rightarrow 0, t \rightarrow \infty$ at rate 
\begin{equation} \frac{t}{\hbar} \sim \frac{1}{ \delta(\hbar)} \end{equation}

\noindent where $\delta(\hbar)$ is the mean-level spacing of the spectrum of the quantum Hamiltonian.
\item \textcolor{gray}{[Stationary Regime]} $t \rightarrow \infty$ first, i.e. $\hbar \rightarrow 0$ for quantum stationary states.
\end{enumerate}
\pagebreak

\noindent Let us now give a precise formulation of what one expects in the stationary regime.  This limit is known to teach us about quantum chaos, as classical ergodic theory teaches us that the classification of classical stationary states reflects the integrable or chaotic nature of classical systems.

\begin{definition} \label{StationaryBohr} \textnormal{\textcolor{gray}{[Stationary Version of Bohr's Correspondence Principle for States]}} Suppose a quantum Hamiltonian $\widehat{\mathcal{T}}^Q(\hbar)$ on $(\mathscr{H}, \langle \cdot, \cdot \rangle_{\hbar})$ is the quantization by $Q$ of a classical Hamiltonian $T$ on $(\mathscr{M}, \omega)$ and has quantum stationary states \begin{equation} \label{YepEigenvalueEquation}\widehat{\mathcal{T}}^Q(\hbar) \Psi_{\lambda}( \cdot | \hbar) = \widehat{T}(\hbar) |_{\lambda} \Psi_{\lambda} \end{equation} 
\noindent with indices $\lambda \in \mathbb{Y}(\mathscr{M})$ in non-compact space $\mathbb{Y}(\mathscr{M})$.  A sequence $\{ \Psi_{\lambda(k)}( \cdot | \hbar)\}_{k = 1}^{\infty}$ of quantum stationary states is \underline{quasi-classical around a classical stationary state $\rho$} if for $k(\hbar)$ in the simultaneous limit $k(\hbar) \rightarrow \infty$ as $\hbar \rightarrow 0$ taken so that \begin{equation} \label{EnergyCorrespondenceRelation} T(\hbar) |_{\lambda(k)} \rightarrow E \end{equation}

\noindent for some $E \in \R$, $\Psi_{\lambda{(k(\hbar))}} ( \cdot | \hbar)$ is quasi-classical around a classical mixed state $\rho$ as in \textnormal{Definition [\ref{MixedStaticBohr}]} for \textbf{some} $T$-invariant $\rho$ supported in the energy surface $T^{-1}(E)$.
\end{definition}

\noindent To relate the classical and quantum notions of energy and stationary states, in practice one needs to work to check that a sequence of stationary states are quasi-classical around a classical mixed state $\rho$ as in Definition [\ref{StationaryBohr}].  Even for two-dimensional systems such as billiards in a domain $\Omega$, whose classical phase spaces $\mathscr{M} \cong T^* \Omega$ are four-dimensional symplectic manifolds, doing so is an active subject full of challenges, conjectures, and mysteries \cite{BourgadeKeating, SarnakQUE}.  A central question in this subject is to identify \textit{\textbf{which}} $T$-invariant classical mixed states $\rho$ occur in such limits.  ``Quantum (\textit{Unique}) Ergodicity'' occurs for classical ergodic systems if along {some} (\textit{all}) subsequences $\lambda(k)$ in $\mathbb{Y}(\mathscr{M})$ so that quantum energy diverges as in (\ref{EnergyCorrespondenceRelation}) one has concentration on $\rho = \rho_{\star |E}$ the Liouville measure on the energy surface $T^{-1}(E)$.  Counterexamples to quantum unique ergodicity are known, featuring ``scarring'' on sparse classical periodic orbits.  Moreover, running this same program for random Schr\"{o}dinger operators in random geometries to model disorder caused by impurities is the setting of the Anderson conjecture and the metal-insulator phase transition.\\
\\
\noindent The asymptotic regime of Definition [\ref{StationaryBohr}] is intimately related to the problem of semi-classical approximation of quantum stationary states as goes by the name of the Wentzl-Kramers-Brillouin (WKB) method and its multivariate generalizations due to Einstein-Brillouin-Keller and Maslov \cite{GuilleminSternbergSemiClassicalAnalysis, MartinezBook, Maslov0}.  Whereas we describe {quantum states} $\Upsilon_v ( \hbar)$ and $\Upsilon_{\rho} (\cdot | \hbar)$ in Definitions [\ref{StaticBohr}] and [\ref{MixedStaticBohr}] as ``quasi-classical'' around classical states $v$ and classical mixed states $\rho$, in WKB theory the ``semi-classical states'' $\Psi_{\lambda} \sim a_0 e^{\frac{\textbf{i}}{\hbar} S_0}$ are good approximations high energy eigenfunctions.  Here $S_0$, $a_0$ solve classical transport and classical Hamilton-Jacobi (eikonal) equations, respectively.  The data of $S_0$ and $a_0$ is associated to a Lagrangian submanifold of $(\mathscr{M}, \omega)$ together with a $T$-invariant half-density, respectively, and may be matched with the $T$-invariant {classical mixed state} $\rho$ on the energy surface in Definition [\ref{StationaryBohr}].

\pagebreak

\section{Coherent States} \label{secColumn2}

\noindent We now specialize the material of section \textbf{[\ref{secColumn1}]} to coherent states.  In the geometric quantization of Hermitian affine spaces $(\mathscr{M}, J, \mathsf{g} , \omega)$, a coherent state $\Upsilon_v ( \cdot | \hbar)$ around a classical state $v \in \mathscr{M}$ is the reproducing kernel in the Fock space $L^2_{\textnormal{J-hol}} (\mathscr{M}, d \rho_{\hbar, \mathsf{g}})$ of $J$-holomorphic functions on $\mathscr{M}$ with Segal-Bargmann weight.  We review Hermitian affine spaces in section \textbf{[\ref{subsecHermitianAffineSpaces}]}, their Fock spaces in section \textbf{[\ref{subsecFockSpaces}]}, and carefully construct a special class of polynomial canonical quantizations of Hermitian affine spaces in Fock spaces in section \textbf{[\ref{subsecCanonicalQuantizations}]}, the $\eta$-quantizations close to Wick quantization.  For $\eta$-quantizations, in Theorems [\ref{LLNColumn2}] and [\ref{CLTColumn2}], in the semi-classical limit $\hbar \rightarrow 0$ \begin{equation} \label{LLNCLTColumn2Formula} \widehat{O}^{\eta}(\hbar) |_{\Upsilon_v(\cdot | \hbar)} \sim O|_v + \hbar^{1/2} \mathbb{G}(O)|_v \end{equation}

\noindent the random value $\widehat{O}^{\eta}(\hbar) |_{\Upsilon_v(\cdot | \hbar)}$ of any quantized observable $\widehat{O}^{\eta}(\hbar)$ in a coherent state $\Upsilon_v( \cdot | \hbar)$ around a classical state $v$, see Definition [\ref{TheModelColumn2Definition}] in \textbf{[\ref{subsecTheModelAppearsColumn2}]}, concentrates to leading order on the non-random value $O|_v$ of the classical observable $O$ at the classical state $v$, independent of $\eta$ \textbf{[\ref{subsecLLNColumn2}]}.  Moreover, quantum corrections are Gaussian $\mathbb{G}(O)|_v$ at scale $\hbar^{1/2}$ with mean $0$ and variance $|| (\nabla O)|_v ||_{\mathsf{g}}^2$ independent of $\eta$ \textbf{[\ref{subsecCLTColumn2}]}.  Theorems [\ref{LLNColumn2}] and [\ref{CLTColumn2}] are confirmations and corrections to the static version of Bohr's Correspondence Principle and follow from our Theorem [\ref{AOEColumn2}], an asymptotic expansion in $\hbar$ of joint cumulants in \textbf{[\ref{subsecAOEColumn2}]}.  The limiting objects in (\ref{LLNCLTColumn2Formula}) are quantities of the type in Proposition [\ref{RespectTheButterflyEffect}], hence do not require the Hamiltonian flow generated by $O$ to be integrable.  We also do not require the quantization $\widehat{O}^{\eta}(\hbar)$ to have discrete spectrum.  Rather, these results follow directly from the fact that at fixed $\hbar >0$ the Segal-Bargmann weight on $\mathscr{M}$ is already Gaussian with covariance kernel $\hbar \mathsf{g}^{-1}$ given by the inverse metric.

\subsection{Classical Phase Spaces as Hermitian Affine Spaces} \label{subsecHermitianAffineSpaces}
\noindent Throughout section \textbf{[\ref{secColumn1}]}, classical phase spaces are arbitrary symplectic manifolds $(\mathscr{M}, \omega)$.  We now consider a simple class of symplectic manifolds: the Hermitian affine spaces.  All definitions above and below hold in the case $(\mathscr{M}, \omega)$ is infinite-dimensional.

\subsubsection{Hermitian Vector Spaces and Compatible Triples}

\noindent Let $\mathscr{M}'$ be a real vector space of dimension $2K$ for $K \in \N \cup \{ \infty\}$.

\begin{definition} \label{ComplexStructureDefinition} A \underline{complex structure} $J: \mathscr{M}' \rightarrow \mathscr{M}'$ on $\mathscr{M}'$ is linear with $J^2 = - \mathbbm{1}$. \end{definition}

\begin{definition} \label{HermitianVectorSpaceDefinition} A \underline{Hermitian inner product} $\mathsf{h} : \mathscr{M}' \times \mathscr{M}' \rightarrow \C$ on $(\mathscr{M}', J)$ is a conjugate symmetric and positive definite $J$-sesquilinear map.  A complex vector space with a Hermitian inner product is called a \underline{Hermitian vector space} $(\mathscr{M}', J, \mathsf{h})$.  \end{definition}

\begin{definition} \label{RiemannianStructureDefinition} A \underline{real inner product} $\mathsf{g}: \mathscr{M}' \times \mathscr{M}' \rightarrow \R$ on a real vector space $\mathscr{M}'$ is a symmetric positive-definite bilinear map. \end{definition}

\begin{definition} \label{SymplecticStructureDefinition} A \underline{symplectic structure} $\omega : \mathscr{M}' \times \mathscr{M}' \rightarrow \R$ on a real vector space $\mathscr{M}'$ is a skew-symmetric non-degenerate bilinear map.
\end{definition}

\begin{proposition} \label{CompatibleDefinition} The real and imaginary parts of a Hermitian inner product $\mathsf{h}$ \begin{eqnarray} \mathsf{g} &=& \textnormal{Re}[ \mathsf{h}] \\  \omega &=& \textnormal{Im}  [ \mathsf{h}  ]\end{eqnarray}

\noindent define a real inner product $\mathsf{g}$ and a symplectic structure $\omega$ on $\mathscr{M}'$.  Conversely, a real inner product $\mathsf{g}$ and a symplectic structure $\omega$ on an even-dimensional real-vector space $\mathscr{M}'$ are real and imaginary parts of a Hermitian inner product $\mathsf{h}$ with respect to a complex structure $J$ on $\mathscr{M}'$ if they are \underline{compatible}, i.e. for all $v^{\textnormal{out}}, v^{\textnormal{in}} \in \mathscr{M}'$
\begin{equation} \omega ( v^{\textnormal{out}}, v^{\textnormal{in}}) = \mathsf{g} ( v^{\textnormal{out}}, - J v^{\textnormal{in}}) \end{equation}
\end{proposition}

\noindent Thus, we may write $(\mathscr{M}', J, \mathsf{h})$ or $(\mathscr{M}', J, \mathsf{g}, \omega)$ to denote a Hermitian vector space.

\subsubsection{Hermitian Affine Spaces as K\"{a}hler Manifolds}

\begin{definition} \label{HermitianAffineSpaceDefinition} A \underline{Hermitian affine space} $(\mathscr{M}, J, \mathsf{g}, \omega)$ is an affine subspace of a Hermitian vector space $\mathscr{M} \subset \mathscr{M}'$ . \end{definition}

\noindent Hermitian affine spaces are very special non-compact K\"{a}hler manifolds.  In light of section \textbf{[\ref{subsecClassicalSystems}]}, we must regard Hermitian affine spaces as manifolds if we are to treat them as classical phase spaces of a classical Hamiltonian system.
  
\begin{proposition} When a real affine space $\mathscr{M}$ is regarded as a real manifold, vector fields $Y$ on $\mathscr{M}$ are just maps $Y : \mathscr{M} \rightarrow \mathscr{M}$.
\end{proposition}
\begin{corollary} A Hermitian affine space $(\mathscr{M}, J, \mathsf{g}, \omega)$ is a K\"{a}hler manifold. \end{corollary}

\begin{definition} \label{GlobalKahlerPotentialDefinition} The \underline{global K\"{a}hler potential} of a Hermitian affine space $(\mathscr{M}, J , \mathsf{h})$ is the restriction of the Hermitian metric $\mathsf{h}$ to the diagonal in $\mathscr{M} \times \mathscr{M}$.  By the conjugate symmetry of $\mathsf{h}$, this potential may be written as either $\mathsf{h}(v, v)$ or $\mathsf{g} (v,v)$.\end{definition}

\begin{definition} \label{GradientVectorFieldDefinition} A \underline{gradient vector field} $\nabla O$ is the vector field on $\mathscr{M}$ defined so \begin{equation} \mathsf{g}( \nabla O, Y) = \partial_Y O \end{equation}
\noindent is the directional derivative of a classical observable $O$ for any vector fields $Y$. \end{definition}

\noindent Comparing with Definitions [\ref{HamiltonianVectorFieldDefinition}] and [\ref{PoissonBracketDefinition}],

\begin{proposition} \label{HamiltonianToGradientConversionCompatibility} In a Hermitian affine space $(\mathscr{M}, J, \mathsf{g}, \omega)$, the gradient $\nabla O$ and Hamiltonian $X_O$ vector fields of a classical observable $O \in C^{\infty}(\mathscr{M})$ are related by \begin{equation} X_O = - J \nabla O .\end{equation} 
\end{proposition}

\begin{corollary} \label{HermitianVectorSpacePoissonBracketCorollary} In a Hermitian affine space $(\mathscr{M}, J, \mathsf{g}, \omega)$, the Poisson bracket $\{ \cdot , \cdot \}$ induced by $\omega$ may instead be written through the metric $\mathsf{g}$ as \begin{equation} \{ O_1, O_2\} = \mathsf{g} ( J \nabla O_1, \nabla O_2).\end{equation}
\end{corollary}

\subsubsection{Welding Operator as Inverse Hermitian Metric}

\noindent The following construction plays a crucial role in this paper:

\begin{definition} \label{WeldingOperatorDefinition} In any Hermitian affine space $(\mathscr{M}, J, \mathsf{g}, \omega)$, the \underline{welding operator} $ \mathscr{B}_1: C^{\infty}(\mathscr{M}, \C) \times C^{\infty}(\mathscr{M}, \C) \longrightarrow C^{\infty}(\mathscr{M}, \C)$ is defined intrinsically by \begin{equation} \mathscr{B}_1 (O_1 , O_2) = \mathsf{h} ( \nabla O_1, \nabla O_2) = \mathsf{h}^{-1} (O_1, O_2) \end{equation}

\noindent the inverse Hermitian metric.

 \end{definition}
 
\noindent By Propositions [\ref{CompatibleDefinition}] and [\ref{HamiltonianToGradientConversionCompatibility}], 

\begin{proposition} \label{PlusMinusPartsOfWeldingOperator} The additive decomposition $\mathscr{B}_1 = \mathscr{B}_1^+ + \mathscr{B}_1^-$ of the welding operator into symmetric and anti-symmetric parts \begin{equation} \mathscr{B}_1(O_1, O_2)^{\pm}  =\tfrac{1}{2} \big (  \mathscr{B}_1 (O_1, O_2) \pm \mathscr{B}_1 (O_2, O_1) \big ) \end{equation}

\noindent coincides with its decomposition into real and imaginary parts \begin{eqnarray} \label{SymmetricPartWeldingIsRiemannian} \mathscr{B}_1(O_1, O_2)^+ &=& \mathsf{g} ( \nabla O_1, \nabla O_2) = \mathsf{g}^{-1} ( O_1, O_2)  \\ \label{SkewSymmetricPartWeldingIsPoisson} \mathscr{B}_1(O_1 , O_2)^- &=& \textnormal{\textbf{i}}  \omega ( \nabla O_1, \nabla O_2) = \textnormal{\textbf{i}} \{ O_1, O_2\} \end{eqnarray}

\noindent as the inverse Riemannian metric $\mathsf{g}^{-1}$ and the Poisson bracket $\{ \cdot, \cdot \}$.
\end{proposition}

\noindent Formula (\ref{SkewSymmetricPartWeldingIsPoisson}) for the \textit{skew-symmetric part} $\mathscr{B}_1^- = \textbf{i} \{ \cdot, \cdot \}$ of the welding operator $\mathscr{B}_1$ coincides with the condition $B_1^Q( \cdot, \cdot )^- = \textbf{i} \{\cdot, \cdot \}$ on the bidifferential operator $B_1^Q$ in the Definition [\ref{QuantizationDefinition}] of any quantization $Q$, the ``Dynamic Version of Bohr's Correspondence Principle for Observables''.  In section \textbf{[\ref{subsecCanonicalQuantizations}]}, we construct the {Wick quantization} of the classical phase spaces $(\mathscr{M}, J, \mathsf{g}, \omega)$ studied in this section whose $B_1^{\text{Wick}} \equiv \mathscr{B}_1$ is the \textit{full} welding operator.  In section \textbf{[\ref{subsecCLTColumn2}]}, the \textit{symmetric part} $\mathscr{B}_1^+$ of the welding operator appears as the variance of the Gaussian in our Theorem [\ref{CLTColumn2}].

\subsubsection{$\sigma$-Canonical Coordinates}

\noindent We now specify a family of coordinates on our Hermitian affine spaces $(\mathscr{M}, J, \mathsf{h})$ which diagonalize the Hermitian inner product $\mathsf{h}$.

\begin{definition} \label{SigmaBasisDefinition} Let $(\mathscr{M}, J, \mathsf{g}, \omega)$ be a Hermitian affine space of complex dimension $K \in \N \cup \{ \infty\}$.  For positive parameters $\{ \sigma_k \}_{k=1}^K$, a \underline{$\sigma$-basis} $\{e_k\}_{k=1}^K$ is a $\C$-basis of the complexification $\mathscr{M} \otimes \C$ so that 

\begin{equation} \mathsf{h} ( e_{k}, e_{k'} ) = \frac{1}{ \sigma_k^2} \cdot \delta(k - k') \end{equation}

\noindent for $\mathsf{h}$ defined by sesquilinear extension so that $\mathsf{g}$ is diagonal with $\mathsf{g}_{kk} = \frac{1}{ \sigma_k^2}$ and $\omega \equiv 0$.
\end{definition}

\begin{definition} \label{SigmaCoordinatesDefinition} The expansion of $v \in \mathscr{M} \subset \mathscr{M} \otimes \C$ in a $\sigma$-basis $\{e_k\}_{k=1}^K$ of $\mathscr{M}$

\begin{equation} v = \sum_{k=1}^K \Big( V_k e_k + \overline{V_k} e_{-k} \Big ) \end{equation} defines $\{ V_k, \overline{V_k}\}_{k=1}^K$ \underline{$\sigma$-holomorphic and $\sigma$-anti-holomorphic coordinates on $\mathscr{M}$.}

\end{definition}

\begin{proposition} \label{WeldingKeyProposition} In a $\sigma$-coordinate system on a Hermitian affine space, using the relation $\mathsf{g}^{kk} = \tfrac{1}{ \mathsf{g}_{kk}} = \sigma_k^2$, the welding operator $\mathscr{B}_1$ of \textnormal{Definition [\ref{WeldingOperatorDefinition}]} has the form

\begin{equation} \mathscr{B}_1 ( O_1, O_2) = \sum_{k=1}^{K} \sigma_k^{2} \frac{ \partial O_1}{ \partial \overline{V_k}} \frac{ \partial O_2}{ \partial V_k }\end{equation}
\noindent written through the Wirtinger derivatives.

\end{proposition}

\noindent $\sigma$-coordinates pick out three complex Poisson subalgebras $\mathsf{A}_{\C } \subset C^{\infty}(\mathscr{M}, \C)$
\begin{definition} \label{GeneralizedPolynomialDefinition} If $\{V_k, \overline{V}_k\}$ are $\sigma$-holomorphic and $\sigma$-anti-holomorphic coordinates on a Hermitian affine space $(\mathscr{M}, J, \mathsf{g}, \omega)$, have Poisson algebras $\mathsf{A}_{\C} \subset C^{\infty}(\mathscr{M}, \C)$:\begin{enumerate}
\item $\mathsf{A}_{\C}^{\textnormal{holpoly}} = \C[ V_{1}, \ldots, V_K ]$ the \underline{holomorphic polynomials}
\item $\mathsf{A}_{\C}^{\textnormal{poly}} = \C[\overline{V_1}, \ldots, \overline{V_k}, V_1, \ldots, V_K]$ the  \underline{polynomials}
\item $\mathsf{A}_{\C}^{\textnormal{genpoly}} = \varprojlim \C[\overline{V_1}, \ldots, \overline{V_N}, V_1, \ldots, V_N]$ the \underline{generalized polynomials} are classical observables $O ( \overline{V_1}, V_1, \overline{V_2}, V_2, \ldots )$ which are ordinary polynomials if all but finitely many $\overline{V_k}, V_{k}$ are set to $0$. 
\end{enumerate} \end{definition}
\noindent Generalized polynomials coincide with polynomials only if $K = \dim_{\C} \mathscr{M} < \infty$.  If a Hermitian affine space $(\mathscr{M}, J, \mathsf{h})$ is complete, it is a {Hilbert space}, but we should not confuse such Hilbert spaces $\mathscr{M}$ with the quantum state spaces $\mathscr{H}$ of section \textbf{[\ref{subsecQuantumSystems}]}.

\subsection{Quantum State Spaces as Fock Spaces} \label{subsecFockSpaces}

\noindent When a classical phase space $(\mathscr{M}, \omega)$ is not just symplectic but also K\"{a}hler, the theory of geometric quantization suggests a preferred choice of quantum state space $(\mathscr{H}, \langle \cdot, \cdot \rangle_{\hbar})$ chosen according to the K\"{a}hler geometry of the classical phase space \cite{Woodhouse}.  In particular, for the non-compact homogeneous K\"{a}hler manifolds $(\mathscr{M}, J, \mathsf{g}, \omega)$ given by the Hermitian affine spaces of subsection \textbf{[\ref{subsecHermitianAffineSpaces}]}, this preferred quantum state space
\begin{equation} (\mathscr{F}_{(\mathscr{M}, J)}, \langle \cdot, \cdot \rangle_{\hbar; \mathsf{h}}) = L^2_{\textnormal{J-hol}} ( \mathscr{M}, d \rho_{\hbar; \mathsf{g}} ) \end{equation}  is the \textit{Fock space} $\mathscr{F}$ of $J$-holomorphic functions on $\mathscr{M}$ square integrable against the Segal-Bargmann weight $d \rho_{\hbar; \mathsf{g}}$, a Gaussian cylinder measure on $\mathscr{M}$ whose covariance is determined by the real inner product $\mathsf{g}$.  In this section we build the Fock spaces and the coherent states $\Upsilon_v ( \cdot | \hbar)$ inside them.

\subsubsection{Reproducing Kernel Hilbert Spaces} \label{subsubsecReproducingKernelHilbertSpaces}

\noindent Fock spaces are special cases of reproducing kernel Hilbert spaces, reviewed in \cite{AronszajnTheoryReproducingKernels}.

\begin{definition} A \underline{reproducing kernel Hilbert space} is a Hilbert space $(\mathscr{H}, \langle \cdot, \cdot \rangle)$ of complex-valued functions $F$ on a space $\mathscr{M}$ so that point-wise evaluation at $v \in \mathscr{M}$ \begin{equation} \textnormal{ev}_v : F \rightarrow F(v) \end{equation} \noindent is a continuous linear functional $\textnormal{ev}_v : \mathscr{H} \rightarrow \C$. \end{definition}

\begin{proposition}Any reproducing kernel Hilbert space has a \underline{reproducing kernel} \begin{equation} \Upsilon : \mathscr{M} \times \mathscr{M} \rightarrow \C \end{equation} \noindent so that for any $v \in \mathscr{M}$, $\Upsilon ( \cdot, v) \in \mathscr{H}$ and  \begin{equation} \textnormal{ev}_v F = \langle \Upsilon_v ( \cdot) , F \rangle \end{equation} \noindent is a realization of the evaluation functional $\textnormal{ev}_v$ by the inner product in $(\mathscr{H}, \langle \cdot , \cdot \rangle)$.
\end{proposition}

\noindent These are spaces of functions, not equivalence classes of functions.

\subsubsection{Weighted Bergman Spaces} \label{subsubsecWeightedBergmanSpaces}

\noindent The material here is also standard and can be found in many places, e.g. \cite{HallBook}.

\begin{definition} \label{WeightedBergmanSpaceDefinition}For $\rho$ a non-negative measure on a complex vector space $(\mathscr{M}, J)$, the \underline{weighted Bergman space} is the space of $J$-holomorphic functions
\begin{equation} L^2_{\textnormal{J-hol}} ( \mathscr{M}, d \rho) \end{equation}

\noindent square integrable with respect to the \underline{weight} $ \rho$ with inner product

\begin{equation} \label{BergmanInnerProduct} \langle F^{\textnormal{out}}, F^{\textnormal{in}} \rangle_{\rho} = \int_{\mathscr{M}} \overline{F^{\textnormal{out}} ( v) } F^{\textnormal{in}} ( v) d \rho (v) .\end{equation}

\noindent \end{definition}

\noindent For $\rho = \rho_{\star | \Omega}$ uniform on a planar domain $\Omega \subset \C$, the weighted Bergman spaces $L^2_{\textnormal{J-hol}} ( \mathscr{M},d \rho_{\star | \Omega})$ are simply called Bergman spaces.

\begin{proposition} If $\dim \mathscr{M} < \infty$, the weighted Bergman spaces $L_{\textnormal{J-hol}}^2(\mathscr{M}, d \rho)$ are closed in $L^2(\mathscr{M}, d \rho)$, hence Hilbert spaces complete with respect to \textnormal{(\ref{BergmanInnerProduct})}.
\end{proposition}

\begin{proposition} \label{WeightedBergmanSpaceBasis} In a $\sigma$-coordinate system on $\mathscr{M}$, the holomorphic polynomials of \textnormal{Definition [\ref{GeneralizedPolynomialDefinition}]} are a dense subspace of the weighted Bergman space $L^2_{\textnormal{J-hol}} ( \mathscr{M}, d \rho) $: \begin{equation} V_{\mu} = V_1^{d_1} V_2^{d_2} \cdots V_k^{d_k} \cdots \end{equation}

\noindent is a basis of $L^2_{\textnormal{J-hol}} ( \mathscr{M}, d \rho)$ for $\mu = (d_1, d_2,\ldots)$ with $d_k \in \N$ and almost all $d_k \equiv 0$. \end{proposition}

\begin{proposition} \label{WeightedBergmanSpacesAreRKHS} The Hilbert space completion of any weighted Bergman space $L^2_{\textnormal{J-hol}} ( \mathscr{M}, d \rho) $ as in \textnormal{Definition [\ref{WeightedBergmanSpaceDefinition}]} is a {reproducing kernel Hilbert space}.  \end{proposition}

\begin{proposition} \label{WeightedBergmanSpacesFormulaRK} For any weighted Bergman space $L^2_{\textnormal{J-hol}} ( \mathscr{M}, d \rho)$, its reproducing kernel $\Upsilon_{\rho}$ may be written in terms of the weight $d \rho$ by the $J$-holomorphic polynomials $P_{\mu}( \cdot | \rho)$ on $\mathscr{M}$ that are orthonormal with respect to the weight $d \rho$ explicitly as

\begin{equation} \Upsilon_{\rho} ( v^{\textnormal{out}}, v^{\textnormal{in}}) = \sum_{\mu} \overline{P_{\mu} ( v^{\textnormal{out}} | \rho )} P_{\mu} (v^{\textnormal{in}} | \rho)  \end{equation}

\noindent for $P_{\mu} ( \cdot | \rho)$ from Gram-Schmidt for $\langle \cdot, \cdot \rangle_{\rho}$ on the basis $V_{\mu}$ of \textnormal{Proposition [\ref{WeightedBergmanSpaceBasis}]}.\end{proposition}

\subsubsection{Segal-Bargmann Gaussian Weights} \label{subsubsecSegalBargmannGaussianWeights}

\noindent In a weighted Bergman space, $\rho$ is arbitrary and does not depend on the geometry of $(\mathscr{M}, J, \mathsf{g}, \omega)$.  We now choose a precious weight $\rho_{\hbar; \mathsf{g}}$ on $\mathscr{M}$ which does.  Informally, \begin{equation} \label{VagueLaw} d \rho_{\hbar; \mathsf{g}}(v) = e^{ - \frac{1}{ \hbar} \mathsf{g} (v , v)} d \text{Leb}_{\mathscr{M}}(v) \end{equation}

\noindent for $\hbar >0$ take the unnormalized Gaussian weight on $\mathscr{M}$ defined via the global K\"{a}hler potential of \textnormal{Definition [\ref{GlobalKahlerPotentialDefinition}]} and the Lebesgue measure on $\mathscr{M}$.  Unfortunately, Lebesgue measure does not exist if $\dim \mathscr{M} = \infty$, our case of interest.  Rigorously, 

\begin{definition} \label{SegalBargmannWeightDefinition} The \underline{Segal-Bargmann weight at scale $\hbar >0$} on a Hermitian affine space $(\mathscr{M}, J, \mathsf{g}, \omega)$ is the Gaussian cylinder set measure on $\mathscr{M}$ with respect to a $\sigma$-coordinate system in the sense of \textnormal{Definition [\ref{SigmaCoordinatesDefinition}]} specified by

\begin{equation} d\rho_{\hbar; \mathsf{g}} (v) = \prod_{k=1}^{\infty} e^{ - \frac{|V_k|^2}{ \hbar \sigma_k^2} } dV_k d\overline{V_k} \end{equation} the law of infinitely-many independent rotation-invariant complex Gaussian random variables indexed by $k=1,2,3,\ldots$ of mean $0$ and complex variance $\hbar \sigma_k^2$.
 \end{definition}
 
 \noindent Under certain conditions, e.g. in the Sazonov theorem, one may check that the Gaussian cylinder set measure $d \rho_{\hbar; \mathsf{g}}$ in Definition [\ref{SegalBargmannWeightDefinition}] extends to a well-defined measure on $\mathscr{M}$.  Such a verification would help give a coordinate-independent approach to the objects that appear below, and would also be necessary if one works with smooth quantizations of Hermitian affine spaces, but we do not pursue such a level of sophistication in our arguments below as we content ourselves with the humble regularity assumptions required of polynomial quantizations.
 
\begin{definition} \label{FockSpaceDefinition} The \underline{Fock space} $(\mathscr{F}_{\mathscr{M}, J}, \langle \cdot, \cdot \rangle_{\hbar, \mathsf{g}})$ is the Hilbert space completion of the weighted Bergman space $L^2_{\textnormal{J-hol}} ( \mathscr{M}, d \rho_{\hbar; \mathsf{g}})$ with the Segal-Bargmann weight $d \rho_{\hbar ; \mathsf{g}}$. \end{definition}

\noindent As may be more familiar to those working in combinatorics and representation theory,
\begin{proposition} The analytic presentation of Fock space in \textnormal{Defintiion [\ref{FockSpaceDefinition}]} is isometric to a symmetric tensor algebra, a ``bosonic Fock space.'' \end{proposition}
\begin{itemize}
\item \textit{Proof:} see Remark 4.4 in \cite{Janson}.
\end{itemize}

\subsubsection{Creation and Annihilation Operators}

\noindent As we have relied on $\sigma$-coordinates to construct the Segal-Bargmann weight $d \rho_{\hbar, \mathsf{g}}$, we must use these coordinates to construct quantum observables in $\mathscr{F}$.

\begin{definition} \label{CreationOperatorDefinition}The \underline{creation operator} $\widehat{\mathcal{V}}_{+k}$ is the unbounded operator on the Fock space $\mathscr{F}$ defined by multiplication by the holomorphic variable $V_k$. \end{definition}

\begin{definition} \label{AnnihilationOperatorDefinition}The \underline{annihilation operator} is the unbounded operator on the Fock space $\mathscr{F}$ defined by differentiation in the holomorphic variable $V_k$ \begin{equation} \widehat{\mathcal{V}}_{-k} ( \hbar) = \hbar \sigma_k^2 \frac{\partial}{\partial V_k} \end{equation}

\noindent scaled by $\hbar \sigma_k^2$ where $V_{k}, \overline{V_k}$ are a $\sigma$-basis of the classical phase space $(\mathscr{M}, J, \mathsf{g}, \omega)$.
\end{definition}
\begin{lemma} \label{MutualAdjointsLemma} The creation and annihilation operators $\widehat{\mathcal{V}}_{\pm k}$ are not self-adjoint on the Fock space $(\mathscr{F}_{\mathscr{M}, J}, \langle \cdot, \cdot \rangle_{\hbar, \mathsf{g}})$ but they are mutual adjoints 
\begin{equation} \widehat{\mathcal{V}}_{\pm k}^{\dagger} = \widehat{\mathcal{V}}_{\mp k} \end{equation} \end{lemma}

\begin{itemize}
\item \noindent \textit{Proof:} see section 14.4 in \cite{HallBook}, where Fock space is ``Segal-Bargmann space.''
\end{itemize}

\noindent The quantum commutation relations of the quantum observables $\widehat{\mathcal{V}}_{k}, \widehat{\mathcal{V}}_k$ are identical to the classical commutation relations of the $\sigma$-coordinates $V_k, \overline{V_k}$.

\begin{proposition} \label{SigmaCoordinatesAreCanonical} $\sigma$-coordinates $(V_k, \overline{V_k})$ on a Hermitian affine space $(\mathscr{M}, J, \mathsf{h})$ are \underline{$\sigma$-canonical coordinates} on $\mathscr{M}$, satisfying classical canonical commutation relations \begin{eqnarray} \{ V_{k} , V_{k'} \} &=& 0 \\ \{ \overline{V}_{k}, \overline{V_{k'}} \} &=& 0 \\  \{ \overline{V_{k}}, V_{k'} \}. &=& \textnormal{\textbf{i}} \sigma_k^2 \delta (k -k')
 \end{eqnarray}
 \end{proposition} 
 
 \begin{itemize}
 \item \textit{Proof:} use Corollary [\ref{HermitianVectorSpacePoissonBracketCorollary}]. $\square$
 \end{itemize}

\begin{proposition} \label{CreationAnnihilationCanonical} Creation and annihilation operators $\widehat{\mathcal{V}}_{\pm k}$ on Fock space $(\mathscr{F}, \langle \cdot, \cdot \rangle_{\hbar})$ satisfy quantum canonical commutation relations \begin{eqnarray} \ [ \widehat{\mathcal{V}}_{k}, \widehat{\mathcal{V}}_{k'} ] &=& 0 \\
\  [ \widehat{\mathcal{V}}_{-k}, \widehat{\mathcal{V}}_{-k'} ]&=& 0 \\
\ [ \widehat{\mathcal{V}}_{-k}, \widehat{\mathcal{V}}_{+k'} ] &=& \hbar \sigma_k^2 \delta(k- k') .\end{eqnarray}
\end{proposition}

\begin{itemize}
\item \textit{Proof:} use Leibniz's rule for the Wirtinger derivatives $\frac{\partial}{\partial V_k}$. $\square$
\end{itemize}

\begin{corollary} \label{OrthogonalBasisFockSpace} The $J$-holomorphic polynomials $V_{\mu}$ of \textnormal{Definition [\ref{GeneralizedPolynomialDefinition}]}, which are a basis of any weighted Bergman space by \textnormal{Proposition [\ref{WeightedBergmanSpaceBasis}]}, are actually an orthogonal basis of the Fock space whose norms may be evaluated explicitly

\begin{equation} || V_{\mu} ||_{\hbar}^2 = \prod_{k=1}^{K} ||V_k^{d_k} ||^2_{\hbar} = \prod_{k=1}^{K} (\hbar \sigma_k^2)^{d_k} d_k!  \end{equation}

 \end{corollary}

\begin{itemize}
\item \textit{Proof:} use Lemma [\ref{MutualAdjointsLemma}], Leibniz's Rule, and Proposition [\ref{CreationAnnihilationCanonical}]. $\square$ \end{itemize}

  \subsubsection{Coherent States}
  
  \noindent In Proposition [\ref{WeightedBergmanSpacesFormulaRK}], we wrote the reproducing kernel $\Upsilon_{\rho}( \cdot , \cdot)$ of a weighted Bergman space $L^2_{\textnormal{J-hol}} ( \mathscr{M}, d\rho)$ in terms of the orthonormal $J$-holomorphic polynomials $P_{\mu} ( \cdot | \rho)$.  For the Segal-Bargmann weight $\rho_{\hbar; \mathsf{g}}$ of Definition [\ref{SegalBargmannWeightDefinition}], Corollary [\ref{OrthogonalBasisFockSpace}] implies
  
  \begin{equation} P_{\mu} ( V_1, V_2, \ldots | \rho_{\hbar, \mathsf{g}} ) = \frac{ V_{\mu} }{ || V_{\mu}||_{\hbar}} \end{equation} 
    
\noindent which enables us to write the reproducing kernel of Fock space in closed form:
  \begin{corollary}  \label{FockReproducingKernelCorollary} The Fock space $\mathscr{F}$ of a Hermitian affine space $(\mathscr{M}, J, \mathsf{h})$ is a reproducing kernel Hilbert space with reproducing kernel (in $\sigma$-coordinates) given by\begin{equation} \label{ReproducingKernelExactFormula}\Upsilon_{\rho_{\hbar; \mathsf{g}}} ( v^{\textnormal{out}}, v^{\textnormal{in}} ) = e^{ \frac{1}{\hbar} \mathsf{h} (v^{\textnormal{out}}, v^{\textnormal{in}})} = \textnormal{exp} \Bigg (  \sum_{k=1}^{K} \frac{\overline{V_k^{\textnormal{out}}} V_k^{\textnormal{in}}} { \hbar \sigma_k^2} \Bigg ). \end{equation}
  \end{corollary} \noindent \begin{itemize}
\item \textit{Proof:} This follows from Propositions [\ref{WeightedBergmanSpacesAreRKHS}], [\ref{WeightedBergmanSpacesFormulaRK}], and Corollary [\ref{OrthogonalBasisFockSpace}]. $\square$
\end{itemize}

\noindent We can finally construct the \textit{coherent states $\Upsilon_v ( \cdot  | \hbar)$ around a classical state $v \in \mathscr{M}$}.
\begin{definition} For a fixed $v \in \mathscr{M}$ in a Hermitian affine space $(\mathscr{M}, J, \mathsf{g}, \omega)$ with Hermitian inner product $\mathsf{h}$, the \underline{coherent state $\Upsilon_v ( \cdot | \hbar)$ around a classical state $v$} is the quantum state in the Fock space $(\mathscr{F}_{\mathscr{M}, J}, \langle \cdot, \cdot \rangle_{\hbar})$ defined by \begin{equation} \Upsilon_{v} ( \cdot | \hbar) = \Upsilon_{d\rho_{\hbar; \mathsf{g}}} ( v, \cdot) \end{equation}

\noindent the reproducing kernel of \textnormal{Corollary [\ref{FockReproducingKernelCorollary}]}.
\end{definition}

\noindent Coherent states take center stage in both theoretical and applied aspects of both mathematics and physics so we cannot hope to give exhaustive references.  Coherent states in Hermitian affine spaces are also called ``Glauber states'', as they were later applied to great effect in quantum optics by Glauber \cite{Glauber1963}, see also \cite{GazeauCoherentBook}, though they were first discovered by Schr\"{o}dinger, playing a key role in the early history of quantum mechanics and Bohr's Correspondence Principle in light of Theorem [\ref{StrongBohrGlauberStates}] below.

\begin{definition} In a Hermitian affine space $(\mathscr{M}, J, \mathsf{g}, \omega)$, for any fixed $1 \leq k \leq K$ the $k$th \underline{classical harmonic oscillator} is the classical system defined by the Hamiltonian \begin{equation} \label{kCHOHamiltonian}T_{2,k}^{\uparrow} |_v = \overline{V_k} V_k = |V_k|^2 \end{equation}

\noindent for $v \in \mathscr{M}$ in the classical phase space $T_{2,k}^{\uparrow} : \mathscr{M} \rightarrow \R_{\geq 0}$.
\end{definition}

\begin{definition} In a Fock space $(\mathscr{F}_{(\mathscr{M}, J)}, \langle \cdot, \cdot \rangle_{\hbar, \mathsf{g}})$, for any fixed $1 \leq k \leq K$ the $k$th \underline{quantum harmonic oscillator} is the quantum system defined by the Hamiltonian \begin{equation} \label{kQHOHamiltonian}\widehat{\mathcal{T}}_{2,k}^{\uparrow}(\hbar) = \widehat{\mathcal{V}}_{+k} \widehat{\mathcal{V}}_{-k} +  \tfrac{1}{2} \hbar \sigma_k^2 \end{equation}
\end{definition}

\noindent Classically, the angular frequency of the $k$th oscillator is $\sigma_k^2$ from our definition of $\sigma$-coodinates, while in quantum mechanics $\tfrac{1}{2} \hbar \sigma_k^2$ is known as the zero-point energy.
\begin{theorem} \label{StrongBohrGlauberStates} For any $k \in \{1,2,\ldots, K\}$, coherent states $\Upsilon_v( \cdot | \hbar)$ are dynamically coherent for the $k$th classical and quantum harmonic oscillator flows in the sense of \textnormal{Definition [\ref{StrongBohr}]} for any quantization $Q$ so that $ \widehat{\mathcal{T}}_{2,k}^{\uparrow}(\hbar)$ is the quantization of $T_{2,k}^{\uparrow}$. \end{theorem}
\begin{itemize} \item \textit{Proof:} use the Baker-Campbell-Hausdorff formula. $\square$
\end{itemize}

\noindent Coherent states for arbitrary Lie groups were defined by Perelomov \cite{Perelomov}, which recovers the coherent states presented here for the case of the Heisenberg group.  From this point of view, the exact formula (\ref{ReproducingKernelExactFormula}) for the reproducing kernel of Fock spaces is attributed to the fact that (I) \textit{translations $Y_{v}$ by $v \in \mathscr{M}$} sending $\varphi \mapsto v +\varphi$ are isometries $Y_{v} : \mathscr{M} \rightarrow \mathscr{M}$, hence the non-compact K\"{a}hler manifold $\mathscr{M}$ is homogeneous, and (II) the translation action lifts to a projective unitary representation of the additive group $\mathscr{M}$ on Fock space via Weyl's displacement operators $\widehat{\mathcal{Y}}_{v}(\hbar)$ also known as vertex operators \cite{Kac0}.  That being said, we do not use vertex operators in this paper.  All we need below is:

\begin{lemma} \label{KeyCoherenceLemma} Annihilation operators and coherent states satisfy an \underline{exchange relation}
\begin{equation} \widehat{\mathcal{V}}_{-k} \Upsilon_{v} ( \cdot | \hbar) = \overline{V}_k \Upsilon_v ( \cdot | \hbar) \end{equation} \end{lemma}

\begin{itemize} \item \textit{Proof:} Follows immediately from Corollary [\ref{FockReproducingKernelCorollary}]. $\square$ \end{itemize}

\subsection{Canonical Quantizations} \label{subsecCanonicalQuantizations}

\noindent Our Definition [\ref{QuantizationDefinition}] of quantizations $Q$ of Poisson algebras $\mathsf{A} \subset C^{\infty} ( \mathscr{M}, \R)$ only required $Q$ to reflect the Poisson geometry of the classical phase space $(\mathscr{M}, \omega)$.  For Hermitian affine spaces $(\mathscr{M}, J, \mathsf{g}, \omega)$, we now define \textit{canonical quantizations}, a class of quantizations that are also required reflect the complex geometry of $\mathscr{M}$ whose definition depends on a fixed system of $\sigma$-coordinates.  We then construct a large family of canonical quantizations of the particular Poisson algebra $\mathsf{A}^{\textnormal{genpoly}}$ of generalized polynomials, the \textit{$\eta$-quantizations close to Wick quantization}.

\subsubsection{Canonical Quantizations of Poisson Algebras} 
\begin{definition} \label{CanonicalQuantizationDefinition} Fix $\sigma$-coordinates on a Hermitian affine space $(\mathscr{M}, J, \mathsf{g}, \omega)$ as in \textnormal{Definition [\ref{SigmaCoordinatesDefinition}]} being $\sigma$-canonical by \textnormal{Proposition [\ref{SigmaCoordinatesAreCanonical}]}.  A \underline{canonical quantization} of $\mathsf{A}_{\C} \subset C^{\infty}(\mathscr{M}, \C)$ including the coordinate functions is a map \begin{equation} Q(\hbar) : \mathsf{A}_{\C} \longrightarrow \mathfrak{gl}(\mathscr{F}) \end{equation}

\noindent so that the restriction of $Q(\hbar)$ to $\mathsf{A} = \mathsf{A}_{\C} \cap C^{\infty}(\mathscr{M}, \R)$ is a quantization in the sense of \textnormal{Definition [\ref{QuantizationDefinition}]} in the Fock space $(\mathscr{F}, \langle \cdot, \cdot \rangle_{\hbar})$ of $\mathscr{M}$ as in \textnormal{Definition [\ref{FockSpaceDefinition}]} so that 
the coordinate functions $V_k, \overline{V_k}$ are quantized to \begin{eqnarray} V_k^Q &=& \widehat{\mathcal{V}}_{+k}  \\
 \overline{V_k}^Q &=& \widehat{\mathcal{V}}_{-k}   \end{eqnarray}

 \noindent the creation and annihilation operators $\widehat{\mathcal{V}}_{\pm k}$ of \textnormal{Definitions [\ref{CreationOperatorDefinition}], [\ref{AnnihilationOperatorDefinition}]}.
 \end{definition}
 
 \noindent We now construct a large family of canonical quantizations of the Poisson algebra
 
 \begin{equation} \mathsf{A}_{\C}^{\textnormal{genpoly}} = \varprojlim \C [ V_1, \ldots,V_N, \overline{V_1}, \ldots, \overline{V_N}] \end{equation}
 
 \noindent of complex generalized polynomials from Definition [\ref{GeneralizedPolynomialDefinition}].

 \subsubsection{$\eta$-Quantizations Close To Wick Quantization}

 \begin{definition} The ring of \underline{non-commutative polynomials} $\C \langle \mathbf{V}_{\pm 1}, \ldots, \mathbf{V}_{\pm K} \rangle$ is the free unital associative $\C$-algebra generated by the $2K$ variables $\{ \mathbf{V}_{\pm k} \}_{k=1}^K$.  Similarly, one defines the ring of \underline{generalized non-commutative polynomials} to be series in \begin{equation} \varprojlim \C \langle \mathbf{V}_{\pm 1}, \ldots, \mathbf{V}_{\pm N} \rangle  \end{equation}
 
 \noindent which are ordinary non-commutative polynomials if all but finitely many $\mathbf{V}_k$ are $0$.

 \end{definition}

\begin{definition} \label{OrderingDefinition} An \underline{ordering} $\eta$ associates a generalized non-commutative polynomial $O^{\eta}$ to every generalized polynomial $O$, i.e. it is a section of the quotient 
\begin{equation} \label{AllowThemToCommute} \varprojlim \C \langle \mathbf{V}_{\pm 1}, \ldots, \mathbf{V}_{\pm N} \rangle \longrightarrow \varprojlim \C[ \mathbf{V}_{\pm 1} ,\ldots, \mathbf{V}_{\pm N} ] \end{equation}

\noindent by the ideal generated by all commutators.\end{definition}

\noindent Here are two distinguished orderings in the theory of canonical quantizations:

\begin{definition} \label{WickOrderingsDefinition}For a generalized polynomial $O$, the \underline{Wick ordering} $O^{\textnormal{Wick}} $ of $O$ is the ordering defined on monomials by
\begin{equation} (\mathbf{V}_1^{p_1} \cdots \mathbf{V}_K^{p_K} {\mathbf{V}_{-1}}^{{d_1}} \cdots {\mathbf{V}_{-K}}^{{d_K}} )^{\textnormal{Wick}}  \ \ =\ \  \mathbf{V}_{+1}^{p_1} \cdots\mathbf{V}_{+K}^{p_K} \mathbf{V}_{-1}^{d_1} \cdots \mathbf{V}_{-K}^{d_K} \end{equation}

\noindent specifying that all $\mathbf{V}_{-k}$ are to the right of all $\mathbf{V}_{+k}$, while \underline{anti-Wick ordering} $O^{\textnormal{anti-Wick}} $ of $O$ is the ordering defined on monomials by
\begin{equation} (\mathbf{V}_1^{p_1} \cdots \mathbf{V}_K^{p_K} {\mathbf{V}_{-1}}^{{d_1}} \cdots {\mathbf{V}_{-K}}^{{d_K}} )^{\textnormal{anti-Wick}}  \ \ =\ \  \mathbf{V}_{-1}^{d_1} \cdots \mathbf{V}_{-K}^{d_K}  \mathbf{V}_{+1}^{p_1} \cdots \mathbf{V}_{+K}^{p_K} \end{equation}
\noindent specifying that all $\mathbf{V}_{-k}$ are to the left of all $\mathbf{V}_{+k}$. \end{definition}

\noindent Consider the linear map \begin{equation} \mathcal{C}(\hbar) :  \varprojlim \C \langle \mathbf{V}_{\pm 1}, \ldots, \mathbf{V}_{\pm N} \rangle \longrightarrow \mathfrak{gl}(\mathscr{F}) \end{equation}

\noindent defined by sending the generators to the creation and annihilation operators
\begin{equation} \mathcal{C}: \mathbf{V}_{\pm k} \longrightarrow \widehat{\mathcal{V}}_{\pm k} ( \hbar). \end{equation}

\noindent We now define a condition on $\eta$ that we use in Theorem [\ref{EtaQuantizationsAreCanonicalQuantizations}] to guarantee that
\begin{equation} \label{CandidateForEtaQuantization} Q_{\eta}(\hbar) = \mathcal{C} ( \hbar) \circ \eta \end{equation}

\noindent is a canonical quantization of the Poisson algebra $\mathsf{A}^{\textnormal{genpoly}}$.

\begin{definition} \label{CloseToWickOrderingDefinition} An ordering $\eta$ is \underline{close to Wick ordering} if for every generalized polynomial $O \in \mathsf{A}_{\C}^{\textnormal{genpoly}}$ there exists a finite sequence $\{O_{\eta, g}\}_{g=0}^{\lfloor \frac{\deg O}{2} \rfloor} \in \mathsf{A}_{\C}^{\textnormal{genpoly}}$ of generalized polynomials of degree $\deg O_{\eta, g} = \deg O- 2g$ beginning with $O_{\eta, 0} = O$ so that the quantum observable \begin{equation} \label{InternalObservableAOEFormula} \widehat{O}^{\eta} ( \widehat{\mathcal{V}}_{\pm 1}, \widehat{\mathcal{V}}_{\pm 2}, \ldots) = \sum_{g=0}^{\lfloor \frac{\deg O}{2} \rfloor} \hbar^g \widehat{O}_{\eta, g}^{\textnormal{Wick}} ( \widehat{\mathcal{V}}_{\pm 1}, \widehat{\mathcal{V}}_{\pm 2} , \ldots) \end{equation}

\noindent which is $Q_{\eta}(\hbar)(O)$ has a finite $\hbar$ expansion in quantum observables $Q_{\textnormal{Wick}}(\hbar)(O_{\eta, g})$, the number of terms being $\lfloor \frac{ \deg O}{2} \rfloor$ at most half the degree of $O$.

\end{definition}

\noindent For general orderings $\eta$, one expects higher and higher order terms $O_{\eta,g}$ in the expansion (\ref{InternalObservableAOEFormula}) to depend in a more and more sensitive way on the choice of $\eta$, in contrast to the independence of $\eta$ of the lowest order term $O_{\eta, 0} = O$.  For special orderings $\eta$, one may have access to explicit formulas for the classical polynomial observables $O_{\eta,g}$ in closed form for all $g=0,1,2,3,\ldots$ in terms of the classical geometry of $O$ and $(\mathscr{M}, \omega)$.\\
\\
\noindent It is easy to check that if $\dim \mathscr{M} < \infty$, so that all generalized polynomials are just polynomials, every ordering $\eta$ is close to Wick ordering, simply by sending all annihilation operators to the right and using Leibniz's Rule. However, for classical phases spaces that are infintie-dimensional, not all orderings are close to Wick ordering.  In particular the anti-Wick ordering is not, as we now describe.  In light of formula (\ref{kCHOHamiltonian}), the total Hamiltonian of a system of infinitely-many independent classical harmonic oscillators indexed by $k=1,2,3,\ldots$ is the generalized polynomial
\begin{equation} T_2^{\uparrow} = \sum_{k=1}^{\infty} V_{k} \overline{V_k} = \sum_{k=1}^{\infty} \overline{V_k} V_k \end{equation}
\noindent in two numerically equivalent ways since $[V_k , \overline{V_k}]=0$.  However, if we formally substitute $V_k \mapsto \widehat{\mathcal{V}}_k$ and $V_{-k} \mapsto \widehat{\mathcal{V}}_{-k}$ in the second expression, which is the prescription of anti-Wick quantization, we get

\begin{equation} \label{KeyBadExample} \sum_{k=1}^{\infty} \widehat{\mathcal{V}}_{-k} \widehat{\mathcal{V}}_{+k}  = \sum_{k=1}^{\infty} \widehat{\mathcal{V}}_{+k} \widehat{\mathcal{V}}_{-k} + \hbar \sum_{k=1}^{\infty} \sigma_k^2 \end{equation}

\noindent after using the $\sigma$-canonical commutation relation $[\widehat{\mathcal{V}}_{-k}, \widehat{\mathcal{V}}_{+k}] = \hbar \sigma_k^2$.  Although (\ref{KeyBadExample}) is evaluation $\mathbf{V}_{\pm k} \rightarrow \widehat{\mathcal{V}}_{\pm k}$ of a well-defined generalized non-commutative polynomial \begin{equation} \sum_{k=1}^{\infty} \mathbf{V}_{-k} \mathbf{V}_{+k} \in \varprojlim_{N} \C \langle \mathbf{V}_{\pm 1}, \ldots, \mathbf{V}_{\pm N} \rangle 
\end{equation}

\noindent the infinite series $\sum_{k=1}^{\infty} \sigma_k^2$ typically diverges.  This is the famous divergence of the ground state energy of a system of infinitely-many quantum harmonic oscillators indexed by $k=1,2,3,\ldots$ each with Hamiltonian (\ref{kQHOHamiltonian}), the infinite energy of assembly of the vacuum of the free massless non-relativistic quantum field theory.  Probabilistically, this divergence simply reflects the fact that when one constructs infinitely-many i.i.d. random variables $\mathbb{O}_1, \mathbb{O}_2, \ldots$ carefully with measure theory, one doesn't always want to assign meaning to the infinite sum $\mathbb{O}_1 + \mathbb{O}_2 + \cdots$.

 \begin{theorem} \label{EtaQuantizationsAreCanonicalQuantizations} Any ordering $\eta$ close to Wick ordering in the sense of \textnormal{Definition [\ref{CloseToWickOrderingDefinition}]} defines a canonical quantization of the Poisson algebra $\mathsf{A}^{\textnormal{genpoly}}$ of generalized polynomials on a Hermitian affine space $(\mathscr{M}, J, \mathsf{g}, \omega)$ in the Fock space $(\mathscr{F}, \langle \cdot, \cdot \rangle_{\hbar})$ by
 \begin{equation} Q_{\eta}(\hbar) : O \longmapsto \widehat{O}^{\eta} ( \widehat{\mathcal{V}}_{\pm 1}, \widehat{\mathcal{V}}_{\pm 2}, \ldots) \end{equation}
 
 \noindent taking the $\eta$-ordered generalized non-commutative polynomial in the creation and annihilation operators $\widehat{\mathcal{V}}_{\pm k}$.  In particular, since $O_{\eta, 0} \equiv O$ from \textnormal{Definition [\ref{CloseToWickOrderingDefinition}]}, 
 
 \begin{equation} \widehat{O}^{\eta}(\hbar) \sim \widehat{O}^{\textnormal{Wick}}(\hbar) \end{equation}
 
 \noindent the resulting $\eta$-quantization agrees with Wick quantization to leading order in $\hbar$.
  \end{theorem}
\begin{itemize}
\item \textit{Proof:} Follows immediately from Definition [\ref{CloseToWickOrderingDefinition}] and the well-known fact \cite{FollandBook, HallBook, MartinezBook} that Wick quantization is a canonical quantization. $\square$
\end{itemize}

\noindent One can relate the bidifferentials $B_g^{\eta}$ of $\eta$-quantizations close to Wick quantization and the sequences $O_{\eta, g}$ of observables in Definition [\ref{CloseToWickOrderingDefinition}].\\
\\
\noindent We emphasize that the asymptotic expansions required by the Definition [\ref{QuantizationDefinition}] of quantizations of Poisson subalgebras $\mathsf{A} \subset C^{\infty}(\mathscr{M}, \R)$ are in the case of Theorem [\ref{EtaQuantizationsAreCanonicalQuantizations}] actually finite polynomial expansions in $\hbar$ because we take $\mathsf{A}^{\textnormal{genpoly}}$ the simpler Poisson subalgebra of generalized polynomials in the $\sigma$-coordinates.\\
\\
\noindent Wick quantization differs in a crucial way from those $\eta$-Quantizations close to it:

\begin{proposition} \label{B1WickIsWeldingOperator} For Wick quantization, the first bidifferential operator $B_1^{\textnormal{Wick}}$ is the welding operator $\mathscr{B}_1$ of \textnormal{Definition [\ref{WeldingOperatorDefinition}]} \begin{equation} B_1^{\textnormal{Wick}} = \mathscr{B}_1.\end{equation} \end{proposition}

\noindent This feature is quite important for us in our probabilistic studies below, and also motivated Karabegov to define a different generalization of Wick quantization:

\begin{definition} A quantization $Q$ of a K\"{a}hler manifold $(\mathscr{M}, J, \mathsf{g}, \omega)$ is said to have \underline{``separation of variables''} if the corresponding bidifferential operators $B_g^{Q}$ are $J$-anti-holomorphic in their first argument, $J$-holomorphic in their second argument, and the first bidifferential operator $B_1^{Q}$ is the welding operator \begin{equation} B_1^{Q} = \mathscr{B}_1\end{equation} given intrinsically by the inverse Hermitian metric.
\end{definition}

\noindent For finite-dimensional compact K\"{a}hler $\mathscr{M}$, deformation quantizations with separation of variables have been classified in \cite{Karabegov}.

\pagebreak

\subsection{Random Value of Quantized Observable in Coherent State} \label{subsecTheModelAppearsColumn2}

\noindent We now specialize Definition [\ref{TheModelColumn1Definition}] to canonical quantizations and to coherent states.
\begin{definition} \label{TheModelColumn2Definition} Under the following structural assumptions,
\begin{itemize}
\item \textnormal{\textcolor{gray}{[Structure of State]}} $\Psi = \Upsilon_v ( \cdot | \hbar)$ is a coherent state
\item \textnormal{\textcolor{gray}{[Structure of Observable]}} $O$ is a classical observable in a Poisson $\mathsf{A} \subset C^{\infty}(\mathscr{M}, \R)$
\item \textnormal{\textcolor{gray}{[Structure of Quantization]}} $Q$ is a canonical quantization of the Poisson $\mathsf{A}$
\end{itemize}

\noindent define $\widehat{O}^Q(\hbar)|_{\Upsilon_v( \cdot | \hbar)}$ the random value of the quantized observable $\widehat{O}^Q(\hbar)$ in the coherent state $\Upsilon_v ( \cdot | \hbar) $ around $v$ using \textnormal{Definition [\ref{TheModelColumn1Definition}]}.
\end{definition}

\subsubsection{Born's Rule: Expectations as Berezin Covariant Symbols}

\noindent Correlation functions of random values $\widehat{\mathcal{O}}(\hbar)|_{\Upsilon_v(\cdot | \hbar)}$ of quantum observables $\widehat{\mathcal{O}}(\hbar)$ in coherent states $\Upsilon_v ( \cdot | \hbar)$, in particular those $\widehat{O}^Q ( \hbar)|_{\Upsilon_v ( \cdot | \hbar)}$ in Definition [\ref{TheModelColumn2Definition}], are well-known objects in semiclassical analysis following the work of Berezin \cite{Berezin1972, Berezin1974, BerezinShubin}:

\begin{definition} \label{BerezinCovariantSymbolDefinition} For any quantum observable $\widehat{\mathcal{O}}(\hbar)$, not necessarily a quantization $\widehat{O}^Q(\hbar)$ of a classical observable $O$, the expected value of the random value $\widehat{O}(\hbar)|_{\Upsilon_v ( \cdot | \hbar)}$ of any quantum observable $\widehat{\mathcal{O}}(\hbar)$ in the coherent state $\Upsilon_v ( \cdot | \hbar) $ around $v \in \mathscr{M}$ is

\begin{equation} \label{BerezinCovariantSymbolFormula} \mathbb{E} [ \widehat{O}(\hbar)|_{\Upsilon_v ( \cdot | \hbar)}] = \frac{ \langle \Upsilon_v ( \cdot | \hbar) | \widehat{\mathcal{O}}(\hbar) | \Upsilon_v ( \cdot | \hbar) \rangle } { \langle \Upsilon_v ( \cdot | \hbar)  | \Upsilon_v ( \cdot | \hbar) \rangle } \end{equation}

\noindent is called the \underline{Berezin covariant symbol} of the operator $\widehat{\mathcal{O}}(\hbar)$. \end{definition}

\noindent In the definition of the Berezin covariant symbol we have not said that $\widehat{\mathcal{O}}(\hbar)$ is realized as the quantization $\widehat{\mathcal{O}}(\hbar) = \widehat{\mathcal{O}}^Q(\hbar)$ of some classical observable $O \in \mathscr{M}$ according to a quantization $Q$.  In the special case that the quantum observable $\widehat{\mathcal{O}}(\hbar) = \widehat{O}^{\textnormal{anti-Wick}}(\hbar)$ is the anti-Wick quantization of a classical observable $O$, the Berezin covariant symbol of $ \widehat{O}^{\textnormal{anti-Wick}}(\hbar)$ is called the \textit{Berezin transform}.  Indeed, anti-Wick quantization of Hermitian affine spaces is a special case of Berezin-Toeplitz quantization of K\"{a}hler manifolds, in which smooth $O$ can be quantized rigorously by taking $\widehat{O}^{\textnormal{anti-Wick}}(\hbar)$ to be a generalized Toeplitz operator with contravariant symbol $O$ \cite{DeMonvelGuillemin, Englis, HallBook}.  We will not use the anti-Wick quantization in this paper.\\
\\
\noindent In the literature, some assume that the Berezin covariant symbol (\ref{BerezinCovariantSymbolFormula}) is defined only for bounded operators, but we allow unbounded self-adjoint $\widehat{\mathcal{O}}(\hbar)$.  In the next section, we give regularity hypothesis on the classical state $v \in \mathscr{M}$ and the quantum observable $\widehat{\mathcal{O}}(\hbar)$ which ensure that the Berezin covariant symbol (\ref{BerezinCovariantSymbolFormula}) is finite.
\pagebreak

\subsubsection{Finite Moments from Regularity Assumptions} \label{FiniteMomentsFromRegularityAssumptionsColumn2}

\begin{proposition} \label{TheModelColumn2RegularityAssumptions} All moments of the random variable $\widehat{O}^{Q}(\hbar)|_{\Upsilon_v ( \cdot | \hbar)}$ in \textnormal{Definition [\ref{TheModelColumn2Definition}]} are finite if the following regularity assumptions are satisfied:

\begin{enumerate}
\item \textcolor{gray}{\textnormal{[Regularity of State]}} In a $\sigma$-coordinate system, the classical state $v \in \mathscr{M}$ has $V_k \equiv 0$ for almost every $1 \leq k \leq K$.
\item \textcolor{gray}{\textnormal{[Regularity of Observable]}} In a $\sigma$-coordinate system, the classical observable $O$ is a generalized polynomial in $\{V_k, \overline{V_k}\}_{k=1}^K$.
\item \textcolor{gray}{\textnormal{[Regularity of Quantization]}} The canonical quantization $Q$ is an $\eta$-quantization $Q_{\eta}$ of the Poisson algebra $\mathsf{A}^{\textnormal{genpoly}}$ of generalized polynomials that is close to Wick quantization of the type constructed in \textnormal{Theorem [\ref{EtaQuantizationsAreCanonicalQuantizations}]}.
\end{enumerate}
\noindent To reflect the third assumption $Q = Q_{\eta}$, we may instead write $\widehat{O}^{\eta}( \hbar)|_{\Upsilon_v ( \cdot | \hbar)}$.
\end{proposition}

\begin{itemize} \item \textit{Proof:} follows from our proof of Theorem [\ref{AOEColumn2}] below. $\square$ \end{itemize}

\noindent Our Theorem [\ref{AOEColumn2}], a polynomial asymptotic $\hbar$-expansion of moments which implies Proposition [\ref{TheModelColumn2RegularityAssumptions}],  relies on the following identity for Berezin covariant symbols of Wick quantized observables which holds for finite $\hbar >0$ and which makes use of all of our regularity assumptions.

\begin{lemma} \label{CovariantBerezinSymbolOfWickKeyLemma} Under the regularity assumptions of \textnormal{Proposition [\ref{TheModelColumn2RegularityAssumptions}]}, for 
$\hbar >0$ fixed, the Berezin covariant symbol of a Wick quantized generalized polynomial $O$ is independent of $\hbar$ and given by

\begin{equation} \mathbb{E}\Big [ \widehat{O}^{\textnormal{Wick}}(\hbar) \Big |_{\Upsilon_v ( \cdot | \hbar)} \Big ] = O|_v \end{equation}

\noindent the value $O|_v$ of the classical observable $O$ at the classical state $v \in \mathscr{M}$. \end{lemma}

\noindent \textit{Proof:} Our proof is in two steps.  We reduce to the case of $O$ polynomial in Step $1$ using our regularity assumptions, and derive the exact formula using the special properties of coherent states in Fock spaces of Hermitian affine spaces in Step $2$.
\begin{itemize}
\item \textit{Step 1:} Our regularity assumption on the classical state $v  = (V_k, \overline{V_k})\in \mathscr{M}$ says that there exists some $K(v) < \infty$ so that $V_{k } \equiv 0$ for all $k \geq K(v)$.  Since $O$ is real valued, all terms in $\widehat{O}^{\textnormal{Wick}}(\hbar)$ contain annihilation operators, and since the only annihilation operators $\widehat{\mathcal{V}}_{-k}$ acting non-trivially on $\Upsilon_v ( \cdot | \hbar)$ are $\widehat{\mathcal{V}}_{-1}, \ldots, \widehat{\mathcal{V}}_{-K}$ and $\widehat{O}^{\textnormal{Wick}}(\hbar)$ is Wick ordered, we may replace the generalized polynomial $O$ by the truncated ordinary polynomial $O(V_1, \ldots, V_{K(v)} , \overline{V_1}, \ldots, \overline{V}_{K(v)})$. 
\item \textit{Step 2:} Since without loss of generality $O$ is polynomial, the desired identity follows from Lemma [\ref{MutualAdjointsLemma}] and Lemma [\ref{KeyCoherenceLemma}]. $\square$
\end{itemize}

\pagebreak

\subsection{Asymptotic Expansion of Cumulants} \label{subsecAOEColumn2}

\noindent Let $W_n ( \mathbb{O})$ denote the $n$th cumulant of a random variable $\mathbb{O}$.

\begin{theorem} \label{AOEColumn2} For the random variable $\widehat{O}^{\eta}(\hbar)|_{\Upsilon_v(\cdot | \hbar)}$ of \textnormal{Definiton [\ref{TheModelColumn2Definition}]} satisfying the regularity assumptions of \textnormal{Proposition [\ref{TheModelColumn2RegularityAssumptions}]}, its cumulants are polynomials in $\hbar$ \begin{equation} \label{AOEColumn2Formula} W_n \Big ( \widehat{O}^{\eta} ( \hbar) \big |_{\Upsilon_v ( \cdot | \hbar) } \Big ) = \sum_{g =0}^{ \lfloor \frac{ n \deg O }{2} \rfloor} \hbar^{n-1+g} W_{\eta, n,g}(O)|_v \end{equation}

\noindent with order of vanishing at least $n-1$ at $\hbar = 0$ and whose coefficients $W_{\eta, n,g}(O)|_v$ are themselves polynomials in the $\sigma$-coordinates $\{V_k , \overline{V_k}\}_{k=1}^{K}$ of $v \in \mathscr{M}$.  Moreover, the leading order coefficient $W_{\eta, n,0}(O)|_v$ is independent of $\eta$.

\end{theorem}

\noindent \textit{Proof:} We prove (\ref{AOEColumn2Formula}) is a polynomial in $\hbar$ whose coefficients are polynomials in the $\sigma$-coordinates of $v$ in Step 1, its independence of $\eta$ to leading order in Step 2, and that its order of vanishing is $\geq n-1$ in Step 3.
\begin{itemize} 
\item \textit{Step 1:} The $n$th cumulant is a polynomial in the $n'$th moments for $0 \leq n' \leq n$, it suffices to prove that the $n'$th moment is a polynomial in $\hbar$ whose coefficients are polynomials in the $\sigma$-coordinates of $v$.  The $n'$th moment is the Berezin covariant symbol of $\widehat{\mathcal{O}}(\hbar) = (\widehat{{O}}^{\eta}(\hbar))^{n'}$ as in Definiton [\ref{BerezinCovariantSymbolDefinition}]:\begin{equation} \label{SubstituteBerezin} \mathbb{E} \Big [ \big ( \widehat{O}^{\eta} ( \hbar) \big |_{\Upsilon_v ( \cdot | \hbar) }\big ) ^{n'} \Big ] = \frac{ \langle \Upsilon_v ( \cdot | \hbar) | \big ( {O}^{\eta}(\widehat{\mathcal{V}}_{\pm 1}, \ldots, \widehat{\mathcal{V}}_{\pm K}) \big )^n | \Upsilon_v ( \cdot | \hbar) \rangle }{ \langle \Upsilon_v ( \cdot | \hbar) | \Upsilon_v ( \cdot | \hbar) \rangle} \end{equation}

\noindent Because our quantization is defined by $\eta$-ordering close to Wick ordering in the sense of Definition [\ref{CloseToWickOrderingDefinition}], there exist generalized polynomials $O_{\eta, g}$ so that \begin{equation} \label{SubstituteMeeee}\widehat{O}^{\eta}(\hbar) = \sum_{g=0}^{\lfloor \frac{ \deg O}{ 2} \rfloor} \hbar^g (\widehat{O}_{\eta, g})^{\textnormal{Wick}} (\hbar). \end{equation}
 \noindent Next, consider the Wick ordering of a product of $n'$ terms of such type: \begin{equation} \label{SubstituteMeTOOO} (\widehat{O}_{\eta, g_1})^{\textnormal{Wick}} (\hbar) \cdots (\widehat{O}_{\eta, g_{n'}})^{\textnormal{Wick}} (\hbar) = \sum_{g' = 0}^{l_*} \hbar^{g'} (\widehat{O}_{\eta, g_1, \ldots, g_n, g'})^{\textnormal{Wick}}(\hbar) \end{equation}

\noindent This is a finite sum up to $l_* = \lfloor \frac{n \deg O}{2} - g_1 - \cdots - g_n - g' \rfloor$.  The expansion comes from the asymptotic expansion axiom in the Definition [\ref{QuantizationDefinition}] of quantization satisfied by Wick quantization, hence $O_{\eta, g_1, \ldots, g_n, g'}$ is itself a polynomial defined in a complicated way through the bidifferential operators $B_g^{\textnormal{Wick}}$ in the phase space star product, together with the $O_{\eta, g_i}$ which themselves depend on $\eta$.  As $O_{\eta, g_i}$ is a generalized polynomial of degree $O - 2g_i$, $\deg O_{\eta, g_1, \ldots, g_n, g'} = n \deg O - 2 ( g_1 + \ldots + g_n + g')$.  Substituting formula (\ref{SubstituteMeeee}) $n'$ times in formula (\ref{SubstituteBerezin}), using (\ref{SubstituteMeTOOO}) and finally in a crucial way the key Lemma [\ref{CovariantBerezinSymbolOfWickKeyLemma}], we get
\begin{equation} \mathbb{E} \Big [ \big ( \widehat{O}^{\eta} ( \hbar) \big |_{\Upsilon_v ( \cdot | \hbar) }\big ) ^{n'} \Big ]  = \sum_{g_1, \ldots, g_n, g'} \hbar^{g_1 + \cdots + g_n + g'} ({O}_{\eta, g_1, \ldots, g_n, g'} ) \big |_v .\end{equation}

\pagebreak

\item \textit{Step 2:} Our regularity assumption on the quantization says that our quantization is defined by an $\eta$-ordering close to Wick ordering as in Definition [\ref{CloseToWickOrderingDefinition}], hence all of our $\eta$-quantized observables agree to leading order with their Wick quantizations, hence the leading order coefficient $W_{\eta, n,0}(O)$ in our $\hbar$-expansion of cumulants (\ref{AOEColumn2Formula}) is independent of $\eta$.
\item \textit{Step 3:} Finally, we verify that the order of vanishing at $\hbar = 0$ of cumulants \begin{equation} W_{n} \big ( \widehat{ O}^{\eta}(\hbar)|_{\Upsilon_v ( \cdot | \hbar)} \big ) \sim W_{n,0}(O)|_v \hbar^{n-1} \end{equation} is at least $n -1$.  By the argument in Step 1 in our proof of Lemma [\ref{CovariantBerezinSymbolOfWickKeyLemma}], it suffices to assume that $O$ is polynomial.  It is important that this step does not use special properties of coherent states, so for the remainder of the proof assume $\Psi$ is an arbitrary quantum state and $O$ polynomial in $\sigma$-coordinates $\{V_k, \overline{V_k}\}_{k=1}^{K}.$  For random value $\widehat{O}^{\eta}(\hbar)|_{\Psi}$ of the quantum observable $\widehat{O}^{\eta}( \widehat{\mathcal{V}}_{\pm 1}, \ldots, \widehat{\mathcal{V}}_{\pm K}) $ in the quantum state $\Psi$, the well-known interpretation of cumulants as ``connected correlators'' is explicitly realized by the ``connections'' made by ``pairings'' \begin{equation} \label{PairingsHere} [ \widehat{\mathcal{V}}_{-k}, \widehat{\mathcal{V}}_{k'}] = \hbar \sigma_k^2 \delta(k-k') \end{equation} \noindent between $\sigma$-canonically conjugate operators in our $\hbar$-expansions in Steps 1-3.  Assuming $|| \Psi||=1$, the variance $W_2  ( \widehat{O}^{\eta}(\hbar)|_{\Psi} , \widehat{O}^{\eta} (\hbar)|_{\Psi}  ) $ is \begin{equation} \label{ThisVariance} \langle \Psi |  {O}^{\eta}(\widehat{\mathcal{V}}_{\pm k}) {O}^{\eta}(\widehat{\mathcal{V}}_{\pm k})  | \Psi \rangle - \langle \Psi | {O}^{\eta}(\widehat{\mathcal{V}}_{\pm k}) | \Psi \rangle \langle \Psi |{O}^{\eta}(\widehat{\mathcal{V}}_{\pm k})| \Psi \rangle .\end{equation}

\noindent The only contributions to the $\hbar$-expansion of (\ref{ThisVariance}) are terms in which inside $\langle \Psi |  {O}_1^{\eta}(\widehat{\mathcal{V}}_{\pm k}) {O_2}^{\eta}(\widehat{\mathcal{V}}_{\pm k})  | \Psi \rangle$ at least one annihilation operator $\widehat{\mathcal{V}}_{-k}$ from the left copy of $O$ pairs with a creation operator $\widehat{\mathcal{V}}_{+k}$ from the right copy of $O$.  More generally, if for $n$ copies of $O$ we draw $n$ ``sites,'' then the $n$th cumulant consists the terms in the $\hbar$-expansion in which the $n$ sites are connected by edges formed by pairings (\ref{PairingsHere}) which can only go from left to right (the Leibniz rule for $\widehat{\mathcal{V}}_{\pm k}$).  The desired order of vanishing thus follows from the basic fact, drawn for $n=4$,

  \begin{figure}[htb]
\centering
\includegraphics[width=0.90 \textwidth]{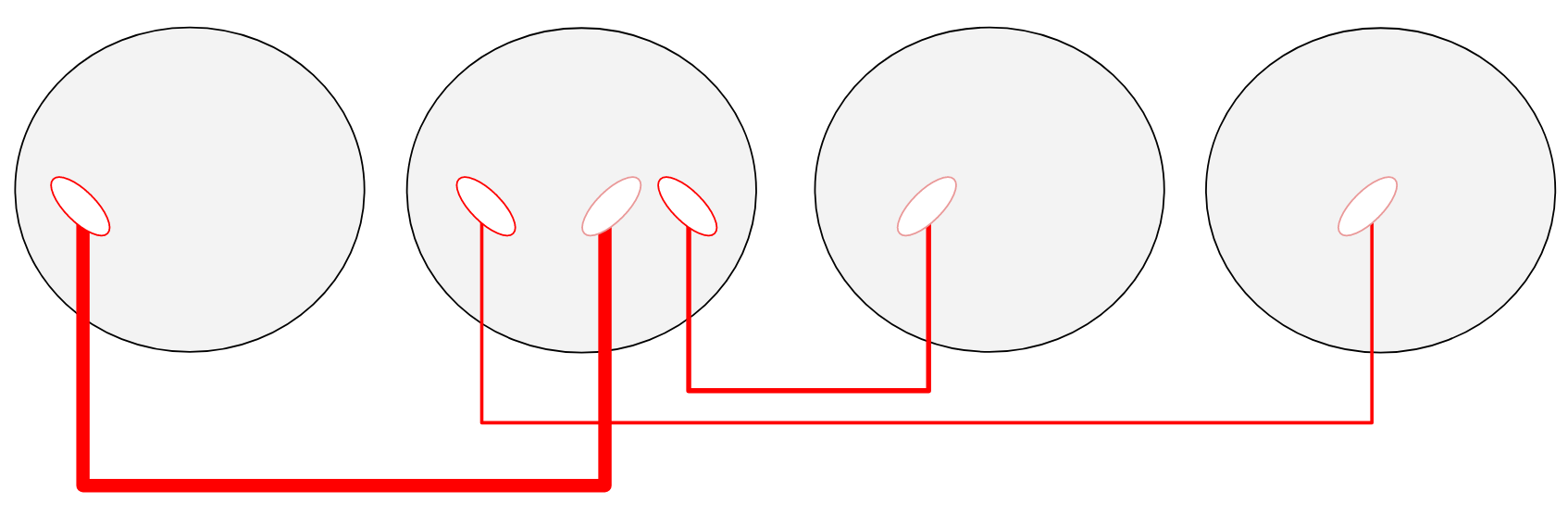}
\end{figure}

\noindent that it requires at least $n-1$ pairings to connect $n$ sites. $\square$

\end{itemize}

\pagebreak

\subsection{Concentration of Measure in Static Semi-Classical Limits} \label{subsecLLNColumn2}

\noindent We now show that coherent states $\Upsilon_v(\cdot | \hbar)$ around a classical state $v \in \mathscr{M}$ are quasi-classical around $v$ as in Definition [\ref{StaticBohr}], namely they obey the Static Version of Bohr's Correspondence Principle for States.  Although coherent states $\Upsilon_v ( \cdot | \hbar)$ are dynamically coherent with respect to the $k$th classical and quantum oscillator flows for every $k=1,2,3,\ldots$ by Theorem [\ref{StrongBohrGlauberStates}], our next result is for general observables.
\begin{theorem} \label{LLNColumn2} For the random variable $\widehat{O}^{\eta}(\hbar)|_{\Upsilon_v(\cdot | \hbar)}$ of \textnormal{Definiton [\ref{TheModelColumn2Definition}]} satisfying the regularity assumptions of \textnormal{Proposition [\ref{TheModelColumn2RegularityAssumptions}]}, in the semi-classical limit $\hbar \rightarrow 0$
\begin{equation} \label{LLNColumn2Formula} \widehat{O}^{\eta} ( \hbar) |_{\Upsilon_v ( \cdot | \hbar) } \rightarrow O|_v \end{equation}

\noindent the random value $\widehat{O}^{\eta}(\hbar)|_{\Upsilon_v ( \cdot | \hbar)}$ of the $\eta$-quantized observable $\widehat{O}^{\eta} (\hbar)$ in a coherent state $\Upsilon_v( \cdot | \hbar)$ around $v$ concentrates on the non-random value $O|_v$ of the classical observable $O$ at $v$ which is independent of the ordering $\eta$ defining the quantization. \end{theorem}

\noindent \textit{Proof:} We prove (\ref{LLNColumn2Formula}) in two steps.  In Step 1, we show why the concentration of measure occurs.  In Step 2, we explain why the limit $O|_v$ has the form it does. 
\begin{itemize}
\item \textit{Step 1:} By Theorem [\ref{AOEColumn2}] for $n=1$, the mean is \begin{equation} W_1 \Big ( \widehat{O}^{\eta}(\hbar) |_{\Upsilon_v ( \cdot | \hbar)} \Big ) = \sum_{g=0}^{\lfloor \frac{\deg O}{ 2} \rfloor} \hbar^{1-1+g} W_{\eta,1,g}(O)|_v \end{equation} \noindent hence in the limit as $\hbar \rightarrow 0$ the mean converges to \begin{equation}W_1 \Big ( \widehat{O}^{\eta}(\hbar) |_{\Upsilon_v ( \cdot | \hbar)} \Big )  \rightarrow W_{1,0}(O)|_v. \end{equation}

\noindent and Theorem [\ref{AOEColumn2}] says $W_{1,0}(O)|_v$ is independent of $\eta$.  By Theorem [\ref{AOEColumn2}] for $n=2$, the variance has order of vanishing at least $2-1 = 1$, so as $\hbar \rightarrow 0$
\begin{equation} W_2 \Big ( \widehat{O}^{\eta}(\hbar) |_{\Upsilon_v ( \cdot | \hbar)} \Big )  \rightarrow 0 \end{equation}

\noindent the variance vanishes, implying concentration of measure occurs at $W_{1,0}(O)|_v$.
\item \textit{Step 2:} The concentration of measure occurs at the classical average $O|_v$ \begin{eqnarray} W_{1,0}(O)|_v &=& \lim_{\hbar \rightarrow 0} \mathbb{E} \Big [ \widehat{O}^{\eta}(\hbar) \Big |_{\Upsilon_v ( \cdot | \hbar) } \Big ] \\ &=&  \lim_{\hbar \rightarrow 0} \mathbb{E} \Big [\widehat{O}^{\textnormal{Wick}}(\hbar) \Big |_{\Upsilon_v ( \cdot | \hbar) } \Big ]  \\ &=& O|_v. \end{eqnarray}

\noindent The first equality is a definition, the second equality uses $\widehat{O}^{\eta}(\hbar) \sim \widehat{O}^{\textnormal{Wick}}(\hbar)$ to leading order in $\hbar$ since the ordering $\eta$ is close to Wick ordering as in Definition [\ref{CloseToWickOrderingDefinition}], and the third equality follows from Lemma [\ref{CovariantBerezinSymbolOfWickKeyLemma}]. $\square$
\end{itemize}

\pagebreak

\subsection{Gaussian Variables as Quantum Fluctuations} \label{subsecCLTColumn2}

\noindent We now determine next-order corrections to Theorem [\ref{LLNColumn2}], the Static Version of Bohr's Correspondence Principle for coherent states.  The relevance of such corrections in dynamical versions of Bohr's Correspondence Principle is discussed in section \textbf{[\ref{subsecDynamicalBohrCP}]}, but we emphasize that the following theorem does not depend on the integrable or chaotic nature of the Hamiltonian flow generated by the classical observable $O$.

\begin{theorem} \label{CLTColumn2} For the random variable $\widehat{O}^{\eta}(\hbar)|_{\Upsilon_v(\cdot | \hbar)}$ of \textnormal{Definiton [\ref{TheModelColumn2Definition}]} satisfying the regularity assumptions of \textnormal{Proposition [\ref{TheModelColumn2RegularityAssumptions}]}, quantum fluctuations of $\widehat{O}^{\eta}(\hbar)|_{\Upsilon_v ( \cdot | \hbar)}$ around its classical limit $O|_v$ occur at scale $\hbar^{1/2}$ and converge in distribution to \begin{equation} \label{CLTColumn2Formula} \frac{1}{\hbar^{1/2}} \Bigg ( \widehat{O}^{\eta} (\hbar)|_{\Upsilon_v ( \cdot | \hbar)} - O|_v \Bigg ) \rightarrow \mathbb{G}(O)|_v .\end{equation}

\noindent a Gaussian random variable $\mathbb{G}(O)|_v $ independent of ordering $\eta$ of mean $0$ and variance $|| (\nabla O)|_v ||_{\mathsf{g}}^2$ given by the square norm of the gradient of $O$ at $v$.
 \end{theorem}

\noindent \textit{Proof:} Our proof is in two steps.  In Step 1, we show why convergence to a Gaussian occurs.  In Step 2, we explain why the Gaussian $\mathbb{G}(O)|_v$ has the form it does.

\begin{itemize}
\item \textit{Step 1:} By Theorem [\ref{AOEColumn2}] for $n=1$, the mean of the centered variable \begin{equation} W_1 \Bigg ( \frac{1}{\hbar^{1/2}} \Bigg ( \widehat{O}^{\eta} (\hbar)|_{\Upsilon_v ( \cdot | \hbar)} - O|_v \Bigg ) \Bigg ) = \frac{1}{\hbar^{1/2}} \sum_{g=1}^{\lfloor \frac{ \deg O}{2} \rfloor} \hbar^{1-1+g} W_{\eta,1,g} (O)|_v \longrightarrow 0 \ \ \ \ \end{equation}

\noindent is not necessarily $0$ but has order of vanishing $1 - \frac{1}{2} = + \frac{1}{2}$ so vanishes as $\hbar \rightarrow 0$.
 By Theorem [\ref{AOEColumn2}] for $n \geq 2$ and scaling properties of cumulants, \begin{eqnarray} W_n \Bigg ( \frac{1}{\hbar^{1/2}} \Bigg ( \widehat{O}^{\eta} (\hbar)|_{\Upsilon_v ( \cdot | \hbar)} - O|_v \Bigg ) \Bigg ) &=& \Bigg ( \frac{1}{ \hbar^{1/2}} \Bigg)^{n} W_n \Bigg (  \widehat{O}^{\eta} (\hbar)|_{\Upsilon_v ( \cdot | \hbar)}\Bigg ) \ \ \ \ \ \ \ \ \\ &=& \Bigg ( \frac{1}{\hbar^{1/2}} \Bigg )^n \sum_{g=0}^{\lfloor \frac{ n \deg O}{2} \rfloor} \hbar^{n-1+g} W_{\eta, n, g} (O)|_v \ \ \ \ \ \ \ \ 
\end{eqnarray}
\noindent has order of vanishing $\geq \tfrac{n}{2} -1$, so as $\hbar \rightarrow 0$ the variance converges to 

\begin{equation} W_2 \Bigg ( \frac{1}{\hbar^{1/2}} \Bigg ( \widehat{O}^{\eta} (\hbar)|_{\Upsilon_v ( \cdot | \hbar)} - O|_v \Bigg ) \Bigg )  \rightarrow W_{2,0}(O)|_v \end{equation}

\noindent independent of $\eta$ by Theorem [\ref{AOEColumn2}], while all higher cumulants $n \geq 3$ vanish

\begin{equation} W_n \Bigg ( \frac{1}{\hbar^{1/2}} \Bigg ( \widehat{O}^{Q} (\hbar)|_{\Upsilon_v ( \cdot | \hbar)} - O|_v \Bigg ) \Bigg ) \rightarrow 0. \end{equation}

\noindent which proves the desired convergence in distribution (\ref{CLTColumn2Formula}) of the quantum fluctuations at scale $\hbar^{1/2}$ to a Gaussian random variable $\mathbb{G}(O)|_v$ of mean $0$ and variance $W_{2,0}(O)|_v$ independent of $\eta$. 

\pagebreak

\item \textit{Step 2:} The limiting variance $W_{2,0}(O)|_v = || (\nabla O)|_v ||_{\mathsf{g}}^2$ has this form because \begin{eqnarray} W_{2,0}(O)|_v &=& \lim_{\hbar \rightarrow 0} \frac{1}{ \hbar } \Bigg ( \mathbb{E} \Big [ \big ( \widehat{O}^{\eta} ( \hbar) \cdot \widehat{O}^{\eta} ( \hbar) \big )  \big |_{\Upsilon_v (\cdot | \hbar)} \Big ] - \mathbb{E} \Big [ \widehat{O}^{\eta} ( \hbar) \big |_{\Upsilon_v ( \cdot | \hbar)} \Big ]^2 \Bigg ) \ \ \ \ \ \ \ \   \\ &=& \lim_{\hbar \rightarrow 0} \mathbb{E} \Bigg [ \widehat{(B_1^{\eta})}(O, O) ^{\eta}(\hbar) |_{\Upsilon_v ( \cdot | \hbar)} \Bigg ] \\ &=& \lim_{\hbar \rightarrow 0} \mathbb{E} \Bigg [ \widehat{(B_1^{\textnormal{Wick}})}(O, O)^{\textnormal{Wick}}(\hbar) |_{\Upsilon_v ( \cdot | \hbar)} \Bigg ] \\ &=& B_1^{\textnormal{Wick}}(O,O)|_v \\ &=& (B_1^{\textnormal{Wick}})^+(O, O) |_v \\ &=& \mathscr{B}_1^+ ( O, O)|_v \\ &=& \mathsf{g}( \nabla O, \nabla O) \end{eqnarray}
\noindent The first equality is a definition, the second is immediate from the definition of the bidifferential $B_1^{\eta}$ of our $\eta$-quantization, and the third uses our assumption that the ordering $\eta$ is close to Wick ordering in Definition [\ref{CloseToWickOrderingDefinition}].  The fourth equality uses Lemma [\ref{CovariantBerezinSymbolOfWickKeyLemma}], the fifth uses $(B_1^Q)^-(O_1, O_2) = \{O_1, O_2\}$ valid for any quantization $Q$ together with the skew-symmetric of the Poisson bracket $\{O, O\}=0$, the sixth uses Proposition [\ref{B1WickIsWeldingOperator}], and the finally the seventh equality uses $\mathscr{B}_1^+(O_1, O_2) = \mathsf{g} ( \nabla O_1, \nabla O_2)$ from Proposition [\ref{PlusMinusPartsOfWeldingOperator}]. $\square$

\end{itemize}

\noindent Let us conclude with three remarks.  First, the variance $\mathscr{B}_1^+(O, O)|_v$ relies on the symmetric part $\mathscr{B}_1^+ = (B_1^{\textnormal{Wick}})^+$ of the first bidifferential of Wick quantization, whereas the symmetric part $(B_1^Q)^+$ of any quantization $Q$ is usually ignored in the deformation quantization of Poisson manifolds as we discussed in section \textbf{[\ref{subsubsecDeformationQuantizationAndComplexStructures}]}.  The variance reflects the complex structure $J$ of our classical phase space $(\mathscr{M}, J, \mathsf{g}, \omega)$.\\
\\
\noindent Second, looking up at the proof of Theorem [\ref{CLTColumn2}], we can now see why
\begin{equation} \mathscr{B}_1 = \sum_{k=1}^K \sigma_k^2 \frac{ \partial }{ \partial \overline{V_k}} \otimes \frac{ \partial }{\partial V_k}  \end{equation}

\noindent defined intrinsically by the inverse Hermitian metric has the name ``welding operator''.  We computed the covariance $W_{2,0}$ of the next-order Gaussian fluctuations simply by applying the welding operator to two independent copies of $W_{1,0}$, the leading-order semi-classical variables $O|_v$ from the concentration of measure in Theorem [\ref{LLNColumn2}].  In terms of the diagrams illustrating the Leibniz Rule in our proof of Theorem [\ref{AOEColumn2}], the covariance $W_{2,0}$ is the result of welding two slit disks $W_{1,0}$, which is a {cylinder}.\\
\\
\noindent Finally, the convergence in Theorem [\ref{CLTColumn2}] to a Gaussian random variable does not at any point invoke the central limit theorem for independent averages.  Rather, the convergence to a Gaussian reflects the fact that at $\hbar >0$ fixed, a coherent state is a unitary translation of the ground state $\Upsilon_0 ( \cdot  | \hbar)$ which is the Gaussian we began with when defining Fock space, i.e. the Segal-Bargmann Gaussian weight $d \rho_{\hbar; \mathsf{g}}$ whose covariance is exactly the inverse Hermitian metric we see in Theorem [\ref{CLTColumn2}].

\pagebreak

\section{Integrable Hierarchies and Conserved Densities} \label{secColumn3}

\noindent Our Theorems [\ref{LLNColumn2}] and [\ref{CLTColumn2}] of the previous section \textbf{[\ref{secColumn2}]}, confirmations of and $\hbar^{1/2}$ Gaussian corrections to the static version of Bohr's Correspondence Principle for coherent states $\Upsilon_v ( \cdot | \hbar)$, do not depend on the integrability of the Hamiltonian flow generated by the classical observable $O$.  In this section, we simply specialize all results of section \textbf{[\ref{secColumn2}]} assuming $O$ does generate a classical integrable Hamiltonian system.  In \textbf{[\ref{subsecColumn3Row1}]}, we define a classical conserved density $dF ( c) |_v$ of a state $v \in \mathscr{M}$ and the classical integrable hierarchy it generates

\begin{equation} \label{JuiceFormula} O_{\phi} = \int_{\mathbb{X}} \phi(c) d F(c) \end{equation}

\noindent for all $\phi : \mathbb{X} \rightarrow \R$.  Likewise, in \textbf{[\ref{subsecColumn3Row2}]} we define quantum {conserved densities} of a state $\Psi \in \mathscr{H}$, and we define integrable quantizations in \textbf{[\ref{subsecIntegrableQuantizations}]}.  Classical conserved densities are non-random signed measures on an auxiliary space $\mathbb{X}$, and by Born's Rule, in \textbf{[\ref{subsecColumn3Row4}]} we reinterpret quantum conserved densities in $\Psi$ as random signed measures $d \widehat{F}(c | \hbar)|_{\Psi}$ on $\mathbb{X}$, finally  specializing to coherent states $\Upsilon_v ( \cdot | \hbar)$.  We give three semi-classical results for the quantum conserved density $d\widehat{F}^{\eta}(c | \hbar)|_{\Upsilon_v ( \cdot | \hbar)}$ of an $\eta$-quantized integrable hierarchy in a coherent state $\Upsilon_v ( \cdot | \hbar)$ around $v$ in any Hermitian affine space $(\mathscr{M}, J, \mathsf{g}, \omega)$.  In Theorems [\ref{LLNColumn3}], [\ref{CLTColumn3}],

\begin{equation} \label{JuiceMe} d\widehat{F}^{\eta}(c | \hbar)|_{\Upsilon_v ( \cdot | \hbar)} \sim dF(c )|_v + \hbar^{1/2} d \mathbb{G}(\{O_l\}_{l=1}^{K}) |_v \end{equation}

\noindent the quantum conserved density concentrates on a limit shape given by the classical conserved density, a confirmation of the static version of Bohr's Correspondence Principle.  Moreover, quantum corrections occur at scale $\hbar^{1/2}$ and approach a Gaussian field on $\mathbb{X}$ of mean $0$ with covariance $W_2(\mathbb{O}_1, \mathbb{O}_2) = \mathbb{E}[ \mathbb{O}_1 \mathbb{O}_2 ] - \mathbb{E}[ \mathbb{O}_1] \mathbb{E}[ \mathbb{O}_2]$ being
\begin{equation}  W_2 \Bigg ( \int_{\mathbb{X} } \phi_1(c_1)d \mathbb{G} ( c) |_v, \int_{\mathbb{X}} \phi_2(c_2) d \mathbb{G} ( c)|_v \Bigg ) = \mathsf{g} \big ( (\nabla O_{\phi_1})|_v, (\nabla O_{\phi_2})|_v \big ). \end{equation}

\noindent for $O_{\phi}$ as in (\ref{JuiceFormula}).  (\ref{JuiceMe}) follows from a polynomial $\hbar$ expansion of joint cumulants  in Theorem [\ref{AOEColumn3}], which itself uses the identical argument for Theorem [\ref{AOEColumn2}].

\subsection{Classical Integrable Hierarchies and Conserved Densities} \label{subsecColumn3Row1}

\noindent Let $(\mathscr{M}, \omega)$ be the classical phase space as in the setting of section \textbf{[\ref{subsecClassicalSystems}]}.

\begin{definition} \label{ClassicalIntegrableHierarchyDefinition} A \underline{classical integrable hierarchy} is a set $\{O_{l}\}_{l = 1}^{\widetilde{K}}$ of $\widetilde{K}$ classical observables $O_l$ on the classical phase space $\mathscr{M}$ so that $\widetilde{K} = K$ where $\dim_{\R} \mathscr{M} = 2K$ and $\{O_{l}\}_{l = 1}^{\widetilde{K}}$ are algebraically independent and pairwise Poisson-commute \begin{equation} \{O_{l}, O_{l' }\} = 0  \end{equation} for all $l', l \in \{1, 2, \ldots, \widetilde{K}\}$.
\end{definition}

\noindent A classical integrable hierarchy generates a Poisson commutative $\mathsf{T} \subset C^{\infty}(\mathscr{M}, \R)$.  The constituents $O_{l'}$ of a hierarchy for $l' \neq l$ are automatically conserved quantities for the $O_{l}$-Hamiltonian flow (\ref{HamiltonianFlow}) since $\frac{ \partial O_{l'}}{ \partial t} = \{ O_{l}, O_{l'} \} = 0$.  However, it is not merely the existence of conservation laws but the location of a Hamiltonian in a hierarchy of pairwise-commuting classical observables that guarantees {Liouville integrability of finite-dimensional Hamiltonian systems}, i.e. the existence of action-angle variables when $K < \infty$ by the Liouville-Arnold theorem.  By contrast, if ${K} = \infty$, existence of a classical integrable hierarchy of size $\widetilde{K} = \infty$ not imply integrability, let alone well-posedness, because an infinite set admits strict subsets of the same cardinality.\\
\\
\noindent Depending on the nature of the classical phase space $(\mathscr{M}, \omega)$ and classical integrable hierarchy $\{O_l\}_{l=1}^K$, it may be possible to simplify the description of the range of the map $v \in \mathscr{M} \rightarrow (O_1 |_v, \ldots, O_K |_v ) \in \R^K$ by locating a conserved density.
 \begin{definition} \label{ClassicalConservedDensityDefinition} A \underline{classical conserved density} is a non-random signed measure $dF(c)|_v$ on an auxiliary space $\mathbb{X}$ parametrized by $c \in \mathbb{X}$ associated to every $v \in \mathscr{M}$ in the classical phase space so that for some collection $\mathcal{C}$ of functions $\phi : \mathbb{X} \rightarrow \R$ one has
 
 \begin{equation} (O_{\phi})|_v = \int_{\mathbb{X}} \phi(c) dF(c) |_v \end{equation} 
 
 \noindent a $\mathcal{C}$-family of Poisson commuting classical observables on $(\mathscr{M}, \omega)$ for $\phi_1, \phi_2 \in \mathcal{C}$
 \begin{equation} \{ O_{\phi_1}, O_{\phi_2}\}= 0 \end{equation} that is big enough, i.e. there are $\{\phi_l\}_{l=1}^K \subset \mathcal{C}$ so that $O_l  = O_{\phi_l}$ is a classical integrable hierarchy in the sense of \textnormal{Definition [\ref{ClassicalIntegrableHierarchyDefinition}]}.\end{definition}
 
\subsection{Quantum Integrable Hierarchies and Conserved Densities} \label{subsecColumn3Row2}
\noindent Let $(\mathscr{H}, \langle \cdot, \cdot \rangle_{\hbar})$ be the quantum state space as in the setting of section \textbf{[\ref{subsecQuantumSystems}]}.

\begin{definition} \label{QuantumIntegrableHierarchyDefinition} A \underline{quantum integrable hierarchy} is a set $\{ \widehat{\mathcal{O}}_l(\hbar) \}_{l = 1}^{\widetilde{K}}$ of $\widetilde{K}$ quantum observables on the quantum state space $\mathscr{H}$ so that $\widetilde{K} = K$ if $\mathscr{H}$ occurs as quantization of $(\mathscr{M}, \omega)$ with $\dim_{\R} \mathscr{M} = 2K$ and $\{\widehat{\mathcal{O}}_{l}(\hbar)\}_{l = 1}^{\widetilde{K}}$ are algebraically independent and that pairwise commute \begin{equation} [ \widehat{\mathcal{O}}_{l}(\hbar), \widehat{\mathcal{O}}_{l'}(\hbar) ] = 0 \end{equation} for $l, l' \in \{1, 2, \ldots, \widetilde{K}\}$.\end{definition}

\noindent A quantum integrable hierarchy generates a commutative subalgebra of $\textbf{i} \mathfrak{u} ( \mathscr{H}, \langle \cdot, \cdot \rangle_{\hbar})$.
\noindent If $K < \infty$, a quantum analog of the Liouville-Arnold theorem for classical integrable Hamiltonian systems holds: the presence of a quantum integrable hierarchy reduces the eigenvalue problem (\ref{EigenvalueProblem}) for a single quantum Hamiltonian to a simultaneous eigenvalue problem (\ref{SimultaneousEigenvalueProblem}), thus reducing what is usually a partial differential equation to an ordinary differential equation.  For a discussion, see chapter 5.2 in \cite{Eti0}.  As in the classical case, this reasoning breaks down in the setting $K= \infty$ of field theory.\\
\\
\noindent The von Neumann spectral theorem, Theorem [\ref{SpectralTheorem}] above, for quantum integrable hierarchies $\{ \widehat{\mathcal{O}}_{l}(\hbar) \}_{l=1}^K$ of commuting self-adjoint operators as in Definition [\ref{QuantumIntegrableHierarchyDefinition}] is:

\begin{theorem} \label{JointSpectralTheorem} For $K \in \N \cup \{ \infty\}$, let $\{\widehat{\mathcal{O}}_l\}_{l=1}^K$ be pairwise-commuting possibly unbounded self-adjoint operators in a Hilbert space $(\mathscr{H}, \langle \cdot, \cdot \rangle_{\hbar})$.  For every $\Psi \in \mathscr{H}$ with $|| \Psi ||=1$ there exists a probability measure $\upmu_{\Psi, \Psi}( \cdot | \widehat{\mathcal{O}})$ on $\R^K$ so that
\begin{equation} \label{JointSpectralMeasureDefiningRelation} { \langle \Psi | \phi(\widehat{\mathcal{O}}_{1} ,\ldots,   \widehat{\mathcal{O}}_K)  | \Psi \rangle } = \int_{- \infty}^{+\infty} \phi(E_1, \ldots, E_K ) d\upmu_{\Psi, \Psi}( E_1, \ldots, E_K | \{\widehat{\mathcal{O}}_l\}_{l = 1}^K)
\end{equation}
\noindent for all bounded continuous $\phi: \R^K \rightarrow \C$, the \underline{joint spectral measure {of} $\{\widehat{\mathcal{O}}_l\}_{l = 1}^K$ {at} $\Psi$.}

\end{theorem}

\noindent Under certain conditions the joint spectrum of commuting self-adjoint operators may admit a simpler description via a joint spectral density. 

\begin{definition} \label{QuantumConservedDensityDefinition} Given a collection $\mathcal{C} = \{ \phi : \mathbb{X} \rightarrow \R\}$ of functions on an auxiliary space $\mathbb{X}$ and a fixed $\Psi \in \mathscr{H}$, a \underline{quantum conserved density} a measure $d \mathsf{\mu}^F(\alpha) |_{\Psi}$ on the dual space $\mathcal{C}^*$ of distributions on $\mathbb{X}$ so that for some $(\phi_1, \ldots, \phi_K) \in \mathcal{C}$ its push forward along the map $\mathcal{C}^* \rightarrow \R^K$ given by \begin{equation}  \alpha \longmapsto \big ( \alpha ( \phi_1), \ldots, \alpha(\phi_K) \big ) \end{equation} \noindent is the joint spectral measure $d \mathsf{\mu}_{\Psi, \Psi}( E_1, \ldots, E_K | \widehat{\mathcal{O}}_1, \ldots, \widehat{\mathcal{O}}_K)$ of a quantum integrable hierarchy in the sense of \textnormal{Definition [\ref{QuantumIntegrableHierarchyDefinition}]}.  \end{definition}

\noindent We reinterpret this crucial definition of quantum conserved density in a probabilistic manner in Proposition [\ref{TheModelColumn3PreDefinition}].

\subsection{Integrable Quantizations} \label{subsecIntegrableQuantizations}

\noindent Our Definition [\ref{QuantizationDefinition}] of quantizations $Q$ does not address whether $Q$ sends \begin{equation} \label{QuantizationPreservesCommutativity}\{O_1, O_2\} = 0 \ \ \ \ \ \Rightarrow \ \ \ \ \ [ \widehat{{O}}_1^{Q}(\hbar), \widehat{{O}}_2^Q(\hbar)] = 0 \end{equation}

\noindent two Poisson commuting classical observables to two commuting quantum observables.

\begin{definition} \label{IntegrableQuantizationDefinition} An \underline{integrable quantization} ${Q}$ of a pair $\mathsf{T} \subset \mathsf{A} \subset C^{\infty}(\mathscr{M},\R)$ of a Poisson algebra $\mathsf{A}$ and a Poisson-commutative subalgebra $\mathsf{T}  \subset \mathsf{A}$ is a quantization $Q(\hbar)$ of $\mathsf{A}$ as in \textnormal{Definition [\ref{QuantizationDefinition}]} so that \textnormal{(\ref{QuantizationPreservesCommutativity})} holds for all $O \in \mathsf{T}$. \end{definition}

\noindent If $\mathsf{T}$ is generated by a classical integrable hierarchy $\{O_l\}_{l=1}^{\widetilde{K}}$, the image of $\mathsf{T}$ under an integrable quantization $Q$ is a quantum integrable hierarchy $\{\widehat{O}_l^{Q}(\hbar)\}_{l=1}^{\widetilde{K}}$.  Well-known quantizations such as Wigner-Weyl, anti-Wick, or Wick quantization will typically not be integrable quantizations of a given classical integrable hierarchy $\{O_l\}_{l=1}^{\widetilde{K}}$.  While existence and uniqueness of integrable quantizations of a given classical integrable hierarchy $\{O_l\}_{l=1}^{\widetilde{K}}$ in $(\mathscr{M}, \omega)$ is far from obvious (phase space star products satisfying \textit{commutativity equations} in addition to the usual \textit{associativity equations} in deformation quantization), the pursuit of explicit integrable quantizations of symplectic or more generally Poisson manifolds is the core of the theory of quantum integrable systems (e.g. discovery of quantum groups and $R$-matrices satisfying Yang-Baxter identities).

\subsection{Random Density of Quantum Hierarchy in Coherent State} \label{subsecColumn3Row4}

\subsubsection{Born's Rule: Correlation from Commutativity}

\noindent In its form due to Robertson, the \textit{Heisenberg Uncertainty Principle} is a lower bound

\begin{equation} \label{UncertaintyPrinciple} W_2 \Big ( \widehat{\mathcal{O}}_1(\hbar) \big |_{\Psi} \Big ) W_2 \Big ( \widehat{\mathcal{O}}_2 (\hbar) \big |_{\Psi} \Big ) \geq \frac{1}{4} W_1 \Big ( [ \widehat{\mathcal{O}}_1(\hbar), \widehat{\mathcal{O}}_2(\hbar)] \big |_{\Psi} \Big )^2 \end{equation}

\noindent for the product of the variances of the random values $\widehat{\mathcal{O}}_1(\hbar)|_{\Psi}$, $\widehat{\mathcal{O}}_2(\hbar)|_{\Psi}$ of two quantum observables $\widehat{\mathcal{O}}_1(\hbar)$ and $\widehat{\mathcal{O}}_2(\hbar)$ in the same quantum state $\Psi \in \mathscr{H}$.  The lower bound is $1/4$ of the square average of the random value of the commutator $[\widehat{\mathcal{O}}_1(\hbar), \widehat{\mathcal{O}_2}(\hbar) ]$ in $\Psi$.\\
\\
\noindent The Uncertainty Principle (\ref{UncertaintyPrinciple}) is a numerical relation that does not require one to specify a \textit{joint law} of the random variables $\widehat{\mathcal{O}}_1(\hbar)|_{\Psi}$, $\widehat{\mathcal{O}}_2(\hbar)|_{\Psi}$.  To specify such a joint law without making any unnatural choices of ordering, one considers the case in which the two quantum observables commute, and can use Theorem [\ref{JointSpectralTheorem}].

\begin{definition}\label{JointRandomValuesDefinition} The \underline{random values} $\{\widehat{\mathcal{O}}(\hbar)|_{\Psi}\}_{l=1}^K$ of a quantum integrable hierarchy $\{\widehat{\mathcal{O}}_l(\hbar)\}_{l=1}^K$ in a quantum state $\Psi \in \mathscr{H}$ are the random variables $\{\widehat{O}_l(\hbar)|_{\Psi}\}_{l=1}^K$ whose joint law is $d \mu_{\Psi, \Psi}( \cdot | \widehat{\mathcal{O}}(\hbar))$ the joint spectral measure of $\widehat{\mathcal{O}}(\hbar)$ at $\Psi$.  For $\phi : \R^K \rightarrow \C$, $\mathbb{E} [ \phi( \widehat{O}_1(\hbar) |_{\Psi}, \ldots, \widehat{O}_K(\hbar)|_{\Psi} )]$ is either side of \textnormal{(\ref{JointSpectralMeasureDefiningRelation})}.
\end{definition} 

\noindent It is absolutely crucial that $[\widehat{\mathcal{O}}_{l_1}, \widehat{\mathcal{O}}_{l_2}] = 0$ does not imply independence of $\widehat{O}_{l_1} |_{\Psi} , \widehat{O}_{l_2} |_{\Psi}$.\\
\\
\noindent By Definition [\ref{JointRandomValuesDefinition}], we arrive at a probabilistic characterization of the ``quantum conserved densities'' of Definition [\ref{QuantumConservedDensityDefinition}] that better matches our Definition [\ref{ClassicalConservedDensityDefinition}]

\begin{definition} \label{TheModelColumn3PreDefinition} A \underline{realized quantum conserved density} for the Schr\"{o}dinger flows generated by a quantum integrable hierarchy $\{\widehat{\mathcal{O}}_l(\hbar)\}_{l=1}^K$ with initial condition $\Upsilon(\hbar)$ is the random signed measure $d\widehat{F}(c | \hbar)|_{\Upsilon(\hbar)}$ on $\mathbb{X}$ whose law is $d \mathsf{\mu}^F(\alpha)|_{\Upsilon(\hbar)}$ from \textnormal{Definition [\ref{QuantumConservedDensityDefinition}]}, i.e.  for appropriate test functions $\chi_l \in \mathcal{C}$ \begin{equation} \widehat{\mathcal{O}}_l(\hbar)|_{\Upsilon(\hbar)} = \int_{\mathbb{X}} \chi_l(c) d \widehat{F}(c | \hbar)|_{\Upsilon(\hbar)} \end{equation}

\noindent are random variables correlated in $l=1,2,3,\ldots$ each defined as in \textnormal{Definition [\ref{TheModelColumn1Definition}]}.

 \end{definition}
  
 \noindent Realized quantum conserved densities are random signed measures $d\widehat{F}(c | \hbar)|_{\Psi}$ on an auxiliary space $\mathbb{X}$ whose law depends on a choice of state $\Psi$ and a choice of quantum integrable hierarchy.  In the special case that $\Psi = \Psi_{\lambda}( \cdot | \hbar)$ is a quantum stationary state, one gets a simpler description of this object.

\begin{definition} A quantum integrable hierarchy $\{\widehat{\mathcal{O}}_l\}_{l=1}^{K}$ has \underline{discrete spectrum} on the quantum state space $\mathscr{H}$ if they are simultaneously diagonalized on a basis of quantum stationary states \begin{equation} \label{SimultaneousEigenvalueProblem} \widehat{\mathcal{O}}_l(\hbar)  \Psi_{\lambda} ( \cdot | \hbar) = \widehat{\mathcal{O}}_l(\hbar) |_{\lambda} \Psi_{\lambda}( \cdot | \hbar )\end{equation} \noindent with eigenvalues $\widehat{\mathcal{O}}_l(\hbar) |_{\lambda} \in \R$ indexed by some set $\lambda \in \mathbb{Y}$. \end{definition}

 \begin{proposition} \label{QuantumConservedDensitiesIfDiscreteSpectrum} If a quantum integrable hierarchy $\{\widehat{\mathcal{O}}_l\}_{l=1}^{K}$ has discrete spectrum, for any state $\Upsilon(\hbar)$, the realization of its quantum conserved density in $\Upsilon(\hbar)$ enjoys
 \begin{equation} d \widehat{F} (c | \hbar) |_{\Upsilon(\hbar)} = d \widehat{F}(c | \hbar) |_{\lambda} \end{equation}
 
 \noindent an equality in law for $\lambda \in \mathbb{Y}$ sampled from the law in \textnormal{Definition [\ref{RandomQuantumStationaryStateIndexDefinition}]}.
 \end{proposition}
 
 \noindent In this chapter we do \textit{not} have to assume that our quantum hierarchy has discrete spectrum.  We now specialize Definition [\ref{TheModelColumn3PreDefinition}] to coherent states.

\begin{definition} \label{TheModelColumn3Definition} Under the following structural assumptions,
\begin{itemize}
\item \textnormal{\textcolor{gray}{[Structure of State]}} $\Psi = \Upsilon_v ( \cdot | \hbar)$ is a coherent state
\item \textnormal{\textcolor{gray}{[Structure of Observables]}} $\{O_l\}_{l=1}^K$ is a classical integrable hierarchy generating a Poisson commutative subalgebra $\mathsf{T} \subset \mathsf{A}$ in a Poisson $\mathsf{A} \subset C^{\infty}(\mathscr{M}, \R)$ with a conserved density $d F_{\star | v}( c)$ on a space $\mathbb{X}$
\item \textnormal{\textcolor{gray}{[Structure of Quantization]}} $Q$ is an integrable canonical quantization as in \textnormal{Definitions [\ref{CanonicalQuantizationDefinition}] and [\ref{IntegrableQuantizationDefinition}]} of the pair $ \mathsf{T} \subset \mathsf{A}$ resulting in a quantum integrable hierarchy $\{\widehat{O}_l^Q(\hbar) \}_{l=1}^K$ with quantum conserved densities $d \widehat{F}(c | \hbar)$ on the same space $\mathbb{X}$.
\end{itemize}

\noindent define a random signed measure $dF_{\lambda}( c | \hbar) |_{\Upsilon_v( \cdot | \hbar)}$ in $\mathbb{X}$, the quantum conserved density of the quantized integrable hierarchy $\{\widehat{O}_l^Q(\hbar)\}_{l=1}^K$ in the coherent state $\Upsilon_v ( \cdot | \hbar) $ around $v$ using \textnormal{Definition [\ref{TheModelColumn3PreDefinition}]}.
\end{definition}

\noindent A challenge is to find explicit realization of these structures.  By definition, the analysis of $dF_{\lambda} ( c | \hbar)|_{\Upsilon_v ( \cdot |\hbar)}$ reduces the study of $K$ correlated random variables of exactly the type studied in the previous section:
\begin{proposition} \label{ReduceColumn3ToColumn2} For $1 \leq l \leq K$, the $\phi_l$-averages
\begin{equation} \widehat{O}_l(\hbar) |_{\Psi} = \int_{\mathbb{X}} \phi_l(c) d \widehat{F}_{\lambda}( c| \hbar )|_{\Psi} \end{equation} 

\noindent of the quantum conserved density $dF_{\lambda} ( c | \hbar)|_{\Upsilon_v ( \cdot |\hbar)}$ of \textnormal{Definition [\ref{TheModelColumn3Definition}]} are correlated  random variables satisfying the structural assumptions of \textnormal{Definition [\ref{TheModelColumn2Definition}]}. \end{proposition}

\subsubsection{Finite Joint Moments from Regularity Assumptions}

\noindent Let us begin with general remarks following section \textbf{[\ref{FiniteMomentsFromRegularityAssumptionsColumn1}]}.  Let $\{\widehat{O}_l(\hbar)|_{\Upsilon(\hbar)} \}_{l=1}^K$ be the random values of a quantum integrable hierarchy $\{ \widehat{\mathcal{O}}_l(\hbar)\}_{l=1}^K$ in a $\hbar$-dependent quantum state $\Psi = \Upsilon(\hbar)$.  By Definition [\ref{JointRandomValuesDefinition}], any bounded function ${\phi} : \R^K \rightarrow \C$ defines a new random variable $\phi(\{\widehat{O}_l(\hbar)|_{\Upsilon(\hbar)} \}_{l=1}^K)$ with finite expectation 
\begin{equation}  \mathbb{E} \big [  {\phi} \big (\{\widehat{O}_l(\hbar)|_{\Upsilon(\hbar)} \}_{l=1}^K \big )  \big ] =   \frac{ \big \langle \Upsilon(\hbar) \big | {\phi} \big ( \{ \widehat{\mathcal{O}}_l ( \hbar) \}_{l=1}^K \big )  \big | \Upsilon(\hbar) \big \rangle_{\hbar} } {\big \langle \Upsilon(\hbar) | \Upsilon(\hbar) \big \rangle_{\hbar}  } .\end{equation}

\noindent It is important that ${\phi}$ may depend on infinitely-many variables if $2K = \dim_{\R} \mathscr{M}= \infty$.  Formally, the Taylor coefficient of bounded $\phi( E_1, E_2, \ldots , E_K)$ around $0 \in \R^K$ at $E_{l_1} \cdots E_{l_n}$ for $l_1, \ldots, l_n \in \{1, \ldots, K\}$ gives the joint moment\begin{equation} \label{JointMoments} \mathbb{E} \big [  \widehat{O}_{l_1}(\hbar)|_{\Upsilon ( \hbar)}  \cdots \widehat{O}_{l_n} ( \hbar) |_{\Upsilon(\hbar)} \big ] = \frac{ \big \langle \Upsilon (\hbar) \big | \ \widehat{\mathcal{O}}_{l_1}(\hbar) \cdots \widehat{\mathcal{O}}_{l_n} ( \hbar)  \ \big | \Upsilon ( \hbar)  \big \rangle_{\hbar} }{ \big \langle \Upsilon( \hbar) \ \big | \ \Upsilon (\hbar) \big \rangle_{\hbar} }\end{equation}

\noindent finite if $\Upsilon(\hbar)$ is in the domain of the possibly unbounded operator $\widehat{\mathcal{O}}_{l_1}(\hbar) \cdots \widehat{\mathcal{O}}_{l_n} ( \hbar)$.  To prove that joint moments of the type (\ref{JointMoments}) are finite for the model of interest, let us take the same regularity assumptions from section \textbf{[\ref{FiniteMomentsFromRegularityAssumptionsColumn2}]}

\begin{proposition} \label{TheModelColumn3RegularityAssumptions} All joint moments \textnormal{(\ref{JointMoments})} of all $\phi_l$-averages $\widehat{O}_l^{Q}(\hbar)|_{\Upsilon_v ( \cdot | \hbar)}$ of the quantum conserved density $dF_{\lambda}( c | \hbar)|_{\Upsilon_v ( \cdot | \hbar)}$ of \textnormal{Definition [\ref{TheModelColumn3Definition}]} are finite if the following regularity assumptions are satisfied:

\begin{enumerate}
\item \textcolor{gray}{\textnormal{[Regularity of State]}} In a $\sigma$-coordinate system, the classical state $v \in \mathscr{M}$ has $V_k \equiv 0$ for almost every $1 \leq k \leq K$.
\item \textcolor{gray}{\textnormal{[Regularity of Observable]}} In a $\sigma$-coordinate system, the classical observables $O_l$ of the classical integrable hierarchy $\{O_l\}_{l=1}^K$ are generalized polynomials in $\{V_k, \overline{V_k}\}_{k=1}^K$ with $\deg O_l = l$.
\item \textcolor{gray}{\textnormal{[Regularity of Quantization]}} $Q$ is an integrable $\eta$-quantization $Q_{\eta}$ of the pair $\mathsf{T} \subset \mathsf{A}^{\textnormal{genpoly}}$ of generalized polynomials for an ordering $\eta$ that is close to Wick quantization of the type constructed in \textnormal{Theorem [\ref{EtaQuantizationsAreCanonicalQuantizations}]}.
\end{enumerate}
\end{proposition}

\begin{itemize} \item \textit{Proof:} follows from our proof of Theorem [\ref{AOEColumn3}] below. $\square$ \end{itemize}

\subsection{Asymptotic Expansions of Joint Cumulants}

\noindent Let $W_n(\mathbb{O}_{l_1}, \ldots, \mathbb{O}_{l_n})$ be the joint cumulant of random variables $\{\mathbb{O}_{l_i}\}_{i=1}^n$, $1 \leq l_i \leq K$.
\begin{theorem} \label{AOEColumn3} For the quantum conserved density $dF_{\lambda}( c | \hbar)|_{\Upsilon_v ( \cdot | \hbar)}$ in $\mathbb{X}$ of \textnormal{Definiton [\ref{TheModelColumn3Definition}]} satisfying the regularity assumptions of \textnormal{Proposition [\ref{TheModelColumn3RegularityAssumptions}]}, all joint cumulants of all of its random $\phi_l$-averages are polynomials in $\hbar$ \begin{equation} \label{AOEColumn3Formula} W_n \Big ( \widehat{O_{l_1}}^{\eta} ( \hbar) \big |_{\Upsilon_v ( \cdot | \hbar) }, \ldots, \widehat{O_{l_n}}^{\eta}(\hbar) \big |_{\Upsilon_v ( \cdot | \hbar)}\Big ) = \sum_{g =0}^{ \lfloor \frac{ l_1 + \cdots + l_n}{2} \rfloor} \hbar^{n-1+g} W_{\eta, n,g}(O_{l_1}, \ldots, O_{l_n})|_v \end{equation}

\noindent with order of vanishing at least $n-1$ at $\hbar = 0$ and whose coefficients $W_{\eta, n,g}(O_{l_1}, \ldots, O_{l_n})$ are themselves polynomials in the $\sigma$-coordinates $\{V_k , \overline{V_k}\}_{k=1}^{K}$ of $v \in \mathscr{M}$.  Moreover, the leading order coefficient $W_{\eta, n,0}(O_{l_1}, \ldots, O_{l_n})$ is independent of $\eta$.

\end{theorem}

\begin{itemize}
\item \textit{Proof:} By Proposition [\ref{ReduceColumn3ToColumn2}], each $\phi_l$-average $\widehat{O_l}^{\eta}(\hbar)|_{\Upsilon_v ( \cdot | \hbar)}$ satisfies the regularity assumptions of Proposition [\ref{TheModelColumn2RegularityAssumptions}], and thus Theorem [\ref{AOEColumn3}] follows by copying the proof of Theorem [\ref{AOEColumn2}] for the correlated random variables $\widehat{O_l}^{\eta}(\hbar)|_{\Upsilon_v ( \cdot | \hbar)}$, replacing everywhere cumulants and moments with joint cumulants and joint moments. Finally, note that the degree of the polynomial expansion is $\lfloor \frac{ l_1 + \cdots + l_n}{2} \rfloor$ because of our assumption $\deg O_l = l$ in Proposition [\ref{TheModelColumn3RegularityAssumptions}].  $\square$
\end{itemize}
\pagebreak

\subsection{Limit Shapes in Static Semi-Classical Limits}

\begin{theorem} \label{LLNColumn3} For the quantum conserved density $d\widehat{F}^{\eta}_{\lambda}( c | \hbar)|_{\Upsilon_v ( \cdot | \hbar)}$ in $\mathbb{X}$ of \textnormal{Definiton [\ref{TheModelColumn3Definition}]} satisfying the regularity assumptions of \textnormal{Proposition [\ref{TheModelColumn3RegularityAssumptions}]}, in the semi-classical limit $\hbar \rightarrow 0$ the quantum conserved density of the $\eta$-quantized integrable hierarchy $\{\widehat{O}_l^Q(\hbar)\}_{l=1}^K$ in a coherent state $\Upsilon_v( \cdot | \hbar)$ around $v$ concentrates on a limit shape

\begin{equation} \label{LLNColumn3Formula} d\widehat{F}^{\eta}_{\lambda}( c | \hbar) \rightarrow dF_{\star |v}(c)  \end{equation}

\noindent  the non-random classical conserved density $dF_{\star |v}(c)$ of the classical hierarchy $\{O_l\}_{l=1}^K$ at $v$, independent of $\eta$, for \textnormal{(\ref{LLNColumn3Formula})} convergence in distribution of $\phi_l$-averages. \end{theorem}

\begin{itemize}
\item \textit{Proof:} For the desired mode of convergence, we have to prove that \begin{equation} \label{LLNColumn3FormulaProof} \widehat{O_l}^{\eta} ( \hbar) |_{\Upsilon_v ( \cdot | \hbar) } \rightarrow (O_l)|_v \end{equation} the random value $\widehat{O_l}^{\eta}(\hbar)|_{\Upsilon_v ( \cdot | \hbar)}$ of the $\eta$-quantized observable $\widehat{O_l}^{\eta} (\hbar)$ in $\Upsilon_v( \cdot | \hbar)$ around $v$ concentrates on the non-random value $(O_l)|_v$ of the classical observable $O_l$ at $v$, independent of $\eta$.  By Proposition [\ref{ReduceColumn3ToColumn2}], use Theorem [\ref{AOEColumn3}] and the same proof for Theorem [\ref{LLNColumn2}] replacing cumulants with joint cumulants. $\square$
\end{itemize}

\subsection{Gaussian Fields as Quantum Fluctuations}

\begin{theorem} \label{CLTColumn3} For the quantum conserved density $d\widehat{F}^{\eta}_{\lambda}( c | \hbar)|_{\Upsilon_v ( \cdot | \hbar)}$ in $\mathbb{X}$ of \textnormal{Definiton [\ref{TheModelColumn3Definition}]} satisfying regularity assumptions \textnormal{Proposition [\ref{TheModelColumn3RegularityAssumptions}]}, quantum fluctuations of $d\widehat{F}^{\eta}_{\lambda}( c | \hbar)|_{\Upsilon_v ( \cdot | \hbar)}$ around its classical limit $dF_{\star | v}(c)$ occur at scale $\hbar^{1/2}$ and converge to\begin{equation} \label{CLTColumn3Formula} \frac{1}{\hbar^{1/2}} \Bigg ( dF_{\lambda}( c | \hbar)|_{\Upsilon_v ( \cdot | \hbar)} - dF_{\star | v}(c) \Bigg ) \rightarrow d\mathbb{G}(O)|_v (c) \end{equation} \noindent a Gaussian field $d \mathbb{G}(O)|_v$ on $\mathbb{X}$ independent of $\eta$, mean $0$, and covariance given by \begin{equation} \iint_{\mathbb{X} \times \mathbb{X}} \chi_{l_1}(c_1) \chi_{l_2}(c_2) W_2 \Big (  d \mathbb{G}(O)|_v (c_1) , d\mathbb{G}(O)|_v ( c_2) \Big )  =  \mathsf{g} ( (\nabla O_1)|_v , (\nabla O_2) |_v )  \end{equation} where \textnormal{(\ref{CLTColumn3Formula})} is convergence in distribution of $\phi_l$-averages.
 \end{theorem}
\begin{itemize}
\item \textit{Proof:} For the desired mode of convergence, we have to prove that quantum fluctuations of the correlated random $\phi_l$-averages $\widehat{O}^{\eta}(\hbar)|_{\Upsilon_v ( \cdot | \hbar)}$ around their classical limits $(O_l)|_v$ occur at scale $\hbar^{1/2}$ and converge in distribution to 

\begin{equation}  \frac{1}{\hbar^{1/2}} \Bigg ( \widehat{O}^{\eta} (\hbar)|_{\Upsilon_v ( \cdot | \hbar)} - O|_v \Bigg ) \rightarrow \mathbb{G}(O)|_v 
\end{equation}

\noindent correlated Gaussians $\{\mathbb{G}(O_l)|_v\}_{l=1}^K$ independent of  $\eta$ of mean $0$ and covariance $W_2 \Big ( \mathbb{G}(O_{l_1})|_v, \mathbb{G}(O_{l_2})|_v \Big ) = \mathsf{g} ( (\nabla O_1)|_v , (\nabla O_2) |_v )$ given by the inner product of the gradients of the classical observables evaluated at the classical state $v$.  By Proposition [\ref{ReduceColumn3ToColumn2}], use Theorem [\ref{AOEColumn3}] and the same proof for Theorem [\ref{CLTColumn2}] replacing cumulants with joint cumulants. $\square$

\end{itemize}

\pagebreak

\section{Constructions for Periodic Benjamin-Ono} \label{secConstructionsForPeriodicBenjaminOno}

\noindent In sections \textbf{[\ref{subsubsecJacobiTW}]-[\ref{subsubsecKMK}]} we review how \textit{spectral shift functions} of Jacobi operators construct a large class of \textit{profiles} in Kerov's Markov-Kre\u{\i}n Correspondence \cite{Ke1}.  In \textbf{[\ref{subsecCBOHConstruction}]}, we apply this construction to Toeplitz operators to construct a classical conserved density for the classical periodic Benjamin-Ono hierarchy as a spectral shift function in Theorem [\ref{CBOHConservedDensityConstruction}], whose existence was announced in Theorem [\ref{CBOHConservedDensityExistence}].  In \textbf{[\ref{subsecQBOHConstruction}]} we do the same for Fock-Block Toeplitz operators to construct a quantum conserved density for the quantum periodic Benjamin-Ono hierarchy as a spectral shift function in Theorem [\ref{QBOHConservedDensityConstruction}], whose existence was announced in Theorem [\ref{QBOHConservedDensityExistence}].  We also identify the quantum conserved densities of quantum stationary states with profiles of anisotropic partitions in Theorem [\ref{QBOHConservedDensitiesForJacksAreAnisotropicPartitionsRevisited}], earlier announced in Theorem [\ref{QBOHConservedDensitiesForJacksAreAnisotropicPartitions}]. Sections \textbf{[\ref{subsecCBOHConstruction}]} and \textbf{[\ref{subsecQBOHConstruction}]} are related by Nazarov-Sklyanin's integrable geometric quantization defined by an ordering $\eta_{NS}$, which we construct in section \textbf{[\ref{subsecNazarovSklyaninQuantizationConstruction}]}. 
\subsubsection{Jacobi Operators and Titchmarsh-Weyl Functions} \label{subsubsecJacobiTW}

\begin{definition} Given a pair $L_{\bullet},L_{+}$ of possibly unbounded self-adjoint operators on a Hilbert space $\mathscr{H}_{\bullet}$ so that $L_{\bullet} - L_+$ is trace class, the \underline{perturbation determinant} 

\begin{equation} \frac{ \det_{\mathscr{H}_{\bullet}} (u - L_+) }{ \det_{\mathscr{H}_{\bullet}} (u -L_{\bullet}) } = : \det\nolimits_{\mathscr{H}_{\bullet}} \big ( \mathbbm{1} - (L_{\bullet}- L_+)( u - L_{\bullet})^{-1} \big ) \end{equation}

\noindent is well-defined by the Fredholm determinant for $u \in \C \setminus \R$. \end{definition}

\noindent We study a very particular rank $2$ perturbation $L_+$ of a generic $L_{\bullet} $.

\begin{definition} Given $\Psi_0 \in \mathscr{H}_{\bullet}$ and an operator $L_{\bullet}$ on $\mathscr{H}_{\bullet}$, the \underline{Krylov subspaces} \begin{equation} \mathscr{H}_{\bullet, \tealN} [  L_{\bullet}; \Psi_0] = \bigoplus_{ \ell = 0 }^{\tealN} \C | L_{\bullet}^{\ell} \Psi_0 \rangle \end{equation} \noindent have a $\tealN \rightarrow \infty$ limit $\mathscr{H}_{\bullet}[L_{\bullet}; \Psi_0]$ called the \underline{orbit} of $\Psi_0$ under $L_{\bullet}$.  Write $\uppi_{L_{\bullet}; \Psi_0}$ for the \underline{projection} from $\mathscr{H}_{\bullet}$ to $\mathscr{H}_{\bullet}[L_{\bullet}; \Psi_0]$.  If its orbit is dense in $\mathscr{H}_{\bullet}$, $\Psi_0$ is \underline{cyclic} for $L_{\bullet}$.
\end{definition} 

\begin{definition} \label{MinorGeneralDefinitions} For $\Psi_0 \in \mathscr{H}_{\bullet}$, $\mathscr{H}_{\Psi_0} = \C | \Psi_0 \rangle$, and $\mathscr{H}_{\bullet} = \mathscr{H}_{\Psi_0} \oplus \mathscr{H}_{\Psi_0}^{\perp}$ with orthogonal projections $\uppi_{\Psi_0}: \mathscr{H}_{\bullet} \rightarrow \mathscr{H}_{\Psi_0}$, $\uppi_{\Psi_0}^{\perp}: \mathscr{H}_{\bullet} \rightarrow \mathscr{H}_{\Psi_0}^{\perp}$.  The \underline{$(\Psi_0, \Psi_0)$-{minor} of $L_{\bullet}$} \begin{equation}  L_+^{\perp} =\uppi_{\Psi_0}^{\perp} L_{\bullet}  \uppi_{\Psi_0}^{\perp}  \end{equation}
\noindent is an operator $L_+^{\perp}:\mathscr{H}_{\Psi_0}^{\perp} \rightarrow \mathscr{H}_{\Psi_0}^{\perp}$ while $L_+  = 0 \oplus L_+^{\perp}$ is block-diagonal $L_{+}:\mathscr{H}_{\bullet} \rightarrow \mathscr{H}_{\bullet}$.
\end{definition}

\noindent The distinction between $L_+^{\perp} =\uppi_{\Psi_0}^{\perp} L_{\bullet}  \uppi_{\Psi_0}^{\perp} $ and $L_+ = 0 \oplus L_+^{\perp}$ is extremely important in this paper.  The next result is well-known in the spectral theory of orthogonal polynomials on the real line \cite{SimonSzego}.  Recall an operator on a pre-Hilbert space is \textit{essentially self-adjoint} if it has a unique self-adjoint extension to the completion. 

\begin{theorem} \label{OPRL} If $\Psi_0 \in \mathscr{H}_{\bullet}$ is {cyclic} for $L_{\bullet}$ and $L_{\bullet} |_{\Psi_0}$ the restriction of $L_{\bullet}$ to its dense orbit is \underline{essentially self-adjoint}, the $(\Psi_0, \Psi_0)$-matrix element of the resolvent
\begin{equation} \label{OPRLformula} \langle \Psi_0 | \frac{1}{u - L_{\bullet}}| \Psi_0 \rangle  = T^{\uparrow}(u) \big |_{L_{\bullet};\Psi_0}  = \frac{1}{u} \cdot \frac{ \det_{\mathscr{H}_{\bullet}} (u -L_+) }{ \det_{\mathscr{H}_{\bullet}} ( u - L_{\bullet}) } \end{equation}

\noindent is $\tfrac{1}{u}$ times the {perturbation determinant} of $L_{\bullet}$ and ${L_+= 0 \oplus \uppi_{\Psi_0}^{\perp} L_{\bullet}  \uppi_{\Psi_0}^{\perp} }$ on $\mathscr{H}_{\bullet}$. \end{theorem}

\noindent Theorem [\ref{OPRL}] is used heavily in this paper, so we provide a quick proof. {\small
\begin{itemize}
\item 
\noindent \textit{Proof:} Let $\uppi_{\bullet, \tealN}$ denote projection from $\mathscr{H}_{\bullet}$ to the Krylov subspace $\mathscr{H}_{\bullet; \tealN} ( L_{\bullet}; \Psi_0)$, $L_{\bullet, \tealN} = \uppi_{\tealN} L_{\bullet} \uppi_{\tealN}$, and $L_{+, \tealN} = \uppi_{+, \tealN} L_{+} \uppi_{+, \tealN}$.  Using cyclicity, have

\begin{equation} \langle \Psi_0 | \frac{1}{u - L_{\bullet, \tealN} }| \Psi_0 \rangle = \frac{1}{u} \cdot \frac{ \det_{\mathscr{H}_{\bullet, \tealN} } (u - L_{+, \tealN}) }{ \det_{\mathscr{H}_{\bullet, \tealN}} ( u - L_{\bullet, \tealN}) } \end{equation}

\noindent since we may write $L_{\bullet, \tealN}$ as a tri-diagonal Jacobi matrix and use Cramer's rule.  To take the limit $\tealN \rightarrow \infty$, essential self-adjointness of $L_{\bullet}$ on the dense orbit $H_{\bullet}[ L_{\bullet}; \Psi_0]$ implies strong resolvent convergence $(u - L_{\bullet, \tealN})^{-1} \rightarrow (u - L_{\bullet})^{-1}$ hence the convergence of matrix elements of resolvents on left side of the formula.  For the limit of ratio of characteristic polynomials on the right-side, recall that if $B_{\tealN}, B$ are bounded operators that strongly converge $B_{\tealN} \rightarrow B$ (not necessarily in operator norm), then if $V$ is trace class we have $VB_{\tealN} \rightarrow VB$ in trace class norm.  Since the perturbation $L_{\bullet} - L_+$ is rank 2, hence trace class, and the Fredholm determinant is continuous in the trace class norm, the right-hand side converges to the perturbation determinant. $\square$

\end{itemize}
}

\noindent To apply Theorem [\ref{OPRL}], one must check that the restriction of the operator $L_{\bullet}$ to the orbit $\mathscr{H}_{\bullet}[L_{\bullet}; \Psi_0]$ of $\Psi_0$ is essentially self-adjoint.  A large class of such $L_{\bullet}$ are the \textit{bounded} self-adjoint operators.  As we saw in the proof of Theorem [\ref{OPRL}], the Galerkin approximation $L_{\bullet, \tealN}$ is a Jacobi matrix, so $L_{\bullet}$ is a \textit{one-sided Jacobi operator}.  

\begin{definition} \label{TWDefinition} The \underline{Titchmarsh-Weyl function} of a Jacobi operator $L_{\bullet}$ with cyclic $\Psi_0$ is the function $T^{\uparrow}(u)|_{L_{\bullet}, \Psi_0}$ of $u \in \C \setminus \R$ defined by either side of formula \textnormal{(\ref{OPRLformula})}. \end{definition}

\noindent $T^{\uparrow}(u)$ is also known as the {Titchmarsh-Weyl ``m-function''} in the oscillation theory and inverse spectral theory of Jacobi operators \cite{SimonTrace}.

\noindent \begin{proposition} \label{FiniteDimSzegoNoGo} If $\dim \mathscr{H}_{\bullet}  < \infty$, with $\mathscr{H}_{\bullet} = \mathscr{H}_{\Psi_0} \oplus \mathscr{H}_{\Psi_0}^{\perp}$ can simplify \textnormal{(\ref{OPRLformula})} by

\begin{equation}\frac{1}{u} \cdot \frac{ \det_{\mathscr{H}_{\bullet}} (u - L_{+}) }{ \det_{\mathscr{H}_{\bullet}} ( u - L_{\bullet}) } = \frac{ \det_{\mathscr{H}_{\Psi_0}^{\perp}} (u - L_+^{\perp}) }{ \det_{\mathscr{H}_{\bullet}}(u - L_{\bullet}) } 
\end{equation}

\end{proposition}

\noindent \textit{Proof:} $\det_{\mathscr{H}_{\bullet}}( u - L_+) = \det_{\mathscr{H}_{\Psi_0} \oplus \mathscr{H}_{\Psi_0^{\perp}}} \begin{bmatrix} u & 0 \\ 0 & u - L_+^{\perp} \end{bmatrix} = u \cdot \det_{\mathscr{H}_{\Psi_0}^{\perp}} ( u - L_+^{\perp})$. $\square$\\
\\
\noindent Such a simplification is not possible for $L_{\bullet}$ in infinite dimensions as is exemplified by Szeg\H{o}'s First Theorem, Theorem [\ref{SzegoFirstTheorem}] below.  We now convert formula (\ref{OPRLformula}), an equality of matrix elements of $L_{\bullet}, L_+$, into a statement about their spectra.

\subsubsection{Measures and Spectral Shift Functions}

\begin{definition} Let $\mathbf{P}$ denote the space of probability measures on $\R$. \end{definition}
\begin{definition} \label{SpectralMeasureDef} The \underline{spectral measure} of $L_{\bullet}$ at $\Psi_0$ is the unique $\tau^{\uparrow} \in \mathbf{P}$ so that

\begin{equation}  \int_{- \infty}^{+\infty} \frac{ d \tau^{\uparrow} (c)}{ u-c} = \langle \Psi_0 | \frac{1}{u - L_{\bullet}} | \Psi_0 \rangle \end{equation}
\noindent for every $u \in \C \setminus \R$.  To emphasize its definition we may write $d \tau^{\uparrow}(c) = d \tau^{\uparrow}_{\Psi_0, \Psi_0}(c | L_{\bullet})$.
\end{definition}
\noindent For trace-class perturbations, there is a relative notion of spectral measure:  
\begin{definition} Given any pair $L_{\bullet}, L_+$ of possibly unbounded self-adjoint operators on a Hilbert space $\mathscr{H}_{\bullet}$ so that $L_{\bullet} - L_+$ is trace class, the \underline{spectral shift function} $\xi(c | L_{\bullet}, L_+)$ is defined for $u \in \C \setminus \R$ by the perturbation determinant

\begin{equation} \frac{ \det_{\mathscr{H}_{\bullet}} ( u - L_+) }{ \det_{\mathscr{H}_{\bullet}} (u - L_{\bullet}) } = \textnormal{exp} \Bigg ( - \int_{- \infty}^{+\infty} \frac{\xi(c | L_{\bullet}, L_+) dc}{u-c } \Bigg ) . \end{equation}
\end{definition}

\begin{theorem} \label{LifshitzKreinTraceFormula} \textnormal{(Lifshitz-Krein Trace Formula)} If $\phi: \R \rightarrow \C$ has $\phi'(c)$ Wiener class, i.e. the Fourier transform of $\phi'(c)$ is in $L^1(\R)$, then for $L_{\bullet}, L_+$ possibly unbounded self-adjoint operators so that $L_{\bullet} - L_+$ is trace class, one has

\begin{equation} \textnormal{Tr}_{\mathscr{H}_{\bullet}}  \Big [ \phi(L_{\bullet}) - \phi(L_+) \Big ] = - \int_{- \infty}^{+\infty} \phi'(c) \xi(c | L_{\bullet}, L_+) dc \end{equation}

\noindent which simplifies to $\int_{- \infty}^{+\infty} \phi(c) d \xi(c | L_{\bullet}, L_+)$ if $\xi$ is bounded variation. \end{theorem}

\noindent For a review of Kre\u{\i}n's theory of the spectral shift function, see \cite{BirPush, BirYaf, SimonTrace}.  We get

\begin{corollary} \label{OPRLcorollary}Under the assumptions of Theorem \textnormal{[\ref{OPRL}]}, the spectral measure $\tau^{\uparrow}$ of $L_{\bullet}$ at $\Psi_0$ determines the spectral shift function of the pair $L_{\bullet}, L_+$ by the formula 

\begin{equation} \label{OPRLcorollaryFormula} \int_{- \infty}^{+\infty} \frac{d \tau^{\uparrow}(c)}{u-c} = \frac{1}{u} \cdot \textnormal{exp} \Bigg ( - \int_{- \infty}^{+\infty} \frac{\xi(c | L_{\bullet}, L_+) dc}{u-c } \Bigg ) \end{equation}

\end{corollary}

\noindent A priori we do not know if every $\tau^{\uparrow} \in \mathbf{P}$ can appear in Corollary [\ref{OPRLcorollary}], since we take the assumption of Theorem [\ref{OPRL}] that the Jacobi operator $L_{\bullet}$ restricted to the dense orbit of its cyclic vector $\Psi_0$ is essentially self-adjoint.  In fact, the exact class of $\tau^{\uparrow} \in \mathbb{P}$ that can appear in this way is known.  For a fixed sequence $\vec{T}^{\uparrow} \in \R^{\infty}$, define \begin{equation} \mathbf{P} [T^{\uparrow}] = \Big \{ \tau^{\uparrow} \in \mathbf{P} \ : \ \forall \ell=0,1,2, \ldots \int_{- \infty}^{+\infty} c^{\ell} d \tau^{\uparrow}(c) = T^{\uparrow}_{\ell} \ \Big \} .\end{equation} \begin{theorem} \label{NevanlinnaTheorem} \textnormal{[Nevanlinna]} The restriction of a Jacobi operator $L_{\bullet}$ to the dense orbit of its cyclic vector $\Psi_0$ is essentially self-adjoint if and only if the spectral measure $\tau^{\uparrow} \in \mathbf{P}$ of $L_{\bullet}$ at $\Psi_0$ is a determinate solution to the Hamburger moment problem, i.e. \begin{equation} \tau^{\uparrow} \in \textbf{P} ( T^{\uparrow}) \ \ \text{with} \ \ \ \# \mathbf{P}(T^{\uparrow})=1. \end{equation}
\end{theorem}

\noindent In general, those $\tau^{\uparrow} \in \mathbf{P}[T^{\uparrow}]$ so that $\Psi_0 = 1 \in L^2 ( \R, \tau^{\uparrow})$ is cyclic for the operator $L_{\bullet} $ of multiplication by $c \in \R$ are known as ``von Neumann solutions'' or ``Akhiezer's extremal solutions''.  For a review of Nevanlinna's parametrization of all $\mathbf{P}[ T^{\uparrow}]$ from the vantage of spectral theory see \cite{SimonCMP}.\\
\\
\noindent Despite the restriction on $\tau^{\uparrow}$ appearing in Corollary [\ref{OPRLcorollary}] due to Theorem [\ref{NevanlinnaTheorem}], a remarkable relation of the form (\ref{OPRLcorollaryFormula}) was found by Kerov for \textit{arbitrary} $\tau^{\uparrow} \in \mathbf{P}$.  To an arbitrary probability measure $\tau^{\uparrow} \in \mathbb{P}$, Kerov's Markov-Krein Correspondence associates a ``{profile}'' $f$ so that its ``{shifted Rayeigh function}'' $\xi$ is related to $\tau^{\uparrow}$ exactly as in the formula (\ref{OPRLcorollaryFormula}).  We define profiles and shifted Rayleigh functions in the next section \textbf{[\ref{subsubsecProfiles}]} and state Kerov's Markov-Kre\u{\i}n Correspondence in section \textbf{[\ref{subsubsecKMK}]}.

\subsubsection{Profiles and Shifted Rayleigh Functions} \label{subsubsecProfiles}

 \noindent The space $\mathbf{P}^{\vee}$ of ``profiles'' $f$ was carefully constructed in \cite{Ke1} as follows.

\begin{definition} A \textit{\underline{profile}} is a function $f : \R \rightarrow \R$ of $c \in \R$ which is $1$-Lipshitz \begin{eqnarray} \label{profileDefLipshitz} |f (c_1) - f(c_2) | & \leq & | c_1 - c_2 |  \ \ \   \ \ \ \ \ \ \ \  \text{for all} \ \ \ \ c_1, c_2 \in \R \end{eqnarray}

\noindent and whose slopes $f'(c) \rightarrow \pm 1$ as $c \rightarrow \pm \infty$ so that \begin{equation} \label{profileDefDecay} \int_{- \infty}^{0} (1 + f'(c))\cdot  \frac{ dc}{1+|c|} < \infty \ \ \ \ \  \text{and}  \ \ \ \ \  \int_{0}^{+\infty} (1- f'(c) ) \cdot \frac{ dc}{ 1+|c|} < \infty . \end{equation}

\noindent Let ${\mathbf{P}^{\vee}}$ denote the space of all profiles.
\end{definition}
\noindent The reason for the decay condition (\ref{profileDefDecay}) will become clear in Theorem [\ref{KMKstatement}].

  \begin{figure}[htb]
\centering
\includegraphics[width=0.9 \textwidth]{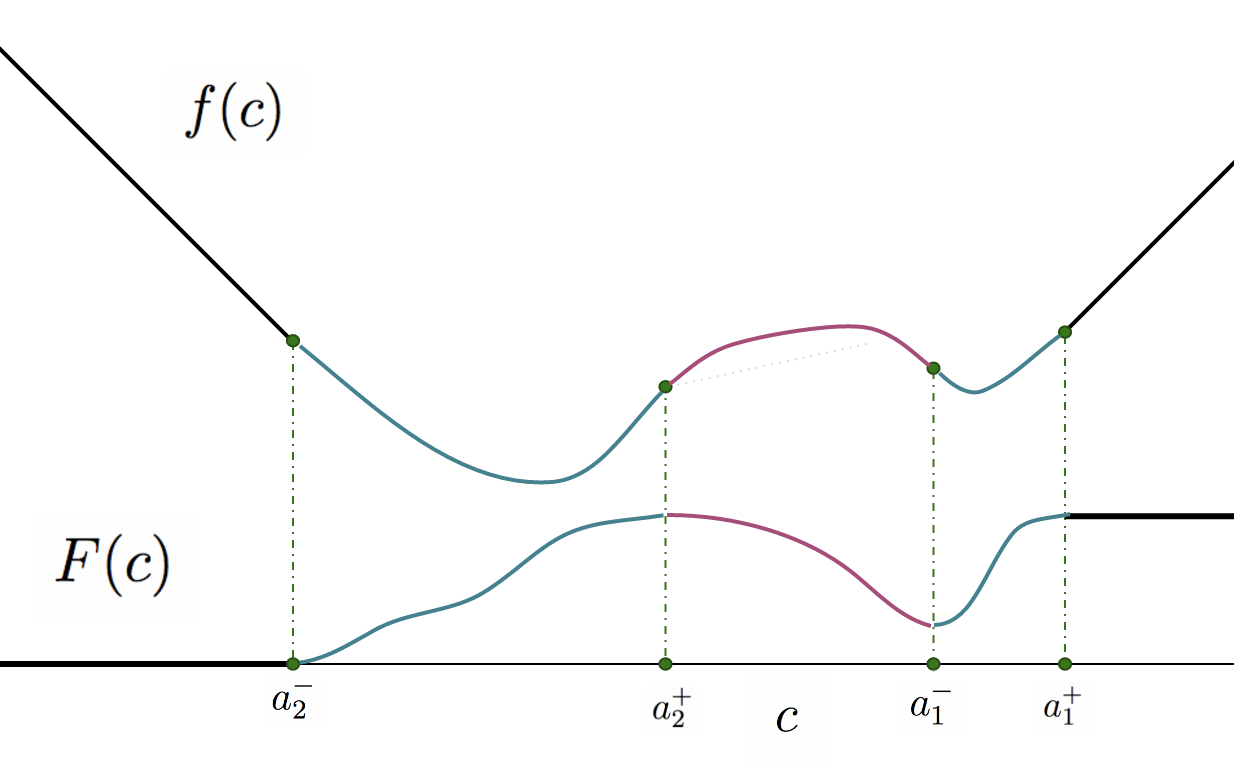}
\end{figure}

\begin{definition} The \underline{Rayleigh function} $F_f$ of a profile $f $ is \begin{equation} F_f(c) := \frac{ 1 + f'(c) }{2} .\end{equation} \end{definition}

\noindent Note that the Rayleigh function $F_f$ takes values in $[0,1]$ since $f$ is $1$-Lipshitz.
\begin{definition} If a Rayleigh function $F_f$ is of bounded variation, it defines a signed measure $dF_f$ on $\R$ we call a \underline{Rayleigh measure}. \end{definition}

\begin{definition} Non-negative measures $dF^{\uparrow}, dF^{\downarrow}$ on $\R$ are \underline{interlacing measures} if their difference $dF = dF^{\uparrow} - dF^{\downarrow}$ is a Rayleigh measure $dF_f$ of some profile $f$. 
\end{definition}

\noindent We say a profile $f$ is of compact support if its Rayleigh measure is.  On the previous page, we depict a generic profile $f$ of compact support $[a_2^-, a_1^+]$ and below its Rayleigh function $F=F_f$.  The points of inflection of $f$ separate regions of convexity and concavity, which correspond to increasing $\uparrow$ or decreasing $\downarrow$ regions of the Rayleigh function $F_f$.  For the profile depicted here, its interlacing measures $dF^{\uparrow}_f$ and $dF^{\downarrow}_f$ are supported on $[a_2^-, a_2^+] \cup [a_1^-, a_1^+]$ and $[a_2^+, a_1]$, respectively.\\
\\
\noindent Kerov's theory of profiles interpolates between two fundamental and very different special cases:  \textit{convex} profiles and profiles of \textit{interlacing sequences} we now describe.

\begin{definition} A \underline{convex profile} is a profile $f$ of bounded variation whose Rayleigh measure $dF_f(c) = dF^{\uparrow}_f(c)$ is an arbitrary probability measure $dF_f \in \mathbf{P}$.
\end{definition}   
  \begin{figure}[htb]
\centering
\includegraphics[width=0.90 \textwidth]{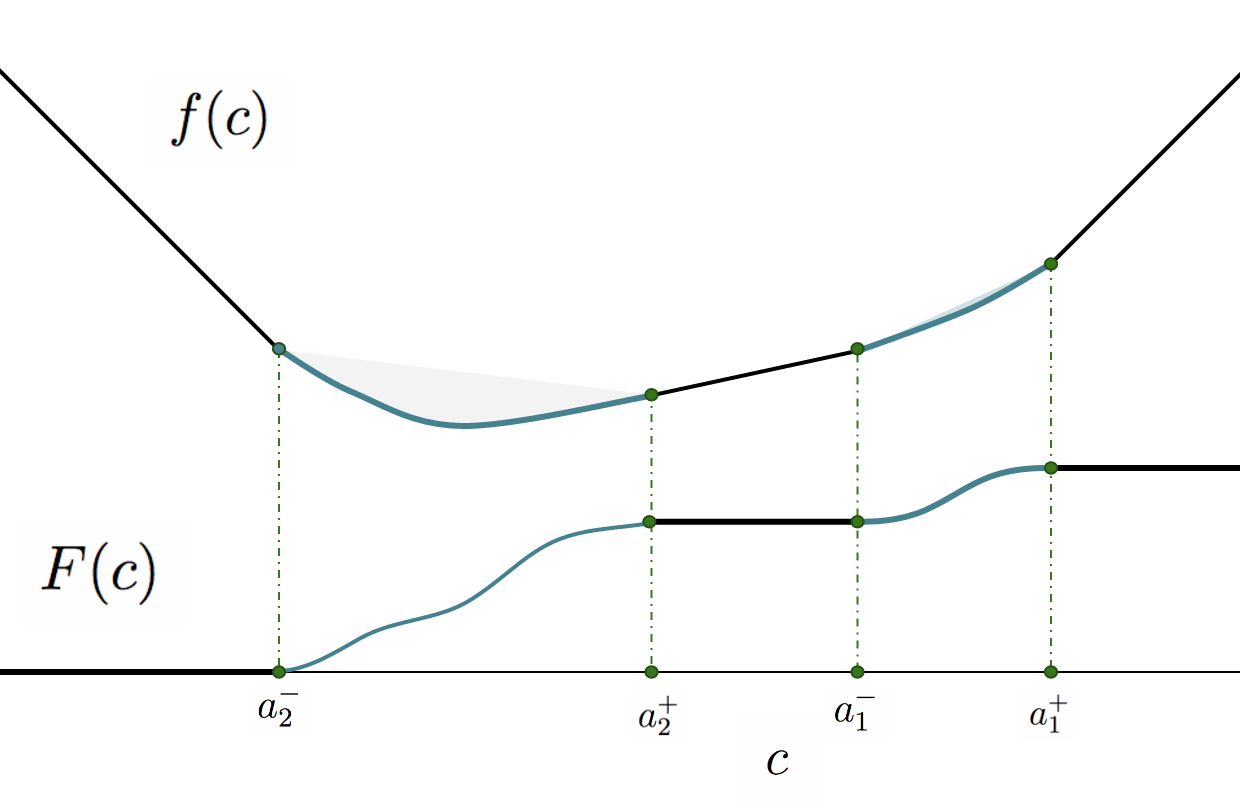}
\end{figure}

\noindent We've drawn a convex profile $f$ whose $dF_f = dF_f^{\uparrow}$ is supported on two disjoint bands $[a_2^-, a_2^+]$ and $[a_1^-, a_1^+]$.  As $dF_f^{\downarrow} \equiv 0$, $f$ forms a linear facet above the gap $[a_2^+, a_1^-]$.

\begin{proposition} \label{CHBDensityProfile}Since $dF_{\star |v}$ of \textnormal{Definition [\ref{CHBHConservedDensityDefinition}]} is a probability measure, there exists a convex profile $f_{\star |v} \in \mathbf{P}^{\vee}$ whose Rayleigh measure $dF_{\star |v}$ is the conserved density of the classical periodic dispersionless Benjamin-Ono equation.  If $v: \mathbb{T} \rightarrow \R$ is continuous, the profile $f_{\star |v}$ is supported on a single connected band $[\min_{\mathbb{T}} v, \max_{\mathbb{T}} v]$. \end{proposition}
\noindent The profiles most unlike convex profiles are the profiles of interlacing sequences:

  \begin{figure}[htb]
\centering
\includegraphics[width=0.90 \textwidth]{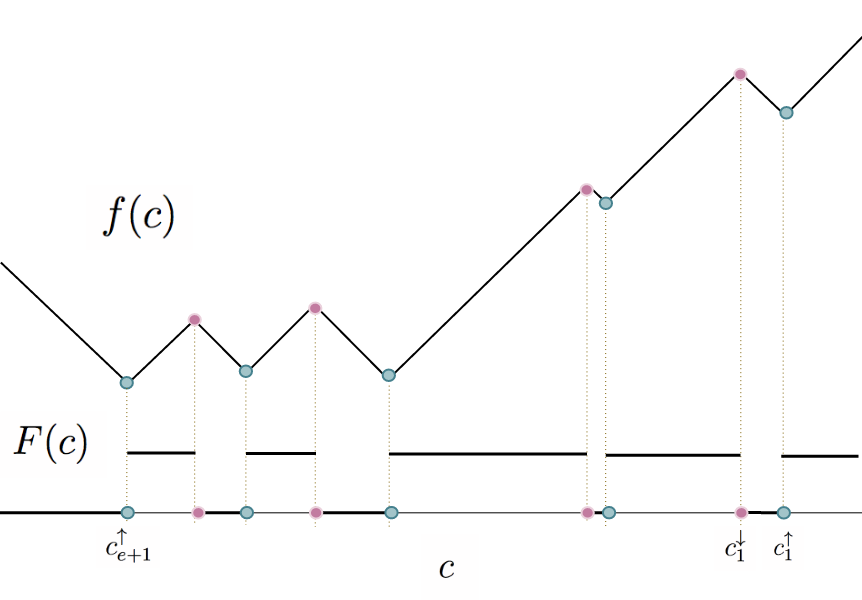}
\end{figure}

 \begin{definition} A profile of a pair $\{c_i^{\uparrow} \}_{i=1}^e$, $\{c_j^{\downarrow} \}_{j=1}^{e-1} \subset \R$ of \underline{interlacing sequences}

\begin{equation} c_e^{\uparrow} < c_{e-1}^{\downarrow} < c_{e-1}^{\uparrow} < \cdots c_2^{\uparrow} < c_1^{\downarrow} < c_1^{\uparrow} \end{equation}
\noindent of real numbers is a profile $f$ whose Rayleigh measure $dF_f$ is given by

\begin{equation} dF_f(c) = \sum_{i=1}^e \delta(c - c_i^{\uparrow}) - \sum_{j=1}^{e-1} \delta( c - c_j^{\downarrow}) \end{equation}

\end{definition}

\noindent Whereas regions of concavity and convexity of a generic profile may be of full Lebesgue measure, for the profile of an interlacing sequence its regions of convexity and concavity are localized at the local minima and maxima of the piecewise-linear profile $f$.  
\begin{proposition} \label{BODensityProfiles} According to \textnormal{Theorems [\ref{CBOHConservedDensityExistence}] and [\ref{QBOHConservedDensitiesForJacksAreAnisotropicPartitions}]}, conserved densities $dF(c | \ebar)|_v$ and $d \widehat{F}^{\eta_{NS}}(c | \ebar, \hbar)|_{\Psi_{\lambda}( \cdot | \hbar)}$ of the classical and quantum periodic Benjamin-Ono systems are Rayleigh measures of profiles of anisotropic partitions are profiles of non-generic interlacing sequences. \end{proposition}

\noindent Our goal will be to verify these claims in Theorems [\ref{CBOHConservedDensityConstruction}], [\ref{QBOHConservedDensitiesForJacksAreAnisotropicPartitionsRevisited}], respectively.\\
\\
\noindent Last but not least, let us turn to the most fundamental family of profiles: the vacua.

  \begin{figure}[htb]
\centering
\includegraphics[width=0.75 \textwidth]{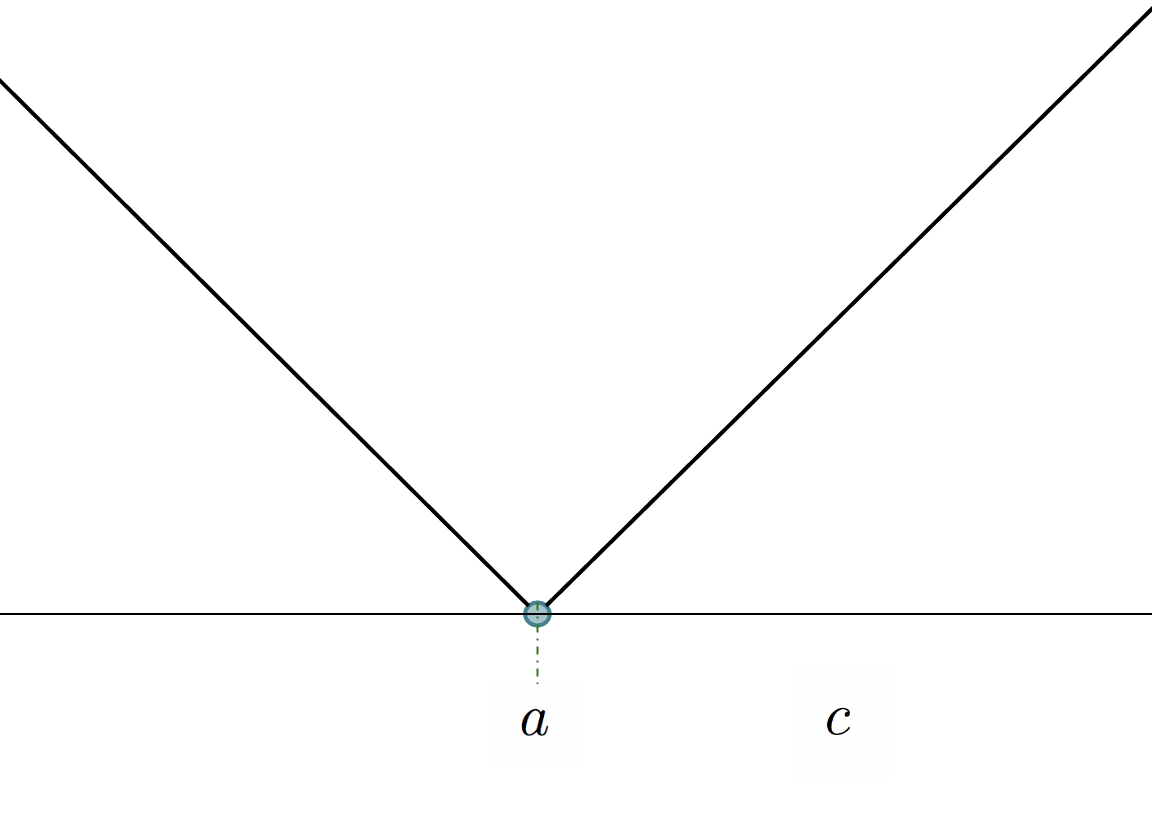}
\end{figure}

\begin{definition} For every $a \in \R$, the \underline{profile of the vacua $\Upsilon_a$} is defined by \begin{equation} f_a( c) = |c- a| .\end{equation} \end{definition}

\noindent Profiles of vacua are both convex and also the profile of the interlacing sequence $\{a\}, \emptyset$.  Our notation is later seen to reflect the fact that $f_a(c)$ describes the random values of the $\widehat{\mathfrak{gl}_1}$ affine Kac-Moody current $\vcurrent ( w | \hbar)$ at level $\hbar$ in the coherent state $\Upsilon_a( \cdot | \hbar)$ around the constant classical state $v(w) \equiv V_0 = a$.

\begin{definition} A profile $f \in \mathbf{P}^{\vee}$ is \underline{centered at $a$} if 

\begin{equation} a = \int_{- \infty}^{+\infty} c \ dF_f(c) \end{equation}

\noindent its Rayleigh measure has mean $a$.
\end{definition}

\noindent The profile of the vacua $\Upsilon_a$ is centered at $a \in \R$.  By comparison, the profiles depicted on the previous pages are drawn to agree outside their support with the profile of some vacua $a$, but we have drawn them without specifying their center $a$.  For all profiles, even those without a center, the value $c=0$ is distinguished since it appears in the limits of integration in the decay condition (\ref{profileDefDecay}), so we study the behavior of profiles $f$ relative to the profile $f_0(c) = |c- 0|$ of the vacua $\Upsilon_0$:

\begin{definition} The \underline{shifted Rayleigh function} $\xi_f$ of a profile $f$ is the difference

\begin{equation} \ \ \ \ \ \ \ \  \xi_f(c) := F_f(c) - F_{f_0}(c) \end{equation}

\noindent of its Rayleigh function from $F_{f_0}(c) = \mathbbm{1}_{[0, \infty)}(c)$ that of the profile of the vacua $\Upsilon_0$. \end{definition}

\noindent Finally, note one can recover $f$ from $\xi_f$ via $f(c) = \int_{- \infty}^c \xi_f(y) dy + \int_c^{+\infty} (1 - \xi_f(y)) dy$.

\subsubsection{Kerov's Markov-Kre\u{\i}n Correspondence} \label{subsubsecKMK}

\noindent Any profile $f \in \mathbf{P}^{\vee}$, not necessarily centered, has a distinguished generating function

\begin{definition} For $u \in \C \setminus \R$, the \underline{$T^{\uparrow}$-observable} of $f \in \mathbf{P}^{\vee}$ is defined through

\begin{equation} \label{TUpObservableDefinition}
 T^{\uparrow}(u) \Big |_{f} = \frac{1}{u} \cdot  \textnormal{exp} \Bigg ( - \int_{- \infty}^{+\infty} \frac{ \xi_f(c) dc}{u-c} \Bigg ) \end{equation}
 
 \noindent the exponential of the Stieltjes transform of the shifted Rayleigh function.

\end{definition}

\begin{proposition} If $f \in \mathbf{P}^{\vee}$ is of bounded variation, its $T^{\uparrow}$-observable 

\begin{equation} T^{\uparrow}(u) \Big |_{f} = \textnormal{exp} \Bigg ( \int_{- \infty}^{+\infty} \log \Bigg [ \frac{1}{u-c} \Bigg ] d F_f(c) \Bigg ) \end{equation}

\noindent can be written through its Rayleigh measure for $u \in \C \setminus \R$. \textnormal{[\textit{Proof:} integrate by parts.]} \end{proposition}

\begin{theorem} \label{KMKstatement} \textnormal{(Kerov's Markov-Kre\u{\i}n Correspondence \cite{Ke1})}  The $T^{\uparrow}$-observable $T^{\uparrow}(u)|_f$ of any profile $f \in \mathbf{P}^{\vee}$ is also the Stieltjes transform  \begin{equation}  \label{KMKgeneral} \int_{- \infty}^{+\infty} \frac{ d \tau_f^{\uparrow}(c) }{ u-c} =  T^{\uparrow}(u)  |_f = \frac{1}{u} \cdot  \textnormal{exp} \Bigg ( - \int_{- \infty}^{+\infty} \frac{ \xi_f(c) dc}{u-c} \Bigg )  \end{equation}

\noindent of a unique probability measure $\tau^{\uparrow}_f \in \mathbf{P}$ called the \underline{transition measure} of $f$.  Moreover, the formula (\ref{KMKgeneral}) defining $f \mapsto \tau_f$ is a bijection $\mathbf{P}^{\vee} \rightarrow \textnormal{\textbf{P}}$.\end{theorem}

\noindent Our notation $\uparrow$ in $T^{\uparrow}(u)$ emphasizes a relationship to $dF^{\uparrow}$ in the Jordan decomposition of the Rayleigh measure $dF$ supported on convex regions of the profile $f$.  For example, if $f$ is the profile of an interlacing sequence $c_{e}^{\uparrow} < \cdots < c_1^{\downarrow} < c_1^{\uparrow}$, equation (\ref{KMKgeneral}) is

\begin{equation} \int_{- \infty}^{+\infty} \frac{ d \tau^{\uparrow}(c)}{u-c} = T^{\uparrow}(u) |_f = \frac{ \prod_{j=1}^{e-1} (u - c_j^{\downarrow}) }{ \prod_{i=1}^e (u - c_i^{\uparrow})} \end{equation}

\noindent Therefore, after taking partial fraction decomposition of the rational function on the right-hand side, one sees that the transition measure $d\tau^{\uparrow}$ is both non-negative and supported on $\{c_e^{\uparrow}, \ldots, c_1^{\uparrow}\}$ the support of $dF^{\uparrow}(c)$.  As discussed in section 2.3 of \cite{Ke1},
\begin{corollary} \label{KMKsupport} In Kerov's Markov-Kre\u{\i}n correspondence \textnormal{(\ref{KMKgeneral})}, the supports of the Rayleigh measure $dF^{\uparrow}$ and the transition measure $d\tau^{\uparrow}$ coincide.  In particular, if their mutual support is bounded, the $T^{\uparrow}$-observable has a convergent expansion 

\begin{equation} \sum_{\ell=0}^{\infty} T_{\ell}^{\uparrow} u^{- \ell -1} = T^{\uparrow}(u) = \textnormal{exp} \Bigg ( \sum_{l = 1}^{\infty} O_l \frac{ u^{-l}}{l} \Bigg ) \end{equation}

\noindent where $T_{\ell}^{\uparrow} = \int_{- \infty}^{+\infty} c^{\ell} d \tau^{\uparrow}$ and $O_l = - l \int_{- \infty}^{+\infty} c^{l-1} \xi(c ) dc$ simplifying to $O_l = \int_{- \infty}^{+\infty} c^l d \xi( c)$ if $\xi$ is bounded variation.  In general, $O_l$ is a polynomial in $T^{\uparrow}_{\ell}$ for $0 \leq \ell \leq l$.  \end{corollary}

\noindent We can finally recognize formula (\ref{OPRLcorollaryFormula}) as a particular case of formula (\ref{KMKgeneral}).

\begin{corollary} \label{KMKstatementII} For $L_{\bullet}$ and $\Psi_0$ satisfying the conditions of Theorem \textnormal{[\ref{OPRL}]}, there exists a profile $f \in \textbf{P}^{\vee}$ so that
\begin{itemize}
\item The $T^{\uparrow}$-observable $T^{\uparrow}(u) |_f$ in \textnormal{(\ref{KMKgeneral})} is the Titchmarsh-Weyl function \textnormal{(\ref{OPRLformula})}
\item The transition measure $\tau_f^{\uparrow}$ of $f$ is the {spectral measure} of $L_{\bullet}$ at $\Psi_0$ and
\item The shifted Rayleigh function $\xi_f$ of $f$ is the \textcolor{black}{\textit{spectral shift function}} $\xi (c | L_{\bullet}, L_+)$. \end{itemize}
\end{corollary}

\noindent Assuming $L_{\bullet}$ bounded, Kerov proved Theorem [\ref{OPRL}] and Corollary [\ref{KMKstatementII}] in chapters 5 and 6 of \cite{Ke1}.  Our distinction between the Rayleigh function $F_f$ and shifted Rayleigh function $\xi_f = F - \mathbbm{1}_{[0, \infty)}$ corrects the statement of Theorem 6.1.3 in \cite{Ke1} by accounting for the difference between the minor $L_+^{\perp}$ on $\mathscr{H}_{+} = \mathscr{H}_{\Psi_0}^{\perp}$ and $L_+ = 0 \oplus L_+^{\perp}$ on $\mathscr{H}_{\bullet}$.

\subsection{Constructing Classical Periodic Benjamin-Ono Densities} \label{subsecCBOHConstruction}
\noindent Our goal in this subsection is to construct in Theorem [\ref{CBOHConservedDensityConstruction}] a conserved density $d F_{ \star |v}( c | \ebar)$ of the classical periodic Benjamin-Ono hierarchy \cite{NaSk2} whose existence was announced in Theorem [\ref{CBOHConservedDensityExistence}].  In section [\textbf{\ref{subsubsecToeplitz}}] we define Toeplitz operators $L_{\bullet}(v)$ on the circle $\mathbb{T}$ with symbol $v$.  In section [\textbf{\ref{subsubsecCBOLax}}], we define the Lax operator for classical Benjamin-Ono \cite{BockKruskal, Nakamura1979} as $L_{\bullet}( v | \ebar) = L_{\bullet}(v ) + \ebar D_{\bullet}$, an elliptic generalized Toeplitz operator of order $1$ where $\ebar>0 $ is the coefficient of dispersion \cite{DeMonvelGuillemin}.  In section \textbf{[\ref{subsubsecNSClassicalCore}]} we recall the classical transfer observables of Nazarov-Sklyanin \cite{NaSk2} to construct $dF_{\star |v}(c | \ebar)$ in section [\textbf{\ref{subsubsecCBOHConservedDensityConstruction}}] via the auxiliary spectral theory of $L_{\bullet}(v | \ebar)$.\\
\\
\noindent To complete the proof of Theorem [\ref{CBOHConservedDensityConstruction}], we must address the dispersionless limit.  In section \textbf{[\ref{subsubsecSZEGO}]}, we check that $\lim_{\ebar \rightarrow 0} dF_{\star |v}(c | \ebar)$ the dispersionless limit of the conserved density agrees with the explicit $dF_{\star |v}(c)$ of Theorem [\ref{CHBHConservedDensityIntro}].  This verification is equivalent to Szeg\H{o}'s First Theorem for Toeplitz operators \cite{DeiftItsKra, SimonSzego}.  We rederive this famous result via Wiener-Hopf factorizations, reviewed in [\textbf{\ref{subsubsecWHFactorizations}}], as we also use Wiener-Hopf factorizations in our treatment of quantum periodic Benjamin-Ono.

\subsubsection{Toeplitz Operators} \label{subsubsecToeplitz}
\noindent Throughout, $\mathbb{T} = \{ w \in \C : |w|=1\}$ is the unit circle and $H$ is the Hilbert space \begin{equation} H := L^2(\mathbb{T}) \end{equation}

\noindent of $\phi:\mathbb{T} \rightarrow \C$ with pairing abbreviated by $\langle \cdot, \cdot \rangle$

\begin{equation} \langle \phi^{\textnormal{out}}, \phi^{\textnormal{in}} \rangle = \oint_{\mathbb{T}} \overline{\phi^{\textnormal{out}}(w)} \cdot \phi^{\textnormal{in}} (w)  \frac{ dw }{ 2 \pi \textbf{i} w}. \end{equation}

\begin{definition} For $\varphi: \mathbb{T} \rightarrow \C$, define the \underline{multiplication operator} $L(\varphi)$ on $H$ by\begin{equation}  (L(\varphi) \phi)(w) = v(w) \cdot \phi(w). \end{equation}
\end{definition} 

\begin{proposition} $D = w \frac{ \partial}{\partial w}$ of \textnormal{Definition [\ref{OperatorsOnCircleDefinition}]} is an unbounded self-adjoint differential operator on $H$ with eigenfunctions $\{w^h \}_{h \in \Z}$ abbreviated by $w^h=|h \rangle$. \end{proposition}

\begin{definition} \label{SzegoProjectionDefinition} The \underline{Szeg\H{o} projection} $\uppi_{\bullet}$ is the bounded singular integral operator

\begin{equation} \label{HardyIntegralOperator} ( \uppi_{\bullet} g)(w_+) = \oint_{\mathbb{T}} \frac{ g(w_-)  }{ w_- - w_+} \frac{ dw_- }{ 2 \pi \textnormal{\textbf{i}} w_-}  \end{equation}

\noindent that projects $\uppi_{\bullet} : H \rightarrow H_{\bullet}$ onto the \underline{Hardy space} $H_{\bullet}$, the closure of $\C[w] \subset H$.

\end{definition}

\begin{definition} The \underline{Toeplitz operator} $L_{\bullet}( \varphi)$ on $H_{\bullet}$ with symbol $\varphi$ is \begin{equation} L_{\bullet}(\varphi) = \uppi_{\bullet} L(\varphi) \uppi_{\bullet} \end{equation}

\noindent a compression of the {multiplication operator} $L(\varphi)$ on $H$. \end{definition}

\noindent Hardy space is $H_{\bullet}$, not $H_+$, because it contains $|h \rangle$ for $h=0,1,2,\ldots$ including $0$.  Such $w^h \in H_{\bullet}$ extend from the unit circle ``$\circ$'' across the origin ``$\cdot$'' so we write ``$\bullet$''.  The distinction between positive and non-negative is particularly sacred in this paper:

\begin{definition} \label{ToeplitzShiftsDefs}While the Szeg\H{o} projection $\uppi_{\bullet}$ projects onto
\begin{equation} H = H_- \oplus H_{\bullet} \end{equation}

\noindent the space $H_{\bullet}$ of non-negative eigenfunctions of $D$, the \underline{shifted Szeg\H{o} projection}
\begin{equation} \uppi_{+} := L(w) \uppi_{\bullet} L(w^{-1}) \end{equation}

\noindent is an operator $\uppi_{+} : H_{\bullet} \rightarrow H_{\bullet}$ block-diagonal $\uppi_+ = 0 \oplus \uppi_+^{\perp}$ where $\uppi^{\perp}_+$ is the projection onto the \underline{shifted Hardy space} $H_{+}$ of positive eigenfunctions of $D$.  We have \begin{equation} H = H_- \oplus H_0 \oplus H_+ \end{equation}

\noindent so $H_{\bullet} = H_0 \oplus H_+$ where $\dim H_0 = 1$ is spanned by the \underline{auxiliary vacuum} $w^0 = |0 \rangle$. 

\end{definition}

\noindent In the basis $|h \rangle = w^h$ of the dense subspace $\C[w] \subset H_{\bullet}$, the Toeplitz operator is
\begin{equation}\label{ToeplitzMatrix} L_{\bullet}(\varphi) \big |_{\C[w]} = \begin{bmatrix} 
\varphi_0 & \varphi_{-1} & \varphi_{-2} & \cdots  \\ 
\varphi_{1} & \varphi_0 & \varphi_{-1} & \ddots  \\
 \varphi_{2} & \varphi_{1} & \varphi_0 & \ddots \\
  \vdots & \vdots & \ddots & \ddots \end{bmatrix} \end{equation}
  
\noindent where $\varphi_k = \oint_{\mathbb{T}} w^{+k} \varphi(w) \frac{ dw}{ 2 \pi \textbf{i} w}$ are the Fourier modes of the symbol $\varphi(x) = \varphi_0 +  \sum_{k=1}^{\infty} \varphi_{-k} e^{+ \textbf{i} k x} + \varphi_k e^{- \textbf{i} k x}$.\\
\\
\noindent We now specialize to symbols $\varphi = v$ that satisfy two crucial assumptions:

\begin{enumerate} 
\item \textcolor{gray}{[Regularity]} \ The symbol $v$ is \underline{bounded}, i.e. $||v||_{\infty} < \infty$
\item \textcolor{gray}{[Reality]}  \ \ \ \ The symbol $v: \mathbb{T} \rightarrow \R$ is \underline{real-valued}, i.e. $V_{-k} = \overline{V_{k}}$
\end{enumerate}

\noindent The multiplication operator $L(v)$ is bounded and self-adjoint on $H$ if $v: \mathbb{T} \rightarrow\R$ is bounded and real.  By Definition [\ref{SpectralMeasureDef}], one has:

\begin{proposition} The spectral measure of the multiplication operator $L(v)$ at the vacuum $\Psi_0 = w^0 = 1 \in H$ is $dF_{\star |v} = v_* \rho_{\star  | \mathbb{T}}$ the push-forward of the normalized uniform measure on the circle $\mathbb{T}$ along the symbol $v: \mathbb{T} \rightarrow \R$ of \textnormal{Definition [\ref{CHBHConservedDensityDefinition}]}, hence also the conserved density of the classical dispersionless periodic Benjamin-Ono by \textnormal{Theorem [\ref{CHBHConservedDensityIntro}]}. \end{proposition}

\noindent In Theorem [\ref{CHBHConstruction}] we will give a different realization of the conserved density $dF_{\star |v}(c)$ for the initial value problem (\ref{CHBE}) via a different auxiliary spectral theory, namely by Szeg\H{o}'s First Theorem for Toeplitz operators $L_{\bullet}(v)$.

\begin{theorem} \label{Toeplitz1911} \textnormal{[Toeplitz 1911]} $L_{\bullet}(v)$ is bounded if and only if $v$ is bounded.  The spectrum of $L_{\bullet}(v)$ coincides with the essential range of $v$ and $||L_{\bullet}(v) ||_{\textnormal{op}} = || v||_{\infty}$.
\end{theorem}

\noindent For a proof, see Theorem 2.7 in \cite{BottcherSilbermann}.  Note that the regularity of symbols is of paramount importance in the theory of Toeplitz operators and their applications in statistical mechanics, as phase transitions in the qualitative nature of their spectral theory occur as families of smooth symbols acquire discontinuities and singularities \cite{DeiftItsKra}.  On the other hand, our reality assumption has immediate consequences:

\begin{proposition} \label{ToeplitzRealNice} For real-valued $v: \mathbb{T} \rightarrow \R$, the Toeplitz operator $L_{\bullet}(v) |_{\C[w]}$ is $\C$-symmetric.  Thus, if $v$ is also bounded, $L_{\bullet}(v)$ is bounded self-adjoint, hence $L_{\bullet}(v)  |_{\C[w]}$ must be essentially self-adjoint on $\C[w]$ and also on the orbit of $\Psi_0 = | 0 \rangle$.  \end{proposition}

\noindent Note that non-symmetric matrices $L_{\bullet}(v) |_{\C[w]}$ cannot be essentially self-adjoint, but one can still ask that the Galerkin approximations $L_{\bullet \tealN}(v)$ are ``stable'' in the sense of strong resolvent convergence $(u - L_{\bullet, \tealN}(v))^{-1} \rightarrow (u-L_{\bullet} (v))^{-1}$ as $u$ varies in the connected components of $\C \setminus v(\mathbb{T})$.  This complement of the range $v( \mathbb{T})$ of the symbol is possibly very complicated for complex and discontinuous symbols $v$.  For real symbols $v: \mathbb{T} \rightarrow \R$, the range $v ( \mathbb{T})$ is an interval if $v$ is continuous no matter how wild the symbol $v$.  For conditions on complex-valued symbols of lower regularity that ensure $L_{\bullet}(v)$ is Fredholm, invertible, or stable, see \cite{BottcherSilbermann, BottcherSilbermannIntro}.

\subsubsection{Lax Operator for Classical Benjamin-Ono} \label{subsubsecCBOLax}

\noindent Recall the differential operator $D$ and $|D| = ( - \Delta)^{1/2}$ the fractional Laplacian on $\mathbb{T}$ with fraction $+\frac{1}{2}$ from Definition [\ref{OperatorsOnCircleDefinition}].

\begin{definition} \label{AuxiliaryDegreeOperatorDefinition} The \underline{auxiliary degree operator} $D_{\bullet}$ is the restriction of either $D$ or $|D|$ to the Hardy space $H_{\bullet} \subset H$. \end{definition}

\noindent $D$ and $|D|$ are both self-adjoint extensions of $D_{\bullet}$ to $H$.
\begin{definition} \label{CBOLaxDefinition} The \underline{Lax operator} for the classical Benjamin-Ono hierarchy is \begin{equation} L_{\bullet}( v| \ebar) = L_{\bullet}(v) + \ebar D_{\bullet} \end{equation} the generalized Toeplitz operator of order $1>0$ on $H_{\bullet}$ with symbol $L(v) + \ebar D$. \end{definition}

\noindent $L_{\bullet}(v  | \ebar)$ unambiguously denotes an unbounded self-adjoint operator on Hardy space $H_{\bullet}$ because both $L_{\bullet}(v)$ and $D_{\bullet}$ are essentially self-adjoint on $\C[w] \subset H_{\bullet}$, i.e.
\begin{proposition} \label{CBOLaxEssentiallySelfAdjoint} For bounded $v$, $L_{\bullet}( v | \ebar)$ is essentially self-adjoint on $\C[w] \subset H_{\bullet}$.\end{proposition}

\noindent The spectral theory of $L_{\bullet}(v | \ebar)$ differs wildly from that of $L_{\bullet}(v)$.  The analytic subtleties underling the limit $\ebar \leftarrow 0$ correspond to the dynamical subtleties in the dispersionless limit of the classical Benjamin-Ono system.

\begin{proposition} \label{ClassicalLaxDiscreteSpectrum} For  $\ebar > 0$, $L_{\bullet}(v | \ebar)$ is elliptic and has discrete spectrum in $H_{\bullet}$ \begin{equation} c_1^{\uparrow}(\ebar)|_v < c_2^{\uparrow}(\ebar)|_v < c_3^{\uparrow}(\ebar)|_v < \cdots \end{equation}

\noindent bounded below with only $+ \infty$ as a point of accumulation.
\end{proposition}
\begin{itemize}
\item \textit{Proof:} use Proposition 2.14 in \cite{DeMonvelGuillemin} for arbitrary elliptic generalized Toeplitz operators on manifolds $\mathbb{X}$ and take $\mathbb{X} = \mathbb{T}$.  See also section 3.1 of \cite{Agnarovich}. $\square$
\end{itemize}
\noindent By contrast, for $\ebar = 0$, Toeplitz operators $L_{\bullet}(v)$ have absolutely continuous spectrum.

\subsubsection{\textcolor{black}{Nazarov-Sklyanin Classical Transfer Observables}} \label{subsubsecNSClassicalCore}

\noindent We now construct the classical periodic Benjamin-Ono hierarchy and its conserved density whose existence was announced in Theorem [\ref{CBOHConservedDensityExistence}].  At the core of what follows is the following result which we state without complete proof:

\begin{theorem} \label{NSClassicalCore} \textnormal{(Nazarov-Sklyanin \cite{NaSk2})} The Titchmarsh-Weyl function 

\begin{equation} {{T}}^{\uparrow}(u ) |_{{{L}}_{\bullet}( \ebar)} = \langle 0 | \frac{1}{ u - L_{\bullet}( v | \ebar)} | 0 \rangle\end{equation}

\noindent of the Lax operator $\widehat{\mathcal{L}}_{\bullet} ( \ebar, \hbar)$ of classical Benjamin-Ono from \textnormal{Definition [\ref{CBOLaxDefinition}]} is a $\C$-valued function on the real $L^2$ Sobolev space $(\mathscr{M}, J, \mathsf{g}_{-1/2}, \omega_{-1/2})$ abbreviated $T^{\uparrow}(u| \ebar) |_{v}$ so that for all $u_1, u_2 \in \C \setminus \R$ we have \begin{equation} \{ T^{\uparrow}(u_1| \ebar) , T^{\uparrow}(u_2 | \ebar) \} = 0 \end{equation}

\noindent Poisson commute for the Gardner-Faddeev-Zakharov bracket \textnormal{(\ref{GFZBracket})}.
\end{theorem}

\begin{definition} The \underline{Nazarov-Sklyanin classical transfer observables} for classical periodic Benjamin-Ono \textnormal{(\ref{CBOE})} are $\widehat{\mathcal{T}}^{\uparrow}(u| \ebar, \hbar) =\widehat{\mathcal{T}}^{\uparrow}(u ) |_{\widehat{\mathcal{L}}_{\bullet}( \ebar, \hbar); \mathscr{H}_0}$ of \textnormal{Theorem [\ref{NSClassicalCore}]}. \end{definition}

\begin{itemize}
\item \textit{Proof:} In \cite{NaSk2}, the authors prove Poisson-commutativity of coefficients in the $1/u$ expansion of the ``cotransfer observable'' ${{T}}^{\downarrow}( u |  \ebar)$ defined via transfer observables \begin{equation} {{T}}^{\uparrow}(u | \ebar) \cdot \Big ( u - {{T}}^{\downarrow}( u | \ebar) \Big ) = \mathbbm{1} \end{equation} \noindent which automatically implies Theorem [\ref{NSClassicalCore}] as stated above.  The conventions in \cite{NaSk2} require a rescaling of the variable $u$ to match the conventions of Definitions [\ref{OmegaVariablesDefinition}] and [\ref{FractionalChargeDefinition}] in the presentation of Jack polynomials in \cite{Mac}. $\square$
\end{itemize}

\subsubsection{\textcolor{black}{Conserved Densities as Spectral Shift Functions}} \label{subsubsecCBOHConservedDensityConstruction}

\noindent We now extend Nazarov-Sklyanin's Theorem [\ref{NSClassicalCore}] as promised in Theorem [\ref{CBOHConservedDensityExistence}].

\begin{theorem}  \label{CBOHConservedDensityConstruction} For bounded real $v: \mathbb{T} \rightarrow \R$ and without loss of generality $\ebar >0$,let $L_{\bullet}(v | \ebar)$ be the Lax operator of classical Benjamin-Ono and define $L_{+}(v | \ebar)$ with respect to $\Psi_0 = | 0 \rangle \in H_{\bullet}$ in the sense of \textnormal{Definition [\ref{MinorGeneralDefinitions}].}  Then $f_{\star |v} ( c | \ebar)$ is the unique profile of an interlacing sequence

\begin{equation} c_1^{\uparrow}(\ebar) |_v <  c_1^{\downarrow}(\ebar)|_v <  c_2^{\uparrow}(\ebar) |_v < \cdots <  c_{e+1}^{\uparrow}(\ebar)|_v < \cdots \leq +  \infty \end{equation}

\noindent with $+ \infty$ as its only accumulation point so that

\begin{itemize}
\item The $T^{\uparrow}$-observable $T^{\uparrow}(u|\ebar ) |_v $ in \textnormal{(\ref{KMKgeneral})} is the Titchmarsh-Weyl function \textnormal{(\ref{OPRLformula})}
\item The transition measure $\tau_{\star |v}^{\uparrow}(c | \ebar)$ is the {spectral measure} of $L_{\bullet}(v|\ebar)$ at $|0 \rangle \in H_{\bullet}$
\item The shifted Rayleigh function $\xi_{\star |v}(c |\ebar)$ is the \textcolor{black}{\textit{spectral shift function}} of the pair $L_{\bullet}(v | \ebar)$, $L_+ (v |\ebar)$.
\end{itemize}

\begin{enumerate}
\item \textnormal{\textcolor{gray}{[Conserved Density}]} The Rayleigh measure is a conserved density $dF_{\star |v}(c| \ebar) = d \xi_{\star |v}(c | \ebar) + \mathbbm{1}_{[0, \infty)}(c)$ of the classical periodic Benjamin-Ono equation \textnormal{(\ref{CBOE})} in the sense of  \textnormal{Definition [\ref{ClassicalConservedDensityDefinition}]}.
\item \textnormal{\textcolor{gray}{[Integrable Hierarchy]}} for $l=1,2,3,\ldots$, the classical observables

\begin{equation} O_l (\ebar) \big |_v = \int_{- \infty}^{\infty} c^l dF_{\star |v} ( c | \ebar) \end{equation}

\noindent pairwise Poisson commute for \textnormal{(\ref{GFZBracket})}

\noindent \begin{equation} \{ O_{l_1}(\ebar)  , O_{l_2}(\ebar) \}_{- \frac{1}{2}} = 0 \end{equation}

\item \textnormal{\textcolor{gray}{[Regularity of Observables]}} $O_l(\ebar)|_v$ is a generalized polynomial of degree $l \in \N$ in the $\sigma$-coordinates $\{V_k, \overline{V_k}\}$ of $v \in \mathscr{M}_0$ as in \textnormal{Definition [\ref{GeneralizedPolynomialDefinition}]}

\item \textnormal{\textcolor{gray}{[Periodic Benjamin-Ono]}} The span of $\{O_l(\ebar)\}_{l=1}^{\infty}$ includes the classical periodic Benjamin-Ono Hamiltonian $T_3^{\uparrow}(\ebar)$ \textnormal{(\ref{CBOHamiltonian})} and also $T_2^{\uparrow}$ \textnormal{(\ref{ClassicalT2})}
 \item \textnormal{{\textcolor{gray}{[``Finite Gap Potentials'']}}} for $\ebar \neq 0$, there are finitely-many interlacing extrema \begin{equation} c_{e+1}^{\uparrow}(\ebar)|_v < \cdots < c_1^{\downarrow}(\ebar)|_v < c_1^{\uparrow}(\ebar)|_v \end{equation}
 \noindent of the profile $f_{\star|v}(c| \ebar)$ centered at $V_0 = a \in \R$ associated to the conserved density $dF_{\star|v}(c|  \ebar)$ by $F(c) = \tfrac{1+ f'(c)}{2}$ if $v$ is a Laurent polynomial in \textnormal{$e^{\pm \textbf{i} k x} $}
 
\item \textnormal{\textcolor{gray}{[Dispersionless Limit]}} as $\ebar \rightarrow 0$ one recovers 
 \begin{equation}  \lim_{\ebar \rightarrow 0} dF_{\star |v} ( c | \ebar) = d F_{\star |v}(c)  \end{equation}
 \noindent the conserved density of \textnormal{Theorem [\ref{CHBHConservedDensityIntro}]}

\end{enumerate}

\end{theorem} 

\noindent We prove the enumerated claims of Theorem [\ref{CBOHConservedDensityConstruction}] in Steps 0-5 below.

\begin{itemize}
\item \textit{Step 0:} We first derive the three bulleted points characterizing the profile $f_{\star |v} ( c | \ebar)$.  Theorem [\ref{NSClassicalCore}] of \cite{NaSk2} implies that the spectral measures $d \tau^{\uparrow}_{\star |v} ( c |\ebar)$ of $L_{\bullet}( v | \ebar)$ at $|0 \rangle$ may be regarded as a conserved density for the classical Benjamin-Ono system.  Our regularity assumption - that $v$ is bounded and real - implies that $L_{\bullet}(v | \ebar)$ is essentially self-adjoint on the pre-Hardy space $\C[w] \subset H_{\bullet}$ by Proposition [\ref{CBOLaxEssentiallySelfAdjoint}], so the desired bullet points follow from Corollary [\ref{KMKstatementII}], taking $f_{\star |v}(c | \ebar)$ to be the Kerov-Markov-Kre\u{\i}n dual of $d \tau^{\uparrow}_{\star |v} c | \ebar)$.  In this step, we can assume without loss of generality that $\Psi_0= |0 \rangle$ is cyclic for $L_{\bullet}(v | \ebar)$, since if not we could just as easily apply Theorem [\ref{OPRL}] not to $\mathscr{H}_{\bullet} = H_{\bullet}$ but to the $L_{\bullet}(v | \ebar)$-orbit of $\Psi_0 = |0 \rangle$ and get the same result in Corollary [\ref{KMKstatementII}].  We also now know that the profile $f_{\star |v}(c)$ is the profile of a generically infinite interlacing sequence and is of bounded variation because there are no accumulation points in the spectrum of $L_{\bullet}(v | \ebar)$ by Proposition [\ref{ClassicalLaxDiscreteSpectrum}].  Note: the difference $L_{\bullet}( v | \ebar) - L_{+}( v | \ebar) = L_{\bullet}(v) - L_{+}(v)$ is independent of $\ebar$, but the spectral shift function $\xi_{\star |v} ( c| \ebar)$ depends on $\ebar$!  
\item \textit{Step 1:} The classical observables $O_l(\ebar)|_v$ are polynomials in the $T^{\uparrow}_{\ell} (\ebar)|_v$ by Corollary [\ref{KMKsupport}], so this result again follows from Theorem [\ref{NSClassicalCore}].

\item \textit{Step 2:} The $T^{\uparrow}_{\ell}(\ebar)|_v$ are easily seen to be generalized polynomials by inspecting formula $\langle 0 | L_{\bullet}(v| \ebar)^{\ell} | 0 \rangle$, so $O_l(\ebar)|_v$ are generalized polynomials since they are ordinary polynomials $T^{\uparrow}_{\ell}(\ebar)|_v$ by Corollary [\ref{KMKsupport}].  $O_l(\ebar)|_v$ are densely-defined because they are finite for Laurent $v$ by our argument in Step 4 below.  Indeed, for such $v$ we will have that 
\begin{equation} O_l ( \ebar) |_v = \text{Tr}_{H_{\bullet}} \Bigg [ L_{\bullet}( v | \ebar)^l - L_{+}(v | \ebar)^l \Bigg ] \end{equation}

\noindent holds even though the test function $\phi(c) = c^l$ does not satisfy the assumptions in the Lifshitz-Kre\u{\i}n Trace Formula, Theorem [\ref{LifshitzKreinTraceFormula}].  This is because Theorem [\ref{LifshitzKreinTraceFormula}] holds for perturbations with $L_{\bullet} - L_+$ trace class, whereas particular operators $L_{\bullet}(v | \ebar), L_+( v | \ebar)$ may have trace formulae for larger classes of $\phi$. 

\item \textit{Step 3:} Direct calculation for $\ell=2,3$.
\item \textit{Step 4:} By Proposition [\ref{ClassicalLaxDiscreteSpectrum}], it suffices to prove that the spectral measure $d \tau^{\uparrow}_{\star |v} ( c | \ebar)$ of the Lax operator $L_{\bullet}(v | \ebar)$ at the auxiliary vacuum $|0 \rangle \in H_{\bullet}$ is bounded if $v$ is a Laurent polynomial in $e^{ \pm \textbf{i} k x}$, since boundedness and discreteness with no accumulation points implies finiteness.  To begin, first prove $d \tau^{\uparrow}_{\star |v} ( c | \ebar)$ has finite $\ell$th moment if $||v||_{\infty}, \ldots, || |D|^{\ell} v ||_{\infty}< \infty$.  For this, Theorem [\ref{Toeplitz1911}] implies $ \sup_{0 \leq m \leq \ell} || L_{\bullet}(D^m v) ||_{\text{op}} = \sup_{0 \leq m \leq \ell} || D^m v ||_{\infty} < \infty$ for fixed $\ell \in \N$.  The fundamental identity $[ D_{\bullet}, L_{\bullet}(v)] = L_{\bullet} (|D| v)$ implies
\begin{equation} \Big | \int_{- \infty}^{\infty} c^{\ell} d \tau^{\uparrow}_{\star |v} (c | \ebar) \Big | = \Big | \langle 0 | \big ( L_{\bullet} (v) + \ebar D_{\bullet} \big )^{\ell} | 0 \rangle \Big | <\infty \end{equation}

\noindent is finite by commuting $D_{\bullet}$ to the right to annihilate the auxiliary vacuum $|0 \rangle$.  Now assuming $v$ Laurent, there is a $K_v \in \N$ so that $V_k \equiv 0$ for all $|k| \geq K_v$.  The compact support of the Fourier transform of $v$ controls supremums of derivatives \begin{equation} ||D^{\ell} v ||_{\infty} \leq K_v^{\ell} \cdot || v||_{\infty} \leq C_v^{\ell+1} \end{equation} \noindent for $C_v = \text{max} ( K_v, || v||_{\infty})$, so by the same argument above $|| L_{\bullet}(D^{\ell} v) ||_{op} \leq C_v^{\ell+1}$ \begin{equation} \Big | \int_{- \infty}^{\infty} c^{\ell} d \tau^{\uparrow}_{\star |v} (c | \ebar) \Big | = \Big | \langle 0 | \big ( L_{\bullet} (v) + \ebar D_{\bullet} \big )^{\ell} | 0 \rangle \Big |  \leq  \sum_{m=0}^{\ell} \ebar^m C_v^{\ell} \binom{l}{m}= \Big ( C_v ( 1 + \ebar ) \Big )^{\ell} \nonumber \end{equation}
\noindent which implies the support of $d\tau^{\uparrow}_{\star |v} (c | \ebar)$ lies in $[ - C_v ( 1 + \ebar), + C_v ( 1+\ebar)]$.  As discussed, this implies the spectral measure $d \tau_{\star |v}^{\uparrow}(c| \ebar)$ of $L_{\bullet}(v | \ebar)$ at $\Psi_0 = |0 \rangle$ is supported on a finite number of points \begin{equation} c_1^{\uparrow}[ v; \ebar] < \cdots < c_{e+1}^{\uparrow}[ v; \ebar]. \end{equation}

\item \textit{Step 5:} Weak convergence of spectral measures $d \tau^{\uparrow}_{\star |v}(c | \ebar) \rightarrow d \tau^{\uparrow}_{\star |v}(c)$ follows from the continuity of the spectral theorem, Theorem [\ref{SpectralTheorem}], in the strong topology, hence we have the desired weak convergence $d \xi_{\star |v}( c| \ebar) \rightarrow d\xi_{\star |v}(c)$ of spectral shift functions, hence also of Rayleigh measures.  This gives an a priori qualitatively different conserved density of the classical \textit{dispersionless} periodic Benjamin-Ono equation than the one we met in Theorem [\ref{CHBHConservedDensityIntro}], but our claim is that they coincide.  To verify this, it remains to compute the spectral shift function $\xi_{\star|v}(c)$ of a Toeplitz operator $L_{\bullet}(v)$ and its minor $L_+(v)$ explicitly and identify it with 
\begin{equation} \label{JelloThrone} d \xi_{\star |v}(c ) + \mathbbm{1}_{[0, \infty}(c)  =  dF_{\star |v}(c) =  (v_* \rho_{\star | \mathbb{T}})(c)\end{equation}

\noindent the push-forward of the uniform measure $\rho_{\star | \mathbb{T}}$ on the circle along $v : \mathbb{T} \rightarrow \R$.  We prove (\ref{JelloThrone}), equivalent to Szeg\H{o}'s First Theorem, in section  \textbf{[\ref{subsubsecSZEGO}]}. $\square$

\end{itemize}

\subsubsection{Wiener-Hopf Factorizations} \label{subsubsecWHFactorizations}

\noindent Assume $\varphi: \mathbb{T} \rightarrow \C$ has Fourier transform in $L^1(\Z)$, i.e. $\varphi$ is in the Wiener algebra.
\begin{definition} Let $\textnormal{wind}_0(\varphi)$ be the \underline{winding number} of $\varphi$ around the origin $0 \in \C$. \end{definition}
\begin{definition} Any Wiener class $\varphi : \mathbb{T} \rightarrow \C^{\times}$ invertible with $\textnormal{wind}_0(\varphi) =0$ has \begin{equation} \varphi = \varphi_{[-]} \cdot \varphi_{[0]}  \cdot \varphi_{[+]} \end{equation}  
 
 \noindent a \underline{Wiener-Hopf factorization}, where $ \varphi_{[0]} \in \C$ is the \underline{geometric mean} of $\varphi$ 
 
 \begin{equation}\label{GeometricMean} \varphi_{[0]} = \textnormal{exp} \Bigg ( \oint_{\mathbb{T}} \log \varphi  \cdot \frac{ dw}{ 2 \pi \textnormal{\textbf{i}} w} \Bigg ) \end{equation} and $\varphi_{[\pm]} : \mathbb{T} \rightarrow \C$ extend to non-vanishing holomorphic functions 
 $\varphi_{[\pm]} : \mathbb{D}_{\pm} \rightarrow \C^{\times}$ on \begin{equation} \mathbb{D}_{\pm } = \{ |w|^{ \pm} < 1 \} \end{equation} in
$\mathbb{P}^1 = \C \cup \{\infty\}$ taking the value $\varphi_{[ \pm]} (0^{\pm})=1$ at $0^{\pm} \in \mathbb{D}_{\pm}$ where $0^- = \infty$.
 \end{definition}
 
 \noindent The square braces in the subscript distinguish the multiplicative factorization of $\varphi$ from our additive factorization $\mathbbm{1} = \uppi_- + \uppi_0 + \uppi_+$, e.g. $\log \varphi_{[0]} = (\log \varphi)_0$ in (\ref{GeometricMean}).
 
\begin{proposition} \textnormal{(Chapter 27.2 in \cite{Lax0})} For Wiener $\varphi$, the Toeplitz operator $L_{\bullet}(\varphi)$ is invertible on Hardy space $H_{\bullet}$ if and only if its symbol $\varphi : \mathbb{T} \rightarrow \C^{\times}$ is invertible with $\textnormal{wind}_0(\varphi) =0$.  In this case, a two-sided inverse of $L_{\bullet}(\varphi)$ is 
\begin{equation} \label{ToeplitzInverse}
L_{\bullet}^{-1}(\varphi) = L_{\bullet}(\varphi_{[+]}^{-1}) L_{\bullet}(\varphi_{[0]}^{-1}) L_{\bullet} (\varphi_{[-]}^{-1} )
\end{equation}
\end{proposition} 

\noindent The factor $L_{\bullet}(\varphi_{[0]}^{-1})$ is $\varphi_{[0]}^{-1} \cdot \mathbbm{1}$.  With formulas (\ref{ToeplitzInverse}) and (\ref{HardyIntegralOperator}), one easily derives:

\begin{proposition}
\label{KCSW}  Let $\varphi: \mathbb{T} \rightarrow \C^{\times}$ be an invertible Wiener function on the circle with $\textnormal{wind}_0(\varphi) =0$ and Wiener-Hopf factorization $\varphi = \varphi_{[-]} \varphi_{[0]} \varphi_{[+]}$.  For $w_{\pm} \in \mathbb{D}_{\pm}$,  
\begin{equation} \label{KCSWformula}
\sum_{h_+ = 0}^{\infty} \sum_{h_- = 0 }^{\infty} w_+^{h_+} \langle h_+ | L_{\bullet}^{-1}(\varphi) | h_- \rangle w_-^{-h_- -1} = \frac{1}{ w_- - w_+} \cdot  \frac{1}{ \varphi_{[0]}} \cdot \frac{1}{ \varphi_{[+]} ( w_+) }  \cdot \frac{1}{ \varphi_{[-]}(w_-) } \end{equation}

 \end{proposition} 

\begin{corollary} \label{KCSWnonneg} As $w_- \rightarrow \infty$ we get 
\begin{equation}
\sum_{h_+ =0 }^{\infty} w_+^{h_+} \langle h_+ | L_{\bullet}^{-1}(\varphi) | 0 \rangle= \frac{1}{\varphi_{[0]}} \cdot \frac{1}{ \varphi_{[+]} ( w_+) }.
\end{equation}
\end{corollary}
\begin{corollary} \label{KCSWnonpos} As $w_+ \rightarrow 0$ we get 
   \begin{equation}
   \sum_{h_- = 0}^{\infty}  \langle 0 | L_{\bullet}^{-1} (\varphi) | h_- \rangle w_-^{-h_-}  = \frac{1}{ \varphi_{[0]}} \cdot \frac{1}{ \varphi_{[-]} ( w_-) }.
   \end{equation}
\end{corollary}

\begin{corollary}\label{KCSWzero} As $w_{\pm} \rightarrow 0^{\pm}$ we get
 \begin{equation}
 \label{zeromode}
  \langle 0| L^{-1}_{\bullet}(\varphi) | 0 \rangle   = \frac{1}{\varphi_{[0]} } = \textnormal{exp} \Bigg (  \oint_{\mathbb{T}} \log \Bigg [\frac{1}{ \varphi(w) } \Bigg ]  \frac{dw}{ 2 \pi \textnormal{\textbf{i}}w}  \Bigg ).
   \end{equation}
   \end{corollary}

 \noindent In the papers of Kre\u{\i}n \cite{Krein1} and Calder\'{o}n-Spitzer-Widom \cite{CSW}, one encounters a version of Proposition [\ref{KCSW}] tailored for the {resolvents} of Toeplitz operators as follows.  Assume $v: \mathbb{T} \rightarrow \R$ is real-valued and bounded, which we recognize as our assumptions on the symbol taken above.  For any $u \in \C \setminus \R$, consider the \textit{family} of symbols 
\begin{equation} \label{symbolFAMILY}
\varphi(w| u ,v) := u - v(w).
\end{equation}

\noindent These $\varphi$ are not real.  Recall $v(w) = V_0 + \sum_{k=1}^{\infty} (\overline{V}_k w^k + V_k w^{-k})$ and $V_0 = a \in \R$.  Thus, all we have done in (\ref{symbolFAMILY}) is introduce a new complex variable $u$ for this same zero mode, though the factors $\varphi_{[ \pm]}(w | u, v)$ may depend in a complicated way on $u$.  With a simple but powerful manipulation \begin{equation}
 u - L_{\bullet}(v) = L_{\bullet}(\varphi ( \cdot | u, v) ),
 \end{equation}
\noindent the {resolvent} of our Toeplitz operator $L_{\bullet}(v)$ becomes 
  \begin{equation}   (u - L_{\bullet}(v))^{-1} = L_{\bullet}(\varphi ( \cdot | u, v))^{-1}
  \end{equation}
  
  \noindent the inverse of the Toeplitz operator with symbol $\varphi ( \cdot |u,v)$.  Moreover, since $v$ takes real values, for each $u \not \in \R$ we know $\varphi (w | u ,v)$ is an invertible function on the circle $\mathbb{T} = \{ |w|=1\}$ with $\textnormal{wind}_0(\varphi) =0$.  This observation allows us to use Proposition [\ref{KCSW}] and its corollaries to express matrix elements of the resolvent in terms of the Wiener-Hopf factors of $\varphi ( w | u, v)$.  In particular, Corollary [\ref{KCSWzero}] for $\varphi = u -v $ gives
  
  \begin{corollary} \label{KCSWzeroSzego} For $v: \mathbb{T} \rightarrow \R$ real-valued in the Wiener algebra and $u \in \C \setminus \R$,
  
  \begin{equation} \langle 0 | \frac{1}{u - L_{\bullet}(v)} | 0 \rangle = T^{\uparrow}(u) |_v = \textnormal{exp} \Bigg ( \oint_{\mathbb{T}} \log \Bigg [ \frac{1}{u - v(w)} \Bigg ] \frac{ dw }{ 2 \pi \textnormal{\textbf{i}} w} \Bigg ) \end{equation} \end{corollary}

 \noindent For further use of Wiener-Hopf factorizations to derive explicit spectral resolutions of self-adjoint Toeplitz operators, see \cite{BottcherGrudsky, Ismagilov, Ros2, Ros3, Ros1}.



\subsubsection{Szeg\H{o}'s First Theorem} \label{subsubsecSZEGO}

\noindent In this section, we complete Step 5 in our proof of Theorem [\ref{CBOHConservedDensityConstruction}] by giving an explicit formula for the conserved density of the classical dispersionless Benjamin-Ono hierarchy already realized as the spectral shift function of the Toeplitz operator $L_{\bullet}(v)$ and its minor $L_+(v)$.  This explicit form is nothing but Szeg\H{o}'s First Theorem for real and bounded symbols $v$, Theorem [\ref{SzegoFirstTheorem}] below, which appears as Theorem 5.10 in section 5.5 of \cite{BottcherSilbermannIntro} and as Theorem 2 in \cite{DeiftItsKra}.  The case stated here is due to Szeg\H{o} in 1920, an improvement on the assumption $v$ continuous and positive in his original proof of 1915.  For versions of this result with complex symbols, see \cite{BottcherSilbermannIntro}, and for less regular symbols (i.e. for symbols $v$ contributing to singular support in the measure $d \mu(x) = v(x) d x$), see the discussion of Verblunsky's approach to Szeg\H{o}'s First Theorem as a Sum Rule in \cite{SimonSzego}.

\begin{theorem} \label{SzegoFirstTheorem} \textnormal{(Szeg\H{o}'s First Theorem \cite{Szego1915})} For bounded real $v: \mathbb{T} \rightarrow \R$, the $\frac{1}{u}$ multiple of the perturbation determinant is the geometric mean of $\frac{1}{u - v}$ \begin{equation} \label{SzegoFirstTheoremFormula}  \frac{1}{u} \cdot \frac{ \det_{H_{\bullet}} (u - L_{+}(v)) }{ \det_{H_{\bullet}} ( u - L_{\bullet}(v)) } = T^{\uparrow}(u ) \big |_v = \textnormal{exp} \Bigg ( \oint_{\mathbb{T}} \log \Bigg [ \frac{1}{u-v(w)} \Bigg ] \frac{ dw}{ 2 \pi \textnormal{\textbf{i}} w} \Bigg ) \end{equation}

\end{theorem}
\begin{itemize}
\item \textit{Proof:} The $\ebar \rightarrow 0$ argument in Step 1 of our proof of Theorem [\ref{CBOHConservedDensityConstruction}] implies that the perturbation determinant is the $(\Psi_0, \Psi_0)$ matrix element of the resolvent of $L_{\bullet}(v)$, so it suffices to prove the identity 

\begin{equation} \label{ToeplitzSpectralKMK} \langle 0 | \frac{1}{ u - L_{\bullet}(v)} | 0 \rangle = T^{\uparrow}(u) |_v = \text{exp} \Bigg ( \oint_{\mathbb{T}} \log \Bigg [\frac{1}{ u - v(w)} \Bigg ]  \frac{dw }{2 \pi \textbf{i} w} \Bigg ) \end{equation}

\noindent Temporarily assuming $v$ is in the Wiener algebra, use Corollary [\ref{KCSWzeroSzego}].  To remove this temporary assumption, follow section 5.4 of \cite{BottcherSilbermannIntro} by checking that the geometric mean is a continuous function of $v$ in the supremum norm. $\square$ \end{itemize}

\noindent The perturbation determinant is the ratio of large characteristic polynomials \begin{equation}  \lim_{\tealN \rightarrow \infty} \frac{ \det_{H_{\bullet, \tealN}} (u - L_{+, \tealN}(v)) }{ \det_{H_{\bullet, \tealN}} ( u - L_{\bullet, \tealN}(v)) }  = \frac{\det_{H_{\bullet}} (u - L_+(v)) }{ \det_{H_{\bullet}} ( u - L_{\bullet}(v))} \end{equation}
because the Toeplitz operator $L_{\bullet}(v)|_{\C[w]}$ is essentially self-adjoint by Proposition [\ref{ToeplitzRealNice}].  In Remark 2 of Theorem 1.6.1 in \cite{SimonSzego}, one learns that Fekete recommended to Szeg\H{o} to first prove Theorem [\ref{SzegoFirstTheorem}], as it implies the limit conjectured by P\'{o}lya $\tfrac{1}{ \textcolor{teal}{N}} \log \det_{H_{\bullet, \textcolor{teal}{N}}} (u - L_{\bullet, \tealN} (v)) \rightarrow \oint_{\mathbb{T}} \log (u - v(w)) \frac{ dw}{ 2 \pi \textbf{i} w}$.  This recasts Szeg\H{o}'s First Theorem as an asymptotic statement about determinants of large Toeplitz matrices.  We prefer to state the theorem directly for perturbation determinants of Toeplitz operators, as this illustrates that an infinite-dimensional version of Proposition [\ref{FiniteDimSzegoNoGo}] is impossible
{\small
\begin{equation} \textnormal{exp} \Bigg ( \oint_{\mathbb{T}} \log \Bigg [ \frac{1}{u-v(w)} \Bigg ] \frac{ dw}{ 2 \pi \textnormal{\textbf{i}} w} \Bigg ) =  \frac{1}{u} \cdot \frac{ \det_{H_{\bullet}} (u - L_{+}(v)) }{ \det_{H_{\bullet}} ( u - L_{\bullet}(v)) } \neq  \frac{ \det_{H_{+}} (u - L_+^{\perp}(v)) }{ \det_{H_{\bullet}}(u - L_{\bullet}(v)) }  \equiv 1
\end{equation}
}

\noindent as the minor $L_+^{\perp}(v)$ on $H_+$ is indistinguishable from $L_{\bullet}(v)$ on $H_{\bullet}$.

\begin{corollary} \label{CHBHConstruction} For bounded real $v$, the profile $f_{\star |v} \in \textnormal{\textbf{P}}^{\vee}$ first introduced via the conserved density $dF_{\star |v} ( c)$ of the classical dispersionless periodic Benjamin-Ono equation in \textnormal{Theorem [\ref{CHBHConservedDensityIntro}]} and later revisited in Proposition \textnormal{[\ref{CHBDensityProfile}]} has a realization via the spectrum of an auxiliary Toeplitz operator $L_{\bullet}(v)$: 
\begin{itemize}
\item The $T^{\uparrow}$-observable $T^{\uparrow}(u) |_v $ defined in formula \textnormal{(\ref{SzegoFirstTheoremFormula})} is the Titchmarsh-Weyl function for all $v$.
\item The transition measure $\tau_{\star |v}^{\uparrow}$ is the {spectral measure} of $L_{\bullet}(v)$ at $|0 \rangle \in H_{\bullet}$ and
\item The shifted Rayleigh function $\xi_{\star |v}$ is the \textcolor{black}{\textit{spectral shift function}} $\xi (c | L_{\bullet}(v), L_+(v))$. \end{itemize}
\end{corollary}

\subsection{\textcolor{black}{Constructing Quantum Periodic Benjamin-Ono Densities}} \label{subsecQBOHConstruction}

\noindent In section \textbf{[\ref{subsubsecBlockJacobi}]} we generalize section \textbf{[\ref{subsubsecJacobiTW}]} by defining the Titchmarsh-Weyl operator of a block Jacobi operator.  In section \textbf{[\ref{subsecFockBlockToeplitz}]} we generalize section \textbf{[\ref{subsubsecToeplitz}]} by defining block Toeplitz operators $L_{\bullet}( \widehat{\boldsymbol{\varphi}})$ with $\mathfrak{gl}(\mathscr{F})$-valued symbols $ \widehat{\boldsymbol{\varphi}}$ as in Definition [\ref{BlockSymbolDefinition}] and also discuss the analytic theory of a particular block Toeplitz operator $L_{\bullet}(\vcurrent (\cdot | \hbar)) $ whose symbol $\vcurrent$ is the affine $\widehat{\mathfrak{gl}}_1$ current at level $\hbar$ in Definition [\ref{KacMoodyCurrentDefinition}].  In section \textbf{[\ref{subsubsecLaxQBO}]}, we define the Lax operator for quantum Benjamin-Ono as 
\begin{equation} \widehat{\mathcal{L}}_{\bullet}( \ebar, \hbar) = L_{\bullet}( \vcurrent ( \cdot | \hbar)) + \ebar \mathbbm{1} \otimes D_{\bullet} \end{equation} for $\ebar \in \R$, $\hbar >0$ the coefficients of dispersion and quantization, respectively.  In section \textbf{[\ref{subsubsecNSorbits}]} we define the Nazarov-Sklyanin quantum transfer operators $\widehat{\mathcal{T}}^{\uparrow}(u | \ebar,\hbar)$ following \cite{NaSk2} as Titchmarsh-Weyl operators of $\widehat{\mathcal{L}}_{\bullet}(\ebar, \hbar)$ and use them in section \textbf{[\ref{subsubsecQBOConservedDensitySSF}]} to construct the quantum periodic Benjamin-Ono hierarchy and its conserved densities $d F_{\Psi}( c | \ebar, \hbar)$ as spectral shift functions in Theorem [\ref{QBOHConservedDensityConstruction}] whose existence was announced in Theorem [\ref{QBOHConservedDensityExistence}].  In \textbf{[\ref{subsubsecAnisotropicSSF}]} we restate Theorem [\ref{QBOHConservedDensitiesForJacksAreAnisotropicPartitions}] as Theorem [\ref{QBOHConservedDensitiesForJacksAreAnisotropicPartitionsRevisited}] to identify the quantum conserved densities of quantum stationary states as $dF_{\lambda} (c | \ee, \e)$, the Rayleigh measures of profiles of anisotropic partitions.

\subsubsection{\textcolor{black}{Block Jacobi Operators and Titchmarsh-Weyl Operators}} \label{subsubsecBlockJacobi}

\noindent The Titchmarsh-Weyl function $T^{\uparrow}(u)|_{L_{\bullet}; \Psi_0}$ of a self-adjoint operator $L_{\bullet}$ on a Hilbert space $\mathscr{H}_{\bullet}$ with cyclic vector $\Psi_0$ from Definition [\ref{TWDefinition}] is a scalar function of $u \in \C \setminus \R$ that one may regard as a linear operator \begin{equation} T^{\uparrow}(u) |_{L_{\bullet}; \Psi_0} : \mathscr{H}_{\Psi_0} \rightarrow \mathscr{H}_{\Psi_0} \end{equation}

\noindent on the one-dimensional space $\mathscr{H}_{\Psi_0} = \C | \Psi_0 \rangle$ spanned by $\Psi_0$ after rewriting

\begin{equation} \label{1by1BlockJacobiTWOperator} T^{\uparrow}(u) =  \uppi_{\Psi_0} \frac{1}{ u - L_{\bullet} } \uppi_{\Psi_0}  \end{equation}

\noindent with projections $\uppi_{\Psi_0}: \mathscr{H}_{\bullet} \rightarrow \mathscr{H}_{\Psi_0}$.  Here is an immediate generalization:

\begin{definition} \label{BlockGeneralDefinitions}Recall $H_{\bullet}$ Hardy space with $\psi_0 = |0 \rangle \in H_{\bullet}$ and $H_0 = \C | 0 \rangle$ so  \begin{equation} \ \ \ H_{\bullet} = H_0 \oplus H_+ .\end{equation}  Given any other Hilbert space $\mathscr{F}$, define \begin{eqnarray} \mathscr{H}_{\bullet} &=& \mathscr{F} \otimes H_{\bullet} \\ \mathscr{H}_{0} &=& \mathscr{F} \otimes H_0 \\ \mathscr{H}_{+} &=& \mathscr{F} \otimes H_+  \end{eqnarray}

\noindent so that \begin{equation} \ \ \  \mathscr{H}_{\bullet} = \mathscr{H}_0 \oplus \mathscr{H}_+ . \end{equation}  If the subspace $\mathscr{H}_0$ is cyclic for a possibly unbounded operator $\widehat{\mathcal{L}}_{\bullet}: \mathscr{H}_{\bullet} \rightarrow \mathscr{H}_{\bullet}$, we say $\widehat{\mathcal{L}}_{\bullet}$ is \underline{one-sided block Jacobi operator}.

\end{definition}

\begin{definition} \label{BlockMinorDefinition} In the setting of \textnormal{Definition [\ref{BlockGeneralDefinitions}]}, the \underline{$(\mathscr{H}_0, \mathscr{H}_0)$-minor} \begin{equation} \widehat{\mathcal{L}}_{+}^{\perp} = (\mathbbm{1} \otimes \uppi_+^{\perp}) \widehat{\mathcal{L}}_{\bullet} ( \mathbbm{1} \otimes \uppi_+^{\perp}) \end{equation}

\noindent of $\widehat{\mathcal{L}}_{\bullet}: \mathscr{H}_{\bullet} \rightarrow \mathscr{H}_{\bullet}$ is an operator $\widehat{\mathcal{L}}_+^{\perp}: \mathscr{H}_+ \rightarrow \mathscr{H}_+$ that appears in $\widehat{\mathcal{L}}_{+} = 0 \oplus \widehat{\mathcal{L}}_{+}^{\perp}$ in the additive decomposition with respect to $\mathscr{H}_{\bullet} = \mathscr{H}_0 \oplus \mathscr{H}_+$ of $\widehat{\mathcal{L}}_{+} : \mathscr{H}_{\bullet} \rightarrow \mathscr{H}_{\bullet}$ given by \begin{equation} \widehat{\mathcal{L}}_{+} = (\mathbbm{1} \otimes \uppi_+) \widehat{\mathcal{L}}_{\bullet} ( \mathbbm{1} \otimes \uppi_+) \end{equation}
\end{definition}

\begin{definition} \label{BlockTWDefinition} For a one-sided block Jacobi operator $\widehat{\mathcal{L}}_{\bullet}$ on $\mathscr{H}_{\bullet} = \mathscr{F}\otimes H_{\bullet}$ with cyclic $\mathscr{H}_0 = \mathscr{F} \otimes H_0$, its \underline{Titchmarsh-Weyl operator} $\widehat{\mathcal{T}}^{\uparrow}(u)|_{\widehat{\mathcal{L}}_{\bullet}; \mathscr{H}_0}: \mathscr{H}_0 \rightarrow \mathscr{H}_0$ is \begin{equation} \label{BlockJacobiTWOperator}  \widehat{\mathcal{T}}^{\uparrow}(u )|_{\widehat{\mathcal{L}}_{\bullet}; \mathscr{H}_0}  = (\mathbbm{1} \otimes \uppi_{\Psi_0}) \cdot \frac{1}{ u - \widehat{\mathcal{L}}_{\bullet}} \cdot (\mathbbm{1} \otimes \uppi_{\Psi_0}) \end{equation}

\noindent the $(\mathscr{H}_0 , \mathscr{H}_0)$ block-entry of the resolvent $\widehat{\mathcal{R}}_{\bullet}(u) = \frac{1}{u - \widehat{\mathcal{L}}_{\bullet}}$ of $\widehat{\mathcal{L}}_{\bullet}$ for $u \in \C \setminus \R$.

\end{definition}

\noindent For $u \in \C \setminus \R$ the Titchmarsh-Weyl operator is a bounded operator on $\mathscr{H}_0 = \mathscr{F} \otimes H_{0}$ because $\widehat{\mathcal{R}}_{\bullet}(u)$ is bounded on $\mathscr{H}_{\bullet} = \mathscr{F} \otimes H_{\bullet}$ for $\widehat{\mathcal{L}}_{\bullet}$ self-adjoint.  For ordinary Jacobi operators, i.e. $\mathscr{F} \cong \C^1$, we know that formula (\ref{1by1BlockJacobiTWOperator}) is also $\frac{1}{u} \cdot \frac{ \det_{H_{\bullet}}(u - L_{+}) }{ \det_{H_{\bullet}} ( u - L_{\bullet})}$ the multiple $\frac{1}{u}$ of the perturbation determinant in the case that the restriction $L_{\bullet} |_{\Psi_0}$ of $L_{\bullet}$ to the orbit of the cyclic vector $\Psi_0$ is essentially self-adjoint, Theorem [\ref{OPRL}] above.  Perhaps a block version of Theorem [\ref{OPRL}] exists for Titchmarsh-Weyl operators $\widehat{\mathcal{T}}^{\uparrow}(u)$ of block Jacobi operators, involving a block perturbation determinant of the pair $\widehat{\mathcal{L}}_{\bullet}$ and $\widehat{\mathcal{L}}_{+}$, but for the purposes of this paper we do not require such a general result to rewrite formula (\ref{BlockJacobiTWOperator}).  Instead, we apply Theorem [\ref{OPRL}] to restrictions of the block Jacobi operator in the following manner:

\begin{definition} Let $\mathscr{H}_{\bullet}[ \widehat{\mathcal{L}}_{\bullet}; \Psi_0]$ denote the $\widehat{\mathcal{L}}_{\bullet}$ orbit of $\Psi_0 \in \mathscr{H}_0$ in $\mathscr{H}_{\bullet}$ and

\begin{equation} \widehat{\mathcal{L}}_{\bullet} \Big |_{\Psi_0} = \uppi_{\widehat{\mathcal{L}}_{\bullet}; \Psi_0}  \ \widehat{\mathcal{L}}_{\bullet} \ \uppi_{\widehat{\mathcal{L}}_{\bullet}; \Psi_0}  
\end{equation}

\noindent  the restriction of $\widehat{\mathcal{L}}_{\bullet}$ to the orbit $\mathscr{H}_{\bullet}[ \widehat{\mathcal{L}}_{\bullet}; \Psi_0 ]$ by projections $\uppi_{\widehat{\mathcal{L}}_{\bullet}; \Psi_0} : \mathscr{H}_{\bullet} \rightarrow \mathscr{H}_{\bullet}[ \widehat{\mathcal{L}}_{\bullet}; \Psi_0]$.

\end{definition}

\noindent Theorem [\ref{OPRL}] restated for $\widehat{\mathcal{L}}_{\bullet} |_{\Psi_0}$ and $\mathscr{H}_{\bullet}[ \widehat{\mathcal{L}}_{\bullet}; \Psi_0]$ and not ``$L_{\bullet}$ and $\mathscr{H}_{\bullet}$'' reads:
\begin{theorem}  \label{OPRLtensororbit}  If $\widehat{\mathcal{L}}_{\bullet} |_{\Psi_0}$ is \underline{essentially self-adjoint}, the $(\Psi_0, \Psi_0)$-element of resolvent
\begin{equation} \label{OPRLformula} \langle \Psi_0 | \frac{1}{u - \widehat{\mathcal{L}}_{\bullet} |_{\Psi_0}}| \Psi_0 \rangle  = T^{\uparrow}(u) \big |_{\widehat{\mathcal{L}}_{\bullet};\Psi_0}  = \frac{1}{u} \cdot \frac{ \det_{\mathscr{H}_{\bullet}[ \widehat{\mathcal{L}}_{\bullet}; \Psi_0 ] } (u -L_+) }{ \det_{\mathscr{H}_{\bullet}[ \widehat{\mathcal{L}}_{\bullet}; \Psi_0]} ( u - \widehat{\mathcal{L}}_{\bullet} |_{\Psi_0}) } \end{equation}

\noindent is $\tfrac{1}{u}$ times the {perturbation determinant} of $\widehat{\mathcal{L}}_{\bullet} |_{\Psi_0}$ and ${L_+= 0 \oplus \uppi_{\Psi_0}^{\perp} \widehat{\mathcal{L}}_{\bullet}  \uppi_{\Psi_0}^{\perp} }$ on $\mathscr{H}_{\bullet}[ \widehat{\mathcal{L}}_{\bullet}; \Psi_0]$. \end{theorem}
\noindent While Theorem [\ref{OPRL}] is stated for ``$\mathscr{H}_{\bullet} = \mathscr{H}_{\Psi_0} \oplus \mathscr{H}_{\Psi_0}^{\perp}$,'' Theorem [\ref{OPRLtensororbit}] is for \begin{equation} \mathscr{H}_{\bullet} [ \widehat{\mathcal{L}}_{\bullet}; \Psi_0] = \mathscr{H}_{\bullet} [ \widehat{\mathcal{L}}_{\bullet}; \Psi_0] _{\Psi_0} \oplus \mathscr{H}_{\bullet} [ \widehat{\mathcal{L}}_{\bullet}; \Psi_0] _{\Psi_0}^{\perp} \end{equation}

\noindent with  $\mathscr{H}_{\bullet} [ \widehat{\mathcal{L}}_{\bullet}; \Psi_0] _{\Psi_0} \cong \mathscr{H}_{\Psi_0} = \C | \Psi_0 \rangle$ one-dimensional.  Two important features of the ingredients in Theorem [\ref{OPRLtensororbit}] are discussed on the next page and depicted here:

  \begin{figure}[htb]
\centering
\includegraphics[width=0.55 \textwidth]{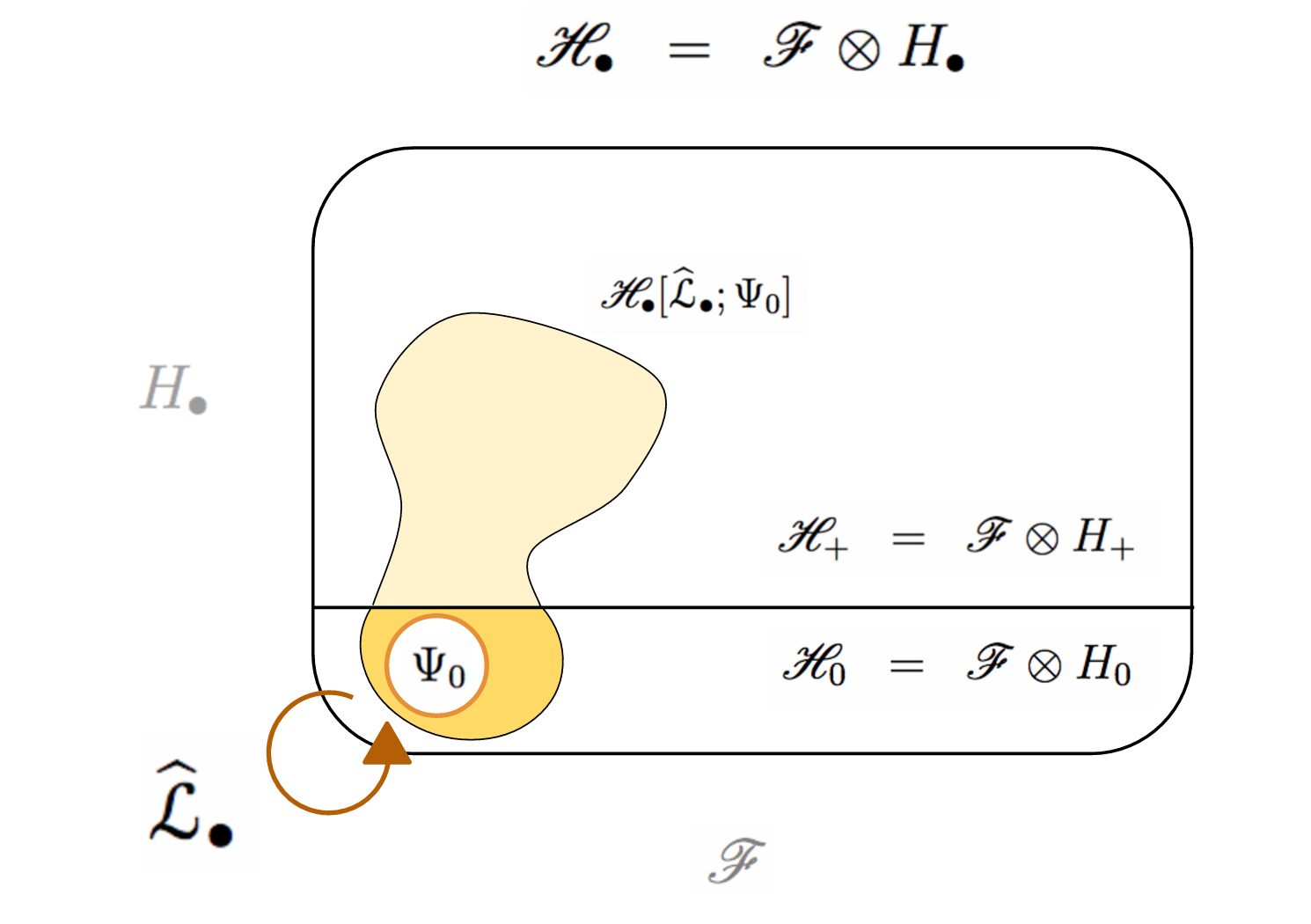}
\end{figure}

\noindent First, the yellow orbit $\mathscr{H}_{\bullet}[ \widehat{\mathcal{L}}_{\bullet}; \Psi_0 ] \subset \mathscr{H}_{\bullet}$ is not necessarily a tensor product of individual subspaces of the two spaces $\mathscr{F}$ and $H_{\bullet}$ that comprise $\mathscr{H}_{\bullet} = \mathscr{F} \otimes H_{\bullet}$.  Second, the orbit $\mathscr{H}_{\bullet}[ \widehat{\mathcal{L}}_{\bullet}; \Psi_0]$ may contain elements of $\mathscr{H}_0$ linearly independent of $\Psi_0$, as is depicted in the dark yellow region of the orbit.  This second feature suggests a condition under which Theorem [\ref{OPRLtensororbit}] can be simplified.

\begin{definition} \label{IsolatedDefinition} An element $\Psi_0 \in \mathscr{H}_0$ in the cyclic subspace $\mathscr{H}_0 \subset \mathscr{H}_{\bullet}$ of one-sided block Jacobi operator $\widehat{\mathcal{L}}_{\bullet}$ is \underline{isolated by $\widehat{\mathcal{L}}_{\bullet}$} if the intersection
\begin{equation} \mathscr{H}_{\bullet}[ \widehat{\mathcal{L}}_{\bullet}; \Psi_0 ] \cap \mathscr{H}_+ = \mathscr{H}_{\Psi_0} = \C | \Psi_0 \rangle \end{equation}

\noindent is one-dimensional and spanned by $\Psi_0$. \end{definition}

\begin{proposition} If $\Psi_0 \in \mathscr{H}_0$ is isolated by a one-sided block Jacobi operator $\widehat{\mathcal{L}}_{\bullet}$, then $L_+ = 0 \oplus \uppi_{\Psi_0}^{\perp} \widehat{\mathcal{L}}_{\bullet}  \uppi_{\Psi_0}^{\perp} $ appearing in \textnormal{Theorem [\ref{OPRLtensororbit}]} is \begin{equation} L_+ = \widehat{\mathcal{L}}_+ |_{\Psi_0} \end{equation} 
\noindent the restriction of $\widehat{\mathcal{L}}_+$ of \textnormal{Definition [\ref{BlockMinorDefinition}]} to the orbit $\mathscr{H}_{\bullet}[ \widehat{\mathcal{L}}_{\bullet}; \Psi_0]$.
\end{proposition}

\begin{theorem} \label{OPRLtensororbitISOLATED} \noindent If $\widehat{\mathcal{L}}_{\bullet} |_{\Psi_0}$ is \underline{essentially self-adjoint} and $\Psi_0 \in \mathscr{H}_0$ is \underline{isolated} by $\widehat{\mathcal{L}}_{\bullet}$,  the $(\Psi_0, \Psi_0)$-element of the resolvent of $\mathcal{L}_{\bullet}|_{\Psi_0}$ is \begin{equation} \label{OPRLformula} \langle \Psi_0 | \frac{1}{u - \widehat{\mathcal{L}}_{\bullet} |_{\Psi_0}}| \Psi_0 \rangle  = T^{\uparrow}(u) \big |_{\widehat{\mathcal{L}}_{\bullet};\Psi_0}  = \frac{1}{u} \cdot \frac{ \det_{\mathscr{H}_{\bullet}[ \widehat{\mathcal{L}}_{\bullet}; \Psi_0 ] } (u - \widehat{\mathcal{L}}_+ |_{\Psi_0}) }{ \det_{\mathscr{H}_{\bullet}[ \widehat{\mathcal{L}}_{\bullet}; \Psi_0]} ( u - \widehat{\mathcal{L}}_{\bullet} |_{\Psi_0}) } \end{equation}

\noindent $\frac{1}{u}$ times the perturbation determinant of the pair $\widehat{\mathcal{L}}_{\bullet} |_{\Psi_0}, \widehat{\mathcal{L}}_{+} |_{\Psi_0}$ in $\mathscr{H}_{\bullet}[ \widehat{\mathcal{L}}_{\bullet}; \Psi_0]$.
\end{theorem}

\noindent To conclude this section, we describe a precious class of one-sided block Jacobi operators $\widehat{\mathcal{L}}_{\bullet}$ with many isolated states $\Psi_0 = \Psi_{\lambda; 0} \in \mathscr{H}_0$ as in Definition [\ref{IsolatedDefinition}].

\begin{definition} \label{TransferOperatorDefinition} The Titchmarsh-Weyl operator $\widehat{\mathcal{T}}^{\uparrow}(u)|_{\widehat{\mathcal{L}}_{\bullet}; \mathscr{H}_0}: \mathscr{H}_0 \rightarrow \mathscr{H}_0$ of a one-sided block Jacobi operator $\widehat{\mathcal{L}}_{\bullet}$ is a \underline{transfer operator} if for any two $u_1, u_2 \in \C \setminus \R$

\begin{equation} [ \widehat{\mathcal{T}}^{\uparrow}(u_1), \widehat{\mathcal{T}}^{\uparrow}(u_2)] = 0.\end{equation}  

\noindent the commutator of two bounded operators $\widehat{\mathcal{T}}^{\uparrow}(u_1), \widehat{\mathcal{T}}^{\uparrow}(u_2)$ is the zero operator. \end{definition}

\noindent Expanding the bounded Tichmarsh-Weyl operator \begin{equation} \widehat{\mathcal{T}}^{\uparrow}(u)  = \sum_{ \ell = 0}^{\infty} u^{- \ell -1} \widehat{\mathcal{T}}^{\uparrow}_{\ell} \end{equation}

\noindent  as an infinite linear combination of unbounded operators
\begin{equation} \widehat{\mathcal{T}}_{\ell}^{\uparrow}  = (\mathbbm{1} \otimes \uppi_0) \  \widehat{\mathcal{L}}_{\bullet} ^{\ell} \ ( \mathbbm{1} \otimes \uppi_0) ,\end{equation}
\noindent if $\widehat{\mathcal{T}}^{\uparrow}(u)$ is a transfer operator, then $[\widehat{\mathcal{T}}^{\uparrow}_{\ell_1}, \widehat{\mathcal{T}}^{\uparrow}_{\ell_2}]=0$ for all $\ell_1, \ell_2 = 0, 1, 2, 3, \ldots$.
\begin{lemma} \label{TransferIsolationLemma} If the Titchmarsh-Weyl operator $\widehat{\mathcal{T}}^{\uparrow}(u)|_{\widehat{\mathcal{L}}_{\bullet}; \mathscr{H}_0}: \mathscr{H}_0 \rightarrow \mathscr{H}_0$ of a one-sided block Jacobi operator $\widehat{\mathcal{L}}_{\bullet}$ is a transfer operator and $\Psi_{\lambda; 0} \in \mathscr{H}_0$ defined for $\Psi_{\lambda} \in \mathscr{F}$ by
\begin{equation} \Psi_{\lambda;0}  = \Psi_{\lambda} \otimes | 0 \rangle \end{equation}

\noindent is a simultaneous eigenfunction of commuting Titchmarsh-Weyl operators 

\begin{equation} \widehat{\mathcal{T}}^{\uparrow} ( u ) \Psi_{\lambda;0 }  = T^{\uparrow}(u)|_{\lambda}  \Psi_{\lambda;0} \end{equation}

\noindent with eigenvalue $T^{\uparrow}(u) |_{\lambda}$, then $\Psi_{\lambda;0} $ is isolated by $\widehat{\mathcal{L}}_{\bullet}$ in the sense of \textnormal{Definition [\ref{IsolatedDefinition}]}.

\end{lemma}

\subsubsection{\textcolor{black}{Fock-Block Toeplitz Operators}} \label{subsecFockBlockToeplitz}

\noindent In section \textbf{[\ref{subsubsecToeplitz}]} we defined Toeplitz operators $L_{\bullet}(\varphi)$ as compressions of multiplication operators $L(\varphi)$ whose symbols $\varphi$ are $\C$-valued distributions on $\mathbb{T}$.  In this section, we define block Toeplitz operators $L_{\bullet}( \widehat{\boldsymbol{\varphi}})$ as compressions of multiplication operators $L(\widehat{\boldsymbol{\varphi}})$ whose symbols are $\mathfrak{gl}(\mathscr{F})$-valued distributions on $\mathbb{T}$ in the sense of Definition [\ref{BlockSymbolDefinition}].  Such $L_{\bullet}( \widehat{\boldsymbol{\varphi}})$ are one-sided block Jacobi operators on $\mathscr{H}_{\bullet} = \mathscr{F} \otimes H_{\bullet}$ where $H_{\bullet}$ is the Hardy space of $\mathbb{T}$ inside $H = L^2(\mathbb{T})$ and $(\mathscr{F}, \langle \cdot, \cdot \rangle)$ is an arbitrary separable $\C$-Hilbert space.  First, as in section \textbf{[\ref{subsubsecToeplitz}]}, we define the two-sided operator:

\begin{definition} For a $\mathfrak{gl}(\mathscr{F})$-valued distribution $\widehat{\boldsymbol{\varphi}}$ on $\mathbb{T}$, the \underline{multiplication operator} $L(\widehat{\boldsymbol{\varphi}})$ is densely defined on $\Psi \otimes w^h \in \mathscr{F} \otimes H$ for $\Psi \in \mathscr{F}$, $H = L^2(\mathbb{T})$, and $h \in \Z$ by \begin{equation} \label{MultiplicationOperatorFormula} L(\widehat{{\boldsymbol{\varphi}}}) ( \Psi \otimes w^h) =  \sum_{k=- \infty}^{+\infty} (\widehat{{\boldsymbol{\varphi}}}_{-k} \Psi ) \otimes w^{k+h}. \end{equation}
\end{definition}

\noindent The domain of definition of $L(\widehat{\boldsymbol{\varphi}})$ on $\mathscr{H} = \mathscr{F} \otimes H$ depends on the regularity of $\widehat{\boldsymbol{\varphi}}$.

\begin{definition} \label{BlockToeplitzOperatorDefinition} For a $\mathfrak{gl}(\mathscr{F})$-valued distribution $\widehat{\boldsymbol{\varphi}}$ on $\mathbb{T}$, the \underline{block Toeplitz operator} $L_{\bullet}(\widehat{\boldsymbol{\varphi}})$ is densely defined on $\mathscr{H}_{\bullet} = \mathscr{F} \otimes H_{\bullet}$ by \begin{equation} L_{\bullet} (\widehat{\boldsymbol{\varphi}}) = (\mathbbm{1} \otimes \uppi_{\bullet}) L(\widehat{\boldsymbol{\varphi}}) (\mathbbm{1} \otimes \uppi_{\bullet}) \end{equation}

\noindent where $\uppi_{\bullet}$ is the Szeg\H{o} projection to Hardy space $H_{\bullet}$ from \textnormal{Definition [\ref{SzegoProjectionDefinition}]} \end{definition}

\noindent Restricted to the dense subspace $\mathscr{F} \otimes \C[w]$ of $\mathscr{H}_{\bullet} = \mathscr{F} \otimes H_{\bullet}$, a block Toeplitz operator

\begin{equation}\label{BlockToeplitzMatrix} L_{\bullet}(\widehat{\boldsymbol{\varphi}}) \big |_{\mathscr{F} \otimes \C[w]} = \begin{bmatrix} 
\widehat{\boldsymbol{\varphi}}_0 & \widehat{\boldsymbol{\varphi}}_{1} & \widehat{\boldsymbol{\varphi}}_{2} & \cdots  \\ 
\widehat{\boldsymbol{\varphi}}_{-1} & \widehat{\boldsymbol{\varphi}}_0 & \widehat{\boldsymbol{\varphi}}_{1} & \ddots  \\
 \widehat{\boldsymbol{\varphi}}_{-2} & \widehat{\boldsymbol{\varphi}}_{-1} & \widehat{\boldsymbol{\varphi}}_0 & \ddots \\
  \vdots & \vdots & \ddots & \ddots \end{bmatrix} \end{equation}
  
\noindent is an infinite block Toeplitz matrix, i.e. an $\N \times \N$ matrix whose $(h_+, h_-)$ entry is a possibly unbounded operator $\widehat{\boldsymbol{\varphi}}_{h_+ - h_-}$ on $\mathscr{F}$ given by $\widehat{\boldsymbol{\varphi}}_{-k} = \oint_{\mathbb{T}} w^{-k} \widehat{\boldsymbol{\varphi}}
(w) \frac{ dw}{ 2 \pi \textbf{i} w}$.\\
\\
\noindent For the rest of this section, take $\widehat{\boldsymbol{\varphi}} (\cdot)= \vcurrent ( \cdot | \hbar)$ the $\widehat{\mathfrak{gl}}_1$ current of Definition [\ref{KacMoodyCurrentDefinition}], a $\mathfrak{gl}(\mathscr{F}(a))$-valued distribution on $\mathbb{T}$ for $(\mathscr{F}, \langle \cdot, \cdot \rangle) = ( \mathscr{F}(a), \langle \cdot, \cdot \rangle_{\hbar; - \frac{1}{2}})$ the Fock-Sobolev space of Definition [\ref{FockSobolevDefinition}] with regularity $s = - \frac{1}{2}$.  Let \begin{equation} \mathscr{H}_{\bullet}(a) = \mathscr{F}(a) \otimes H_{\bullet} \end{equation} be the Hilbert space completion of\begin{equation} \mathcal{H}_{\bullet} (a)= \mathcal{F}(a) \otimes \C[w]\end{equation}

\noindent the tensor product of $\mathcal{F}(a) \cong \C[V_1, V_2, \ldots] \cdot \Upsilon_a( \cdot | \hbar)$ and pre-Hardy space $\C[w] \subset H_{\bullet}$.  We now discuss properties of the Fock-block Toeplitz operator $L_{\bullet}( \vcurrent ( \cdot | \hbar))$ on $\mathscr{H}_{\bullet}(a)$.

\begin{definition} The \underline{total degree operator} $\widehat{\mathcal{D}}_{\bullet}$ is the sum of the momentum operator \begin{equation} \widehat{\mathcal{D}}_{\bullet} = \frac{1}{ \hbar} \widehat{\mathcal{T}}_2^{\uparrow} \otimes \mathbbm{1} + \mathbbm{1} \otimes D_{\bullet} \end{equation}

\noindent of formula \textnormal{(\ref{MomentumOperatorFormula})} and the auxiliary degree operator 
of \textnormal{Definition [\ref{AuxiliaryDegreeOperatorDefinition}]}.\end{definition}

\begin{proposition}The total degree operator $\widehat{\mathcal{D}}_{\bullet}$ is unbounded and self-adjoint on $\mathscr{H}_{\bullet}(a)$ preserving the dense subspace $\mathcal{H}_{\bullet}(a)$ with basis of eigenfunctions the monomial basis $V_{\mu} \otimes |h \rangle$ with eigenvalue $\deg \mu + h$ where $\deg \mu $ is the size of the partition $\mu$
\end{proposition}
\begin{itemize}
\item \textit{Proof:} follows from Proposition [\ref{FockSobolevBosonicBasisProposition}]. $\square$
\end{itemize}

\begin{definition} Let $\mathcal{H}_{\bullet}(a)[d]$ denote the eigenspace of $\widehat{\mathcal{D}}_{\bullet}$ with eigenvalue $d \in \N$. \end{definition}

\begin{proposition} $\dim \mathcal{H}_{\bullet}(a) [d] = \sum_{h=0}^d \dim \mathbb{Y}_{d-h} < \infty$ is finite-dimensional. \end{proposition}

\noindent The next fact is crucial and can be easily checked from the formula for $\vcurrent ( \cdot | \hbar)$:

\begin{proposition} The Fock-block Toeplitz operator $L_{\bullet}( \vcurrent ( \cdot | \hbar))$ whose symbol is the $\widehat{\mathfrak{gl}_1}$ current on $\mathbb{T}$ at level $\hbar$ commutes with the total degree operator on $\mathcal{H}_{\bullet}(a)$
\begin{equation} [ \widehat{\mathcal{D}}_{\bullet}, L_{\bullet}( \vcurrent ( \cdot | \hbar)) ] = 0 \end{equation}

\noindent the dense subspace of $\mathscr{H}_{\bullet}(a)$, hence preserves $\mathcal{H}_{\bullet}(a)$. \end{proposition}

\begin{corollary} \label{BlockFockToeplitzFinite} The Fock-block Toeplitz operator $L_{\bullet}( \vcurrent ( \cdot | \hbar))$ is a densely-defined $\C$-symmetric operator on $\mathscr{H}_{\bullet}(a)$ preserving finite-dimensional subspaces $\mathcal{H}_{\bullet}(a)[d]$. \end{corollary}

\begin{proposition} $L_{\bullet}( \vcurrent ( \cdot | \hbar))$ is essentially self-adjoint on pre-Hilbert space $\mathcal{H}_{\bullet}(a)$. \end{proposition}
{\small
\begin{itemize} \item \textit{Proof:} We say $\Psi \in \mathscr{H}_{\bullet}(a)$ is a \underline{state of uniqueness} for a possibly unbounded self-adjoint operator $L_{\bullet}$ if $L_{\bullet}^{\ell} \Psi $ lies in the domain of $L_{\bullet}$ for every $\ell \in \N$.  If the domain of $L_{\bullet}$ contains a set of states of uniqueness for $L_{\bullet}$ whose linear span is dense, then by Nussbaum's lemma (Proposition 5.46 in \cite{MorettiBook}), $L_{\bullet}$ is essentially self-adjoint on their linear span.  Thus, it suffices to check that $V_{\mu} \otimes | h \rangle$ are states of uniqueness for $L_{\bullet}( \vcurrent ( \cdot | \hbar))$, which holds immediately because the $L_{\bullet}( \vcurrent ( \cdot | \hbar))$ orbit of $V_{\mu} \otimes |h \rangle$ is a subspace of $\mathcal{H}_{\bullet}(a)[d]$ for $d = \deg \mu + h$ and $\dim \mathcal{H}_{\bullet}(a) [d] < \infty$. $\square$
\end{itemize}
}

\begin{corollary} $L_{\bullet}( \vcurrent ( \cdot | \hbar))$ is a well-defined self-adjoint operator on $\mathscr{H}_{\bullet}(a)$, hence 
\begin{equation} \widehat{\mathcal{R}}_{\bullet}(u | \hbar) = \frac{1}{u - L_{\bullet}( \vcurrent ( \cdot | \hbar))}  \end{equation} \noindent the resolvent is a bounded operator on $\mathscr{H}_{\bullet}(a)$ for $u \in \C \setminus \R$. \end{corollary}

\noindent Finally, using Corollary [\ref{BlockFockToeplitzFinite}], it is simple to check:
\begin{corollary} As $\hbar \rightarrow 0$, have strong resolvent convergence $L_{\bullet}( \vcurrent ( \cdot | \hbar)) \rightarrow 0$. \end{corollary}

\subsubsection{\textcolor{black}{Lax Operator for Quantum Benjamin-Ono}} \label{subsubsecLaxQBO}

\begin{definition} \label{QBOLaxDefinition}The \underline{Lax operator} for the quantum Benjamin-Ono hierarchy is \begin{equation} \widehat{\mathcal{L}}_{\bullet}( \ebar, \hbar) = L_{\bullet}( \vcurrent ( \cdot | \hbar)) +\ebar  (\mathbbm{1} \otimes D_{\bullet} ) \end{equation}

\noindent the perturbation of $L_{\bullet}( \vcurrent( \cdot | \hbar))$ the Fock-block Toeplitz operator with symbol $\vcurrent (\cdot | \hbar)$ the $\widehat{\mathfrak{gl}_1}$ current at level $\hbar$ by an amount $\ebar$ of the auxiliary degree operator $D_{\bullet}$, where $\ebar$ is the coefficient of dispersion for classical Benjamin-Ono.\end{definition}

\noindent For $|h \rangle = w^h$ in dense subspace $( \mathcal{F}(a) \otimes \C[w]) \cong \mathcal{H}_{\bullet}(a) \subset \mathscr{H}_{\bullet}(a)$, this Lax operator is

\begin{equation} \label{QuantumLaxMatrixVisualize} \widehat{\mathcal{L}}_{\bullet}( \ebar, \hbar)  \Big |_{\mathcal{H}_{\bullet} (a)} = \begin{bmatrix} a+ \ebar \cdot {0} & {\widehat{\mathcal{V}}}_1& \widehat{\mathcal{V}}_2 & \widehat{\mathcal{V}}_3 & \cdots & \widehat{\mathcal{V}}_{h} & \cdots  \\ 
    \widehat{\mathcal{V}}_{-1} (\hbar)& a+ \ebar \cdot 1& \widehat{\mathcal{V}}_1 & \widehat{\mathcal{V}}_2 & \ddots & \widehat{\mathcal{V}}_{h-1} & \ddots  \\ 
    \widehat{\mathcal{V}}_{-2} (\hbar)& \widehat{\mathcal{V}}_{-1}(\hbar) &a + \ebar \cdot 2 & \widehat{\mathcal{V}}_1& \ddots & \ddots & \ddots \\ 
    \widehat{\mathcal{V}}_{-3} (\hbar)& \widehat{\mathcal{V}}_{-2} (\hbar)& \widehat{\mathcal{V}}_{-1}(\hbar) &a+ \ebar \cdot 3 & \ddots & \ddots & \ddots  \\
    \vdots & \ddots & \ddots & \ddots & \ddots & \ddots & \ddots \\
    \widehat{\mathcal{V}}_{-h} (\hbar)& \widehat{\mathcal{V}}_{-(h-1)}(\hbar) & \ddots  & \ddots & \ddots & a+\ebar \cdot h & \ddots \\
     \vdots & \ddots & \ddots & \ddots & \ddots & \ddots & \ddots 
    \end{bmatrix} \nonumber
    \end{equation}
    
    \noindent a $\C$-symmetric $\N \times \N$ matrix whose matrix elements are themselves in $\mathfrak{gl}(\mathscr{F}(a))$: \begin{equation}  \widehat{\mathcal{L}}_{\bullet}( \ebar, \hbar) _{h_+, h_-} = \widehat{\mathcal{V}}_{h_- - h_+} (\hbar)+ (a+ \ebar h )\delta( h_+ - h_-). \end{equation}
    
    \noindent The $(h,h)$-diagonal block entry act by the scalar $a + \ebar \cdot h$ in $\mathscr{F}(a) \otimes |h \rangle$, independent of $\hbar$, while the off-diagonal block entries $\widehat{\mathcal{V}}_{h_- - h_+}(\hbar)$ are either creation operators $\widehat{\mathcal{V}}_{+k}$ or annihilation operators $\widehat{\mathcal{V}}_{-k}$ depending on the sign of $h_- - h_+$ for $h_{\pm } \geq 0$.  Moreover, the diagonal is constant in the dispersionless limit $\ebar \rightarrow 0$.\\
 \\
 \noindent  Since the auxiliary degree operator $\mathbbm{1} \otimes D_{\bullet}$ commutes $[ \widehat{\mathcal{D}}_{\bullet}, \mathbbm{1} \otimes D_{\bullet}] = 0$ with the total degree operator $\widehat{\mathcal{D}}_{\bullet}$, identical arguments from the previous section \textbf{[\ref{subsecFockBlockToeplitz}]} imply:

\begin{proposition} The quantum Lax operator $\widehat{\mathcal{L}}_{\bullet}( \ebar, \hbar) $ commutes
\begin{equation} [ \widehat{\mathcal{D}}_{\bullet}, \widehat{\mathcal{L}}_{\bullet}( \ebar, \hbar)  ] = 0 \end{equation}

\noindent with the total degree operator on $\mathcal{H}_{\bullet}(a)$ the dense subspace of $\mathscr{H}_{\bullet}(a)$. \end{proposition}

\begin{corollary} \label{QuantumLaxFinite} $\widehat{\mathcal{L}}_{\bullet}( \ebar, \hbar) $ is a densely-defined $\C$-symmetric operator on $\mathscr{H}_{\bullet}(a)$ that preserves the finite-dimensional subspaces $\mathcal{H}_{\bullet}(a)[d]$. \end{corollary}

\begin{proposition} $\widehat{\mathcal{L}}_{\bullet}( \ebar, \hbar)$ is essentially self-adjoint on the pre-Hilbert space $\mathcal{H}_{\bullet}(a)$. \end{proposition}

\begin{corollary} $\widehat{\mathcal{L}}_{\bullet}( \ebar, \hbar) $ is a self-adjoint operator on $\mathscr{H}_{\bullet}(a)$, hence the resolvent 
\begin{equation} \widehat{\mathcal{R}}_{\bullet}(u | \ebar, \hbar) = \frac{1}{u - \widehat{\mathcal{L}}_{\bullet}( \ebar, \hbar) }  \end{equation} \noindent is a bounded operator on $\mathscr{H}_{\bullet}(a)$ for $u \in \C \setminus \R$. \end{corollary}

\begin{corollary} As $\hbar \rightarrow 0$, strong resolvent convergence $\widehat{\mathcal{L}}_{\bullet}( \ebar, \hbar)  \rightarrow \ebar (\mathbbm{1} \otimes D_{\bullet})$. \end{corollary}

\noindent With these results in place, we may apply the finite-dimensional spectral theorem to the restriction $\widehat{\mathcal{L}}_{\bullet}(\ebar, \hbar)|_{\mathscr{H}_{\bullet}(a)[d]}$ of $\widehat{\mathcal{L}}_{\bullet}( \ebar, \hbar)$ to the finite-dimensional subspace $\mathcal{H}_{\bullet}(a)[d]$:

\begin{theorem} \label{SpectralTheoremQuantumLax} There is a basis $\Phi_{(\lambda, h)} ( V | \ebar, \hbar, a)$ of $\mathcal{H}_{\bullet}(a) \cong \C[V_1, V_2, \ldots] \otimes \C[w]$ of eigenstates of the Lax operator $\widehat{\mathcal{L}}_{\bullet}( \ebar, \hbar)$ of quantum Benjamin-Ono that are polynomials in $V_1, V_2, \ldots$ and $w$ indexed by the data $(\lambda, h)$ of a partition $\lambda$ and $h \in \N$.  \end{theorem}

\noindent Essential self-adjointness of $\widehat{\mathcal{L}}_{\bullet}( \ebar, \hbar)$ on $\mathcal{F}(a) \otimes \C[w]$ implies that all such eigenstates $\Phi_{(\lambda, h)} ( V_1, V_2, \ldots | \ebar, \hbar,a)$ are necessarily in the dense subspace $\mathcal{H}_{\bullet}(a)$ of $\mathscr{H}_{\bullet}(a)$ and are solutions to the fundamental system \begin{equation} \label{FundamentalSystemQuantumLax} \widehat{\mathcal{L}}_{\bullet} (\ebar, \hbar) \Phi ( \cdot | \ebar, \hbar ) = u \ \Phi( \cdot | \ebar, \hbar) \end{equation}
\noindent for some very particular eigenvalues $u = u_{\lambda, h,a}(\ebar, \hbar)$ depending on the data $\lambda, h,a$ and dynamical coefficients $\ebar, \hbar$ of dispersion and quantization.  $\widehat{\mathcal{L}}_{\bullet}( \ebar, \hbar)$ is self-adjoint for $\ebar \in \R$, $\hbar >0$, $a \in \R$, which ensures that its eigenvalues are real $u =c$ for $c \in \R$, but we do not find $c_{\lambda, h,a}(\ebar, \hbar)$ in this paper.

\subsubsection{\textcolor{black}{Nazarov-Sklyanin Transfer Operators and Jack-Lax Orbits}} \label{subsubsecNSorbits}

\noindent The content of this paper hinges on the following fact which we state without proof:

\begin{theorem} \label{NSQuantumCore} \textnormal{(Nazarov-Sklyanin \cite{NaSk2})} The Titchmarsh-Weyl operator 

\begin{equation}\widehat{\mathcal{T}}^{\uparrow}(u ) |_{\widehat{\mathcal{L}}_{\bullet}( \ebar, \hbar); \mathscr{H}_0} = (\mathbbm{1} \otimes \uppi_0) \frac{1}{ u - \widehat{\mathcal{L}}_{\bullet}(\ebar, \hbar) } ( \mathbbm{1} \otimes \uppi_0) \end{equation}

\noindent of the Lax operator $\widehat{\mathcal{L}}_{\bullet} ( \ebar, \hbar)$ of quantum Benjamin-Ono from \textnormal{Definition [\ref{QBOLaxDefinition}]} is a transfer operator in the sense of \textnormal{Definition [\ref{TransferOperatorDefinition}]}, i.e. for $u_1, u_2 \in \C \setminus \R$
\begin{equation} [ \widehat{\mathcal{T}}^{\uparrow}(u_1  | \ebar, \hbar) , \widehat{\mathcal{T}}^{\uparrow}(u_2 | \ebar, \hbar) ] = 0 \end{equation}

\noindent are bounded pairwise-commuting operators on $\mathscr{H}_0(a) = \mathscr{F}(a) \otimes H_0$ for $H_0 = \C| 0 \rangle$.
\end{theorem}

\begin{definition} The \underline{Nazarov-Sklyanin quantum transfer operators} for quantum periodic Benjamin-Ono are $\widehat{\mathcal{T}}^{\uparrow}(u| \ebar, \hbar) =\widehat{\mathcal{T}}^{\uparrow}(u ) |_{\widehat{\mathcal{L}}_{\bullet}( \ebar, \hbar); \mathscr{H}_0}$ of \textnormal{Theorem [\ref{NSQuantumCore}]}. \end{definition}
\begin{itemize}
\item \textit{Proof:} In \cite{NaSk2}, the authors prove commutativity of the coefficients in the $1/u$ expansion of a ``cotransfer operator'' $\widehat{\mathcal{T}}^{\downarrow}( u |  \ebar, \hbar)$ related to $\widehat{\mathcal{T}}^{\uparrow}(u | \ebar,\hbar)$ by

\begin{equation} \widehat{\mathcal{T}}^{\uparrow}(u | \ebar, \hbar) \cdot \Big ( u - \widehat{\mathcal{T}}^{\downarrow}( u | \ebar, \hbar) \Big ) = \mathbbm{1} \end{equation}

\noindent which automatically implies Theorem [\ref{NSQuantumCore}] as stated above.  Our conventions differ from \cite{NaSk2} by identifications in Definitions [\ref{OmegaVariablesDefinition}] and [\ref{FractionalChargeDefinition}]. $\square$
\end{itemize}

\noindent By direct computation, one can check that Nazarov-Sklyanin's quantum transfer operators $\widehat{\mathcal{T}}^{\uparrow}( u |\ebar, \hbar)$ are relevant for the quantum periodic Benjamin-Ono system:
\begin{proposition} \label{LaxtoQBO} For $\ell = 2,3$, the commuting self-adjoint operators $\widehat{\mathcal{T}}_{\ell}^{\uparrow} ( \ebar, \hbar)$ in \textnormal{Theorem [\ref{NSQuantumCore}]} are the momentum operator $\widehat{\mathcal{T}}_2^{\uparrow}$ of \textnormal{formula (\ref{MomentumOperatorFormula})} and the quantum periodic Benjamin-Ono Hamiltonian $\widehat{\mathcal{T}}^{\uparrow}_3 ( \ebar, \hbar)$ of \textnormal{Definition [\ref{QBOHamiltonian}]} after \begin{equation} \mathscr{F}(a) \cong \mathscr{F}(a) \otimes H_0  =\mathscr{H}_{0}(a) \end{equation}

\noindent identifying the state space $\mathscr{F}(a)$ of quantum periodic Benjamin-Ono with the subspace $\mathscr{H}_{0}(a) = \mathscr{F}(a) \otimes | 0 \rangle$ of $\mathscr{H}_{\bullet}(a) = \mathscr{F}(a) \otimes H_{\bullet}$, so that $H_{\bullet}$ is an auxiliary Hardy space.\end{proposition}

\noindent We didn't find the classical stationary states of the classical periodic Benjamin-Ono equation in this paper, but we did identify the quantum stationary states of the quantum periodic Benjamin-Ono equation as Jack polynomials in Theorem [\ref{QBOStationaryStatesAreJacks}].  Proposition [\ref{LaxtoQBO}] implies
\begin{corollary} The Nazarov-Sklyanin quantum transfer operators in $\mathscr{F}(a) \cong \mathscr{H}_0(a)$ for $a \in \R$ are $\widehat{\mathcal{T}}^{\uparrow}( u | \ebar, \hbar): \mathscr{H}_{0}(a) \rightarrow \mathscr{H}_{0}(a)$ are simultaneously diagonalized on 
\begin{equation}\Psi_{\lambda,a} ( \cdot | \ebar, \hbar) \otimes |0 \rangle \end{equation}

\noindent where $\Psi_{\lambda,a}( V_1, V_2, \ldots | \ebar, \hbar ) \in \mathscr{F}(a)$ are Jack polynomials as in \textnormal{Theorem [\ref{QBOStationaryStatesAreJacks}]}. \end{corollary}

\begin{definition} The \underline{Jack-Lax orbits} are the orbits \begin{equation} \mathscr{H}_{\bullet, \lambda} (a; \ebar, \hbar) =\mathscr{H}_{\bullet} [ \widehat{\mathcal{L}}_{\bullet}( \ebar, \hbar) , \Psi_{\lambda,a}( \cdot | \ebar, \hbar) \otimes | 0 \rangle ] \end{equation}

\noindent of $\Psi_{\lambda,a} ( \cdot | \ebar, \hbar) \otimes |0 \rangle$ under $\widehat{\mathcal{L}}_{\bullet}( \ebar, \hbar)$.  

\end{definition}

\noindent We abbreviate restrictions to the Jack-Lax orbit $\mathscr{H}_{\bullet, \lambda} (a; \ebar, \hbar)$ by \begin{eqnarray} \widehat{\mathcal{L}}_{\bullet}( \ebar, \hbar)|_{\lambda, a} &=&  \widehat{\mathcal{L}}_{\bullet}( \ebar, \hbar) |_{\mathscr{H}_{\bullet, \lambda} (a; \ebar, \hbar) }  \\ \widehat{\mathcal{L}}_{+} ( \ebar, \hbar) |_{\lambda,a}&=& \widehat{\mathcal{L}}_{+}( \ebar, \hbar)|_{\mathscr{H}_{\bullet, \lambda} (a; \ebar, \hbar) }   \end{eqnarray}

\noindent Note: \underline{$\Psi_{\lambda,a}( \cdot | \ebar, \hbar) \otimes | 0 \rangle$ are \textit{not} eigenfunctions of $\widehat{\mathcal{L}}_{\bullet}$!}  The Jack-Lax orbit decomposes

\begin{equation} \mathscr{H}_{\bullet, \lambda} (a; \ebar, \hbar)  = \bigoplus_{(\nu, h) \sim \lambda } \C | \Phi_{\nu, h}( \cdot | \ebar, \hbar) \rangle \end{equation}

\noindent into one-dimensional eigenspaces of $\widehat{\mathcal{L}}_{\bullet}(\ebar, \hbar)$ for fundamental solutions $\Phi_{\nu, h, a} ( \cdot | \ebar, \hbar)$ indexed by some pair $(\nu, h)$ of a partition $\nu$ and $h \in \N$ with $\deg \nu + h = \deg \lambda$.  

\begin{proposition} The Jack-Lax orbits $\mathscr{H}_{\bullet, \lambda} (a; \ebar, \hbar) $ are finite-dimensional. \end{proposition}

\begin{itemize}
\item \textit{Proof:} $\mathscr{H}_{\bullet, \lambda} (a; \ebar, \hbar) $ are subspaces of the finite-dimensional eigenspace $\mathscr{H}_{\bullet}(a)[d]$ of the total degree operator $\widehat{\mathcal{D}}_{\bullet}$ for $d = \deg \lambda$ by Corollary [\ref{QuantumLaxFinite}]. $\square$\end{itemize}

\subsubsection{\textcolor{black}{Conserved Densities as Spectral Shift Functions}} \label{subsubsecQBOConservedDensitySSF}

\noindent We now extend Nazarov-Sklyanin's Theorem [\ref{NSQuantumCore}] as promised in Theorem [\ref{QBOHConservedDensityExistence}].

\begin{theorem} \label{QBOHConservedDensityConstruction} For any $\Psi \in \mathscr{F}(a)$, define a {\textit{random}} Rayleigh function \begin{equation} d \widehat{F}(c | \ebar, \hbar) |_{\Psi} \end{equation} \noindent of a profile $\widehat{f}_{\Psi} ( c | \ebar, \hbar) \in \mathbf{P}^{\vee}$ on the real line $\mathbb{X} = \R$ by specifying the characteristic functions of its random $\phi$-averages for special

\begin{equation} \label{SmartChoice} \phi(c) = \frac{1}{ \textnormal{\textbf{i}} t} \log \Big [ \frac{1}{u-c} \Big ]. \end{equation}
 
\noindent to be
\begin{equation} \label{SuperCool} \mathbb{E} \Big [ e^{ \textnormal{\textbf{i}} t \int_{- \infty}^{+\infty} \phi(c) d \widehat{F}(c | \ebar, \hbar)|_{\Psi} } \Big ] = \langle \Psi \otimes 0 | \frac{1}{ u- \widehat{\mathcal{L}}_{\bullet}( \ebar, \hbar)} | \Psi \otimes 0 \rangle \end{equation}
\noindent the diagonal matrix element of the resolvent of the quantum Lax operator $\widehat{\mathcal{L}}_{\bullet}( \ebar, \hbar)$.

\begin{enumerate}
\item \textnormal{\textcolor{gray}{[Conserved Density]}} The law of $d \widehat{F}(c | \ebar, \hbar)|_{\Psi}$ does not change if $\Psi$ evolves by the quantum periodic Benjamin-Ono equation \textnormal{(\ref{QBOE})}, i.e. $d \widehat{F}(c | \ebar, \hbar)|_{\Psi}$ is a (realized) quantum conserved density in the sense of \textnormal{Definitions [\ref{QuantumConservedDensityDefinition}], [\ref{TheModelColumn3PreDefinition}]}.
\item \textnormal{\textcolor{gray}{[Integrable Hierarchy]}} for $l=1,2,3,\ldots$, the quantum observables $\{\widehat{\mathcal{O}}_{l}( \ebar, \hbar)\}_{l=1}^{\infty}$

\begin{equation} \sum_{l=1}^{\infty} u^{- l-1} \widehat{\mathcal{O}}_{l}( \ebar, \hbar) := \frac{ \partial}{\partial u } \log \widehat{\mathcal{T}}^{\uparrow}(u | \ebar, \hbar) \end{equation}

\noindent defined to be coefficients of the logarithmic derivative of the Nazarov-Sklyanin quantum transfer operator commute \noindent \begin{equation} [ \widehat{\mathcal{O}}_{l_1}(\ebar, \hbar)  , \widehat{\mathcal{O}}_{l_2}(\ebar, \hbar) ]= 0  \end{equation}\noindent and their random values in a state $\Psi$ are identically \begin{equation} \widehat{\mathcal{O}}_l (\ebar, \hbar) \big |_{\Psi} = \int_{- \infty}^{\infty} c^l d\widehat{F} ( c | \ebar, \hbar)|_{\Psi} \end{equation}

\item \textnormal{\textcolor{gray}{[Regularity of Observables]}} $\widehat{\mathcal{O}}_l(\ebar, \hbar)$ is a generalized non-commutative polynomial of degree $l \in \N$ in creation and annihilation operators $\widehat{\mathcal{V}}_{\pm k}$ as \textnormal{Definition [\ref{GeneralizedPolynomialDefinition}]}

\item \textnormal{\textcolor{gray}{[Periodic Benjamin-Ono]}} The span of $\{\widehat{\mathcal{O}}_l(\ebar, \hbar)\}_{l=1}^{\infty}$ includes the quantum periodic Benjamin-Ono Hamiltonian $\widehat{\mathcal{T}}_3^{\uparrow}(\ebar, \hbar)$ \textnormal{(\ref{QBOHamiltonian})} and also $\widehat{\mathcal{T}}_2^{\uparrow}(\hbar)$ \textnormal{(\ref{MomentumOperatorFormula})}
 
\end{enumerate}
\end{theorem}

\noindent We prove the enumerated claims of Theorem [\ref{QBOHConservedDensityConstruction}] in Steps 1-4 below.

\begin{itemize}
\item \textit{Step 1:} The smart choice of $\phi(c)$ in formula (\ref{SmartChoice}) is enough to characterize the law of the random signed measure $d \widehat{F}(c | \ebar, \hbar)|_{\Psi}$ due to our ability to vary the spectral parameter $u \in \C \setminus \R$, so the desired equality follows from the definition of the Nazarov-Sklyanin quantum transfer operator $\widehat{\mathcal{T}}^{\uparrow}(u | \ebar, \hbar)$ and the desired time invariance from Theorem [\ref{NSQuantumCore}].
\item \textit{Step 2:} Write Hamiltonians $O_l(\ebar)|_v$ of classical periodic Benjamin-Ono hierarchy as polynomials in the $T_{\ell}^{\uparrow}(\ebar)|_v$ as in Corollary [\ref{KMKsupport}], define $\widehat{\mathcal{O}}_l( \ebar, \hbar)$ to be the result of substituting $T_{\ell}^{\uparrow}(\ebar)|_v \rightarrow \widehat{\mathcal{T}}^{\uparrow}_{\ell}( \ebar, \hbar)$, and use Theorem [\ref{NSQuantumCore}].
\item \textit{Step 3:} Follows from the previous step and Step 2 of Theorem [\ref{CBOHConservedDensityConstruction}].
\item \textit{Step 4:} Follows from Proposition [\ref{LaxtoQBO}].
\end{itemize}

\noindent Let us conclude with remarks about the proof.  First, formula (\ref{SuperCool}) carries a non-trivial relationship between two distinct quantities:
\begin{itemize}
\item The Kerov-Markov-Kre\u{\i}n transform of the random Rayleigh measure $d \widehat{F}( c | \ebar, \hbar)|_{\Psi}$ on the left-hand side of formula (\ref{SuperCool}) is \begin{equation} \label{LeftSecret} d \widehat{\tau}^{\uparrow}( c | \ebar, \hbar)|_{\Psi} \end{equation} \noindent a \textit{random} probability measure on $\R$, while
\item The spectral measure of the quantum Benjamin-Ono Lax operator $\widehat{\mathcal{L}}_{\bullet}(\ebar, \hbar)$ at $\Psi \otimes | 0 \rangle$ whose Stieltjes transform is the right-hand side of formula (\ref{SuperCool}) is \begin{equation} \label{RightSecret} d \tau_{\Psi \otimes 0 , \Psi \otimes 0 } ( c | \widehat{\mathcal{L}}(\ebar, \hbar ) )\end{equation} \noindent a \textit{non-random} probability measure on $\R$.
\end{itemize}

\noindent Second, whereas for our classical periodic Benjamin-Ono hierarchy it is the case that

\begin{equation} O_l ( \ebar) |_v = \text{Tr}_{H_{\bullet}} \Bigg [ L_{\bullet}( v | \ebar)^l - L_{+}(v | \ebar)^l \Bigg ] \end{equation}
\noindent over the auxiliary Hardy space, the quantum periodic Benjamin-Ono hierarchy is \textbf{not}
\begin{equation} \label{WatchOutEveryone} \widehat{\mathcal{O}}_{l} ( \ebar, \hbar) \neq \text{Tr}_{H_{\bullet}} \Bigg [ \widehat{\mathcal{L}}_{\bullet}(\ebar, \hbar)^{l} - \widehat{\mathcal{L}}_+ ( \ebar, \hbar)^l \Bigg ] \end{equation}

\noindent the result of substituting $(V_k, \overline{V_k} ) \rightarrow ( \widehat{\mathcal{V}}_{+k}, \widehat{\mathcal{V}}_{-k})$ into the Lifshitz-Kre\u{\i}n trace formula.  The operator defined by the right-hand side is plagued by divergences.  Instead, there is a distinguished ordering, Nazarov-Sklyanin's ordering $\eta_{NS}$ of Theorem [\ref{NazarovSklyaninQuantizationConstruction}], so

\begin{equation} \widehat{\mathcal{O}}_{l} ( \ebar, \hbar) = \Bigg (  \text{Tr}_{H_{\bullet}} \Bigg [ L_{\bullet}( v | \ebar)^l - L_{+}(v | \ebar)^l \Bigg ] \Bigg )^{\eta_{NS}}  \end{equation}

\noindent We will come back to this subtle issue once more in the next section, when we consider the joint spectrum of the operators $\widehat{\mathcal{O}}_l(\ebar, \hbar)$ at a Jack polynomial.

\subsubsection{\textcolor{black}{Anisotropic Partitions as Spectral Shift Functions}} \label{subsubsecAnisotropicSSF}

\noindent We now describe the quantum conserved densities $d\widehat{F}(c | \ebar, \hbar)|_{\Psi}$ of quantum periodic Benjamin-Ono from Theorem [\ref{QBOHConservedDensityConstruction}] in the case of quantum stationary states as promised in Theorem [\ref{QBOHConservedDensitiesForJacksAreAnisotropicPartitions}].  To do so, we use another black-boxed result from \cite{NaSk2}:

\begin{theorem} \label{NSQuantumSpectrumCore} \textnormal{(Nazarov-Sklyanin \cite{NaSk2})} The eigenvalue $T^{\uparrow}(u | \ebar, \hbar) |_{\lambda}$ of the quantum transfer operator $\widehat{\mathcal{T}}^{\uparrow}(u | \ebar ,\hbar) $ at a Jack $\Psi_{\lambda}( \cdot | \ebar, \hbar)$ is the rational function of $u$ given by \begin{equation} \label{OPRLformula} \langle \Psi_{\lambda} \otimes 0  | \frac{1}{u - \widehat{\mathcal{L}}_{\bullet}(\ebar, \hbar) |_{\lambda}}|  \Psi_{\lambda} \otimes 0 \rangle   = T^{\uparrow}(u | \ebar, \hbar)|_{\lambda}  = \frac{1}{u} \cdot \frac{ \det_{\mathscr{H}_{\bullet}[ \lambda; \ebar, \hbar ]  } (u - \widehat{\mathcal{L}}_+ (\ebar, \hbar) |_{\lambda}) }{ \det_{\mathscr{H}_{\bullet}[\lambda; \ebar, \hbar]} ( u - \widehat{\mathcal{L}}_{\bullet}(\ebar, \hbar) |_{\lambda}) } \nonumber \end{equation}

\noindent the Titchmarsh-Weyl function of $\widehat{\mathcal{L}}_{\bullet} ( \ebar, \hbar) |_{\lambda}$ with cyclic $\Psi_{\lambda,a}( \cdot | \ebar, \hbar) \otimes 0$ in $\mathscr{H}_{\bullet}[ \lambda; \ebar, \hbar]$. \end{theorem} \begin{itemize}
\item \textit{Proof:} The eigenvalue statement follows from \cite{NaSk2} in conventions of Definitions [\ref{OmegaVariablesDefinition}] and [\ref{FractionalChargeDefinition}].  To complete the proof, the restriction $\widehat{\mathcal{L}}_{\bullet}( \ebar, \hbar)|_{\lambda}$ of the unbounded self-adjoint quantum Lax operator $\widehat{\mathcal{L}}_{\bullet} (\ebar, \hbar)$ to a finite-dimensional Jack-Lax orbit is bounded and so essentially self-adjoint.  Moreover, Theorem [\ref{NSQuantumCore}] implies $\Psi_{\lambda} \otimes |0 \rangle$ is isolated by Lemma [\ref{TransferIsolationLemma}], hence the rest follows from Theorem [\ref{OPRLtensororbitISOLATED}] for $\widehat{\mathcal{L}}_{\bullet} = \widehat{\mathcal{L}}_{\bullet} ( \ebar, \hbar)$ and $\Psi_0 = \Psi_{\lambda;0} ( \cdot | \ebar, \hbar)= \Psi_{\lambda} ( \cdot | \ebar, \hbar) \otimes | 0 \rangle$. $\square$
\end{itemize}

\begin{corollary} \label{QBOConservedDensitiesSSFExist} There exists a profile $f_{\lambda} ( c| \ee, \e) \in \mathbf{P}^{\vee}$ so that
\begin{itemize}
\item The $T^{\uparrow}$-observable of $f_{\lambda}( c | \ee, \e)$ is the Titchmarsh-Weyl function \begin{equation} T^{\uparrow}(u) |_{f_{\lambda}( \cdot | \ee, \e)} = T^{\uparrow}(u )|_{\widehat{\mathcal{L}}_{\bullet} ( \ebar, \hbar) |_{\lambda}; \Psi_{\lambda; 0 } ( \cdot | \ebar, \hbar)} \end{equation} of $\widehat{\mathcal{L}}_{\bullet}( \ebar, \hbar) |_{\lambda}$ with cyclic $\Psi_{\lambda;0} ( \cdot | \ebar, \hbar)$, and hence also {the eigenvalue of the Nazarov-} {Sklyanin transfer operator $\widehat{\mathcal{T}}^{\uparrow}(u | \ebar, \hbar)$ at $\Psi_{\lambda;0} ( \cdot | \ebar, \hbar) \in \mathscr{H}_{\bullet}$}.
\item The transition measure $d\tau_f^{\uparrow}( c | \ee, \e)$ of $f_{\lambda}( c | \ee, \e)$ is the {spectral measure} of the restriction $\widehat{\mathcal{L}}_{\bullet} ( \ebar, \hbar) |_{\lambda}$ at $\Psi_{\lambda;0} ( \cdot | \ebar, \hbar)$ and hence also {the spectral measure of the non-} {restricted $\widehat{\mathcal{L}}_{\bullet} ( \ebar, \hbar)$ Lax operator at $\Psi_{\lambda;0} ( \cdot | \ebar, \hbar) \in \mathscr{H}_{\bullet}$}.
\item The shifted Rayleigh function $\xi_f(c | \ee, \e)$ of $f_{\lambda}( c | \ee, \e)$ is the \textcolor{black}{\textit{spectral shift function}} of $\widehat{\mathcal{L}}_{\bullet} ( \ebar, \hbar)|_{\lambda}, \widehat{\mathcal{L}}_{+} ( \ebar, \hbar) |_{\lambda}$ on the Jack-Lax orbit $\widehat{\mathscr{H}}_{\bullet}[ \lambda; \ebar, \hbar]$.
 \end{itemize}
 
 \noindent using the conventions of \textnormal{Definition [\ref{OmegaVariablesDefinition}]}.

\end{corollary}

\noindent With this corollary in place, we have:

\begin{theorem}\label{QBOHConservedDensitiesForJacksAreAnisotropicPartitionsRevisited} Under the identifications in \textnormal{Definitions [\ref{OmegaVariablesDefinition}] and [\ref{FractionalChargeDefinition}]}, in the case of quantum stationary states $\Psi = \Psi_{\lambda, a} ( \cdot | \ebar, \hbar)$ which we identified with Jack polynomials in \textnormal{Theorem [\ref{QBOStationaryStatesAreJacks}]}, quantum conserved densities of \textnormal{Theorem [\ref{QBOHConservedDensityConstruction}]} are \begin{equation} d \widehat{F}_{\Psi_{\lambda, a}( \cdot | \ebar, \hbar)} ( c | \ebar, \hbar) = d F_{\lambda} ( c - a | \ee, \e) \end{equation}

\noindent Rayleigh measures of profiles $f_{\lambda} (c - a | \ee, \e)$ of anisotropic partitions. \end{theorem}
\begin{itemize}
\item \textit{Proof:} In \cite{NaSk2}, Nazarov-Sklyanin find joint spectrum of $\widehat{\mathcal{T}}^{\uparrow}(u | \ebar, \hbar)$ to be
\begin{equation} T^{\uparrow}(u | \ebar, \hbar) |_{\lambda,a} =  T^{\uparrow}(u) |_{f_{\lambda}( c-a | \ee, \e)}\end{equation}

\noindent the $T^{\uparrow}$ observable of the profile $f_{\lambda}( c -a | \ee, \e)$.  The formula in \cite{NaSk2} involves the rescaled and shifted row lengths $ \ee ( i-1) + \e \lambda_i $ of a partition $\lambda$ but can be converted to a product over contents and hence of this form.  The desired claim follows from Theorem [\ref{QBOHConservedDensityConstruction}] and by applying the general results for block Jacobi operators from section \textbf{[\ref{subsubsecBlockJacobi}]} to the quantum transfer operator $\widehat{\mathcal{T}}^{\uparrow}(u  | \ebar, \hbar)$. $\square$
\end{itemize}

\noindent A priori, interlacing extrema of $f_{\lambda}(  c-a | \ee, \e)$ only agree with interlacing eigenvalues of $\widehat{\mathcal{L}}_{\bullet}( \ebar, \hbar) |_{\lambda,a}, \widehat{\mathcal{L}}_{+} ( \ebar, \hbar) |_{\lambda,a}$ in the Jack-Lax orbit $\mathscr{H}_{\bullet, \lambda} (a; \ebar, \hbar)$ up to multiplicities.  A remark is in order about these interlacing extrema, in light of the fundamental distinction in formula (\ref{WatchOutEveryone}) from the previous section.  The Lifshitz-Krein Trace Formula of Theorem [\ref{LifshitzKreinTraceFormula}] applied to $\widehat{\mathcal{L}}_{\bullet}( \ee, \e) |_{\lambda}, \widehat{\mathcal{L}}_+ ( \ee, \e) |_{\lambda}$ in finite-dimensional Jack-Lax orbits $\mathcal{H}_{\bullet}[ \lambda; \ee, \e]$ says

\begin{equation}  \widehat{{O}}_l ( \ee, \e) \big |_{\lambda} =  \text{Tr}_{\mathscr{H}_{\bullet}[ \lambda; \ee, \e]} \Bigg [ \Big ( \widehat{\mathcal{L}}_{\bullet}( \ee, \e) |_{\lambda} \Big )^{l} - \Big (  \widehat{\mathcal{L}}_{+}( \ee, \e)|_{\lambda} \Big )^l  \Bigg ] \end{equation}
\noindent the eigenvalues of commuting $\widehat{\mathcal{O}}_l( \ee, \e)$ may be written as a finite-dimensional trace.  This {{does not mean}} that the commuting operators $\widehat{\mathcal{O}}_l(\ee, \e)$ themselves are traces over an auxiliary pre-Hardy space as in formula (\ref{WatchOutEveryone}) of the non-restricted operators $\widehat{\mathcal{L}}_{\bullet}(\ee, \e), \widehat{\mathcal{L}}_+ ( \ee, \e)$ without normal ordering.  This same issue arose in section \textbf{[\ref{subsubsecBlockJacobi}]}, our inability a priori to write the Titchmarsh-Weyl operator as a block perturbation determinant.  Note that the restricted $\widehat{\mathcal{L}}_{\bullet}( \ee , \e)|_{\lambda}, \widehat{\mathcal{L}}_{+}(\ee, \e)|_{\lambda}$ and the unrestricted $\widehat{\mathcal{L}}_{\bullet}( \ee, \e)$ preserve Jack-Lax orbits $\mathscr{H}_{\bullet}[\lambda; \ee, \e]$ by definition but $\widehat{\mathcal{L}}_+(\ee, \e)$ does not.\\
\\
\noindent Let us contrast Theorem [\ref{QBOHConservedDensitiesForJacksAreAnisotropicPartitionsRevisited}] to what we have seen in the classical case.  For classical periodic Benjamin-Ono, we did not write the Titchmarsh-Weyl function $T^{\uparrow}(u | \ebar) |_v$ of the classical Lax operator $L_{\bullet}(v  | \ebar) = L_{\bullet} + \ebar D_{\bullet}$ explicitly in terms of $v$ and $0< \ebar$, but at least checked that it is a meromorphic function with isolated poles that do not accumulate except at infinity.  We did better for the classical dispersionless periodic Benjamin-Ono system, computing explicitly the Titchmarsh-Weyl function $T^{\uparrow}( u )|_v$ of the classical dispersionless Lax operator $L_{\bullet}(v)$, a Toeplitz operator with symbol $v$, explicitly in terms of $v$ as the geometric mean of $\frac{1}{u-v(w)}$ in Szeg\H{o}'s First Theorem, Theorem [\ref{SzegoFirstTheorem}].

\begin{corollary} 
\label{SpectralCorollary} The spectral measure of the quantum Benjamin-Ono Lax operator $\widehat{\mathcal{L}}_{\bullet}( \ebar, \hbar)$ at $ \Psi_{\lambda,a} ( {V} | \ebar, \hbar) \otimes |0 \rangle \in \mathscr{F}(a) \otimes H_{\bullet}$ is the transition measure $d\tau^{\uparrow}_{\lambda}(c -a| \ee, \e)$ of the profile of the anisotropic partition $\lambda \in \mathbb{Y}(a; \ee, \e)$.

\end{corollary}

\noindent Compare Corollary [\ref{SpectralCorollary}] to Biane's realization of transition measures $\tau^{\uparrow}_{\lambda}(c | 0 , \hbar)$ in the isotropic case $\ebar = \eblue = 0$ via Jucy-Murphy elements in irreducible symmetric group modules, in which $\C  [ S(d+1) / S(d)  ]$ is the $(d+1)$-dimensional auxiliary space \cite{Bi1}.  Similarly, cotransfer operator $\widehat{\mathcal{T}}^{\downarrow}(u | \ebar, \hbar)$ yields cotransition measures \cite{Ke4, Ol1}.\\
\\
\noindent Finally, since the quantum periodic Benjamin-Ono hierarchy has discrete spectrum in the Fock space $\mathscr{F}(a)$, by Definition [\ref{RandomQuantumStationaryStateIndexDefinition}] and Proposition [\ref{DiscreteSpectrumImpliesRandomLambda}], we have:

\begin{corollary} \label{QuantumConservedDensitiesOfQBOAreRandomPartitions} For any $\Upsilon(\hbar) \in \mathscr{F}(a)$, the random profile $\widehat{f}(c | \ebar, \hbar)|_{\Upsilon(\hbar)}$ associated to the quantum conserved density of the quantum periodic Benjamin-Ono hierarchy of \textnormal{Theorem [\ref{QBOHConservedDensityConstruction}]} is identical in law to the random profile $f_{\lambda} ( c-a | \ee, \e)$ of the anisotropic partition $\lambda$ sampled from the law

\begin{equation} \textnormal{Prob} ( \lambda) = \frac{1}{ || \Upsilon(\hbar) ||_{- \frac{1}{2}, \hbar}^2 } \cdot \Big | \Big \langle \Upsilon(\hbar) | \Psi^{\textnormal{norm}}_{\lambda, a}( \cdot | \ebar, \hbar) \Big \rangle_{- \frac{1}{2}, \hbar} \Big |^2 \end{equation}

\noindent where $\Psi_{\lambda,a}^{\textnormal{norm}}( \cdot | \ebar, \hbar)$ is the Jack polynomial normalized with respect to the inner product $\langle \cdot, \cdot \rangle_{- \frac{1}{2}, \hbar}$ on the Fock space $\mathscr{F}(a)$ of the Sobolev leaf $(\mathscr{M}(a), J, \mathsf{g}_{-1/2}, \omega_{-1/2})$.
\end{corollary}

\noindent One of the bright strands connecting Kerov's diverse work is his realization that the Kerov-Markov-Kre\u{\i}n  transform $\tau_{\lambda}^{\uparrow}(c | 0 , \hbar)$ of the profile $f_{\lambda}(c| 0, \hbar)$ of an \textit{isotropic} Young diagram made from squares of area $2 \hbar$ is actually the \textit{transition measure} $\tau^{\uparrow}_{\lambda}(c | 0, \hbar)$ of $\lambda$ with respect to the \textit{Plancherel growth process}, a Gibbs measure the space of infinite Young tableaux corresponding to the regular representation of the infinite symmetric group $S(\infty)$ \cite{Ke0}.  This is a model for a growing discrete profile $f_{\lambda} (c | 0, \hbar)$ whose marginal at time $d$ is the Plancherel measure of $S(d)$.  In \cite{Ke4}, Kerov extended this observation: the Kerov-Markov-Kre\u{\i}n transform of the profile of the anisotropic partition is yet again a transition measure $\tau^{\uparrow}_{\lambda} ( c| \ee, \e)$, this time for the \textit{Jack-Plancherel growth process} on Young's lattice with Jack edge multiplicities \cite{Ke4, KeOkOl}.  Kerov realized that the algebraic theory of Jack polynomials comes from a larger, analytic theory of interlacing sequences when specialized to a precious case $c_i^{\uparrow}, c_j^{\downarrow} \in \ee \N + \e \N$ of interlacing extrema in a degenerate quarter lattice.  In \cite{Ke1}, Kerov did show that Jacobi operator spectra gave another example of the Kerov-Markov-Kre\u{\i}n correspondence, but he did not realize the case of anisotropic partitions in this way.  We do so via the spectral theory of the quantum Benjamin-Ono Lax operator, an elliptic generalized Fock-block Toeplitz operator of order $1$ \cite{DeMonvelGuillemin}, a contribution to Kerov's legacy made possible by the work of Nazarov-Sklyanin \cite{NaSk2}.

\subsection{\textcolor{black}{Constructing Nazarov-Sklyanin's Quantization}} \label{subsecNazarovSklyaninQuantizationConstruction}
\noindent We now construct Nazarov-Sklyanin's integrable geometric quantization in Theorem [\ref{NazarovSklyaninQuantizationConstruction}], previously announced as Theorem [\ref{NazarovSklyaninQuantizationExistence}].

\begin{definition} Let $\mathsf{T} \subset \mathsf{A}^{\textnormal{genpoly}}(\mathscr{M}(a))$ be the Poisson-commutative subalgebra of the Poisson algebra of generalized polynomials on Sobolev leaves $(\mathscr{M}(a), J, \mathsf{g}_{-1/2}, \omega_{-1/2})$ generated by the classical periodic Benjamin-Ono hierarchy $\{T_{\ell}^{\uparrow}(\ebar)\}_{l=1}^{\infty}$ as in \textnormal{Definition [\ref{GeneralizedPolynomialDefinition}], Corollary [\ref{SobolevLeavesAreHermitianAffineSpaces}], and {Theorem [\ref{NSClassicalCore}]}}.  The \underline{Nazarov-Sklyanin quantization} $Q_{NS}^{\circ}$ of $\mathsf{T}$ in the Fock-Sobolev space $(\mathscr{F}(a), \langle \cdot, \cdot \rangle_{- \frac{1}{2}, \hbar})$ is defined for polynomials $O$ by
\begin{equation} O( T_1^{\uparrow}(\ebar), T_{2}^{\uparrow}(\ebar), \ldots)^{Q_{NS}^{\circ}} = O ( \widehat{\mathcal{T}}_1^{\uparrow}(\ebar, \hbar), \widehat{\mathcal{T}}_2^{\uparrow}(\ebar, \hbar), \ldots ) \end{equation} \noindent for $\{\widehat{\mathcal{T}}_{\ell}(\ebar, \hbar) \}_{\ell=1}^{\infty}$ the quantum periodic Benjamin-Ono hierarchy of \textnormal{Theorem [\ref{NSQuantumCore}]}.
\end{definition}

\noindent It is important to extend $Q_{NS}^{\circ}$ to a quantization of the full Poisson algebra $\mathsf{A}^{\textnormal{genpoly}}$.

\begin{proposition} \label{NSExtensions} There exist extensions of $Q_{NS}$ of Nazarov-Sklyanin's quantization $Q_{NS}^{\circ}$ of the Poisson-commutative $\mathsf{T}$ to the Poisson algebra $\mathsf{A}^{\textnormal{genpoly}}$ so that

\begin{itemize}
\item \textcolor{gray}{\textnormal{[Ordering for Quantization]}} $Q_{NS}$ are defined by orderings $\eta_{NS}$ in the sense of \textnormal{Definition [\ref{OrderingDefinition}]}, which we call \underline{Nazarov-Sklyanin orderings $\eta_{NS}$}

\item {\textcolor{gray}{\textnormal{[Canonical Quantizations]}}} $Q_{NS}$ are canonical quantizations as in \textnormal{Definition [\ref{CanonicalQuantizationDefinition}]} of the Sobolev leaf $(\mathscr{M}(a), J, \mathsf{g}_{-1/2}, \omega_{-1/2})$ as a Hermitian affine space.\end{itemize}

\noindent but such extensions are certainly not unique.
\end{proposition}

\begin{itemize}
\item \textit{Proof:} Follows from the fact that the quantum periodic Benjamin-Ono hierarchy $\widehat{\mathcal{T}}_{\ell}^{\uparrow}(\ebar, \hbar)$ of Theorem [\ref{NSQuantumCore}] is defined from the classical periodic Benjamin-Ono hierarchy $T_{\ell}^{\uparrow}(\ebar)$ simply by substituting the classical symbol $v(x)$ for the affine Kac-Moody $\widehat{\mathfrak{gl}_1}$ current $\vcurrent ( x | \hbar)$ in the formula for the Lax operator. $\square$
\end{itemize}

\noindent For the purposes of this paper, we do not have to single out the ``correct'' extension $Q_{NS}$, as we only make use of the values of an extension $Q_{NS}$ on $\mathsf{T}$, i.e. the original $Q_{NS}^{\circ}$, but it is important that such extensions exist so that we can apply the arguments from section \textbf{[\ref{secColumn2}]}.  We now collect two remarkable facts about the Nazarov-Sklyanin quantization which is implicit in their original article, so we attribute it to them:

\begin{theorem} \label{NazarovSklyaninQuantizationConstruction} \textnormal{(Nazarov-Sklyanin \cite{NaSk2})} For any $Q_{NS}$ as in \textnormal{Proposition [\ref{NSExtensions}]}, 
\begin{itemize}
\item \textcolor{gray}{\textnormal{[Structure of Quantization]}} $Q_{NS}$ is an integrable quantization of $\mathsf{T} \subset \mathsf{A}^{\textnormal{genpoly}}$ as in \textnormal{Definition [\ref{IntegrableQuantizationDefinition}]}
\item \textcolor{gray}{\textnormal{[Regularity of Quantization]}} $Q_{NS}$ is an $\eta$-quantization of the pair $\mathsf{T} \subset \mathsf{A}^{\textnormal{genpoly}}$ of generalized polynomials for an ordering $\eta$ that is close to Wick quantization of the type constructed in \textnormal{Theorem [\ref{EtaQuantizationsAreCanonicalQuantizations}]}.
\end{itemize}

\noindent hence $Q_{NS}$ satisfies the structural assumptions of \textnormal{Definition [\ref{TheModelColumn3Definition}]} and the regularity assumptions of \textnormal{Proposition [\ref{TheModelColumn3RegularityAssumptions}]}.
\end{theorem}

\begin{itemize}
\item \textit{Proof:} Integrability follows from Theorem [\ref{NSQuantumCore}] while the regularity follows from the very particular form of the quantized observables $\widehat{\mathcal{T}}_{\ell}^{\uparrow}(\ebar, \hbar)$ as written through Fock-block Toeplitz operators. $\square$
\end{itemize}

{\footnotesize

\bibliographystyle{plain}
\bibliographystyle{amsalpha}

\bibliography{Bbib2017seas1AUG}}

{\small

\begin{quote}
$ \ $\\
$ \ $\\
ALEXANDER MOLL\\
{Institut des Hautes \'{E}tudes Scientifiques\\
35 Route de Chartres\\
Bures-sur-Yvette 91440 France}\\
$ \ $\\
\textbf{e-mail address:} {moll@ihes.fr}\\
\end{quote}

}

\end{document}